\documentclass[aps,prb,onecolumn,nofootinbib,citeautoscript,10pt]{revtex4-2}

\usepackage{amsmath,amssymb}
\usepackage{comment}

\usepackage[tight]{subfigure}

\usepackage[dvipsnames]{xcolor}
\usepackage[papersize={8.5in,11in}]{geometry}
\usepackage[colorlinks=true]{hyperref}
\hypersetup{
    bookmarks=true,
    unicode=false,
    pdftoolbar=true,
    pdfmenubar=true,
    pdffitwindow=false,
    pdfstartview={FitH},
    pdfkeywords={plasmons} {nodal-ring semimetal} {random phase approximation},
    pdfnewwindow=true,
    colorlinks=true,
    linkcolor=magenta,
    citecolor=blue,
    filecolor=magenta,
    urlcolor=blue
}

\usepackage{dcolumn}
\usepackage{color}
\usepackage{amssymb,amsmath}
\usepackage{tabularx,graphicx,makecell}
\usepackage{epstopdf}
\usepackage{latexsym}
\usepackage{colortbl}
\usepackage{psfrag}
\usepackage{bbm,bm,array,physics}
\usepackage{dsfont}
\usepackage{float, mathrsfs}
\newcommand{\bs}[1]{\boldsymbol{#1}}

\def \nn{\nonumber \\}
\def\*#1{\boldsymbol{#1}}

\begin{document}

\title{Plasmon modes in tilted three-dimensional nodal-ring semimetals. I. $\mathcal{PT}$-symmetric nodal ring and its gapped cousin}

\author{Ipsita Mandal}
\email{ipsita.mandal@snu.edu.in}

\affiliation{Department of Physics, Shiv Nadar Institution of Eminence (SNIoE), Gautam Buddha Nagar, Uttar Pradesh 201314, India}

\begin{abstract}
We study collective charge excitations in three-dimensional $\mathcal{PT}$-symmetric nodal-ring semimetals (PTNRs) and their gapped counterparts (GNRs) in the presence of band tilt, together with Weyl semimetals (WSMs). Within a low-energy two-band description, we consider untilted systems as well as in-plane and axial tilts. Using the random-phase approximation, we obtain the density of states, Drude weights, dielectric functions, and plasmon dispersions at low doping and in the intrinsic limit. The mass gap and the two inequivalent tilt directions generate distinct Fermi-surface topologies and thereby control the formation and damping of plasmons. By comparing the collective modes with the particle-hole continuum, we identify the regimes in which plasmons remain well defined. We also calculate the response of untilted and tilted WSMs and contrast it with that of PTNRs and GNRs. The two classes of semimetals differ in the carrier-density dependence and isotropy of the plasmon frequency as well as in interband screening and their response to tilt. These differences reflect their distinct Fermi-surface geometries and interband structures. Our results establish how mass and tilt govern collective charge dynamics through their interplay with intraband spectral weight and interband particle-hole excitations.
\end{abstract}

\maketitle

\tableofcontents

\section{Introduction}

Over the last decade, there have been intensive efforts to understand the transport properties of semimetals, which are materials with nodal points in their band structure. At these points, two or more bands cross and the density-of-states (DOS) vanishes. Among the three-dimensional (3d) semimetals with twofold band-crossings, the best-known examples are the Weyl semimetals (WSMs)~\cite{burkov11_weyl, yan17_topological}, whose Brillouin zone (BZ) forms a manifold with nontrivial topology due to the Berry phase. The band-crossing points of a WSM are protected by the point-group symmetries of the crystal lattice. Each node acts as a source or sink of Berry curvature (BC) flux in momentum space, and thus mimics a magnetic monopole. The monopole charge equals the Chern number arising from the Berry connection, and its sign is the chirality $\chi$ of the node. Obeying the Nielsen-Ninomiya theorem~\cite{nielsen}, the nodes appear in pairs of opposite chiralities. The BC flux emanating from one partner of a pair disappears into the other. Examples of WSMs include TaAs~\cite{huang15_observation, lv_Weyl, yang_Weyl} and HgTe-class materials~\cite{ruan_Weyl}. A node can also be tilted in generic settings \cite{marcus-tilt_magneto, yadav23_magneto, ips-tilted}. For a tilt below the critical value, the node is of type I and has a closed Fermi surface (FS) at finite doping, which is the case we use for comparison in Sec.~\ref{secweyl}. Band-crossings can also occur along curves in the BZ instead of at isolated points, as we describe below, and those are the main subject of this paper.

Nodal-ring semimetals (NRSs) host symmetry-protected band-crossings along one-dimensional (1d) curves in the Brillouin zone (BZ)~\cite{balents-nodal, fu_nlsm, yang_review_nlsm}. Experimental realisations of NRSs span a broad range of material and artificial platforms, including SrAs$_3$ \cite{arpes-nlsm}, Ca$_3$P$_2$ \cite{expt1_nlsm}, the hexagonal pnictides CaAgP and CaAgAs \cite{expt2_nlsm}, alkaline-earth metals such as Ca, Sr, and Yb \cite{alkaline_nlsm}, Fe$_2$MnX compounds \cite{claudia_nlsm}, Co$_3$Sn$_2$S$_2$ \cite{enke}, photonic metamaterials \cite{biao_nodal}, and acoustic crystals \cite{nlr-acoustic}. Among these systems, first-principles calculations predict that CuTeO$_3$ \cite{cuteo_nlsm} hosts an almost ideal nodal ring: the nodal loop lies close to the Fermi level, exhibits negligible energy dispersion, is confined approximately to the $k_xk_y$ plane, and is well described by a circular contour, free from interference by extraneous bands. Furthermore, the weak spin--orbit coupling in CuTeO$_3$ preserves the nodal-ring physics over an appreciable energy scale, making it an excellent realisation of the low-energy model Hamiltonian considered in this work. These material characteristics provide concrete justification for our idealised description of a circular nodal curve. The low-energy excitations in the vicinity of the ring are characterised by a toroidal FS when the chemical potential is finite, and by a ring-shaped zero-energy contour in the intrinsic limit. NRSs exhibit a variety of unconventional transport and optical properties, including anisotropic magnetoconductivity~\cite{ips-nlsm-ph, phe_nlsm, gnr-hall-ah, ips-gnr-strain} and Friedel oscillations~\cite{rahmipoor-2d-nlsm, rhim-kim-friedel}. The collective excitations of NRSs have been studied in both two-dimensional (2d) and 3d systems~\cite{rahmipoor-2d-nlsm, rhim-kim-friedel}. For the intrinsic case, where the chemical potential lies exactly at the nodal-ring energy, the polarisability exhibits a singular structure that is qualitatively distinct from that of nodal-point semimetals. The plasmon dispersion is highly anisotropic, reflecting the toroidal geometry of the FS. The doped regimes bring in more distinctive features.

Recent work on 2d tilted NRSs has shown that a finite tilt enhances the plasmon frequency and leads to beat-patterns in the Friedel oscillations~\cite{rahmipoor-2d-nlsm}. In that study, the polarisability was evaluated at zero chemical potential, and the plasmon mode was obtained from the zeros of the dielectric function. Motivated by these findings, we investigate the analogous problem for 3d NRSs. In 3d, the tilt can lie either in the plane of the ring or along its axis. Together with a mass gap, these two inequivalent directions open a much richer variety of FS topology, and hence of intraband and interband contributions to the charge response, than in the untilted 3d case~\cite{rhim-kim-friedel} or in 2d tilted NRSs~\cite{rahmipoor-2d-nlsm}.
We consider the minimal two-band model that captures a single circular nodal-ring in the BZ, with two distinct symmetry settings. When the combined presence of inversion ($\mathcal{P}$) and time-reversal ($\mathcal{T}$) symmetries protects the band crossing, we refer to the resulting gapless system as a $\mathcal{PT}$-symmetric nodal ring (PTNR). When a mass term is introduced, breaking $\mathcal{P}$ and opening a gap of $2\,\Delta$ at the ring, we refer to the resulting gapped system as a gapped nodal ring (GNR). The PTNR has a vanishing Berry curvature everywhere except on the nodal ring itself, which makes it an ideal platform for studying the interplay between topology and collective excitations without the complications of a finite Berry curvature in the bulk or a gap-term~\cite{schnyder_nodal, rhim-kim-friedel, ips-dipole-vnr, lpge-ptnr}. The GNR, in turn, allows us to address how the mass-gap modifies the topological properties \cite{ips-nlsm-ph, ips-gnr-strain}, polarisability, the Drude weight, and the plasmon dispersion. Both settings have been studied in related transport contexts~\cite{ips-magnus, ips-nlsm-ph, ips-gnr-strain, yang1, ips-sanskar-2dgnr, gnr-hall-ah, ips-sandip-fano}, but their collective modes have not been analysed in a unified way, especially in the context of tilting.

In this paper, our focus is on the intrinsic limit and small-chemical-potential scenarios. For each of the PTNR and GNR, we treat the untilted configuration, the in-plane tilt along $k_x$, and the axial tilt along $k_z$. In every case we compute the non-interacting density-density response function at zero temperature, following the approach of Ref.~\cite{rahmipoor-2d-nlsm}. The polarisability is expressed as an integral over the toroidal FS, and we extract the plasmon dispersion within the random-phase approximation (RPA) from the long-wavelength limit of the dielectric function. To identify which features are specific to the toroidal FS, we also compare with a type-I WSM node.

The paper is organised as follows. In Sec.~\ref{secmodel}, we introduce the model Hamiltonian for the NRS, define the toroidal coordinates, and summarise the density-response formalism and the FS geometry. Sec.~\ref{secungapped} is devoted to the PTNR, with subsections on the untilted configuration, the in-plane tilt along $k_x$, and the axial tilt along $k_z$. Sec.~\ref{secgapped} repeats this analysis for the GNR, treating the untilted, $k_x$-tilted, and $k_z$-tilted configurations in turn. In Sec.~\ref{secweyl}, we contrast the long-wavelength response of the nodal ring, with and without tilt, with that of a Weyl node. The derivations for the latter are given in Appendix~\ref{app:weyl}. We conclude with a summary and outlook in Sec.~\ref{secsum}, where the main results are also collected in tables. Appendix~\ref{appdirac} gives the mapping of the NRS systems onto 2d Dirac cones. The remaining appendices are devoted to providing concrete details about various intermediate steps and derivations. We use natural units and, hence, set $\hbar= c = k_B = e = 1$. However, we retain the symbol $e$ in the analytical expressions for the sake of book-keeping.

\section{Model}
\label{secmodel}

The minimal two-band model of a GNR with a single circular nodal loop 
lying in the $k_x k_y$-plane is given by~\cite{balents-nodal, yang1}
\begin{align}
\mathcal{H}_0(\bs{k}) = \bs{d}_0(\bs{k})\cdot \boldsymbol{\sigma}\,, \quad
\bs{d}_0(\bs{k}) = \left\{\lambda\left(k_\perp^2 - k_0^2\right),\,
v_z\,k_z,\,\Delta\right\}, \quad
k_\perp = \sqrt{k_x^2 + k_y^2}\,,
\end{align}
where $\boldsymbol{\sigma} = \{\sigma_x, \sigma_y, \sigma_z\}$ denotes 
the vector of Pauli matrices. The parameters $\lambda$ and $k_0$ are 
material-dependent, while $\Delta$ encodes the small gap induced by 
symmetry breaking, for instance through inversion-breaking uniaxial
strain, pressure, or an external electric field~\cite{schnyder_nodal}. 
Thus, $\Delta$ is the mass term, which opens a gap of $2\, \Delta$ at the nodal ring. The gapless limit corresponds to a PTNR, when the two bands cross along the locus 
$k_\perp^2 - k_0^2 = 0$, which defines a nodal ring of radius $k_0$. 

When the chemical potential satisfies $\mu \ll \lambda\,k_0^2$ and the gap is small, $\Delta \ll \lambda\,k_0^2$, the low-energy excitations are confined to the neighbourhood of 
the nodal ring. For $\mu > \Delta$ they lie on the FS encircling the ring. To characterise the 
transport signatures of these quasiparticles, it is convenient to 
linearise $\mathcal{H}_0$ in the momentum deviation from the nodal curve 
\cite{linearize-nlsm}, which is most naturally accomplished by 
introducing toroidal coordinates:
\begin{align}
\label{eqtrs}
k_x = \left(k_0 + \kappa\cos\gamma\right)\cos\phi\,, \quad
k_y = \left(k_0 + \kappa\cos\gamma\right) \sin\phi\,, \quad
k_z = \frac{\kappa\sin\gamma}{\alpha}\,, \quad
\alpha = \frac{v_z}{v_0}\,, \quad v_0 = 2 \,\lambda\,k_0\,.
\end{align}
The Jacobian of this coordinate transformation is 
$J = \kappa\left(k_0 + \kappa\cos\gamma\right)/\alpha$. Inverting the 
transformation gives $k_0 + \kappa\cos\gamma = \pm\,k_\perp$, and 
since $\kappa \ll k_0$ in the low-energy limit, we have
$ \kappa \cos  \gamma   = k_\perp - k_0 $.
This leads to
\begin{align}
& \mathcal{H}_0 (\bs k ) = 
\mathcal{H} (\delta \bs k) + \order{\kappa^2}\,, \quad 
\mathcal{H} (\delta \bs k) = 
{\bs d} ( \delta \bs k) \cdot \boldsymbol{\sigma} \,,
\quad
\delta \bs k =
\kappa  \left \lbrace \cos  \gamma  \cos  \phi , \,
\cos  \gamma  \sin  \phi , \,
\frac{\sin  \gamma } {\alpha} 
 \right \rbrace, \nn 
& \bs{d}( \delta \bs{k}) 
= \left \lbrace v_0 \, \kappa \cos  \gamma  ,
\, v_0 \, \kappa \sin  \gamma ,\, \Delta \right\rbrace
=
\left \lbrace v_0  \left( k_\perp - k_0 \right) ,
\, v_z \, k_z ,\, \Delta \right\rbrace .
\end{align}
In the toroidal coordinate system, $k_0$ represents the major radius 
of the torus, which is the distance from the centre of the tube to the centre 
of the nodal ring. The variable $\kappa$ plays the role of the minor radius, 
measuring the cross-sectional extent of the torus. The toroidal angle 
$\phi$ and the poloidal angle $\gamma$, each ranging over $[0, 2\pi)$, 
describe rotations around the nodal ring and around the torus's axis 
of revolution, respectively. The parameter $\alpha$ captures the 
anisotropy of the system, encoding the ratio of the velocity along the 
$k_z$-axis to that within the $k_x k_y$-plane. Setting $v_z = v_0$ for simplicity and adding a tilt term, the linearised Hamiltonian that we use in the rest of this paper is
\begin{align}
\label{eq:ham3d}
\mathcal{H} (\bs k) = {\bs d}  (\bs k) \cdot \boldsymbol{\sigma}
+ v_0 \,{\bs \eta} \cdot {\bs k}\; \mathbb{I}_{2\times 2} \,,
\quad {\bs d} (\bs k) = \left \lbrace v_0 \left( k_\perp - k_0 \right) , 
\, v_0 \, k_z ,\, \Delta \right \rbrace \,.
\end{align}
Here $k_0$ is the radius of the nodal ring, and $\boldsymbol{\sigma}$ is the vector of Pauli matrices. The tilt is parametrised by the dimensionless vector $\bs \eta = \eta_x \, \bs{\hat x} + \eta_y \, \bs{\hat y} + \eta_z \, \bs{\hat z}$.

The eigenvalues are
\begin{align}
\label{eq:ev3d}
\varepsilon_s (\bs k) = (-1)^s\, E_k + v_0\, \bs \eta \cdot \bs k \,,
\quad E_k = \sqrt{ v_0^2 \, \kappa^2 + \Delta^2 } \,,
\quad \kappa = \sqrt{ \left( k_\perp - k_0 \right)^2 + k_z^2 } \,,
\end{align}
with $s \in \lbrace 1, 2 \rbrace$. The band-velocity has the components $V_i^{(s)} = \partial_i \varepsilon_s$, where the indices $i, j \in \{x, y, z\}$ label Cartesian components here and in the rest of the paper. Using $\partial_x \kappa = \cos \gamma \cos \phi$, $\partial_y \kappa = \cos \gamma \sin \phi$, and $\partial_z \kappa = \sin \gamma$, we obtain
\begin{align}
\label{eq:velocity3d}
V_x^{(s)} = (-1)^s \, \frac{v_0^2\, \kappa}{E_k}\, \cos \gamma \cos \phi + v_0\, \eta_x \,, \quad
V_y^{(s)} = (-1)^s \, \frac{v_0^2\, \kappa}{E_k}\, \cos \gamma \sin \phi + v_0\, \eta_y \,, \quad
V_z^{(s)} = (-1)^s \, \frac{v_0^2\, \kappa}{E_k}\, \sin \gamma + v_0\, \eta_z \,.
\end{align}
For $\Delta = 0$, we have $v_0\, \kappa/E_k = 1$. The velocity of the $s=2$ band is then $\bs V = v_0 \left( \cos \gamma \cos \phi + \eta_x ,\, \cos \gamma \sin \phi + \eta_y ,\, \sin \gamma + \eta_z \right)$. The eigenstates are two-component spinors,
\begin{align}
|\psi_+ (\bs k) \rangle = \begin{pmatrix} 
\cos (\theta/2) \\ \sin (\theta/2)\, e^{i \, \gamma (\bs k)} \end{pmatrix} , \quad
|\psi_- (\bs k) \rangle = \begin{pmatrix} \sin (\theta/2) \\ 
-\cos (\theta/2)\, e^{i \, \gamma (\bs k)} \end{pmatrix} ,
\end{align}
where the polar and azimuthal angles are defined by
\begin{align}
\label{eq:gamma3d}
\cos \theta = \frac{\Delta}{E_k} \,, \quad \sin \theta 
= \frac{v_0 \, \kappa}{E_k} \,, \quad e^{i \, \gamma (\bs k)} 
= \frac{ \left( k_\perp - k_0 \right) + i\, k_z }{ \kappa } \,.
\end{align}
For $\Delta = 0$, we have the PTNR with $\cos \theta = 0$ and $\sin \theta = 1$.
When $\bs \eta = 0$, the Hamiltonian of Eq.~\eqref{eq:ham3d} is invariant under continuous rotations about the $ k_z $-axis and under reflection through the $k_z = 0$ plane. A finite in-plane tilt $\bs \eta = \eta_x \bs{\hat x} $ breaks the rotational symmetry while preserving the mirror one. An axial tilt $\bs \eta = \eta_z \bs{\hat z} $ does the reverse.

\subsection{Density response, Drude weight, and plasmon dispersion}

The noninteracting density-density response function at temperature $T$ and chemical potential $\mu$ is given by the retarded bare bubble diagram, evaluated with the eigenstates of Eq.~\eqref{eq:ham3d}~\cite{giuliani-vignale}. It is the 3d version of the expression used in Ref.~\cite{rahmipoor-2d-nlsm} for a 2d tilted nodal-line semimetal. Explicitly,
\begin{align}
\label{eq:polgeneral}
& \Pi (\bs q, \omega) = g \int \frac{d^3 \bs k}{(2\, \pi)^3} 
\sum_{s,s'} \mathcal{F}_{ss'} (\bs k, \bs k')
\frac{ f_0 (\varepsilon_s (\bs k)) - f_0 (\varepsilon_{s'} (\bs k')) }
{ \omega + \varepsilon_s (\bs k) - \varepsilon_{s'} (\bs k') + i\, 0^+ } \,,
\nn & \mathcal{F} (\bs k, \bs k') = 
\frac{ 1 - \cos \theta \cos \theta' - \sin \theta \sin \theta' \cos (\gamma - \gamma')} {2} \, ,
\quad \bs k' = \bs k + \bs q\,,
\end{align}
where $g$ is the degeneracy factor, $f_0$ is the Fermi-Dirac distribution, and $\mathcal{F}_{ss'}(\bs k, \bs k')$ is the band overlap factor, namely the squared overlap between the initial state at $\bs k$ in band $s$ and the final state at $\bs k'$ in band $s'$, with $\mathcal{F} \equiv \mathcal{F}_{12} = \mathcal{F}_{21}$ the interband form-factor entering the second line above. For the spinless model considered here, $g=1$, but we keep the $g$ factor to keep the analytical expressions general. The form-factor reduces to $[1 - \cos (\gamma - \gamma')]/2$ when $\Delta = 0$. The complex response function splits as $\Pi = \Re \Pi + i\, \Im \Pi$, with $\Re \Pi$ the physical polarisability entering the RPA dielectric function and $\Im \Pi$ governing the particle-hole damping. We perform all calculations in this paper at $T = 0$. The Fermi-Dirac functions then reduce to step functions, and the sum over $s, s'$ in Eq.~\eqref{eq:polgeneral} splits into two parts. The intraband part, $\Pi_{\rm intra}$, collects the terms with $s = s'$. The interband part, $\Pi_{\rm inter}$, collects the terms with $s \neq s'$. Hence, $\Pi = \Pi_{\rm intra} + \Pi_{\rm inter}$. Depending on the value of $\mu$, the polarisability receives contributions from one or both of these.

For both PTNR and GNR, we consider low but nonzero doping, $0 < \mu \ll v_0\, k_0$. We will need to evaluate the band-dependent DOS,
\begin{align}
\label{eqdos}
\rho_s (\mu) = g \int \frac{d^3\bs k}{(2\pi)^3}\, \delta (\varepsilon_s - \mu) \,.
\end{align}
It will feed into the Drude-weight tensor, defined as
\begin{align}
\label{eqdrude}
D_{ij} = \pi\, g\, e^2 \sum_s \int \frac{d^3 \bs k}{(2\, \pi)^3}\, 
\delta (\varepsilon_s - \mu)\, V_i^{(s)} \, V_j^{(s)} \,,
\end{align}
which also fixes the Drude peak of the optical conductivity, $\Re \sigma_{ij} (\omega) = D_{ij}\, \delta (\omega)$. Taking the $q \to 0$ limit first and thereafter $\omega \to 0$, the intraband part of Eq.~\eqref{eq:polgeneral} gives~\cite{giuliani-vignale}
\begin{align}
\label{eqintra}
\Pi_{\rm intra} (\bs q, \omega) \simeq 
q_i\, q_j\, D_{ij}/(\pi\, e^2\, \omega^2)\,.
\end{align} 
The opposite order of limits, in which $\omega \to 0$ is taken first, gives instead the static long-wavelength result $\Pi (\bs q \to 0, 0) = -\, \rho (\mu)$, where $\rho = \sum_s \rho_s$ is the total DOS. Within the RPA, the dielectric function is given by
\begin{align}
\label{eq:epsrpa}
\varepsilon^{\rm RPA} (\bs q, \omega) = 1 - v (q)\, \Pi (\bs q, \omega) \,, \quad
v (q) = \frac{4\, \pi\, e^2}{q^2} \,,\quad \Pi = \Pi_{\rm intra} + \Pi_{\rm inter}\,,
\end{align}
where $\Pi $ is the full complex polarisability and $v (q)$ is the Fourier transform of the 3d Coulomb interaction $e^2/r$. This kernel differs from the 2d one, $v (q) = 2\, \pi\, e^2/q$, used in Ref.~\cite{rahmipoor-2d-nlsm}. The additional factor of $1/q$ changes the long-wavelength plasmon from the $\sqrt{q}$ behaviour of 2d systems to a finite frequency, as we show below. The dielectric function is therefore complex as well. Its imaginary part, $-\, v (q)\, \Im \Pi$, describes the damping into the particle-hole continuum. Undamped collective modes lie outside the continuum, where $\Im \Pi = 0$. They are the zeros of $\Re \varepsilon^{\rm RPA} = 1 - v (q)\, \Re \Pi$. At long wavelengths, $\Re \Pi$ is dominated by the real intraband part of Eq.~\eqref{eqintra}. We therefore keep only $\Pi_{\rm intra}$ in the plasmon condition. The interband part is real and negative below the interband threshold and it, therefore, enters as a background dielectric constant, which we estimate for the GNR in Sec.~\ref{secgapped}. Appendix~\ref{app:monotone} shows that $\Re \varepsilon^{\rm RPA}$ then increases monotonically with $\omega$ below the onset. Its zero is therefore unique, and the static screened frequency is an upper bound of it. In many expressions that follow, we will write $\varepsilon^{\rm RPA} (\bs q, \omega)$ simply as $\varepsilon^{\rm RPA}_{i} $, where $i \in \lbrace x, y, z \rbrace $ will indicate the Cartesian direction of $\bs q$.

Inserting $\Re \Pi (\bs q, \omega) \to \Pi_{\rm intra} (\bs q, \omega) = q_i\, q_j\, D_{ij}/(\pi\, e^2\, \omega^2)$ into $\Re \varepsilon^{\rm RPA} = 0$ yields the square of the plasmon dispersion along the principal direction $i$,
\begin{align}
\label{eqplasmon}
\omega_{p, i}^2 = 4\, D_{ii} \,,
\end{align}
where no sum over $i$ is implied. The interband part lowers these bare frequencies through the static background dielectric constant $\epsilon_{b, i}$. The screened frequencies are $\tilde\omega_{p, i}^2 = \omega_{p, i}^2/\epsilon_{b, i}$, and we use this notation throughout. In every configuration we consider, $D_{ij}$ is diagonal in the Cartesian frame set by the tilt direction. For the axial tilt and the untilted case, this follows from the residual rotational symmetry about $k_z$. For the in-plane tilt $\bs \eta = \eta_x\, \bs{\hat x}$, no such rotational symmetry survives, yet $D_{ij}$ remains diagonal, because two discrete symmetries persist. The energy $\varepsilon_s (\bs k)$ depends on $\phi$ only through $\cos \phi$, and on $\gamma$ only through $\kappa$ and $\cos \gamma$. It is therefore invariant under $\phi \to -\phi$ and, independently, under $\gamma \to -\gamma$. The first is a reflection through the $xz$-plane, which leaves $V_x^{(s)}$ and $V_z^{(s)}$ unchanged while flipping the sign of $V_y^{(s)}$, as follows from Eq.~\eqref{eq:velocity3d}. Since the FS and the measure are even under this reflection, the integrand of $D_{xy} = \pi\, g\, e^2 \sum_s \int d^3\bs k/(2\,\pi)^3\, \delta (\varepsilon_s - \mu)\, V_x^{(s)}\, V_y^{(s)}$ is odd, and $D_{xy}$ vanishes identically, despite $D_{xx} \neq D_{yy}$. The second symmetry, $\gamma \to -\gamma$, leaves $V_x^{(s)}$ and $V_y^{(s)}$ unchanged while flipping the sign of $V_z^{(s)}$, and by the same argument kills $D_{xz}$ and $D_{yz}$. The three diagonal entries need not all be equal, however. For the untilted configuration and for the axial tilt, the rotational symmetry about $k_z$ additionally enforces $D_{xx} = D_{yy}$. Only two independent plasmon frequencies then appear, $\omega_{p,z}$ and $\omega_{p,x} = \omega_{p,y}$. The in-plane tilt breaks this rotational symmetry explicitly, and $D_{xx} \neq D_{yy}$ in general, even though $D_{xy}$ stays zero. Correspondingly, $\omega_{p,x} \neq \omega_{p,y}$, and the plasmon dispersion has three distinct principal frequencies. We give the value of $D_{yy}$ alongside $D_{xx}$ wherever the two differ.

The interband part of Eq.~\eqref{eq:polgeneral} is treated separately. We first consider the situation in which band $s=1$ is completely filled and band $s=2$ is completely empty, as is the case when $\mu$ lies in the gap or at the nodal ring. The two terms with $s \neq s'$ then give
\begin{align}
\label{eq:polinter}
\Pi_{\rm inter} (\bs q, \omega) = g \int \frac{d^3 \bs k}{(2\, \pi)^3}\, \mathcal{F} (\bs k, \bs k')
\left ( \frac{1}{\omega - E_k - E_{k'} - v_0\, \bs \eta \cdot \bs q + i\, 0^+}
- \frac{1}{\omega + E_k + E_{k'} - v_0\, \bs \eta \cdot \bs q + i\, 0^+} \right ).
\end{align}
Since the tilt term is proportional to the identity in band space, it leaves the eigenstates and the form-factor unchanged. It enters any response only through the shift $v_0\, \bs \eta \cdot \bs q$ of the transition energies, as shown in Appendix~\ref{app:tiltshift}. For $q \to 0$, the form-factor vanishes as $\mathcal{F} \simeq q_i\, q_j\, g_{ij} (\bs k)$, where $g_{ij}$ is the quantum metric of the two bands. In the static limit, the two energy shifts cancel to leading order in $q$, and we obtain
\begin{align}
\label{eq:polinterstatic}
\Pi_{\rm inter} (\bs q, 0) \simeq -\, g \int \frac{d^3 \bs k}{(2\, \pi)^3}\, \frac{q_i\, q_j\, g_{ij} (\bs k)}{E_k}\, \left[ f_0 (\varepsilon_1) - f_0 (\varepsilon_2) \right] .
\end{align}
The occupation difference equals unity for a filled lower band and an empty upper band. It vanishes inside any electron- or hole-pocket, which is then excluded from the integral. Eq.~\eqref{eq:polinterstatic} is used in Sec.~\ref{secgapped} to account for the interband screening of the plasmon in the regime where both pockets contribute.

\subsection{Validity of the low-energy description}

\begin{table}[t!]
\centering
\begin{tabular}{|c|c|}
\hline
Parameter & Natural units \\ \hline
$v_0$ from Ref.~\cite{phe_nlsm} & $0.0004$ \\ \hline
$k_0$ from Ref.~\cite{phe_nlsm} & $160$ eV \\ \hline
$\Delta$ from Ref.~\cite{phe_nlsm} & $\sim 0.02$ eV \\ \hline
$\mu$ from Ref.~\cite{phe_nlsm} & $\sim 0.01$--$0.04$ eV \\ \hline
$g\, e^2\, k_0$ & $160$ eV \\ \hline
\end{tabular}
\caption{\label{tab-params} Representative parameters in natural units, $\hbar = c = 1$. The coupling $g\, e^2\, k_0$ follows from $e^2 = 1$ and $g=1$.}
\end{table}

Two restrictions apply throughout this paper. First, we work at low doping, $\mu \ll v_0\,k_0 $ (equivalently, $\kappa_F \ll k_0 $). In the gapped case we additionally require a small gap, $\Delta \ll v_0\,k_0$. For the in-plane tilt of the GNR, we also assume that $\eta_x\,\Delta$ is small compared with $v_0\,k_0$. This condition controls the width of the pockets in the narrow windows of Sec.~\ref{secfermisurface}, which then occupy only small arcs of the ring. This is the regime in which the linearised Hamiltonian of Eq.~\eqref{eq:ham3d} is valid, since the low-energy expansion about the nodal ring requires the Fermi momentum to stay well below the ring radius. The condition on $\Delta$ ensures the same for the insulating regimes $\mu < \Delta$, where no Fermi momentum exists and the relevant momenta are set by $\Delta/v_0$. Second, we do not compute the Friedel oscillations. Their determination requires the static polarisability at the threshold wavevector $q = 2\, \kappa_F $, whereas the local mapping onto a 2d Dirac cone that underlies the whole calculation is valid only for $q \ll k_0 $. For realistic values of material parameters \cite{ips-gnr-strain, ips-dipole-vnr, ips-firdous-vnr}, $2\, \kappa_F $ is comparable to $k_0 $, and in the tilted cases the effective threshold is comparable to or larger than $k_0 $ once the angular average over the FS is taken. The threshold therefore lies outside the window in which the present approach applies, and any numerical estimate of the real-space decay obtained by continuing the low-$q$ polarisability beyond its range of validity would not be a controlled result. The numerical plots use the representative parameter set of Table~\ref{tab-params}.

\subsection{Fermi surface under chemical potential and tilt}
\label{secfermisurface}

\subsubsection{PTNR case}

The FS follows from $\varepsilon_s (\bs k) = \mu$. Without tilt and for $\Delta = 0$, it is a torus with major radius $k_0$ and minor radius $\kappa_F = \mu/v_0$. The low-doping condition $\mu \ll v_0\, k_0$ ensures that this is a ring torus. The tilt term $v_0\, \bs{\eta} \cdot \bs k$ shifts both bands rigidly at each $\bs k$, and it deforms the FS, even for the PTNR. The nature of the deformation depends on the direction of the tilt, and it is summarised below following Ref.~\cite{ahn-tilted-nl, ips-dipole-vnr}.

\begin{figure*}[t!]
\centering
\includegraphics[width= 0.9 \textwidth]{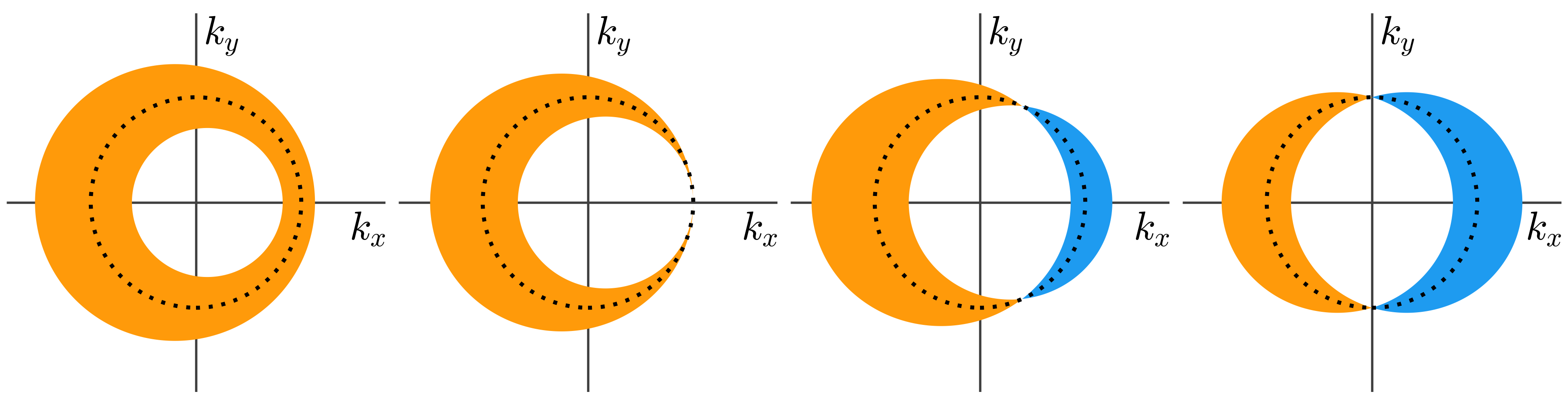}
\caption{\label{figfs} Evolution of the FS of a PTNR under an in-plane tilt, shown as the slice in the $k_x k_y$-plane at $k_z = 0$. The panels show, from left to right, the ring-cyclide regime, the boundary case $\mu = \tilde{\eta}$, the horn-cyclide regime, and the symmetric horn-cyclide regime ($\mu = 0$), with $(\tilde{\eta}, \mu) = (0.15, 0.30)$, $(0.20, 0.20)$, $(0.30, 0.12)$, and $(0.30, 0)$, respectively. Here $\tilde{\eta} = v_0\, \eta_x\, k_0$, and the units are $v_0 = k_0 = 1$. The orange and blue colours depict the $s=2$ (electron) and $s=1$ (hole) pockets, respectively. The dotted black circle marks the original PTNR with $\mu = \eta = 0$. The parameter values are chosen only to make the regimes clearly visible.}
\end{figure*}

An in-plane tilt, e.g., $\bs{\eta} = \eta_x\, \bs{\hat x}$, shifts the energy by $v_0\, \eta_x\, k_x$. On the nodal ring, this shift equals $\tilde{\eta} \cos \phi$, where $\tilde{\eta} = v_0\, \eta_x\, k_0$. The ring is therefore no longer at a constant energy. The resulting evolution is shown in Fig.~\ref{figfs}. At $\mu = 0$, the FS consists of an electron-pocket and a hole-pocket, which touch each other at the two points $\phi = \pm \pi/2$ where the shift vanishes. The resulting surface is a symmetric horn-cyclide. The dotted black circle in Fig.~\ref{figfs} is a reference contour, marking the original PTNR at $\mu = 0$ and $\bs{\eta} = 0$. The coloured surfaces are the actual FSs, with the orange and blue colours denoting the $s=2$ and $s=1$ bands, respectively. For $0 < \mu < \tilde{\eta}$, the two pockets coexist on a horn-cyclide. The contact points move away from the symmetry axis as $\mu$ increases. The hole-pocket shrinks to a single point at $\mu = \tilde{\eta}$ and disappears for larger $\mu$. For $\mu > \tilde{\eta}$, the FS is a single closed ring-cyclide enclosing the nodal ring. When $\mu$ approaches $v_0\, k_0$, the central hole closes and the surface turns into a spindle-cyclide, a regime that we exclude, as stated above.

An axial tilt, $\bs{\eta} = \eta_z\, \bs{\hat z}$, shifts the energy by $v_0\, \eta_z\, k_z$. This shift vanishes on the nodal ring, which therefore stays at zero energy. For $|\eta_z| < 1$, no pocket of the opposite carrier type forms. The FS remains a single closed surface of revolution about the $k_z$-axis for all $\mu > 0$. At fixed $\phi$, its cross-section is an ellipse with eccentricity $|\eta_z|$, and the nodal ring sits at one of its foci. For $|\eta_z| > 1$ the factor $1 + \eta_z \sin\gamma$ passes through zero, and the FS opens into two unbounded hyperboloid-like sheets that extend to $k_z \to \pm\infty$. This type-II or over-tilted regime lies outside the scope of the present work, and we restrict to $|\eta_z| < 1$ throughout.

\subsubsection{GNR case}


The mass $\Delta$ of the GNR adds a gap on top of the features elucidated for PTNR. For an in-plane tilt the classification is less clear-cut than for the PTNR, and we list all cases here. Those are summarised in Table~\ref{tab:fsclass} of Sec.~\ref{sec:gaptiltx}.

At fixed $\phi$, the cross-section of the GNR maps exactly onto a gapped Dirac cone with the effective chemical potential $\mu_{\rm eff} (\phi) = \mu - \tilde{\eta} \cos \phi$ and a radial tilt $t (\phi) = \eta_x \cos \phi$. For $|t| < 1$ the band minimum of this cone lies at the energy $\Delta_{\rm eff} (\phi) = \Delta \sqrt{1 - t^2}$, and it is displaced from the nodal ring. The local FS is an ellipse. It is absent for $|\mu_{\rm eff}| < \Delta_{\rm eff} (\phi)$. For $\Delta_{\rm eff} (\phi) \le \mu_{\rm eff} < \Delta$ it is an electron ellipse that does not enclose the nodal ring, so that a ray from the ring meets it twice, as in the axial case of Sec.~\ref{sec:gaptiltz}. For $\mu_{\rm eff} \ge \Delta$ the ellipse encloses the ring. The same three classes hold for holes with $\mu_{\rm eff} \to -\mu_{\rm eff}$. As $\phi$ runs from $0$ to $\pi$, $\mu_{\rm eff}$ rises monotonically from $a \equiv \mu - \tilde{\eta}$ to $b \equiv \mu + \tilde{\eta}$. The thresholds $\pm \Delta_{\rm eff} (\phi)$ vary with $\phi$ by at most $\Delta \left( 1 - \nu_x \right)$, where $\nu_x = \sqrt{1 - \eta_x^2}$. This is much slower than the variation of $\mu_{\rm eff}$ as long as $\eta_x\, \Delta \ll \nu_x\, v_0\, k_0$. Each class then occupies a single arc, and the regime is fixed by the classes at the two endpoints $\phi = 0$ and $\phi = \pi$, where $\Delta_{\rm eff} = \Delta\, \nu_x$. Taking $\mu \ge 0$, since the case $\mu < 0$ follows by exchanging electrons and holes, there are eight cases. Their derivation is given in Appendix~\ref{app:fsclass}.

\begin{figure*}[t!]
\centering
\includegraphics[width=0.9\textwidth]{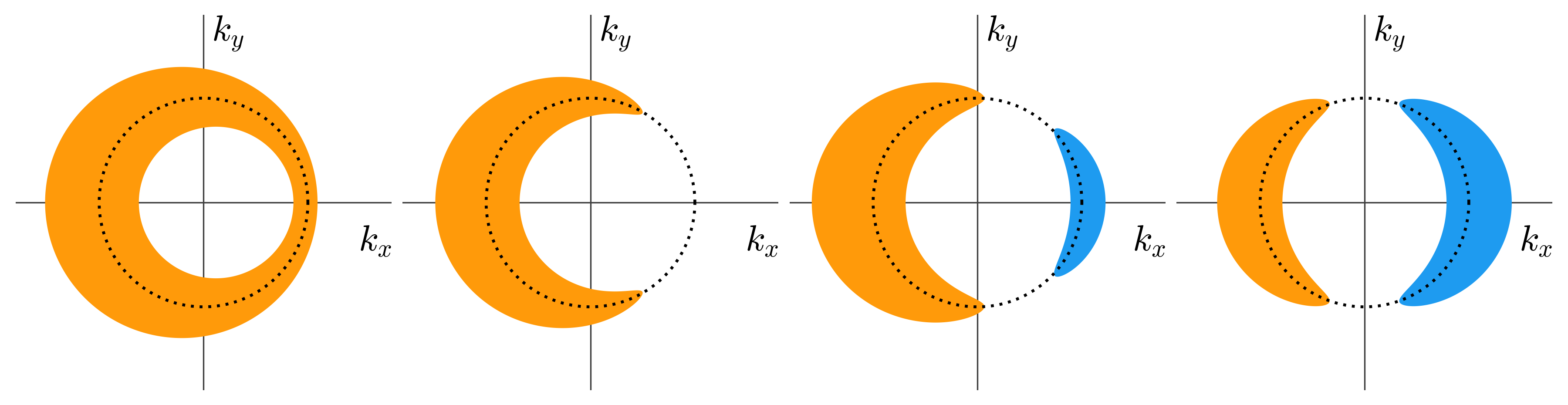}
\caption{\label{fig:fsgnrx} Evolution of the FS of a GNR under an in-plane tilt, shown as the slice in the
$k_x k_y$-plane at $k_z = 0$. The panels show, from left to right, the fully-conducting, crescent-cyclide, twin-crescent, and symmetric twin-crescent ($\mu = 0$) regimes, with $(\tilde{\eta}, \mu) = (0.15, 0.30)$, $(0.20, 0.20)$, $(0.30, 0.12)$, and $(0.30, 0)$, respectively. The fully-insulating regime has no FS and is therefore not shown. The orange and blue colours depict the $s=2$ (electron) and $s=1$ (hole) pockets, respectively. The dotted black circle marks the original ring. The units are $v_0 = k_0 = 1$, and $\Delta = 0.1$. The parameter values are chosen only to make the regimes clearly visible.}
\end{figure*}

\begin{figure*}[t!]
\centering
\subfigure[]{\includegraphics[width=0.28\textwidth]{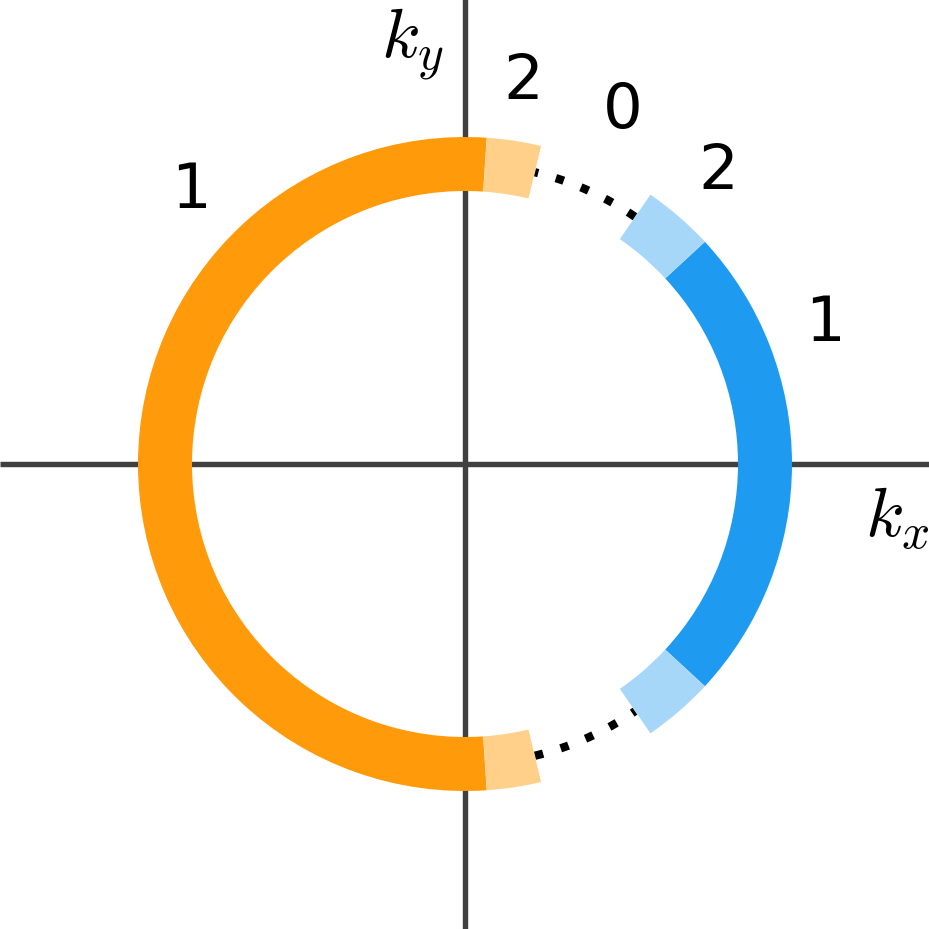}}
\subfigure[]{\includegraphics[width=0.28\textwidth]{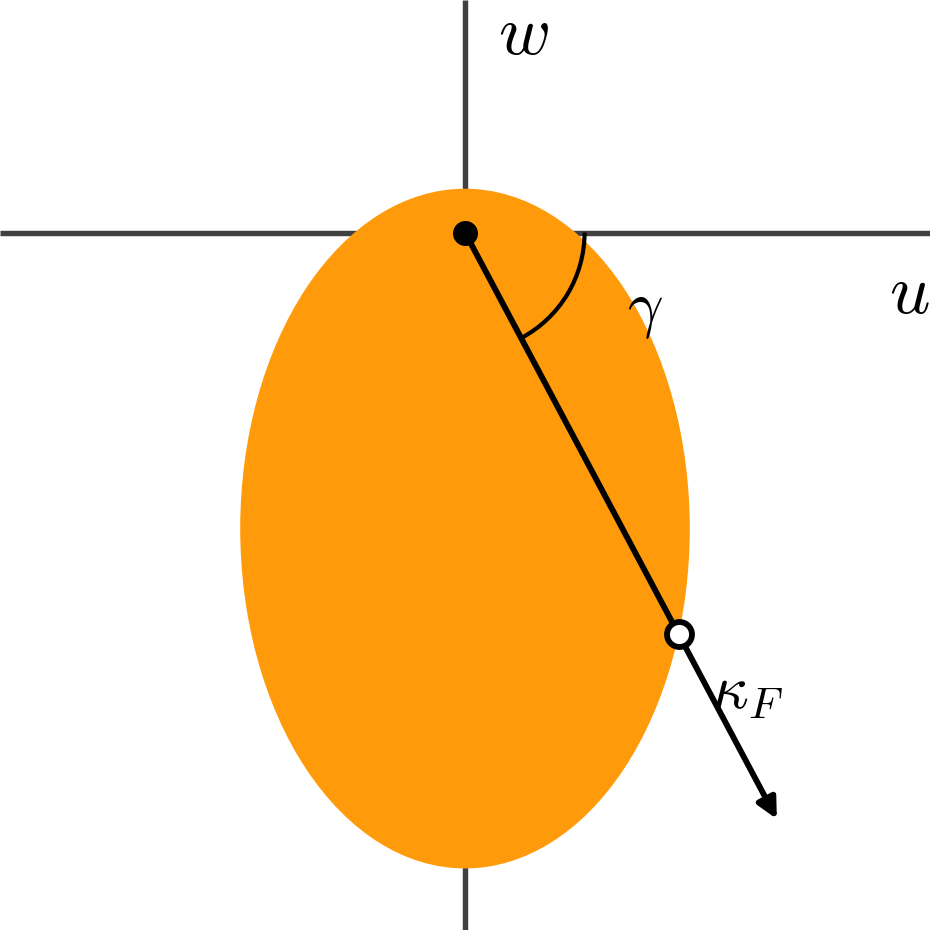}}
\subfigure[]{\includegraphics[width=0.28\textwidth]{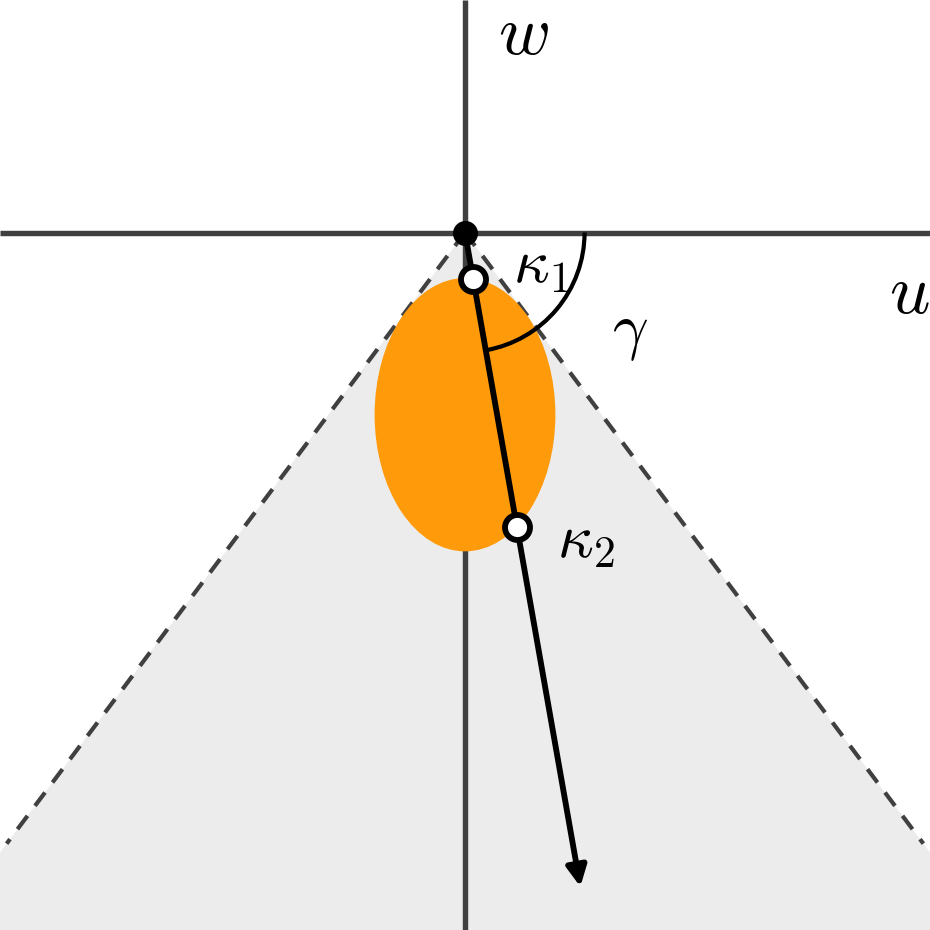}}
\subfigure[]{\includegraphics[width=1\textwidth]{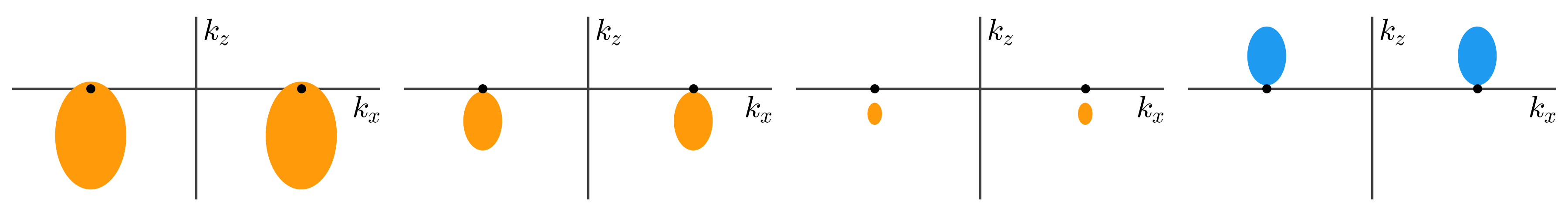}}
\subfigure[]{\includegraphics[width= 0.9\textwidth]{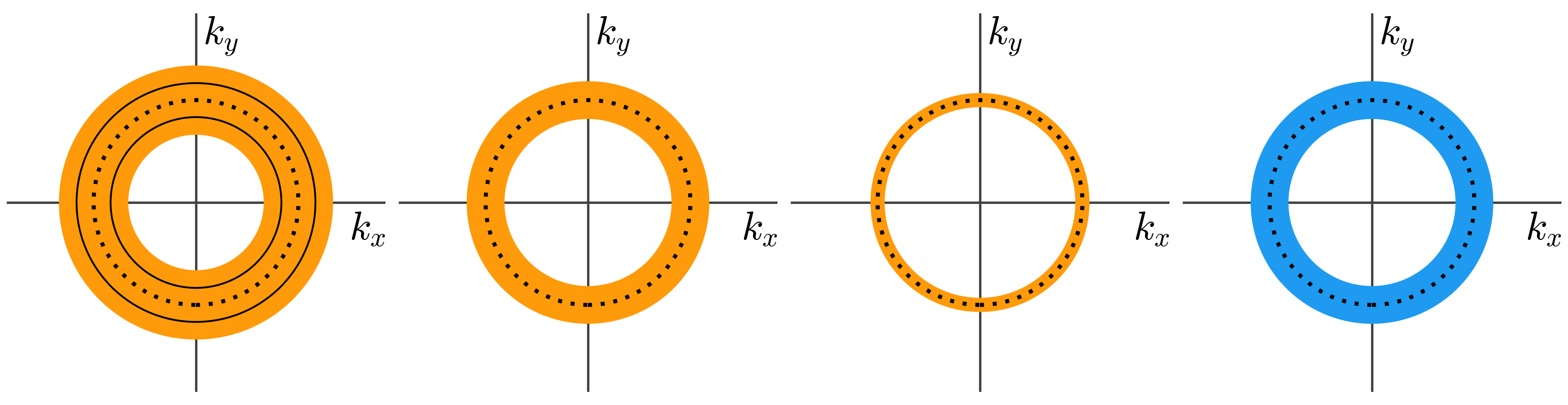}}
\caption{\label{fig:fsgnrz} FS of the GNR under in-plane and axial tilts. (a) Projection onto the $k_x\,k_y$-plane for the in-plane tilt in the twin-crescent regime (Sec.~\ref{secgtxbp}), not drawn to scale, with $\Delta = 0.1$, $\tilde{\eta} = 0.3$, and $\mu = 0.12$. The numbers count the intersections of a ray from the ring with the local FS. The dotted arcs are gapped, defined here by $|\mu_{\rm eff}| < \Delta/2$, and the light shading marks the arcs with two intersections, widened for clarity. (b), (c) Local FS of the axially-tilted GNR (Sec.~\ref{sec:gaptiltz}) in the $(u, w) = (k_\perp - k_0, k_z)$ plane, with the ring at the black dot. For $\mu = 0.26$ in (b), the FS encloses the ring and each ray meets it once. For $\mu = 0.16$ in (c), the FS is non-enclosing. Rays inside the shaded window between the dashed tangent rays meet it at the roots $\kappa_1$ and $\kappa_2$ of Eq.~\eqref{eq:gaptiltz-quad} (open circles). (d), (e) Cut at $k_y = 0$ and projection onto the $k_x\,k_y$-plane for $\mu = 0.26$, $0.18$, $0.14$, and $-0.18$ from left to right. In (d), the electron ($s=2$) and hole ($s=1$) pockets are orange and blue, displaced along $-k_z$ and $+k_z$, respectively. In (e), the ring is dotted and the solid circles mark the $k_z = 0$ slice, which exists for $|\mu| > \Delta$. In (b) to (e), $\Delta = 0.2$, $\eta_z = 0.75$, and $v_0 = k_0 = 1$. The values are chosen for visibility of the regimes.}
\end{figure*}

For $b < \Delta\, \nu_x$, no FS exists. For $\Delta\, \nu_x \le b < \Delta$ only a small electron-pocket exists around $\phi = \pi$, and it is non-enclosing everywhere. It is accompanied by a non-enclosing hole-pocket around $\phi = 0$ if $a \le -\Delta\, \nu_x$, which requires $\mu < \Delta \left( 1 - \nu_x \right)/2$ and $\tilde{\eta} \simeq \Delta$. For $b \ge \Delta$ and $-\Delta\, \nu_x < a < \Delta\, \nu_x$ an arc of the ring is gapped. The electron-pocket ends on this arc in a non-enclosing rim, and it encloses the ring further away from the gap. For $b \ge \Delta$ and $-\Delta < a \le -\Delta\, \nu_x$ this arc is bounded by an additional non-enclosing hole-pocket around $\phi = 0$. The pocket then has the shape of a crescent, which tapers to a tip at each end of the gapped arc. We call this the crescent-cyclide regime, and it is the partially-gapped regime in the sense that an arc of the ring is gapped. For $a \le -\Delta$ and $b \ge \Delta$ electron- and hole-pockets coexist, each ending on the gapped arc between them in a non-enclosing rim, as illustrated in Fig.~\ref{fig:fsgnrz}(a). Each of the two pockets is then a crescent, and we call this the twin-crescent regime, which is also the disjoint-pocket regime. Unlike in the horn-cyclide of the PTNR, the two pockets do not touch. For $\Delta\, \nu_x \le a < \Delta$ there is no gapped arc and the electron-pocket runs around the whole ring, but it does not enclose the ring near $\phi = 0$. For $a \ge \Delta$ it encloses the ring everywhere. The non-enclosing rims have the width $\Delta\left( 1 - \nu_t \right)/(\tilde{\eta}\, |\sin\phi|)$ in $\phi$, which is at most $\Delta\left( 1 - \nu_x \right)/(\tilde{\eta}\, |\sin\phi|)$ and equals $\Delta\, \eta_x^2 \cos^2\phi/(2\, \tilde{\eta}\, |\sin\phi|)$ at leading order in $\eta_x$. They vanish where the crossing $\mu_{\rm eff} = \pm\Delta$ occurs at $\phi = \pi/2$, as shown in Appendix~\ref{app:fsclass}. The windows in $\mu$ or $\tilde{\eta}$ that separate the eight cases from the four regimes of Sec.~\ref{sec:gaptiltx} have the width $\Delta \left( 1 - \nu_x \right) \simeq \Delta\, \eta_x^2/2$. To leading order in $\eta_x$ they close and leave the fully-insulating, twin-crescent, crescent-cyclide, and fully-conducting regimes, with the boundaries $b = \Delta$, $a = \pm\Delta$, and $\mu + \tilde{\eta} = \Delta$. All four windows are treated in Appendix~\ref{app:windows}. In the first two, only a small electron-pocket exists or it is accompanied by a small hole-pocket. In the third, the electron-pocket of the crescent-cyclide regime is accompanied by a small non-enclosing hole-pocket around $\phi = 0$. In the fourth, the non-enclosing part of the electron-pocket closes the gapped arc around $\phi = 0$. The fully-conducting, crescent-cyclide, and twin-crescent regimes, together with the symmetric case $\mu = 0$, are shown in Fig.~\ref{fig:fsgnrx}.

For an axial tilt, the nodal ring stays at zero energy, and the local plane is the same for every $\phi$. The FS of the GNR is the ellipse of Sec.~\ref{sec:gaptiltz}, displaced along $k_z$, and the band minimum lies at $\Delta_{\rm eff} = \Delta\, \sqrt{1 - \eta_z^2}$. For $\mu \ge 0$, there is no FS for $\mu < \Delta_{\rm eff}$. It is a displaced torus that does not enclose the ring for $\Delta_{\rm eff} \le \mu < \Delta$, and it encloses the ring for $\mu \ge \Delta$. These regimes are shown in Fig.~\ref{fig:fsgnrz}.

\section{PTNR}
\label{secungapped}

In this section we study the PTNR, setting $\Delta = 0$ in the Hamiltonian of Eq.~\eqref{eq:ham3d}. The chemical potential may lie exactly at the nodal ring, $\mu = 0$, or be finite but small, $\mu \ll v_0\,k_0$. The latter condition keeps the minor radius of the toroidal FS, $\kappa_F = \mu/v_0$, well below the major radius $k_0$. The FS is then a ring torus. For $\mu \geq v_0\,k_0$, it would turn into a horn torus and eventually a spindle torus, which self-intersects. Such values of $\mu$ lie outside the range of validity of the linearised Hamiltonian. The essential simplification is that at each fixed toroidal angle $\phi$ the low-energy Hamiltonian reduces to a 2d massless Dirac cone in the local cross-sectional plane spanned by $u = k_\perp - k_0$ and $w = k_z$. The ring then supplies an overall multiplicity $k_0$ through the Jacobian weight, a mapping that we make explicit in Sec.~\ref{secuntiltednr} and use throughout. It is exact to leading order in $q/k_0$ and lets us import known results for undoped and doped graphene~\cite{hwang-dassarma-graphene}, adapted to the toroidal geometry. We first treat the untilted case, where the residual rotational symmetry about $k_z$ keeps the response isotropic in the cross-section. We then add an in-plane tilt with respect to $k_x$-axis and an axial with respect to $k_z$-axis. The results are summarised in Tables~\ref{tab:plasmon-summary} and \ref{tab:anisotropy-onsets}.


\subsection{Untilted PTNR}
\label{secuntiltednr}

\begin{figure}[t!] 
\centering 
\includegraphics[width=0.4 \textwidth]{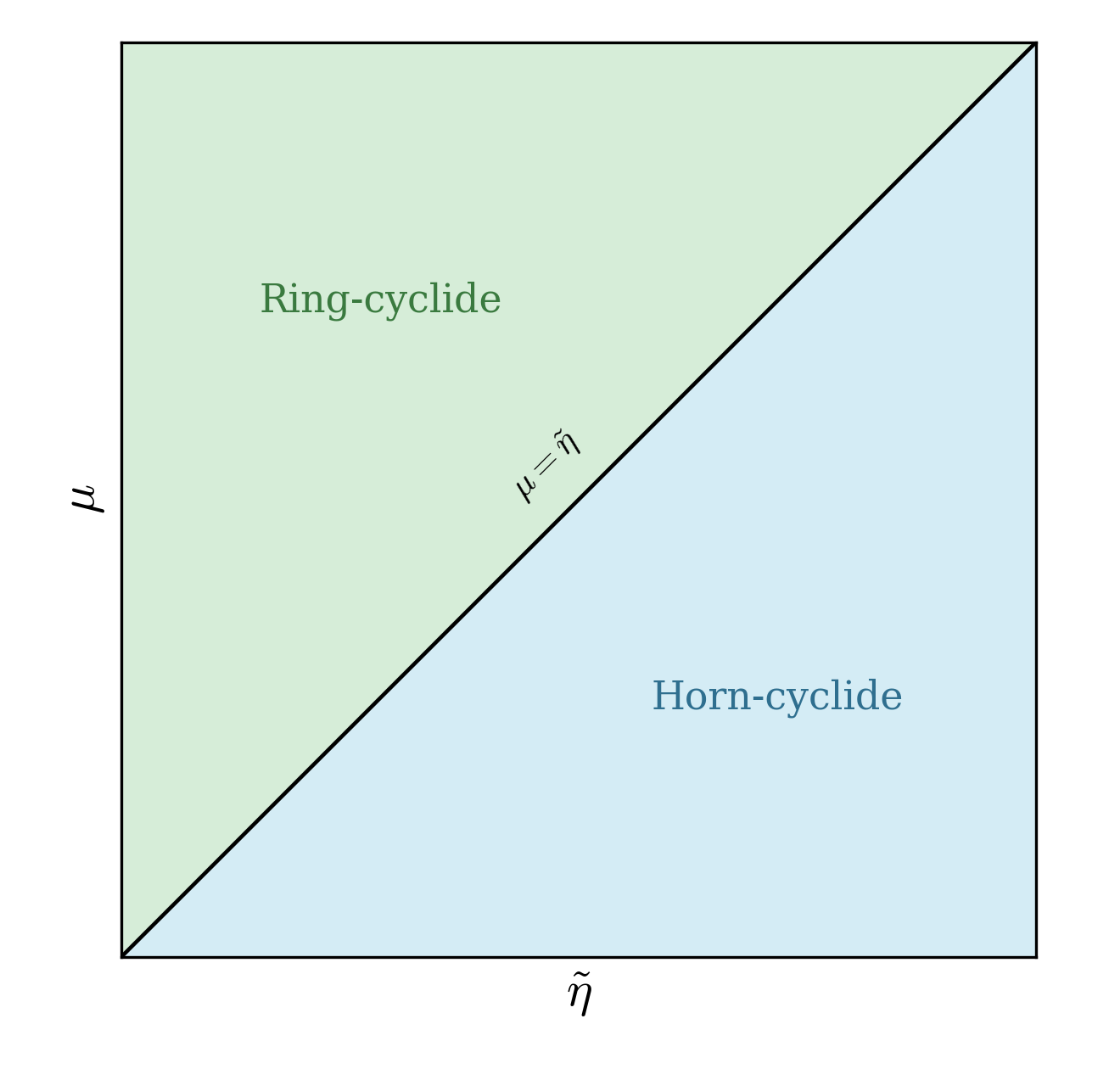} 
\caption{Phase diagram of an in-plane-tilted PTNR in the $\mu$-$\tilde{\eta}$-plane. The line $\mu=\tilde{\eta}$ separates the ring-cyclide and horn-cyclide regimes, and only the ratio $\tilde{\eta}/\mu$ determines the regime. Only $\mu \ge 0$ and $\tilde{\eta} \ge 0$ are shown. The regimes for $\mu < 0$ follow by exchanging electrons and holes, as explained in Sec.~\ref{secfermisurface}, and those for $\tilde{\eta} < 0$ follow from $\tilde{\eta} \to |\tilde{\eta}|$.\label{fig:phaseptnr}}
\end{figure}

We first consider the untilted case $\bs \eta = 0$. To leading order in $q/k_0$, the polarisability can be obtained easily by noting that, near the ring, the Hamiltonian of Eq.~\eqref{eq:ham3d} reduces at each fixed $\phi$ to a single isotropic 2d massless Dirac cone, with velocity $v_0$, in the local cross-sectional plane spanned by $u = k_\perp - k_0$ and $w = k_z$ (see also Appendix~\ref{appdirac}). The Jacobian weight $k_0 + \kappa \cos \gamma \to k_0$ is $\phi$-independent in this limit. A wavevector $\bs q = q\, \bs{\hat z} $ therefore couples the ring to a continuum of decoupled copies of this 2d system, one for each $\phi $. The ring supplies an overall multiplicity through $\int_0^{2\pi} k_0\, d\phi/(2\pi) = k_0 $. For $\bs q = q\, \bs{\hat x} $ the copy at angle $\phi$ sees only the projection $q \cos\phi $ of $\bs q$ onto the local radial direction. The component along the ring changes neither the energies nor the form-factor at leading order in $q/k_0$, and the copies remain decoupled. Consequently,
\begin{align}
\label{eq:ringmap}
\Pi (q\, \bs{\hat z}, \omega) = k_0\, \Pi_{2\rm D} (q, \omega) \text{ and }
\Pi (q\, \bs{\hat x}, \omega) = k_0 \int_0^{2\pi} 
\frac{d\phi}{2\pi}\, \Pi_{2\rm D} \big( q\, | \cos \phi |, \omega \big) \text{ for } q \ll k_0 \,,
\end{align}
where $\Pi_{2\rm D} (q,\omega)$ is the exact noninteracting polarisability of a single massless Dirac cone. This mapping is exact to leading order in $q/k_0 $ and reduces the whole problem to results already established for undoped and doped graphene~\cite{hwang-dassarma-graphene}. The derivation is given in Appendix~\ref{appdirac}.

At the intrinsic limit, $\mu = 0$, the FS collapses to the nodal ring itself and only the valence band is occupied. Both bands are then filled or empty everywhere at $T=0$. Only interband transitions between them contribute, and $\Pi = \Pi_{\rm inter}$ of Eq.~\eqref{eq:polinter}, with $\bs \eta = 0$, $E_k = v_0\, \kappa$, and $\mathcal{F} = [1 - \cos (\gamma' - \gamma)]/2$. This is precisely $k_0$ copies of the corresponding undoped-graphene expression. Its known closed form gives, for $\bs q$ in the local cross-section and below the interband threshold,
\begin{align}
\label{eq:polmu03dexact}
\Pi (\bs q, \omega) =  \frac{ -\,g\, k_0\, q^2}
{16\, \sqrt{v_0^2\, q^2 - \omega^2}} 
\text{ for } 0 \le \omega < v_0\, q \text{ and } q \ll k_0 \,.
\end{align}
For $\omega > v_0\, q$, a pair of gapless quasiparticles can be created directly and $\Pi (\bs q, \omega)$ acquires an imaginary part. This signals Landau damping into the interband continuum. Deep inside the continuum, i.e., for $\omega \gg v_0\, q$, Eq.~\eqref{eq:polmu03dexact} continues analytically to
\begin{align}
\label{eq:imfinal3d}
\Im \Pi (\bs q, \omega) \simeq \frac{ -\, g\, k_0\, q^2}{16\, \omega} 
\text{ for } v_0\, q \ll \omega \ll v_0\, k_0 \,.
\end{align}
This holds for $\bs q = q\, \bs{\hat z} $. For $\bs q = q\, \bs{\hat x} $ the local wavevector is $q \cos \phi $. Since $\Im \Pi_{2\rm D} \propto q^2 $ deep in the continuum, the angular average gives $\langle \cos^2 \phi \rangle = 1/2 $ and $\Im \Pi (q\, \bs{\hat x}, \omega) \simeq -\, g\, k_0\, q^2/(32\, \omega) $. The intrinsic damping is thus anisotropic by a factor of two. The static limit of Eq.~\eqref{eq:polmu03dexact}, viz.
\begin{align}
\label{eq:pi0final3d}
\Pi (\bs q, 0) =  \frac{ -\, g\, k_0\, q} {16\, v_0} \,,
\end{align}
for $\bs q \parallel \bs{\hat z} $ vanishes linearly with $q$. For $\bs q \parallel \bs{\hat x} $ the same average gives $\Pi (q\, \bs{\hat x}, 0) = -\, g\, k_0\, q/(8\, \pi\, v_0) $, which is also linear in $q$. This reflects the 1d character of the nodal ring, as found in Ref.~\cite{rhim-kim-friedel}.

For $\bs q \parallel \bs{\hat z}$, Eq.~\eqref{eq:polmu03dexact} shows that $\Pi (\bs q, \omega)$ is real and strictly negative throughout the window $0 \le \omega < v_0\, q$ below the continuum, the only place an undamped collective mode could appear. Substituting into Eq.~\eqref{eq:epsrpa} gives
\begin{align}
\label{eq:epsintrinsic}
\varepsilon^{\rm RPA} (\bs q, \omega) = 1 + \frac{\pi\, g\, e^2\, k_0}{4\, \sqrt{v_0^2 \,q^2 - \omega^2}} 
\text{ for } 0 \le \omega < v_0\, q \,,
\end{align}
which is strictly greater than unity for every $q$ and every $\omega$ in this range. Hence $\varepsilon^{\rm RPA}$ has no zero below the continuum. For $\bs q \parallel \bs{\hat x}$ the interband continuum extends down to $\omega = 0$, since the local threshold $v_0\, q\, |\cos \phi|$ vanishes at $\phi = \pi/2$. No undamped mode exists there either. At exactly $\mu = 0$, the PTNR therefore supports no undamped RPA plasmon, in the same way that undoped graphene does not. With no FS there is no Drude weight to balance the repulsive sign of $-\,v(q)\, \Pi$. Extrapolating the large-$\omega$ form of Eq.~\eqref{eq:imfinal3d} below the threshold at $\omega = v_0\, q$ does not produce a plasmon either, since that form does not apply there. A genuine plasmon only emerges once a nonzero $\mu$ is applied.

For $0 < \mu \ll v_0\, k_0$, the FS is a torus of minor radius $\kappa_F = \mu/v_0 $. Since $ \kappa_F\, (k_0 + \kappa_F \cos \gamma) \to \kappa_F \,k_0 $ for $ \kappa_F \ll k_0$, the DOS reduces to
\begin{align}
\label{eq:dos}
\rho (\mu) = \frac{g}{(2\, \pi)^3\, v_0} \int_0^{2\pi} \! d\phi \int_0^{2\pi} \! d\gamma\, \kappa_F\, k_0
= \frac{g\, k_0\, \kappa_F}{2\, \pi\, v_0} = \frac{g\, k_0\, \mu}{2\, \pi\, v_0^2} \,,
\end{align}
which is linear in $\mu$. It equals $k_0$ times the DOS, $g\,\mu/(2 \,\pi \,v_0^2)$, of a single doped Dirac cone, consistent with Eq.~\eqref{eq:ringmap}. Evaluating the Drude weight of Eq.~\eqref{eqdrude} for the $s=2$ band, with the velocity of Eq.~\eqref{eq:velocity3d} at $\bs \eta = 0$ and $\Delta = 0$, gives the nonzero components
\begin{align}
\label{eq:dzz}
D_{zz} = \frac{g\, e^2\, k_0\, \mu}{4} \,, \quad
D_{xx} = D_{yy} = \frac{g\, e^2\, k_0\, \mu}{8} \,.
\end{align}
The ratio $D_{zz}/D_{xx} = 2 $ results purely from the relative weight of $\langle \sin^2 \gamma \rangle $ and $\langle \cos^2 \gamma \cos^2 \phi \rangle $ over the ring. It is the same factor of two as in the intrinsic damping. 

\begin{figure}[t!]
\centering
\includegraphics[width=0.7\textwidth]{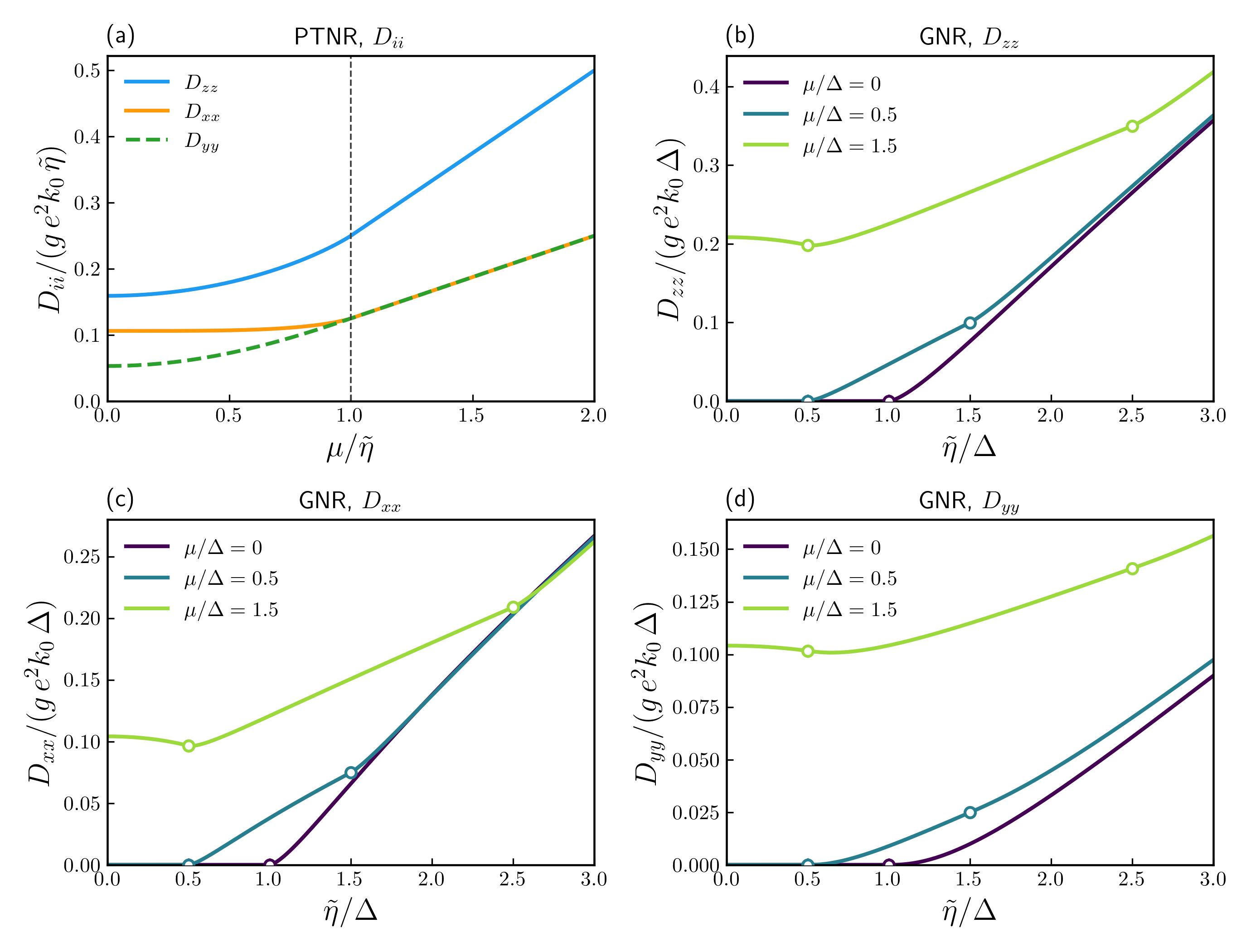}
\caption{\label{fig:drudex} Drude weights of the in-plane-tilted PTNR and GNR. Panel (a) shows $D_{zz}$, $D_{xx}$, and $D_{yy}$ of the PTNR in units of $g\, e^2\, k_0\, \tilde{\eta}$ as functions of $\mu/\tilde{\eta}$. The dashed vertical line marks $\mu = \tilde{\eta}$, which separates the horn-cyclide and ring-cyclide regimes. Panels (b), (c), and (d) show $D_{zz}$, $D_{xx}$, and $D_{yy}$ of the GNR as functions of $\tilde{\eta}/\Delta$, for $\mu/\Delta = 0$, $0.5$, and $1.5$. The open circles mark the boundaries between the fully-insulating, crescent-cyclide, twin-crescent, and fully-conducting regimes. The $\order{\eta_x^2} $ corrections are neglected throughout.}
\end{figure}

Using Eq.~\eqref{eqintra}, we get
\begin{align}
\Pi_{\rm intra} (q\, \bs{\hat z}, \omega) \simeq \frac{g\, k_0\, \mu\, q^2}{4\, \pi\, \omega^2} \,, \quad
\Pi_{\rm intra} (q\, \bs{\hat x}, \omega) \simeq \frac{g\, k_0\, \mu\, q^2}{8\, \pi\, \omega^2} \,.
\end{align}
The plasmon frequencies along these two principal directions are
\begin{align}
\label{eq:wpz}
\omega_{p, z}^2 = g\, e^2\, k_0\, \mu \,,
\quad \omega_{p, x}^2 = \omega_{p, y}^2 = \frac{g\, e^2\, k_0\, \mu}{2} \,,
\end{align}
with anisotropy ratio $ {\omega_{p, z}} /{\omega_{p, x}} = \sqrt{2} $. This is a distinctive signature of the toroidal FS. Unlike the 2d case of Ref.~\cite{rahmipoor-2d-nlsm}, where the long-wavelength Drude weight is independent of the carrier density below a threshold, here $D_{ij} \propto \mu $ throughout the low-doping window.

For $\mu > 0$ both bands contribute to the response, and the interband part must be kept alongside $\Pi_{\rm intra}$. It follows from the mapping of Eq.~\eqref{eq:ringmap}, in which only the states with $E_k > \mu$ are unblocked. To leading order in $q$,
\begin{align}
\label{eq:pinterdopedptnr}
\Pi_{\rm inter} (q\, \bs{\hat z}, \omega) = -\, \frac{g\, k_0\, q^2}{16\, \pi\, \omega} \left[ \ln \!\left| \frac{\omega + 2\, \mu}{\omega - 2\, \mu} \right| + i\, \pi\, \Theta \left( \omega - 2\, \mu \right) \right] ,
\end{align}
and $\Pi_{\rm inter} (q\, \bs{\hat x}, \omega)$ is half of this, because $\langle \cos^2 \phi \rangle = 1/2$. For $\omega > 2\, \mu$ the imaginary part is $-\, g\, k_0\, q^2/(16\, \omega)$, which is Eq.~\eqref{eq:imfinal3d}. It now sets in at the onset $\omega_{\rm th} = 2\, \mu$ instead of at $\omega = v_0\, q$. For $\omega < 2\, \mu$ the polarisability is real and negative. Its static value is $-\, g\, k_0\, q^2/(16\, \pi\, \mu)$. Together with the Drude term of Eq.~\eqref{eq:wpz} it gives the dielectric function
\begin{align}
\label{eq:epsptnrdoped}
\varepsilon^{\rm RPA} (q\, \bs{\hat z}, \omega) = 1 - \frac{g\, e^2\, k_0\, \mu}{\omega^2} + \frac{g\, e^2\, k_0}{4\, \omega} \left[ \ln \!\left| \frac{\omega + 2\, \mu}{\omega - 2\, \mu} \right| + i\, \pi\, \Theta \left( \omega - 2\, \mu \right) \right] ,
\end{align}
for $v_0\, q \ll \omega \ll v_0\, k_0$. The $\bs{\hat x}$ channel follows by halving the prefactor $g\, e^2\, k_0$ in all three terms. Below the onset the interband part acts as a background dielectric constant, $\epsilon_{b,z} = 1 + g\, e^2\, k_0\, \text{arctanh} (\omega/2\, \mu)/(2\, \omega)$, with the static values $\epsilon_{b,z} = 1 + g\, e^2\, k_0/(4\, \mu)$ and $\epsilon_{b,x} = \epsilon_{b,y} = 1 + g\, e^2\, k_0/(8\, \mu)$. With $\lambda = g\, e^2\, k_0/(4\, \mu)$, the $\bs{\hat z}$ channel reads $\varepsilon^{\rm RPA} = 1 + \lambda \left[ \text{arctanh} (\omega/(2\, \mu))/(\omega/(2\, \mu)) - (\omega/(2\, \mu))^{-2} \right]$. It tends to $-\infty$ for $ \omega/(2\, \mu) \to 0$ and to $+\infty$ for $ \omega/(2\, \mu) \to 1^-$, and hence it has a zero below the onset for every coupling. The same holds along $\bs{\hat x}$, with $\lambda \to \lambda/2$. The bare Drude frequencies of Eq.~\eqref{eq:wpz} are recovered for $\lambda \ll 1$. In the static approximation for $\epsilon_b$, the screened frequencies are $\tilde\omega_{p,z}^2 = 4\, \mu^2\, \lambda/(1 + \lambda)$ and $\tilde\omega_{p,x}^2 = 2\, \mu^2\, \lambda/(1 + \lambda/2)$. They always lie below $\omega_{\rm th}$, and the anisotropy ratio becomes $\tilde\omega_{p,z}/\tilde\omega_{p,x} = \sqrt{2\, (1 + \lambda/2)/(1 + \lambda)}$, which is slightly below $\sqrt{2}$. The steps behind Eq.~\eqref{eq:pinterdopedptnr} and the static values are given in the $\eta_z \to 0$ limit of Appendix~\ref{app:interband}. The behaviour of $\Re \varepsilon^{\rm RPA}$ at the onset is derived in Appendix~\ref{app:onset}.

\begin{figure*}[t!]
\centering
\includegraphics[width=0.95\textwidth]{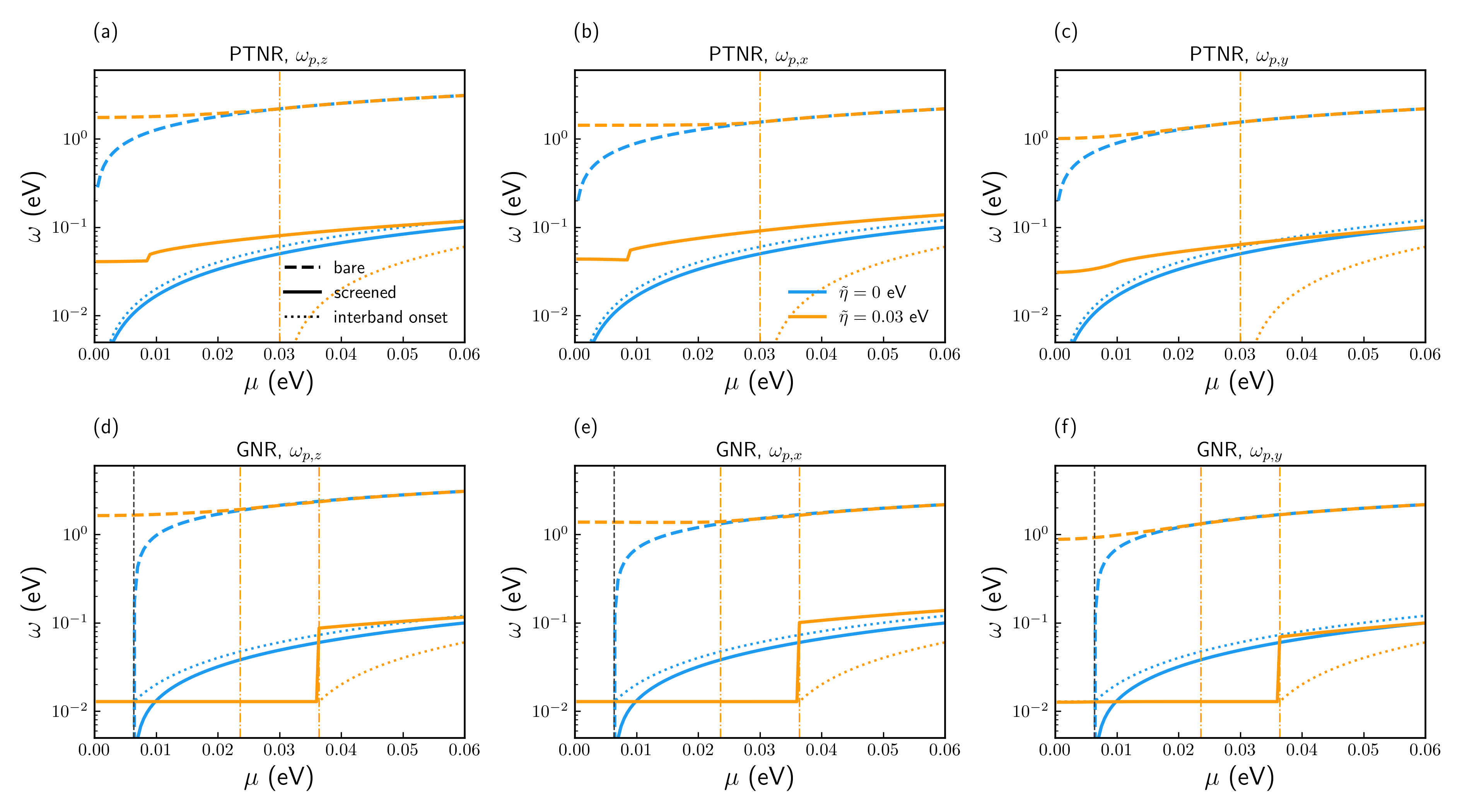}
\caption{\label{fig:plasmax} Plasmon frequencies of the in-plane-tilted PTNR and GNR for $\tilde{\eta} = 0$ (blue) and $\tilde{\eta} = 0.03$ eV (orange), for the parameters of Table~\ref{tab-params}. Panels (a), (b), and (c) show $\omega_{p, z}$, $\omega_{p, x}$, and $\omega_{p, y}$ of the PTNR as functions of $\mu$, and panels (d), (e), and (f) show those of the GNR with $\Delta = 0.02$ eV. The dashed curves are the bare frequencies $\omega_{p, i}$. The solid curves are the screened frequencies $\tilde\omega_{p, i}$, defined as the lowest zero of $\Re \varepsilon^{\rm RPA}$ with the full frequency dependence of the interband part. The dotted curves are the interband onset $\omega_{\rm th}$, which vanishes in the horn-cyclide regime of the PTNR and is not drawn there. The in-plane tilt $\tilde{\eta} = 0.03$ eV corresponds to $\eta_x \approx 0.47$, and the plotted results are the leading order in $\eta_x$. The dash-dotted orange vertical lines mark the regime boundaries of the tilted system, namely $\mu = \tilde{\eta}$ in (a), (b), and (c), and $\mu = \tilde{\eta} - \Delta$ and $\mu = \tilde{\eta} + \Delta$ in (d), (e), and (f). The dashed grey vertical line in (d), (e), and (f) marks $\mu = \Delta$.}
\end{figure*}

\subsection{Tilt with respect to the \texorpdfstring{$k_x$}{kx}-axis}
\label{sectiltx}

We now include an in-plane tilt along the $ k_x $-axis by setting $\bs \eta = \eta_x\, \bs{\hat x} $. The tilt term becomes $v_0\, \eta_x\, k_x = v_0\, \eta_x\, (k_0 + \kappa \cos \gamma) \cos \phi $, and the eigenvalues are $\varepsilon_s (\bs k) = (-1)^s\, v_0\, \kappa + v_0\, \eta_x\, \left( k_0 + \kappa \cos \gamma \right) \cos \phi $. Setting $\varepsilon_s = \mu $ gives the Fermi momentum
\begin{align}
\label{eq:cyclide}
\kappa_s (\phi, \gamma) = 
\frac{ \mu - \tilde{\eta} \cos \phi }{ v_0\, \zeta_s (\phi, \gamma) } \,,
\quad \zeta_s (\phi, \gamma) = (-1)^s + \eta_x \cos \gamma \cos \phi \,,
\quad \tilde{\eta} = v_0\, \eta_x\, k_0 \,,
\end{align}
leading to
\begin{align}
\label{eq:jacobian}
\delta \big( \varepsilon_s (\bs k) - \mu \big) = 
\frac{ \delta \big( \kappa - \kappa_s (\phi, \gamma) \big) }{ v_0\, | \zeta_s (\phi, \gamma) | } \,.
\end{align}
Depending on the ratio $\tilde{\eta}/\mu$, the FS takes the form of a ring-cyclide (for $0 < \tilde{\eta} < \mu $ ) or a horn-cyclide (for $0 < \mu < \tilde{\eta} $), as shown in Fig.~\ref{fig:phaseptnr}. We first focus on the ring-cyclide regime, where the FS is a single closed surface containing the nodal ring. The horn-cyclide regime is treated at the end of this subsection. We assume the type-I condition $|\eta_x| < 1$ throughout. The classification depends only on $|\tilde{\eta}|$, since $\eta_x \to -\eta_x$ is equivalent to $\phi \to \phi + \pi$. For $\eta_x < 0$ the conditions above and below are therefore read with $\tilde{\eta} \to |\tilde{\eta}|$. Quantities even in $\eta_x$ are unchanged. The part of $\Im \Pi (q\, \bs{\hat x}, \omega)$ odd in $\eta_x$ changes sign together with $q \to -q$.

\begin{figure}[t!]
\centering
\includegraphics[width=0.7\textwidth]{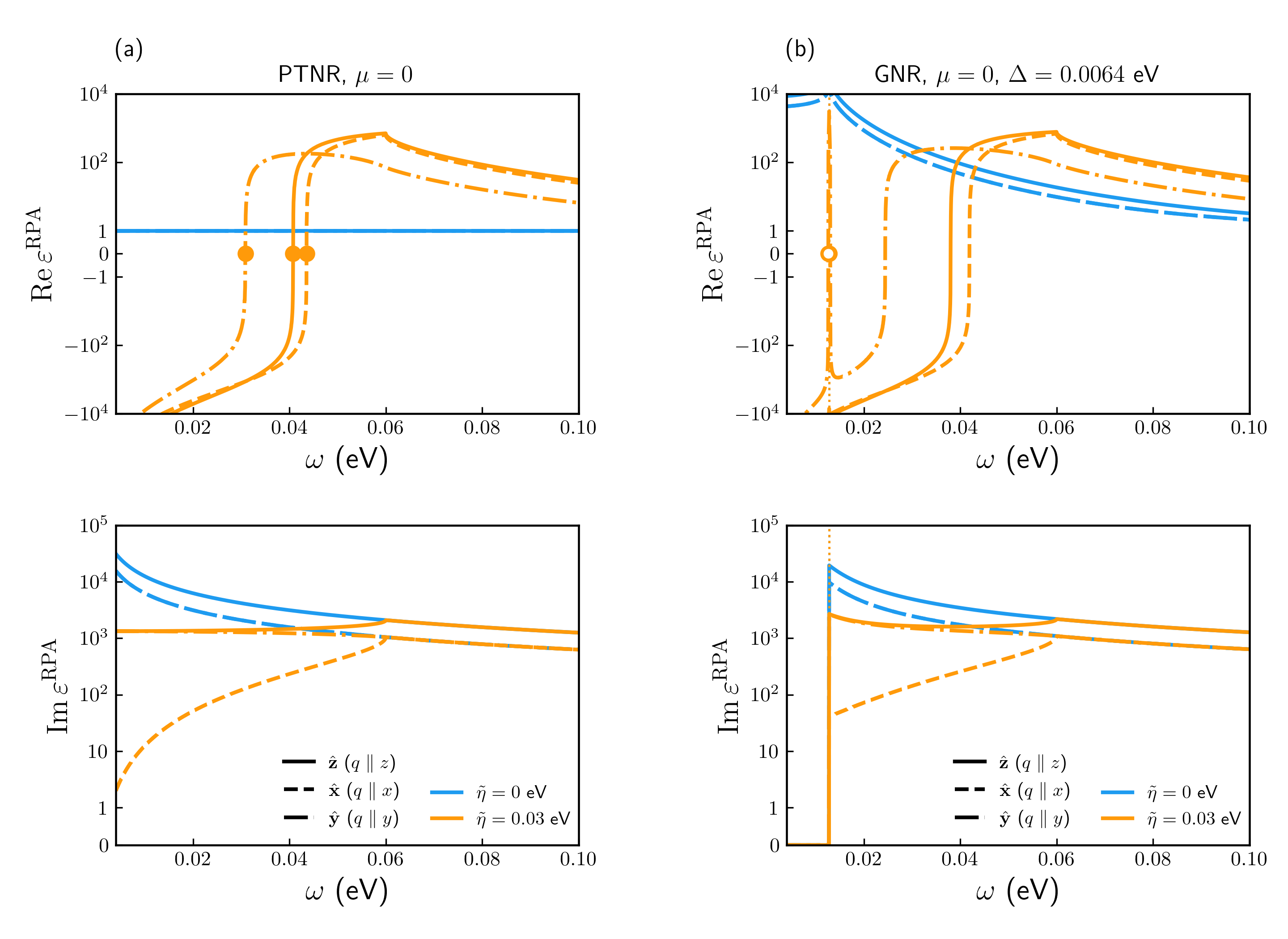}
\caption{\label{fig:epsx} RPA dielectric function of the in-plane-tilted (a) PTNR and (b) GNR at $\mu = 0$, with $\Delta = 0.02$ eV for the GNR, for $\tilde{\eta} = 0$ (blue) and $\tilde{\eta} = 0.03$ eV (orange). Since $\tilde{\eta} > \Delta$, the GNR lies in the twin-crescent regime. The other parameters are taken from Table~\ref{tab-params}. Upper panels: $\Re \varepsilon^{\rm RPA}$ (symmetric-logarithmic, linear for $|\Re \varepsilon^{\rm RPA}| < 1$). Lower panels: $\Im \varepsilon^{\rm RPA}$. Solid, dashed, and dash-dotted curves are for $\bs q \parallel \bs{\hat z}$, $\bs q \parallel \bs{\hat x}$, and $\bs q \parallel \bs{\hat y}$, respectively. The dotted vertical line in (b) marks the interband onset $\omega_{\rm th} = 2\,\Delta$. Filled circles in (a) mark the damped plasmons of the PTNR, and open circles in (b) mark the undamped, screened plasmons of the GNR. All curves are for $\omega \gg v_0\, q$.}
\end{figure}

At the intrinsic point, $\mu = 0 $, the FS is the nodal ring only for $\tilde{\eta} \to 0$. At finite $\tilde{\eta}$ it is the symmetric horn-cyclide, whose static properties are given at the end of this subsection. Here we consider the interband response at frequencies well above the pocket scale. At fixed $\phi$ and in the local plane $(u, w) = (\kappa \cos \gamma, k_z)$, the two bands read $\varepsilon_\pm = \pm\, v_0 \sqrt{u^2 + w^2} + v_0\, \eta_x \cos \phi\, u + \tilde{\eta} \cos \phi $. This is a tilted Dirac cone with the tilt along $u$ and a constant offset $\tilde{\eta} \cos \phi $. The offset induces small local pockets. These do not overlap the region of phase space with $\kappa + \kappa' \sim \omega/v_0 $ that the delta function selects, provided $\omega \gg v_0\, q $ and $\omega \gg \tilde{\eta} $. For $\bs q = q\, \bs{\hat z} $ the momentum transfer has no $x$-component. Then $k_x' = k_x $ and the tilt contribution to the interband transition energy cancels exactly. The pole condition and the form-factor are identical to those of the untilted ring, and
\begin{align}
\label{eq:imintrinsicxtilt}
\Im \Pi (q\, \bs{\hat z}, \omega) \simeq -\, \frac{g\, k_0\, q^2}{16\, \omega} \,,
\end{align}
with no dependence on $\eta_x $. The channel $\bs q = q\, \bs{\hat x} $ is different, because $\bs q $ then has a component along the tilt. The tilt contribution to the transition energy is $v_0\, \eta_x\, \left( k_x' - k_x \right) = v_0\, \eta_x\, q $ exactly. The displacement along the ring enters only through this term and not through $\kappa $ or the form-factor. The resonance condition is therefore $ v_0\, \left( \kappa + \kappa' \right) = \omega - v_0\, \eta_x\, q \equiv \Omega $, and the result is the untilted one with $\omega \to \Omega $. Averaging the local wavevector $q \cos \phi $ over $\phi $ gives
\begin{align}
\Im \Pi (q\, \bs{\hat x}, \omega) = -\, \frac{g\, k_0\, q^2}{16}\, \int_0^{2\pi} \frac{d\phi}{2\pi}\, \frac{ \cos^2 \phi\, \Theta \left( \Omega - v_0\, q\, | \cos \phi | \right) }{ \sqrt{ \Omega^2 - v_0^2\, q^2 \cos^2 \phi } } \simeq -\, \frac{g\, k_0\, q^2}{32\, \omega}\, \left( 1 + \frac{ v_0\, \eta_x\, q }{ \omega } \right) ,
\end{align}
where the last step uses $\omega \gg v_0\, q $. In contrast to the $\bs{\hat z}$ channel, the damping along $\bs{\hat x}$ therefore acquires a correction that is odd in $\eta_x $ and of first order in $v_0\, \eta_x\, q/\omega $. The steps behind the frequency shift $\Omega$ and this expansion are given in Appendix~\ref{app:tiltshift}. For $\bs q = q\, \bs{\hat y}$ the momentum transfer is again perpendicular to the tilt, so the shift vanishes. The local transfer is $q \sin\phi$, and the angular average $\langle \sin^2 \phi \rangle = 1/2$ gives $\Im \Pi (q\, \bs{\hat y}, \omega) \simeq -\, g\, k_0\, q^2/(32\, \omega)$, with no dependence on $\eta_x$.

The counterpart of Eq.~\eqref{eq:epsintrinsic} follows in the same way. For $\bs q = q\, \bs{\hat z}$ the tilt drops out of the pole condition. In the window $\omega < v_0\, q - 2\, |\tilde{\eta}|$, with $\tilde{\eta} \ll v_0\, q \ll v_0\, k_0$ and to leading order in $\eta_x$, the small pockets add the correction
\begin{align}
\label{eq:dpipocketptnr}
\delta \Pi (q\, \bs{\hat z}, \omega) = -\, \frac{4\, g\, k_0\, q^2\, |\tilde{\eta}|^3}{9\, \pi^2 \left( v_0^2\, q^2 - \omega^2 \right)^2}
\end{align}
to Eq.~\eqref{eq:polmu03dexact}. It is of relative order $(\tilde{\eta}/v_0\, q)^3$ and is derived in Appendix~\ref{app:pockets}. The correction is negative, as is Eq.~\eqref{eq:polmu03dexact}, and it raises $\varepsilon^{\rm RPA}$ above the value of Eq.~\eqref{eq:epsintrinsic}. Hence $\varepsilon^{\rm RPA} > 1$ has no zero there. For $\bs q = q\, \bs{\hat x}$ the shifted frequency $\Omega = \omega - v_0\, \eta_x\, q$ enters, and the local threshold $v_0\, q\, | \cos \phi |$ vanishes at $\phi = \pi/2$. The interband continuum then reaches down to $\Omega = 0$, and $\varepsilon^{\rm RPA}$ is complex at all frequencies, as in Sec.~\ref{secuntiltednr}. At small tilt the interband response thus gives no undamped plasmon in either channel.

\subsubsection{Ring-cyclide regime}

For low doping, $0 < \mu \ll v_0\, k_0 $, the FS is a ring-cyclide, with $\kappa_F (\phi, \gamma)$ given by Eq.~\eqref{eq:cyclide}. Since $\tilde{\eta} = v_0\, \eta_x\, k_0 $, the condition $\tilde{\eta} < \mu $ is equivalent to $\eta_x < \mu/(v_0\, k_0) $. This is far tighter than the type-I condition $0 < \eta_x < 1 $ of nodal-point semimetals and it forces the tilt to be small at low doping. The DOS follows from Eq.~\eqref{eq:jacobian} and the exact Jacobian $\kappa\, (k_0 + \kappa \cos \gamma)$ as
\begin{align}
\rho (\mu) = \frac{g}{(2\, \pi)^3\, v_0} \int_0^{2\pi} \! d\phi \int_0^{2\pi} \! d\gamma\, \frac{ \kappa_F \left( k_0 + \kappa_F \cos \gamma \right) }{ \zeta_2 } \,.
\end{align}
We expand $1/\zeta_2^2$ to $\mathcal{O} (\eta_x^2)$ in the term proportional to $k_0\, \kappa_F$. The piece linear in $\tilde\eta$ averages to zero, and no factor of $\tilde\eta^2/\mu$ survives. This term contributes $3\, \eta_x^2/4$. The term proportional to $\kappa_F^2 \cos \gamma$ is smaller by $\kappa_F/k_0 $. Its cross-term with the offset $\tilde\eta \cos \phi $ in $\kappa_F$ is nevertheless of the same order in $\eta_x$, since $\tilde\eta/k_0 = v_0\, \eta_x $. It contributes a further $3\, \eta_x^2/2$, and
\begin{align}
\label{eq:dostiltx}
\rho (\mu) = \frac{g\, k_0\, \mu}{2\, \pi\, v_0^2}\, \left( 1 + \frac{9\, \eta_x^2}{4} \right).
\end{align}
The compressibility sum rule fixes the static polarisability at small $q$ as
\begin{align}
\label{eq:pistatictiltx}
\Pi (\bs q, 0) \to 
-\, \frac{g\, k_0\, \mu}{2\, \pi\, v_0^2}\, \left( 1 + \frac{9\, \eta_x^2}{4} \right)
\text{ for } q \to 0 \,.
\end{align}
It is independent of whether $\bs q$ points along $\bs{\hat z}$ or $\bs{\hat x}$, since the response must reduce to the isotropic thermodynamic compressibility as $q \to 0$. A genuine split between the two directions only shows up at the next order in $q/\kappa_F$, through the momentum dependence of the full Lindhard function, and requires numerical evaluation.

Setting $\Delta = 0$ and $\bs \eta = \eta_x\, \bs{\hat x}$ in Eq.~\eqref{eq:velocity3d}, the components of the $s=2$ band-velocity are $V_x^{(2)} = v_0\, \left( \cos \gamma \cos \phi + \eta_x \right)$, $V_y^{(2)} = v_0\, \cos \gamma \sin \phi$, and $V_z^{(2)} = v_0\, \sin \gamma$. As explained in Sec.~\ref{secmodel}, the in-plane tilt breaks the rotational symmetry that enforces $D_{xx} = D_{yy}$ in the untilted case, while the reflection symmetries $\phi \to -\phi$ and $\gamma \to -\gamma$ keep $D_{ij}$ diagonal. The Drude-weight components are therefore
\begin{align}
\label{eq:dzzT}
D_{zz} &= \frac{ \pi\, g\, e^2\, v_0^2}{(2\, \pi)^3} \int d\phi \int d\gamma\, \frac{ \kappa_2 \left( k_0 + \kappa_2 \cos \gamma \right) }{ v_0\, \zeta_2 }\, \sin^2 \gamma \,,
\nn  D_{xx} &= \frac{ \pi\, g\, e^2\, v_0^2}{(2\, \pi)^3} 
\int d\phi \int d\gamma\, \frac{ \kappa_2 \left( k_0 + \kappa_2 \cos \gamma \right) }
{ v_0\, \zeta_2 }\, \left( \cos \gamma \cos \phi + \eta_x \right)^2 ,
\nn  D_{yy} &= \frac{ \pi\, g\, e^2\, v_0^2}{(2\, \pi)^3} 
\int d\phi \int d\gamma\, \frac{ \kappa_2 \left( k_0 + \kappa_2 \cos \gamma \right) }
{ v_0\, \zeta_2 }\, \cos^2 \gamma \sin^2 \phi \,.
\end{align}
Expanding to $\mathcal{O} (\eta_x^2)$ in the ring-cyclide regime, with the $\kappa_2^2 \cos \gamma$ term retained as in the DOS, gives (see Appendix~\ref{app:xtiltdrude})
\begin{align}
\label{eq:dzzx}
D_{zz} \simeq \frac{ g\, e^2\, k_0\, \mu }{ 4 }\, \left ( 1 + \frac{ 9\, \eta_x^2 }{ 8 } \right ) ,
\quad
D_{xx} \simeq \frac{ g\, e^2\, k_0\, \mu }{ 8 }\, 
\left( 1 + \frac{ 17\, \eta_x^2 }{ 16 } \right) , \quad
D_{yy} \simeq \frac{ g\, e^2\, k_0\, \mu }{ 8 }\, 
\left( 1 + \frac{ 27\, \eta_x^2 }{ 16 } \right) .
\end{align}
Unlike $D_{zz}$ and $D_{xx}$, $D_{yy}$ receives no contribution from the offset $\eta_x$ appearing directly in the velocity of Eq.~\eqref{eq:velocity3d}, since $V_y^{(2)} = v_0\, \cos \gamma \sin \phi$ does not involve $\eta_x$. Its $\mathcal{O} (\eta_x^2)$ correction comes entirely from the tilt-induced deformation of the FS itself, through $\kappa_2 (\phi, \gamma)$ and $\zeta_2 (\phi, \gamma)$. The three principal Drude weights obey
\begin{align}
\label{eq:anisotropyx}
\frac{ D_{zz} }{ D_{xx} } \simeq 2\, \left( 1 + \frac{ \eta_x^2 }{ 16 } \right ), \quad
\frac{ D_{xx} }{ D_{yy} } \simeq 1 - \frac{ 5\, \eta_x^2 }{ 8 } , \quad
\frac{ D_{zz} }{ D_{yy} } \simeq 2\, \left( 1 - \frac{ 9\, \eta_x^2 }{ 16 } \right ).
\end{align}
While $D_{zz}$ and $D_{xx}$ both grow with the tilt, $D_{yy}$ grows faster still, and the in-plane tilt therefore reduces both $D_{xx}/D_{yy}$ and $D_{zz}/D_{yy}$ at quadratic order, even though it increases $D_{zz}/D_{xx}$.
Using Eq.~\eqref{eqintra}, we get
\begin{align}
\Pi_{\rm intra} (q\, \bs{\hat z}, \omega) \simeq \frac{g\, k_0\, \mu\, q^2}{4\, \pi\, \omega^2}\, 
\left ( 1 + \frac{9\, \eta_x^2}{8} \right ), \quad
\Pi_{\rm intra} (q\, \bs{\hat x}, \omega) \simeq 
\frac{g\, k_0\, \mu\, q^2}{8\, \pi\, \omega^2}\, 
\left ( 1 + \frac{17\, \eta_x^2}{16} \right ) , \quad
\Pi_{\rm intra} (q\, \bs{\hat y}, \omega) \simeq 
\frac{g\, k_0\, \mu\, q^2}{8\, \pi\, \omega^2}\, 
\left ( 1 + \frac{27\, \eta_x^2}{16} \right ) ,
\end{align}
and the plasmons are identified from
\begin{align}
\omega_{p,z}^2  = g\, e^2\, k_0\, \mu\, \left( 1 + \frac{ 9\, \eta_x^2 }{ 8 } \right ) ,
\quad
\omega_{p,x}^2 = \frac{ g\, e^2\, k_0\, \mu }{ 2 }\, 
\left( 1 + \frac{ 17\, \eta_x^2 }{ 16 } \right) ,\quad
\omega_{p,y}^2 = \frac{ g\, e^2\, k_0\, \mu }{ 2 }\, 
\left( 1 + \frac{ 27\, \eta_x^2 }{ 16 } \right) .
\end{align}
At $\eta_x = 0$, $\omega_{p,x} = \omega_{p,y}$, as required by the residual rotational symmetry. To quadratic order in the tilt,
\begin{align}
\frac{ \omega_{p,z} }{ \omega_{p,x} } \simeq 
\sqrt{2}\, \left ( 1 + \frac{ \eta_x^2 }{ 32 } \right ), \quad
\frac{ \omega_{p,x} }{ \omega_{p,y} } \simeq 1 - \frac{ 5\, \eta_x^2 }{ 16 } , \quad
\frac{ \omega_{p,z} }{ \omega_{p,y} } \simeq 
\sqrt{2}\, \left ( 1 - \frac{ 9\, \eta_x^2 }{ 32 } \right ) .
\end{align}
The in-plane tilt therefore splits the two originally-degenerate transverse plasmon frequencies, $\omega_{p,x}$ and $\omega_{p,y}$, by an amount of order $\eta_x^2$, in addition to enhancing all three frequencies relative to their untilted values.

The dielectric function follows from Eqs.~\eqref{eq:epsrpa} and \eqref{eqintra}. For $v_0\, q \ll \omega$ it is $\varepsilon^{\rm RPA} (q\, \bs{\hat e}_i, \omega) \simeq 1 - \omega_{p,i}^2/\omega^2$ along each principal axis, and its zero is the plasmon above. At $q \to 0$ the interband transitions are Pauli-blocked for $\omega < \omega_{\rm th}$, where $\omega_{\rm th} = 2\, \left( \mu - |\tilde{\eta}| \right)/\left( 1 + |\eta_x| \right)$. This threshold is set by the point $\phi = \gamma = 0$ of the FS, at which $\kappa_F$ is smallest. The threshold vanishes as $|\tilde{\eta}| \to \mu$, as the ring-cyclide gives way to the horn-cyclide, in accordance with the absence of an interband threshold in the latter. Below the threshold the interband part is real and screens the plasmon. Above it the plasmon is damped.

Both bands contribute here, and the full interband response follows from the same local mapping. At fixed $\phi$ the local cone has the chemical potential $\mu_{\rm eff} (\phi) = \mu - \tilde{\eta} \cos \phi$, and Pauli blocking removes the states with $E_k < |\mu_{\rm eff}|$. Dropping the tilt terms of relative order $\eta_x$ inside the local cone, as for the Drude weights, Eq.~\eqref{eq:pinterdopedptnr} generalises to
\begin{align}
\label{eq:pintertiltxptnr}
\Pi_{\rm inter} (q\, \bs{\hat e}_i, \omega) \simeq -\, \frac{g\, k_0\, q^2}{16\, \pi\, \omega} \int_0^{2\pi} \frac{d\phi}{2\pi}\, \check w_i (\phi) \left[ \ln \!\left| \frac{\omega + 2\, |\mu_{\rm eff}|}{\omega - 2\, |\mu_{\rm eff}|} \right| + i\, \pi\, \Theta \left( \omega - 2\, |\mu_{\rm eff}| \right) \right] ,
\end{align}
with the weights $\check w_z = 1$, $\check w_x = \cos^2 \phi$, and $\check w_y = \sin^2 \phi$. It holds for $v_0\, q \ll \omega \ll v_0\, k_0$ and in both cyclide regimes. The imaginary part reproduces Eq.~\eqref{eq:imhorn} below. In the ring-cyclide regime, $\mu_{\rm eff} > 0$ everywhere, and for $\omega \ll 2\, (\mu - |\tilde{\eta}|)$ the logarithm reduces to $\omega/|\mu_{\rm eff}|$. With $\tilde w \equiv \sqrt{\mu^2 - \tilde{\eta}^2}$, the static background dielectric constants are then
\begin{align}
\label{eq:epsbtiltxptnr}
\epsilon_{b,z} = 1 + \frac{g\, e^2\, k_0}{4\, \tilde w} \,, \quad
\epsilon_{b,x} = 1 + \frac{g\, e^2\, k_0\, \mu}{4\, \tilde w \left( \mu + \tilde w \right)} \,, \quad
\epsilon_{b,y} = 1 + \frac{g\, e^2\, k_0}{4 \left( \mu + \tilde w \right)} \,.
\end{align}
They are the $\Delta \to 0$ limit of Eq.~\eqref{eq:epsb-xtilt}, and they reduce to Eq.~\eqref{eq:epsptnrdoped} for $\tilde{\eta} \to 0$. The real part of the dielectric function is $\Re \varepsilon^{\rm RPA} \simeq \epsilon_{b,i} - \omega_{p,i}^2/\omega^2$ for $\omega_{p,i}$ well below the threshold, where the static value applies. This gives the screened frequencies $\tilde\omega_{p,i}$ defined in Sec.~\ref{secmodel}. By Appendix~\ref{app:monotone}, the zero below the threshold is unique, and the static screened frequency is an upper bound of it. Unlike in the untilted case, $\Re \varepsilon^{\rm RPA}$ stays finite at the threshold, because the logarithmic singularity of Eq.~\eqref{eq:pintertiltxptnr} is integrable over the ring. The approach to the threshold is derived in Appendix~\ref{app:onset}. The zero exists below the threshold if $\Re \varepsilon^{\rm RPA} > 0$ there, and this is guaranteed if the static screened frequency lies below the threshold. Since $\epsilon_{b,i}$ grows as $|\tilde{\eta}| \to \mu$, the screening becomes stronger as the ring-cyclide approaches the horn-cyclide.

\subsubsection{Horn-cyclide regime}

In the horn-cyclide regime, $0 \le \mu < \tilde{\eta}$, the FS splits into an electron-pocket ($s=2$) and a hole-pocket ($s=1$), as shown in Fig.~\ref{figfs}. Both bands cross the chemical potential. The local doping $\mu_{\rm eff} (\phi) = \mu - \tilde{\eta} \cos \phi$ is positive on the electron arc, $\phi \in (\phi_b, 2\,\pi - \phi_b)$, and negative on the hole arc, $|\phi| < \phi_b$, with $\cos \phi_b = \mu/\tilde{\eta}$. The two pockets touch at $\phi = \pm \phi_b$. The Drude tensor therefore receives contributions from $s=1$ and $s=2$, and it must be evaluated with the corresponding roots of $\varepsilon_s (\bs k) = \mu$. The velocity enters the Drude weight quadratically, and hence the hole arc adds to the electron arc. We first drop the $\eta_x^2$ terms, which change the results by a relative amount of order $\eta_x^2$, and restore them below. Then $\kappa_F = |\mu_{\rm eff}|/v_0$, and the angular integrals of Appendix~\ref{app:horn} give
\begin{align}
\label{eq:horncyclide}
\rho (\mu) &= \frac{g\, k_0}{\pi^2\, v_0^2} 
\left[ \mu\, \arcsin \left(  {\mu}/{\tilde{\eta}} \right) + \bar{w} \right ], \quad
D_{zz} = \frac{g\, e^2\, k_0}{2\, \pi} \left( \mu\, \arcsin \left(  {\mu}/{\tilde{\eta}} \right) + \bar{w} \right) , \nn
D_{xx} &= \frac{g\, e^2\, k_0}{4\, \pi} \left[ \mu\, \arcsin \left(  {\mu}/{\tilde{\eta}} \right) + \frac{\bar{w} \left( 4\, \tilde{\eta}^2 - \mu^2 \right)}{3\, \tilde{\eta}^2} \right] , \quad
D_{yy} = \frac{g\, e^2\, k_0}{4\, \pi} \left[ \mu\, \arcsin \left(  {\mu}/{\tilde{\eta}} \right) + \frac{\bar{w} \left( 2\, \tilde{\eta}^2 + \mu^2 \right)}{3\, \tilde{\eta}^2} \right] ,
\end{align}
where $\bar{w} = \sqrt{\tilde{\eta}^2 - \mu^2}$. These coincide with the $\Delta \to 0$ limit of the corresponding regime of the tilted GNR in Sec.~\ref{secgapped}. The intraband polarisability follows from Eq.~\eqref{eqintra} as $\Pi_{\rm intra} (q\, \bs{\hat e}_i, \omega) \simeq q^2\, D_{ii}/(\pi\, e^2\, \omega^2)$, and the plasmon frequencies are $\omega_{p,i}^2 = 4\, D_{ii}$. The anisotropy ratio $D_{zz}/D_{xx}$ increases from $3/2$ at $\mu = 0$ to $2$ at $\mu = \tilde{\eta}$. The ratio $D_{yy}/D_{xx}$ increases from $1/2$ to $1$. Fig.~\ref{fig:drudex}(a) shows the three weights in units of $g\, e^2\, k_0\, \tilde{\eta}$ as functions of $\mu/\tilde{\eta}$. At $\mu = \tilde{\eta}$, the hole-pocket shrinks to a point, and Eq.~\eqref{eq:horncyclide} joins continuously onto the ring-cyclide values, $\rho = g\, k_0\, \mu/(2\, \pi\, v_0^2)$, $D_{zz} = g\, e^2\, k_0\, \mu/4$, and $D_{xx} = D_{yy} = g\, e^2\, k_0\, \mu/8$.

The $\eta_x^2$ corrections follow from the exact integrals over the two pockets, with the Jacobian $\kappa\, (k_0 + \kappa \cos \gamma)$, the factors $\zeta_s$, and the tilt term of the velocity retained, as shown in Appendix~\ref{app:horn}. The arcs bounded by $\phi_b$ are not modified, because $\zeta_s$ does not change the sign of $\kappa_s$. With $\varsigma = \bar{w}/\tilde{\eta}$ and $\Lambda = \left[ \bar{w} + \mu \arcsin (\mu/\tilde{\eta}) \right]/\tilde{\eta}$, we find
\begin{align}
\label{eq:horncorr}
&\rho (\mu) = \frac{g\, k_0\, \tilde{\eta}}{\pi^2\, v_0^2} \left[ \left( 1 + \frac{9\, \eta_x^2}{4} \right) \Lambda - \frac{\eta_x^2\, \varsigma^3}{4} \right] , \quad
D_{zz} = \frac{g\, e^2\, k_0\, \tilde{\eta}}{2\, \pi} \left[ \left( 1 + \frac{9\, \eta_x^2}{8} \right) \Lambda - \frac{\eta_x^2\, \varsigma^3}{8} \right] , \nn
& D_{xx} = \frac{g\, e^2\, k_0\, \tilde{\eta}}{4\, \pi} \left[ \Lambda + \frac{\varsigma^3}{3} + \eta_x^2 \left( \frac{17\, \Lambda}{16} - \frac{3\, \varsigma^3}{16} - \frac{3\, \varsigma^5}{40} \right) \right] , \quad
D_{yy} = \frac{g\, e^2\, k_0\, \tilde{\eta}}{4\, \pi} \left[ \Lambda - \frac{\varsigma^3}{3} + \eta_x^2 \left( \frac{27\, \Lambda}{16} - \frac{9\, \varsigma^3}{16} + \frac{3\, \varsigma^5}{40} \right) \right] .
\end{align}
At $\eta_x = 0$ these reduce to Eq.~\eqref{eq:horncyclide}. At $\mu = \tilde{\eta}$, where $\varsigma = 0$ and $\Lambda = \pi/2$, they reproduce the ring-cyclide corrections of Eqs.~\eqref{eq:dostiltx} and \eqref{eq:dzzx}. The relations $\Pi_{\rm intra} (q\, \bs{\hat e}_i, \omega) \simeq q^2\, D_{ii}/(\pi\, e^2\, \omega^2)$ and $\omega_{p,i}^2 = 4\, D_{ii}$ remain valid with these $D_{ii}$.

The interband response in this window follows from the same local mapping. For $v_0\, q \ll \omega$, Pauli blocking removes the transitions with $\omega < 2\, | \mu_{\rm eff} (\phi) |$. We again drop the tilt terms of relative order $\eta_x$ inside the local cone, and find
\begin{align}
\label{eq:imhorn}
\Im \Pi (q\, \bs{\hat z}, \omega) \simeq -\, \frac{g\, k_0\, q^2}{16\, \omega}\, F_0 (\omega) \,,
\quad
\Im \Pi (q\, \bs{\hat x}, \omega) \simeq -\, \frac{g\, k_0\, q^2}{16\, \omega}\, F_2 (\omega) \,, \nn
\Im \Pi (q\, \bs{\hat y}, \omega) \simeq -\, \frac{g\, k_0\, q^2}{16\, \omega} \left[ F_0 (\omega) - F_2 (\omega) \right] ,
\end{align}
where $F_n (\omega) = \int_0^{2\pi} \frac{d\phi}{2\pi}\, \cos^n \phi\, \Theta \left( \omega - 2\, | \mu - \tilde{\eta} \cos \phi | \right)$. The $\bs{\hat y}$ channel carries the weight $\sin^2 \phi = 1 - \cos^2 \phi$, because the local transfer is $q \sin\phi$ in that case. The functions $F_n$ are elementary,
\begin{align}
F_0 = \frac{ \arcsin b_\omega - \arcsin a_\omega }{ \pi } \,,
\quad
F_2 = \frac{ \arcsin b_\omega - \arcsin a_\omega + a_\omega \sqrt{ 1 - a_\omega^2 } - b_\omega \sqrt{ 1 - b_\omega^2 } }{ 2\, \pi } \,,
\end{align}
with $a_\omega = \max \left[ -1, (\mu - \omega/2)/\tilde{\eta} \right]$ and $b_\omega = \min \left[ 1, (\mu + \omega/2)/\tilde{\eta} \right]$. For $\omega > 2\, (\tilde{\eta} + \mu)$ they reduce to $F_0 = 1$ and $F_2 = 1/2$, which recovers the intrinsic-limit expressions. The horn-cyclide has no interband threshold, $\omega_{\rm th} = 0$, because the local Dirac point crosses the chemical potential at $\cos \phi_b = \mu/\tilde{\eta}$. At small frequencies $F_0 \simeq \omega/(\pi\, \bar{w})$, as shown in Appendix~\ref{app:horn}, and $\Im \Pi (q\, \bs{\hat z}, \omega)$ tends to the constant $-\, g\, k_0\, q^2/(16\, \pi\, \bar{w})$. The plasmons in this window are therefore always damped by interband processes.

The leading corrections in $\eta_x$ and $\omega/(v_0\, k_0)$ to Eq.~\eqref{eq:imhorn} are given in Eq.~\eqref{eq:imhorncorr} of Appendix~\ref{app:horn}. They vanish for $\omega > 2\, (\tilde{\eta} + \mu)$, and the $\bs{\hat x}$ channel then retains only the factor $1 + v_0\, \eta_x\, q/\omega$ found in the intrinsic limit. The term linear in $q$ is the only one odd in $\eta_x$. The factors $(1 - u^2)^{-3/2}$ in the coefficients show that the expansion fails in a narrow window around the thresholds $\omega = 2\, (\tilde{\eta} \pm \mu)$.

The Drude weights and the interband response combine into the dielectric function. For $v_0\, q \ll \omega$, Eqs.~\eqref{eq:epsrpa}, \eqref{eqintra}, and \eqref{eq:imhorn} give
\begin{align}
\label{eq:epshorn}
\varepsilon^{\rm RPA} (q\, \bs{\hat z}, \omega) &\simeq 1 -
\frac{4\, D_{zz}}{\omega^2} + i\, \frac{\pi\, g\, e^2\, k_0}{4\, \omega}\, F_0 (\omega) \,, \nn
\varepsilon^{\rm RPA} (q\, \bs{\hat x}, \omega) &\simeq
1 - \frac{4\, D_{xx}}{\omega^2} + i\, \frac{\pi\, g\, e^2\, k_0}{4\, \omega}\, F_2 (\omega) \,, \nn
\varepsilon^{\rm RPA} (q\, \bs{\hat y}, \omega) &\simeq
1 - \frac{4\, D_{yy}}{\omega^2} + i\, \frac{\pi\, g\, e^2\, k_0}{4\, \omega}\, \left[ F_0 (\omega) - F_2 (\omega) \right] .
\end{align}
Eq.~\eqref{eq:epshorn} is the analogue of Eq.~\eqref{eq:epsintrinsic} for the horn-cyclide, with the real part of the interband response left out. The real part follows from Eq.~\eqref{eq:pintertiltxptnr} and adds the term $\left( g\, e^2\, k_0/(4\, \omega) \right) R_i (\omega)$ to $\Re \varepsilon^{\rm RPA}$, where
\begin{align}
R_i (\omega) = \int_0^{2\pi} \frac{d\phi}{2\pi}\, \check w_i (\phi)\, \ln \!\left| \frac{\omega + 2\, |\mu_{\rm eff}|}{\omega - 2\, |\mu_{\rm eff}|} \right| .
\end{align}
The integrand has an integrable logarithmic singularity at $\phi = \pm \phi_b$, where the local Dirac point sits at the chemical potential. In contrast to the ring-cyclide, the interband part is therefore not a constant background. For $\omega \ll \tilde{\eta}$ the two crossings give $R_z \simeq 2\, \omega \left[ \ln \!\left( \tilde{\eta}/\omega \right) + c \right]/(\pi\, \bar{w})$, with a real constant $c$ of order unity, as derived in Appendix~\ref{app:horn}. Together with $F_0 \simeq \omega/(\pi\, \bar{w})$ this yields
\begin{align}
\label{eq:epsinterhorn}
\varepsilon^{\rm RPA}_{\rm inter} (q\, \bs{\hat z}, \omega) 
\simeq \frac{g\, e^2\, k_0}{2\, \pi\, \bar{w}} 
\left[ \ln \!\left( \tilde{\eta}/ \omega \right)+ c + i\, \frac{\pi}{2} \right] ,
\end{align}
and the $\bs{\hat x}$ and $\bs{\hat y}$ channels carry the extra factors $\mu^2/\tilde{\eta}^2$ and $\bar{w}^2/\tilde{\eta}^2$, that is, the weights at $\phi_b$. This agrees with the logarithm found for the small-gap limit of the GNR in Sec.~\ref{secgapped}, with $\Delta$ replaced by $\omega$. The Drude term supplies a zero of $\Re \varepsilon^{\rm RPA}$ at $\tilde\omega_{p,i}^2 = \omega_{p,i}^2/\epsilon_{b,i} (\omega_{p,i})$, where $\epsilon_{b,i} (\omega) = 1 + g\, e^2\, k_0\, R_i (\omega)/(4\, \omega)$. Since $F_0 (\omega)$, $F_2 (\omega)$, and $F_0 (\omega) - F_2 (\omega)$ are positive for all $\omega > 0$, the imaginary part never vanishes. The mode is therefore always damped. Neglecting the logarithmic derivative of $\epsilon_{b,i}$, its damping rate is $\Gamma_i = \left( \pi\, g\, e^2\, k_0/(8\, \epsilon_{b,i}) \right) F_n (\omega_{p,i})$, where $F_n$ is replaced by $F_0$ for $i = z$, by $F_2$ for $i = x$, and by $F_0 - F_2$ for $i = y$.

At the symmetric point $\mu = 0$, we have $\phi_b = \pi/2$, $\bar{w} = \tilde{\eta}$, and $\varsigma = \Lambda = 1$. The FS is the symmetric horn-cyclide, which consists of an electron-pocket and a hole-pocket of equal size that touch at $\phi = \pm \pi/2$, as shown in the last panel of Fig.~\ref{figfs}. Eq.~\eqref{eq:horncyclide} then gives a finite Drude weight and a finite DOS, $\rho = g\, k_0\, \tilde{\eta}/(\pi^2\, v_0^2)$, $D_{zz} = g\, e^2\, k_0\, \tilde{\eta}/(2\, \pi)$, $D_{xx} = g\, e^2\, k_0\, \tilde{\eta}/(3\, \pi)$, and $D_{yy} = g\, e^2\, k_0\, \tilde{\eta}/(6\, \pi)$. This is in contrast to the untilted PTNR at $\mu = 0$, which has neither. Eq.~\eqref{eq:horncorr} multiplies these quantities by $1 + 2\, \eta_x^2$ for $\rho$, $1 + \eta_x^2$ for $D_{zz}$, $1 + 3\, \eta_x^2/5$ for $D_{xx}$, and $1 + 9\, \eta_x^2/5$ for $D_{yy}$.

\begin{figure}[t!]
\centering
\includegraphics[width=0.7\textwidth]{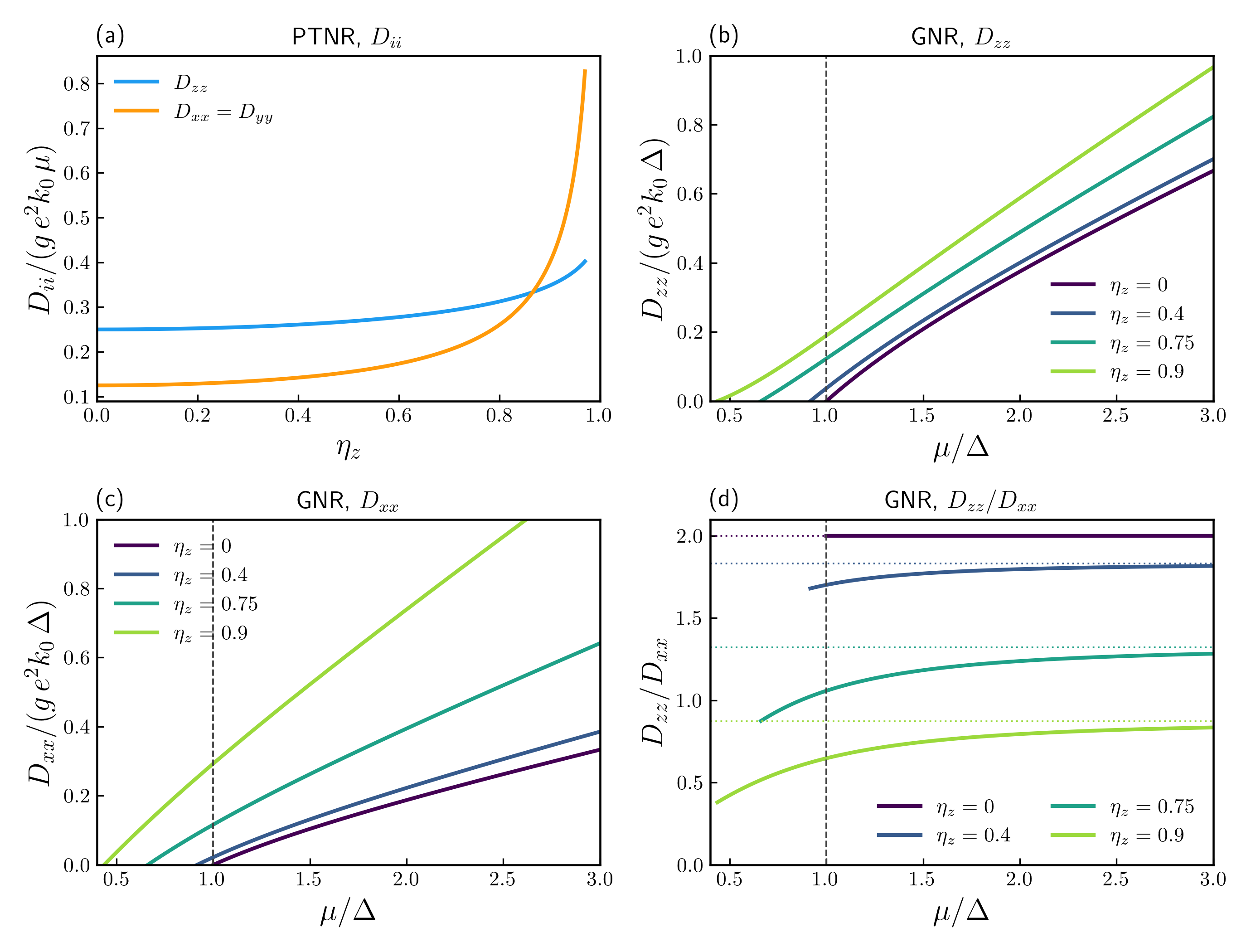}
\caption{\label{fig:drude} Drude weights of the axially-tilted PTNR and GNR, from Eqs.~\eqref{eq:dzzz} and \eqref{eq:dtiltgapped}. Panel (a) shows $D_{zz}$ and $D_{xx} = D_{yy}$ of the PTNRs functions of $\eta_z$. Panels (b) and (c) show $D_{zz}$ and $D_{xx}$ of the GNR as functions of $\mu/\Delta$, for $\eta_z = 0$, $0.4$, $0.75$, and $0.9$. Each curve starts at $\mu = \Delta_{\rm eff}$, and the dashed vertical line marks $\mu = \Delta$. Panel (d) shows the anisotropy ratio, $D_{zz}/D_{xx}$. The dotted horizontal lines there correspond to $2\, \sqrt{1 - \eta_z^2}$, which are the PTNR-case values.}
\end{figure}

The interband response follows from Eq.~\eqref{eq:imhorn} with $\mu = 0$, for which $a_\omega = -\, b_\omega$. Then $F_0 = (2/\pi)\, \arcsin b_\omega$ and $F_2 = \left[ \arcsin b_\omega - b_\omega \sqrt{1 - b_\omega^2} \right]/\pi$, with $b_\omega = \min \left[ 1, \omega/(2\, \tilde{\eta}) \right]$. Both approach their intrinsic values for $\omega > 2\, \tilde{\eta}$. The analogue of Eq.~\eqref{eq:epsintrinsic} at this point is Eq.~\eqref{eq:epshorn} with $\mu = 0$, namely
\begin{align}
& \varepsilon^{\rm RPA} (q\, \bs{\hat z}, \omega) \simeq 1 
- \frac{2\, g\, e^2\, k_0\, \tilde{\eta}}{\pi\, \omega^2}
 + i\, \frac{g\, e^2\, k_0}{2\, \omega}\, \arcsin b_\omega \,, \nn &
\varepsilon^{\rm RPA} (q\, \bs{\hat x}, \omega) \simeq 1 -
\frac{4\, g\, e^2\, k_0\, \tilde{\eta}}{3\, \pi\, \omega^2}
+ i\, \frac{g\, e^2\, k_0}{4\, \omega}
\left ( \arcsin b_\omega - b_\omega \,\sqrt{1 - b_\omega^2} \right) , \nn &
\varepsilon^{\rm RPA} (q\, \bs{\hat y}, \omega) \simeq 1 -
\frac{2\, g\, e^2\, k_0\, \tilde{\eta}}{3\, \pi\, \omega^2}
+ i\, \frac{g\, e^2\, k_0}{4\, \omega}
\left ( \arcsin b_\omega + b_\omega \,\sqrt{1 - b_\omega^2} \right) .
\end{align}
In contrast with the untilted case, $\Re \varepsilon^{\rm RPA}$ now has a zero, at $\omega_{p,z}^2 = 2\, g\, e^2\, k_0\, \tilde{\eta}/\pi$, $\omega_{p,x}^2 = 4\, g\, e^2\, k_0\, \tilde{\eta}/(3\, \pi)$, and $\omega_{p,y}^2 = 2\, g\, e^2\, k_0\, \tilde{\eta}/(3\, \pi)$, to leading order in $\eta_x$. For $g\, e^2\, k_0 \ll \tilde{\eta}$ these lie well below $2\, \tilde{\eta}$, where $\arcsin b_\omega \simeq \omega/(2\, \tilde{\eta})$. The damping rate of the $\bs{\hat z}$ mode then reduces to $\Gamma_z = g\, e^2\, k_0\, \omega_{p,z}/(8\, \tilde{\eta})$, and the ratio $\Gamma_z/\omega_{p,z} = g\, e^2\, k_0/(8\, \tilde{\eta})$ is small.

The real part of the interband response is finite along $\bs{\hat x}$ at this point, because the weight $\cos^2 \phi$ vanishes at the touching points $\phi = \pm \pi/2$. Eq.~\eqref{eq:pintertiltxptnr} gives $\epsilon_{b,x} = 1 + g\, e^2\, k_0/(2\, \pi\, \tilde{\eta})$ in the static limit, and $\omega_{p,x}^2$ is reduced by this factor. Along $\bs{\hat z}$ the real part is logarithmically enhanced, as in Eq.~\eqref{eq:epsinterhorn} with $\bar{w} = \tilde{\eta}$. The same holds along $\bs{\hat y}$, because the weight $\sin^2 \phi$ equals unity at the touching points. It is small compared with unity only for $g\, e^2\, k_0\, \ln \!\left( \tilde{\eta}/\omega_{p,z} \right) \ll \tilde{\eta}$ along $\bs{\hat z}$, and for $g\, e^2\, k_0\, \ln \!\left( \tilde{\eta}/\omega_{p,y} \right) \ll \tilde{\eta}$ along $\bs{\hat y}$.

Figs.~\ref{fig:plasmax}(a)--\ref{fig:plasmax}(c) compare the bare and screened plasmon frequencies of the PTNR with the interband onset for the parameters of Table~\ref{tab-params} and $\tilde{\eta} = 0.03$ eV. As for the axial tilt, the bare frequencies lie far above the onset for the strong coupling of Table~\ref{tab-params}, and the interband screening lowers them substantially. Without tilt, the screened frequencies stay below the onset, and the plasmon is undamped. With tilt, the screened frequency lies above the onset wherever the onset exists, and the plasmon is therefore damped for all $\mu$. This is illustrated by the curves in Figs.~\ref{fig:plasmax}(a)--\ref{fig:plasmax}(c).

\subsection{Tilt with respect to the \texorpdfstring{$ k_z $}{kz}-axis}
\label{sectiltz}

We next consider the axial tilt $\bs \eta = \eta_z\, \bs{\hat z}$. The tilt term becomes $v_0\, \eta_z\, k_z = v_0\, \eta_z\, \kappa \sin \gamma$, and the eigenvalues are $\varepsilon_s (\bs k) = v_0\, \kappa\, \left[ (-1)^s + \eta_z \sin \gamma \right]$. For $|\eta_z| < 1$, we have $\varepsilon_1 < 0 < \varepsilon_2$ for all allowed values of $\gamma$. The FS is rotationally symmetric about the $ k_z $-axis.

\begin{figure}[t!]
\centering
\includegraphics[width=0.7\textwidth]{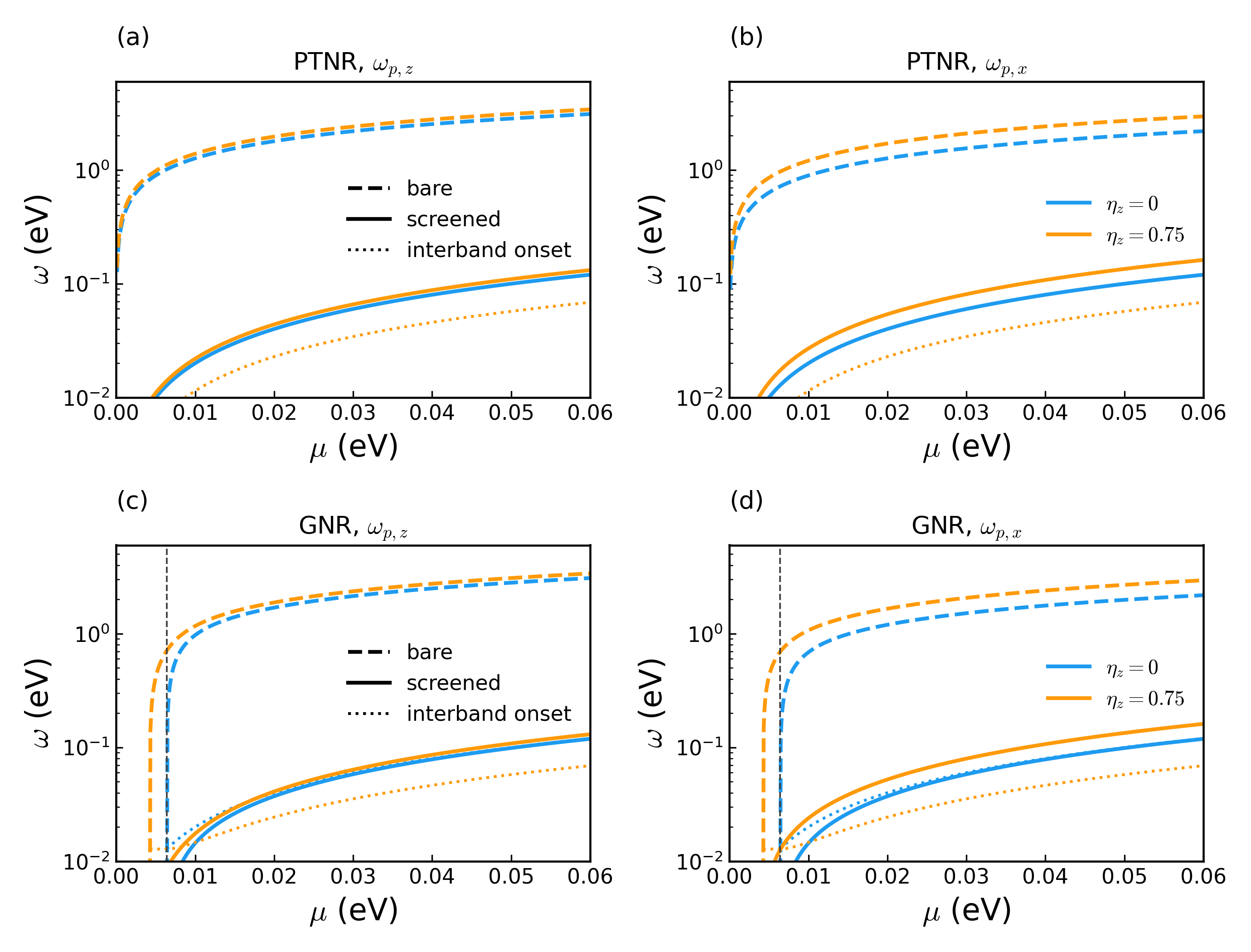}
\caption{\label{fig:plasma} Plasmon frequencies of the axially-tilted PTNR and GNR for $\eta_z = 0$ (blue) and $\eta_z = 0.75$ (orange), for the parameters of Table~\ref{tab-params}. Panels (a) and (b) show $\omega_{p, z}$ and $\omega_{p, x}$ of the PTNR as functions of $\mu$, and panels (c) and (d) show those of the GNR. The dashed curves are the bare frequencies $\omega_{p, i}$ of Eq.~\eqref{eq:wpgappedztilt}, and the PTNR curves are its $\Delta \to 0$ limit. The solid curves are the screened frequencies $\tilde\omega_{p, i}$ obtained with Eq.~\eqref{eq:epsb-ztilt}. The dotted curves are the interband onset $\omega_{\rm th}$. The dashed vertical line in (c) and (d) marks $\mu = \Delta$. The vertical axes are logarithmic.}
\end{figure}

At the intrinsic limit, $\mu = 0$, the FS coincides with the nodal ring. In the local plane $(u, w) = (\kappa \cos \gamma, k_z)$ at fixed $\phi$, the tilt is along $w$ and produces no energy offset, since $k_z = w$. The tilt only shifts the interband transition energy by $v_0\, \eta_z\, q_z$. For $\bs q = q\, \bs{\hat x} $ the momentum transfer has no $z$-component and the shift vanishes. The result is then that of the untilted ring, including its factor of $1/2$ relative to the $\bs{\hat z}$ channel. For $\bs q = q\, \bs{\hat z} $ the momentum transfer lies along the tilt, and the resonance condition becomes $ v_0\, \left( \kappa + \kappa' \right) = \omega - v_0\, \eta_z\, q $. The phase space is that of the untilted ring at the shifted frequency $\omega - v_0\, \eta_z\, q $. For $\omega \gg v_0\, q $, this gives
\begin{align}
\Im \Pi (q\, \bs{\hat z}, \omega) &= -\, \frac{g\, k_0\, q^2}
{16\, \sqrt{ \left( \omega - v_0\, \eta_z\, q \right)^2 - v_0^2\, q^2 }} 
\simeq -\, \frac{g\, k_0\, q^2}{16\, \omega}\, 
\left( 1 + \frac{ v_0\, \eta_z\, q }{ \omega } \right) , \quad
\Im \Pi (q\, \bs{\hat x}, \omega) \simeq -\, \frac{g\, k_0\, q^2}{32\, \omega} \,.
\end{align}
The correction along $\bs{\hat z}$ is odd in $\eta_z$ and of first order in $v_0\, \eta_z\, q/\omega $, which is small in the regime $\omega \gg v_0\, q $. The frequency shift and this expansion are derived in Appendix~\ref{app:tiltshift}. The axial tilt also moves the onset of Landau damping along $\bs{\hat z}$ to $\omega = v_0\, q\, \left( 1 + \eta_z \right)$.
Below this onset $\Pi (q\, \bs{\hat z}, \omega)$ is real, because the same frequency shift applies to Eq.~\eqref{eq:polmu03dexact}. The analogue of Eq.~\eqref{eq:epsintrinsic} is therefore
\begin{align}
\label{eq:epsintrinsicztilt}
\varepsilon^{\rm RPA} (q\, \bs{\hat z}, \omega) = 1 + \frac{\pi\, g\, e^2\, k_0}{4 \sqrt{v_0^2\, q^2 - \left( \omega - v_0\, \eta_z\, q \right)^2}}
\text{ for } 0 \le \omega < v_0\, q\, \left( 1 + \eta_z \right) \,.
\end{align}
For $|\eta_z| < 1$ the argument of the square root is positive throughout this window, and the second term is positive as well. Hence $\varepsilon^{\rm RPA} > 1$ has no zero, and the axially-tilted PTNR at $\mu = 0$ supports no undamped plasmon, as in the untilted case. Unlike the in-plane tilt, the axial tilt creates no pockets at $\mu = 0$ and gives no Drude weight. For $\bs q = q\, \bs{\hat x}$ the shift vanishes, the continuum reaches down to $\omega = 0$, and $\varepsilon^{\rm RPA}$ is complex at all frequencies.

For low doping, $0 < \mu \ll v_0\, k_0 $, only the $ s=2 $ band crosses the Fermi level. The Fermi momentum and the DOS are
\begin{align}
\label{eq:kappaFztilt}
\kappa_F (\gamma) = \frac{ \mu }{ v_0\, \left( 1 + \eta_z \sin \gamma \right) } \text{ and }
\rho (\mu) = \frac{ g\, k_0\, \mu }{ 2\, \pi\, v_0^2 \, (1 - \eta_z^2)^{3/2} } \,,
\end{align}
respectively. The DOS is even in $\eta_z$, increases with $|\eta_z|$, and diverges as $|\eta_z| \to 1 $. Setting $\Delta = 0$ and $\bs \eta = \eta_z\, \bs{\hat z}$ in Eq.~\eqref{eq:velocity3d}, the components of the $s=2$ band-velocity are $V_x^{(2)} = v_0\, \cos \gamma \cos \phi$, $V_y^{(2)} = v_0\, \cos \gamma \sin \phi$, and $V_z^{(2)} = v_0\, \left( \sin \gamma + \eta_z \right)$. The Drude-weight components evaluate to
\begin{align}
\label{eq:dzzz}
D_{zz} = \frac{ g\, e^2\, k_0\, \mu }{ 2\,  \eta_z^2 } 
\left( 1 - \sqrt{1 - \eta_z^2} \right) ,\quad
D_{xx} = D_{yy} = \frac{ g\, e^2\, k_0\, \mu }{ 4\, \eta_z^2 } 
\left( \frac{ 1 }{ \sqrt{1 - \eta_z^2} } - 1 \right ).
\end{align}
For $\eta_z \to 0$, these reduce to the untilted results $g\, e^2\, k_0\, \mu /4$ and $g\, e^2\, k_0\, \mu /8$. The anisotropy ratio is $ D_{zz} / D_{xx} = 2\, \sqrt{1 - \eta_z^2} $. The steps behind Eq.~\eqref{eq:dzzz} follow from the $\Delta \to 0$ limit of Appendix~\ref{app:gnrztilt}.
Using Eq.~\eqref{eqintra}, we get
\begin{align}
\Pi_{\rm intra} (q\, \bs{\hat z}, \omega) \simeq \frac{g\, k_0\, \mu\, q^2}{2\, \pi\, \eta_z^2\, \omega^2}\, \left( 1 - \sqrt{1 - \eta_z^2} \right) , \quad
\Pi_{\rm intra} (q\, \bs{\hat x}, \omega) \simeq \frac{g\, k_0\, \mu\, q^2}{4\, \pi\, \eta_z^2\, \omega^2}\, \left( \frac{ 1 }{ \sqrt{1 - \eta_z^2} } - 1 \right) .
\end{align}
The plasmon dispersion is given by
\begin{align}
\omega_{p,z}^2 = \frac{ 2\, g\, e^2\, k_0\, \mu }{ \eta_z^2 } 
\left ( 1 - \sqrt{1 - \eta_z^2} \right ) ,
\quad
\omega_{p,x}^2 = \omega_{p,y}^2 = \frac{ g\, e^2\, k_0\, \mu }{ \eta_z^2 } 
\left ( \frac{ 1 }{ \sqrt{1 - \eta_z^2} } - 1 \right) ,\quad
\frac{ \omega_{p,z} }{ \omega_{p,x} } = \sqrt{2\,\sqrt{1 - \eta_z^2}}\,.
\end{align}
The axial tilt enhances both Drude weights and hence both plasmon frequencies, as shown in Fig.~\ref{fig:drude}(a) for the prefactors of the weights. The anisotropy ratio decreases from its untilted value $\sqrt{2}$ at quadratic order in $\eta_z$. The ratio $D_{zz}/D_{xx}$ reaches unity at $\eta_z = \sqrt{3}/2$, beyond which $D_{xx}$ exceeds $D_{zz}$.

The dielectric function follows from Eqs.~\eqref{eq:epsrpa} and \eqref{eqintra}. For $v_0\, q \ll \omega$ it reads $\varepsilon^{\rm RPA} (q\, \bs{\hat e}_i, \omega) \simeq 1 - \omega_{p,i}^2/\omega^2$, with its zero at the plasmon frequencies above. At $q \to 0$ the interband transitions at wavevector $\bs k$ have the energy $2\, v_0\, \kappa$, independent of the tilt, and they are Pauli-blocked for $\kappa < \kappa_F (\gamma)$. By Eq.~\eqref{eq:kappaFztilt} the interband onset therefore lies at $\omega_{\rm th} = 2\, \mu/\left( 1 + |\eta_z| \right)$. The plasmon is free of interband damping, up to corrections of order $v_0\, q$, if $\omega_{p,i} < \omega_{\rm th}$.

The static real part follows from the $\Delta \to 0$ limit of Eq.~\eqref{eq:epsb-ztilt}, and it is independent of the tilt, $\epsilon_{b,z} = 1 + g\, e^2\, k_0/(4\, \mu)$ and $\epsilon_{b,x} = \epsilon_{b,y} = 1 + g\, e^2\, k_0/(8\, \mu)$. The reason is that on the FS $E_F (\gamma) = \mu/\left( 1 + \eta_z \sin \gamma \right)$, and the term odd in $\sin \gamma$ drops out of the $\gamma$-average of $1/E_F$.

Figs.~\ref{fig:plasma}(a) and \ref{fig:plasma}(b) compare the bare and screened plasmon frequencies of the PTNR with the interband onset $\omega_{\rm th}$ for the parameters of Table~\ref{tab-params}. For $g\, e^2\, k_0 \gg \mu$, the static background dielectric constants $\epsilon_{b,i}$ are much larger than unity, as seen from the expressions above. The interband screening is therefore essential. It lowers the frequencies more strongly along $\bs{\hat z}$ than along $\bs{\hat x}$, because $\epsilon_{b, z} > \epsilon_{b, x}$. Without tilt, the screened frequencies stay below the onset, which they approach closely, and the plasmon is undamped. For the tilt shown by the orange curves, the screened frequencies lie above the onset throughout, and the plasmon is damped. The untilted (blue) and tilted (orange) cases are shown in Figs.~\ref{fig:plasma}(a) and \ref{fig:plasma}(b). The plotted range of $\mu$ is limited by the condition $\mu \ll v_0\, k_0$ of the low-doping description.

\begin{figure}[t!]
\centering
\includegraphics[width=0.7\textwidth]{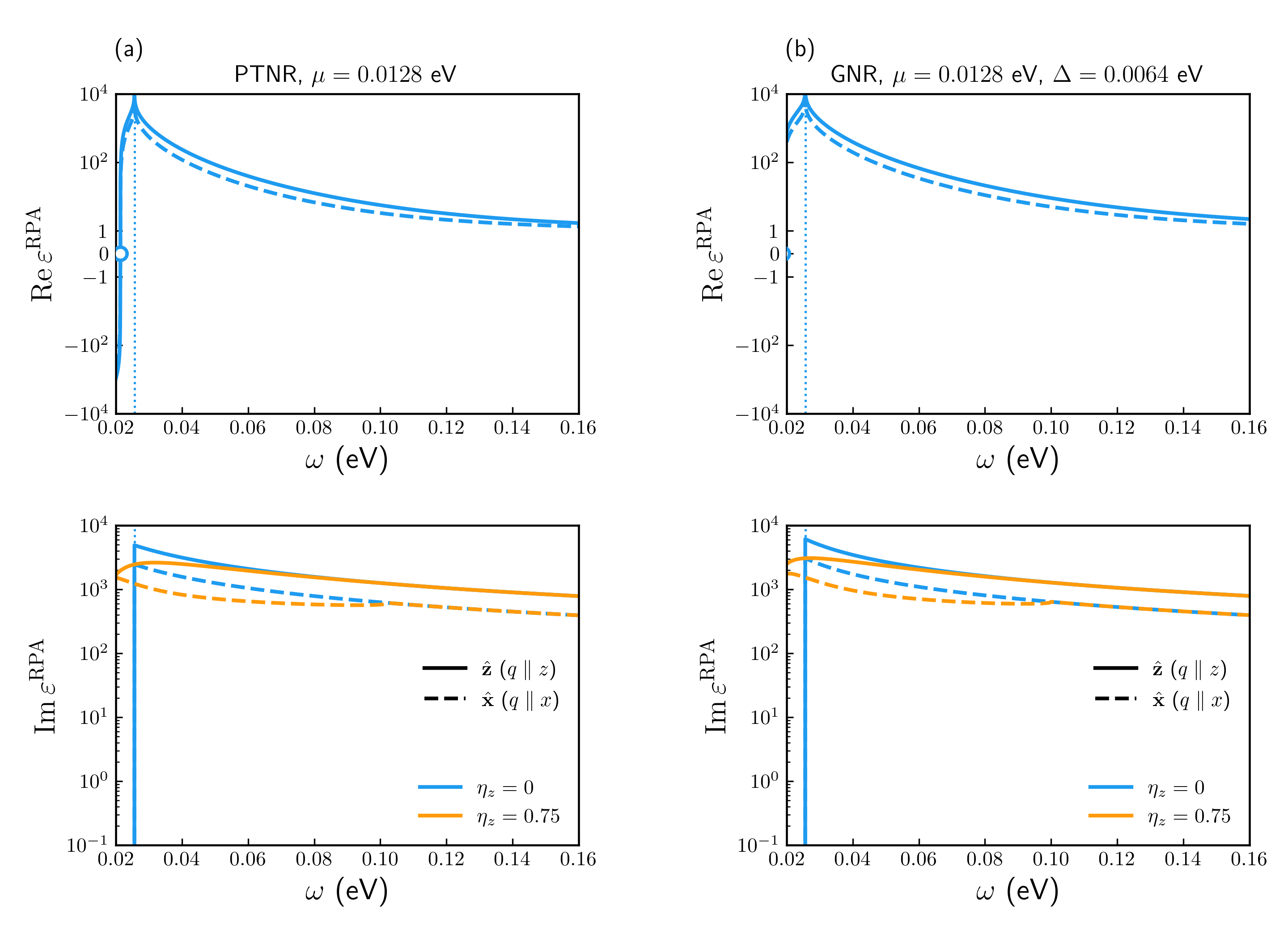}
\caption{\label{fig:eps} RPA dielectric function of the axially-tilted (a) PTNR and (b) GNR for $\eta_z = 0$ (blue) and $\eta_z = 0.75$ (orange). The other parameters are taken from Table~\ref{tab-params}. Upper panels: $\Re \varepsilon^{\rm RPA}$ (symmetric-logarithmic, linear for $|\Re \varepsilon^{\rm RPA}| < 1$); lower panels: $\Im \varepsilon^{\rm RPA}$. Solid and dashed curves are for $\bs q \parallel \bs{\hat z}$ and $\bs q \parallel \bs{\hat x}$, respectively. Dotted vertical lines mark the interband onset $\omega_{\rm th}$, and open circles mark the screened plasmons. The tilted $\Re \varepsilon^{\rm RPA}$ is shown only below $\omega_{\rm th}$, where the closed-form results hold. All curves are for $\omega \gg v_0\, q$.}
\end{figure}

The frequency dependence of the real part follows from the radial integral over $\kappa > \kappa_F (\gamma)$, which is elementary. The derivation of the following results is given in Appendix~\ref{app:interband}. With $A_\pm = 2\, \mu \pm \omega$, $r_\pm = \sqrt{A_\pm^2 - \eta_z^2\, \omega^2}$, and $\varrho_\pm = |\eta_z|\, \omega/(A_\pm + r_\pm)$, for $\omega < 2\, \mu/(1 + |\eta_z|)$, we obtain
\begin{align}
\label{eq:pinterztiltptnr}
& \Pi_{\rm inter} (q\, \bs{\hat z}, \omega) = -\, \frac{g\, k_0\, q^2}{16\, \pi\, \omega} 
\left[ \ln \!\left( \frac{A_+ + r_+}{A_- + r_-} \right) + \frac{\varrho_+^2 - \varrho_-^2}{2} \right] , \quad
\Pi_{\rm inter} (q\, \bs{\hat x}, \omega) = -\, \frac{g\, k_0\, q^2}{32\, \pi\, \omega} 
\left[ \ln \!\left( \frac{A_+ + r_+}{A_- + r_-} \right) - \frac{\varrho_+^2 - \varrho_-^2}{2} \right] ,\nn
& \Pi_{\rm inter} (q\, \bs{\hat y}, \omega) = \Pi_{\rm inter} (q\, \bs{\hat x}, \omega)\,,
\quad \Pi_{\rm inter} (q\, \bs{\hat y}, \omega) = \Pi_{\rm inter} (q\, \bs{\hat x}, \omega)\,.
\end{align}
The angular integrals are done with the Fourier series of $\ln \!\left( A + B \sin\gamma \right)$. For $\eta_z \to 0$ these reduce to Eq.~\eqref{eq:pinterdopedptnr}. At $\omega \to 0$ they give $-\, g\, k_0\, q^2/(16\, \pi\, \mu)$, independent of $\eta_z$, in agreement with the static constants above. The real part of the dielectric function is $\Re \varepsilon^{\rm RPA} = 1 - \omega_{p,i}^2/\omega^2 - 4\, \pi\, e^2\, \Pi_{\rm inter}/q^2$. By Appendix~\ref{app:monotone}, the zero is unique and the static screened frequency is an upper bound of it. At the onset $r_- \to 0$, and $\Re \varepsilon^{\rm RPA}$ stays finite. In the $\bs{\hat x}$ and $\bs{\hat y}$ channels it approaches this value with an infinite slope, through a $\sqrt{\omega_{\rm th} - \omega}$ term, whereas the $\bs{\hat z}$ channel has the weaker exponent $3/2$. The coefficients are given in Appendix~\ref{app:onsetz}. Above the onset the imaginary part follows from the delta function $\delta (\omega - 2\, E_k)$ restricted to the unblocked arc of the circle $E_k = \omega/2$. It reads
\begin{align}
\Im \varepsilon^{\rm RPA} (q\, \bs{\hat z}, \omega) = \frac{\pi\, g\, e^2\, k_0}{4\, \omega}\, 
\mathcal{G}_+ (\tau) \,, \quad
\Im \varepsilon^{\rm RPA} (q\, \bs{\hat x}, \omega) = \frac{\pi\, g\, e^2\, k_0}{8\, \omega}\, 
\mathcal{G}_- (\tau) \,, \quad
 \tau = \frac{ 2\, \mu/\omega - 1 } {|\eta_z| }\,,
\end{align}
with 
\begin{align}
\mathcal{G}_\pm (\tau) =
\begin{cases}
1/2 - \left[ \arcsin \tau \pm \tau \sqrt{1 - \tau^2} \right]/\pi & \text{ for }
|\tau| < 1 \\
0 & \text{ for }  \tau  \geq 1 \\
1 & \text{ for } \tau  \leq -1 
\end{cases}.
\end{align}
The condition $ \tau  = 1$ reproduces the onset $2\, \mu/(1 + |\eta_z|)$, and for $ \tau \leq -1$ we recover the untilted values.

Fig.~\ref{fig:eps}(a) shows the dielectric function of the PTNR at $\mu = 0.04$ eV. Without tilt, $\Im \varepsilon^{\rm RPA}$ jumps at $\omega = 2\, \mu$, and $\Re \varepsilon^{\rm RPA}$ has a zero below the onset, which is the screened plasmon. The axial tilt moves the onset down to $\omega_{\rm th}$ and replaces the jump by a continuous rise. For the tilt shown by the orange curves in Fig.~\ref{fig:eps}(a), the onset falls below the untilted plasmon. $\Re \varepsilon^{\rm RPA}$ is then negative throughout the window $\omega < \omega_{\rm th}$ and has no zero. Hence no undamped plasmon exists at this $\mu$.

\section{GNR}
\label{secgapped}

In this section we study the GNR, keeping $\Delta $ nonzero in the Hamiltonian of Eq.~\eqref{eq:ham3d}. We take $\mu\geq0$, since all results are even in $\mu$, and we consider $\mu \ll v_0\,k_0$ together with $\Delta \ll v_0\,k_0$. The second condition keeps the linearised Hamiltonian valid for $\mu < \Delta$ as well as for $\mu > \Delta$. For $\mu > \Delta$, the FS is a torus of minor radius $\kappa_F = \sqrt{\mu^2-\Delta^2}/v_0$. For $\mu < \Delta$, the untilted GNR has no FS. The essential simplification is the same as in Sec.~\ref{secungapped}: at each fixed toroidal angle $\phi$ the low-energy Hamiltonian reduces to a 2d massive Dirac cone with mass $\Delta$ in the local cross-sectional plane spanned by $u = k_\perp - k_0$ and $w = k_z$. The ring then supplies an overall multiplicity $k_0$ through the Jacobian weight, exactly as in Sec.~\ref{secuntiltednr}. It is exact to leading order in $q/k_0$ and lets us import known results for gapped graphene~\cite{pyatkovskiy-gapped-graphene}, adapted to the toroidal geometry. We first treat the untilted case, where the residual rotational symmetry about $k_z$ keeps the response isotropic in the cross-section. We then add an in-plane tilt with respect to $k_x$-axis and an axial with respect to $k_z$-axis. The results are summarised in Tables~\ref{tab:plasmon-summary} and \ref{tab:anisotropy-onsets}.

\begin{figure}[t!]
\centering
\subfigure[]{\includegraphics[width=0.42 \textwidth]{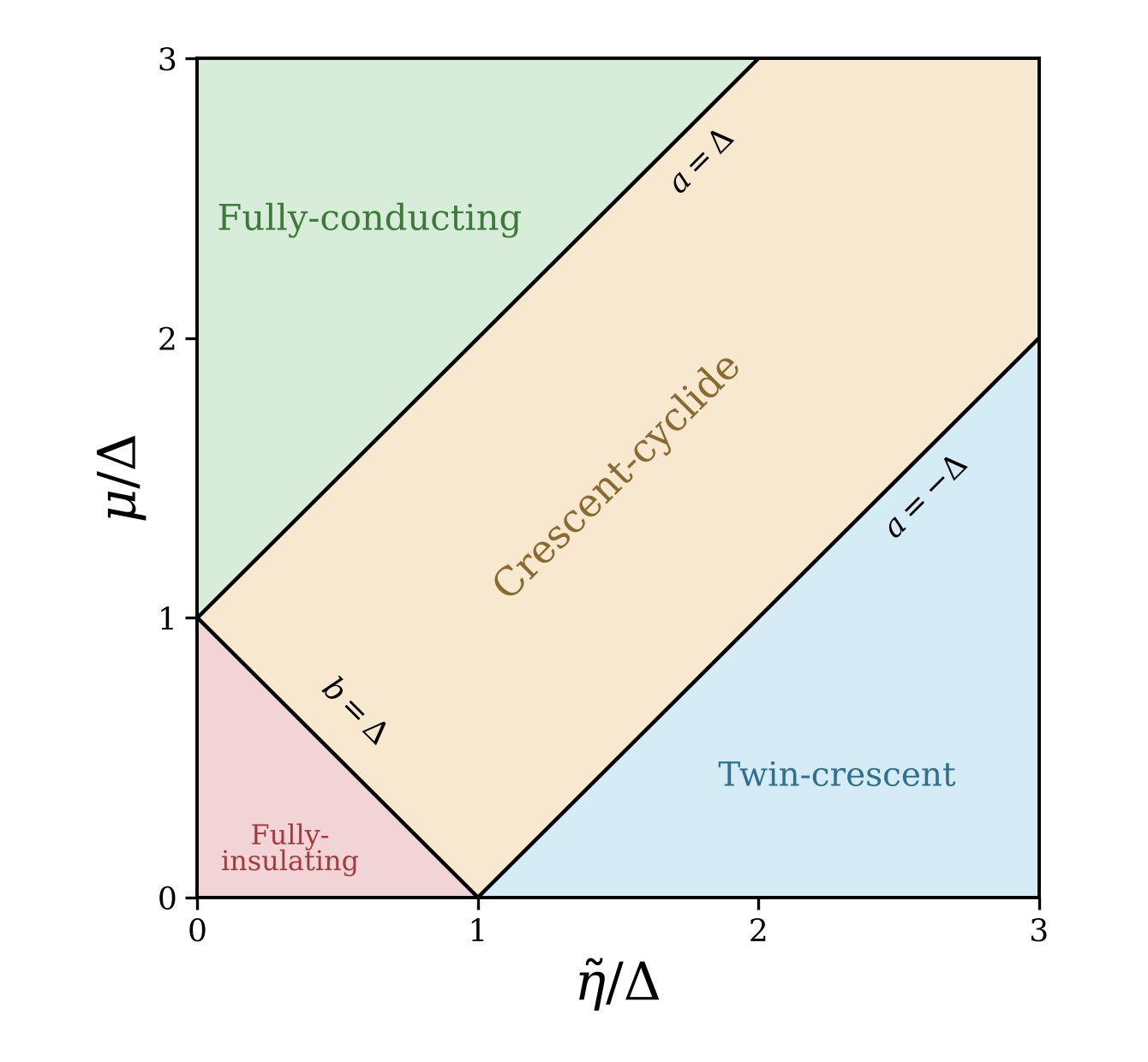}}\quad
\subfigure[]{\includegraphics[width=0.4\textwidth]{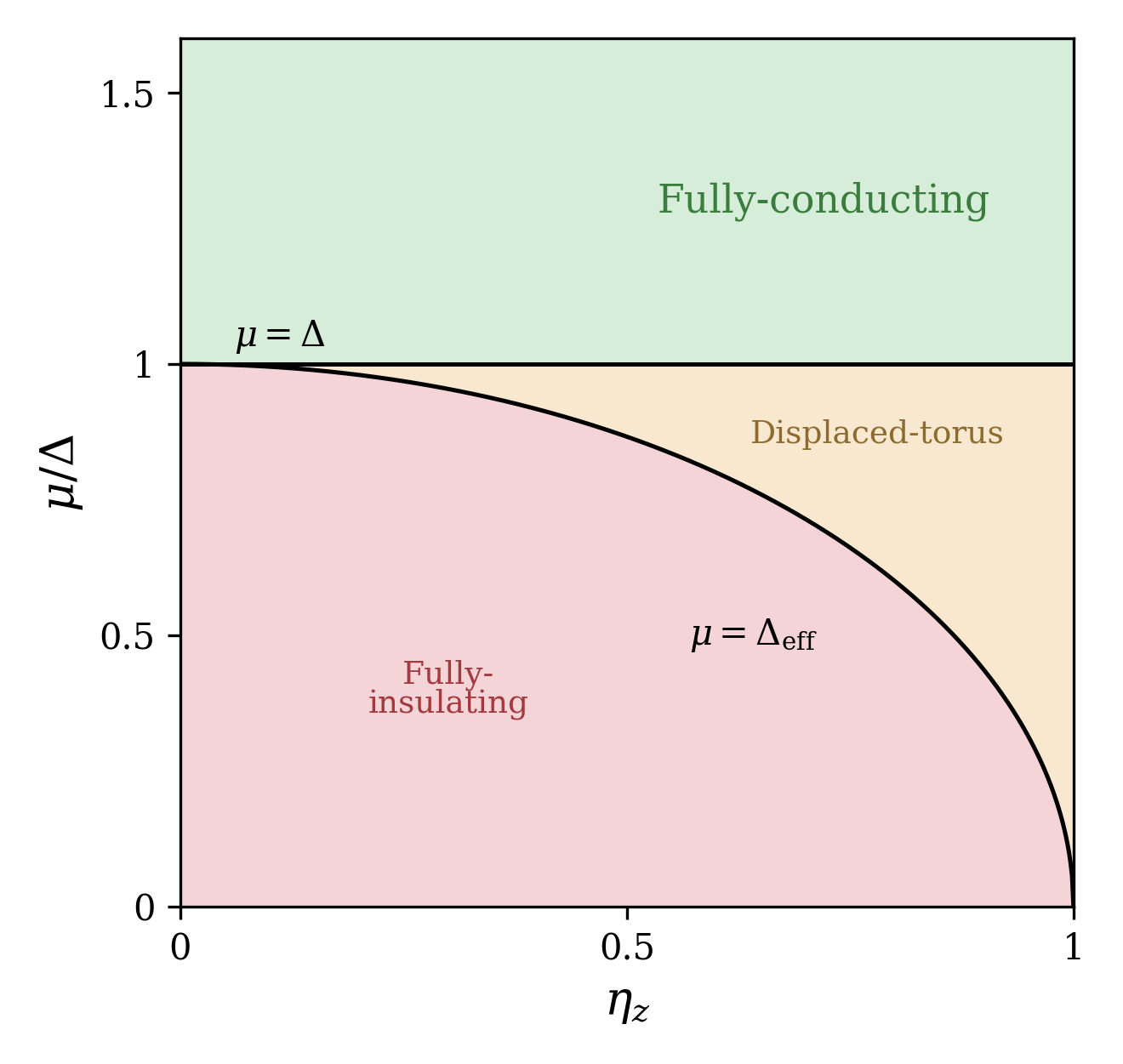}}
\caption{Phase diagrams of the tilted GNR. Panel (a) is for the in-plane tilt, in the $\mu/\Delta$-$\tilde{\eta}/\Delta$-plane. To leading order in $\eta_x$, the boundary lines $b=\Delta$, $a=-\Delta$, and $a=\Delta$, with $a$ and $b$ defined in Table~\ref{tab:fsclass}, delineate the fully-insulating, twin-crescent, crescent-cyclide, and fully-conducting regimes. Panel (b) is for the axial tilt, in the $\eta_z$-$\mu/\Delta$-plane. The curve $\mu=\Delta_{\rm eff}$ and the line $\mu=\Delta$ delineate the fully-insulating, displaced-torus, and fully-conducting regimes. Only $\mu \ge 0$ is shown in both panels, together with $\tilde{\eta} \ge 0$ in (a) and $0 \le \eta_z < 1$ in (b). The regimes for $\mu < 0$ follow by exchanging electrons and holes, as explained in Sec.~\ref{secfermisurface}. Those for $\tilde{\eta} < 0$ in (a) follow from $\tilde{\eta} \to |\tilde{\eta}|$, as explained in Sec.~\ref{sectiltx}, and those for $\eta_z < 0$ in (b) follow from the evenness of all results in $\eta_z$. Reflecting each diagram about both axes therefore gives the full phase diagram.\label{fig:phase}}
\end{figure}

\subsection{Untilted GNR}
\label{secgapnotilt}

We start with an untilted GNR. For $\mu < \Delta$, which includes the intrinsic limit $\mu=0$, the system is a band insulator and there is no FS. The linearisation about the nodal ring nevertheless requires $\Delta \ll v_0\,k_0$ here, because no $\kappa_F$ is available to bound the relevant momenta. The static polarisability is dominated by interband transitions. Equation~\eqref{eq:ringmap} applies verbatim with $\Pi_{2\rm D}$ replaced by the polarisability of a gapped 2d Dirac cone~\cite{pyatkovskiy-gapped-graphene}, whose static limit is $\Pi_{2\rm D} (q, 0) = -\, g\, q^2/(12\, \pi\, \Delta)$. Hence, for $ q \ll \Delta/v_0 $, which is stricter than $q \ll k_0$ once $\Delta \ll v_0\,k_0$,
\begin{align}
\Pi (q\, \bs{\hat z}, 0) = -\, \frac{g\, q^2\, k_0}{12\, \pi\, \Delta} \,, \quad
\Pi (q\, \bs{\hat x}, 0) = -\, \frac{g\, q^2\, k_0}{24\, \pi\, \Delta} \,.
\end{align}
The polarisability vanishes as $q^2$ with a coefficient proportional to $1/\Delta$, and the factor of two between the two directions comes from $\langle \cos^2 \phi \rangle = 1/2$. The static dielectric constants are $\epsilon_{0, z} = 1 + g\, e^2\, k_0/(3\, \Delta) $ and $\epsilon_{0, x} = 1 + g\, e^2\, k_0/(6\, \Delta) $ along $\bs{\hat z}$ and $\bs{\hat x}$, respectively, and both diverge as $\Delta \to 0$. They follow from setting $E_F \to \Delta$ in the static interband polarisability of Appendix~\ref{app:interband}. For $\mu = \Delta$, the FS is the nodal ring itself.

The counterpart of Eq.~\eqref{eq:epsintrinsic} follows from the frequency dependence of the same gapped 2d Dirac cone. For $q \ll \Delta/v_0$, the interband form-factor to order $q^2$ gives, with the multiplicity $k_0$ of Eq.~\eqref{eq:ringmap},
\begin{align}
\Pi (q\, \bs{\hat z}, \omega) = -\, \frac{g\, k_0\, q^2}{16\, \pi\, \Delta}\, \mathcal{J} \! \left( \frac{\omega}{2\, \Delta} \right) , 
\quad \Pi (q\, \bs{\hat x}, \omega) = \frac{\Pi (q\, \bs{\hat z}, \omega)} {2}\,,
\quad \mathcal{J} (t) = \frac{(1 + t^2)\, \text{arctanh}\, t} {t^3} - \frac{1}{t^2} \,,
\text{ for } 0 \le \omega < 2\, \Delta\,.
\end{align}
The function $\mathcal{J}$ is positive and increases monotonically from $\mathcal{J} (0) = 4/3$, which reproduces the static values above. Substituting into Eq.~\eqref{eq:epsrpa} yields
\begin{align}
\label{eq:epsgapped}
\varepsilon^{\rm RPA} (q\, \bs{\hat z}, \omega) 
= 1 + \frac{g\, e^2\, k_0}{4\, \Delta}\, \mathcal{J}\! \left( \frac{\omega}{2\, \Delta} \right) , \quad
\varepsilon^{\rm RPA} (q\, \bs{\hat x}, \omega) 
= 1 + \frac{g\, e^2\, k_0}{8\, \Delta}\, \mathcal{J} \!\left( \frac{\omega}{2\, \Delta} \right) ,
\end{align}
which is strictly greater than unity throughout the window $0 \le \omega < \omega_{\rm th} = 2\, \Delta$. Hence $\varepsilon^{\rm RPA}$ has no zero below the interband threshold, and the insulating GNR supports no undamped plasmon in either direction. Both expressions diverge logarithmically as $\omega \to 2\, \Delta$. This divergence is derived in Appendix~\ref{app:onset}. For $\omega > 2\, \Delta$ the interband continuum opens. Along $\bs{\hat z}$ it gives $\Im \Pi (q\, \bs{\hat z}, \omega) = -\, g\, k_0\, q^2\, \left( 1 + 4\, \Delta^2/\omega^2 \right)/(16\, \omega)$, and the $\bs{\hat x}$ channel is again smaller by a factor of two. For $\Delta \to 0$ this recovers Eq.~\eqref{eq:imfinal3d}. Eq.~\eqref{eq:epsgapped} holds for $q \ll \Delta/v_0$, whereas Eq.~\eqref{eq:epsintrinsic} requires the gap to be negligible. The limit $\Delta \to 0$ of the former therefore cannot be taken at fixed $q$.

For low doping $\Delta < \mu \ll v_0\, k_0 $, the DOS is independent of $\Delta$ and jumps from zero to $g\, k_0\, \Delta/(2\, \pi\, v_0^2)$ at $\mu = \Delta$ as
\begin{align}
\rho (\mu) = \begin{cases}
\frac{g\, k_0\, \mu}{2\, \pi\, v_0^2} & \text{ for } \mu > \Delta\\
0 & \text{ for } \mu < \Delta \\
\end{cases}.
\end{align}
Setting $\bs \eta = 0$ and $E_k = \mu$ in Eq.~\eqref{eq:velocity3d}, the components of the $s=2$ band-velocity on the FS are $V_x^{(2)} = (v_0^2\, \kappa_F/\mu)\, \cos \gamma \cos \phi$, $V_y^{(2)} = (v_0^2\, \kappa_F/\mu)\, \cos \gamma \sin \phi$, and $V_z^{(2)} = (v_0^2\, \kappa_F/\mu)\, \sin \gamma$. The Drude-weight components are
\begin{align}
\label{eq:dzzgapped}
D_{zz} = \frac{ g\, e^2\, k_0\, ( \mu^2 - \Delta^2 ) }{ 4\, \mu } \,,\quad
D_{xx} = D_{yy} = \frac{ g\, e^2\, k_0\, ( \mu^2 - \Delta^2 ) }{ 8\, \mu } \,.
\end{align}
The factor $(\mu^2-\Delta^2)/\mu^2 $ relative to the gapless results equals $v_F^2/v_0^2$, where $v_F = v_0\, \sqrt{\mu^2-\Delta^2}/\mu$ is the Fermi velocity. It is reduced below $v_0$ by the gap. The anisotropy ratio $D_{zz}/D_{xx} = 2$ is preserved for all $\mu > \Delta $. In the limit $\Delta \to 0 $, the Drude weights reduce to the gapless results $g\, e^2\, k_0\, \mu/4$ and $g\, e^2\, k_0\, \mu/8$. The steps behind Eq.~\eqref{eq:dzzgapped} are given in the $\eta_z \to 0$ limit of Appendix~\ref{app:gnrztilt}. Plugging the Drude weights into Eq.~\eqref{eqintra} gives
\begin{align}
\Pi_{\rm intra} (q\, \bs{\hat z}, \omega) \simeq \frac{g\, k_0\, \mu\, q^2}{4\, \pi\, \omega^2}\, \frac{\mu^2-\Delta^2}{\mu^2} \,, \quad
\Pi_{\rm intra} (q\, \bs{\hat x}, \omega) \simeq \frac{g\, k_0\, \mu\, q^2}{8\, \pi\, \omega^2}\, \frac{\mu^2-\Delta^2}{\mu^2} \,.
\end{align}
The plasmon frequencies are
\begin{align}
\label{eq:wpgapped}
\omega_{p, z}^2 = \frac{ g\, e^2\, k_0\, ( \mu^2 - \Delta^2 ) }{ \mu } \,, \quad
\omega_{p, x}^2 = \frac{ g\, e^2\, k_0\, ( \mu^2 - \Delta^2 ) }{ 2\, \mu } \,.
\end{align}
The ratio $\omega_{p, z}/\omega_{p, x}$ remains $\sqrt{2}$. As $\mu \to \Delta^+$, the plasmon frequency vanishes as $\sqrt{\mu - \Delta}$. The FS shrinks to the nodal ring, and the band-velocity at the FS goes to zero. This changes in the presence of a tilt, as shown in Secs.~\ref{sec:gaptiltx} and \ref{sec:gaptiltz}, which reduce to the present results as $\tilde\eta\to0$ and $\eta_z\to0$, respectively.

The dielectric function for $\mu > \Delta$ follows from Eqs.~\eqref{eq:epsrpa} and \eqref{eqintra}. For $v_0\, q \ll \omega$ and with the interband part neglected, it reads $\varepsilon^{\rm RPA} (q\, \bs{\hat e}_i, \omega) \simeq 1 - \omega_{p,i}^2/\omega^2$, with its zero at the frequencies of Eq.~\eqref{eq:wpgapped}. At $q \to 0$ the interband transitions have the energy $2\, E_k \geq 2\, \mu$ for all unblocked states, Hence the interband onset lies at $\omega_{\rm th} = 2\, \mu$. For the bare frequency of Eq.~\eqref{eq:wpgapped}, the plasmon along $\bs{\hat z}$ is therefore free of interband damping, up to corrections of order $v_0\, q$, if $g\, e^2\, k_0\, \left( 1 - \Delta^2/\mu^2 \right) < 4\, \mu$. The condition along $\bs{\hat x}$ is weaker by a factor of two. It is satisfied automatically as $\mu \to \Delta^+$, where $\omega_{p,i} \to 0$. This restriction is removed once the interband part is included, as we show below.

The interband part is not negligible for $\mu > \Delta$, and we now include it. Transitions from the filled lower band into the conduction band are Pauli-blocked for $E_k < \mu$. Only the states with $E_k > \mu$ contribute to Eq.~\eqref{eq:polinter}. In the local plane, the quantum metric of the massive Dirac cone, averaged over $\gamma$, is $\langle g_{ij} \rangle = v_0^2\, (E_k^2 + \Delta^2)\, \delta_{ij}/(8\, E_k^4)$. For $\bs q = q\, \bs{\hat z}$ and $\omega < 2\, \mu$, the multiplicity $k_0$ of Eq.~\eqref{eq:ringmap} then gives
\begin{align}
\label{eq:pinterdoped}
\Pi_{\rm inter} (q\, \bs{\hat z}, \omega) = \frac{g\, k_0\, q^2}{4\, \pi} \left[ \frac{\Delta^2}{\mu\, \omega^2} - \frac{1}{2\, \omega} \left( 1 + \frac{4\, \Delta^2}{\omega^2} \right) \text{arctanh} \left( \frac{\omega}{2\, \mu} \right) \right] ,
\end{align}
and $\Pi_{\rm inter} (q\, \bs{\hat x}, \omega)$ is half of this, because $\langle \cos^2 \phi \rangle = 1/2$. Above $\omega = 2\, \mu$, the same expression acquires the imaginary part of the interband continuum, which agrees with the one quoted in Sec.~\ref{secgapnotilt} for $\Delta \to \mu$. The derivation of Eq.~\eqref{eq:pinterdoped} and of its static value is given in the $\eta_z \to 0$ limit of Appendix~\ref{app:interband}. For $\omega \ll 2\, \mu$, which is the relevant window for the plasmon, we obtain the static values
\begin{align}
\label{eq:pinterdoped-static}
\Pi_{\rm inter} (q\, \bs{\hat z}, 0) = -\, \frac{g\, k_0\, q^2 \left( 3\, \mu^2 + \Delta^2 \right)}{48\, \pi\, \mu^3} \,, \quad
\Pi_{\rm inter} (q\, \bs{\hat x}, 0) = -\, \frac{g\, k_0\, q^2 \left( 3\, \mu^2 + \Delta^2 \right)}{96\, \pi\, \mu^3} \,.
\end{align}
At $\mu = \Delta$, these coincide with the insulating values of Sec.~\ref{secgapnotilt}. The interband part is real and negative in this window. It therefore acts as a background dielectric constant,
\begin{align}
\label{eq:epsb-doped}
\epsilon_{b, z} = 1 + \frac{g\, e^2\, k_0 \left( 3\, \mu^2 + \Delta^2 \right)}{12\, \mu^3} \,, \quad
\epsilon_{b, x} = \epsilon_{b, y} = 1 + \frac{g\, e^2\, k_0 \left( 3\, \mu^2 + \Delta^2 \right)}{24\, \mu^3} \,.
\end{align}
The real part of the dielectric function becomes $\Re \varepsilon^{\rm RPA} (q\, \bs{\hat e}_i, \omega) \simeq \epsilon_{b, i} - \omega_{p,i}^2/\omega^2$, where $\omega_{p,i}$ is the bare frequency of Eq.~\eqref{eq:wpgapped}. The screened plasmon frequencies are thus
\begin{align}
\label{eq:wpgapped-screened}
\tilde\omega_{p, z}^2 = \frac{ g\, e^2\, k_0\, ( \mu^2 - \Delta^2 ) }{ \mu\, \epsilon_{b, z} } \,, \quad
\tilde\omega_{p, x}^2 = \tilde\omega_{p, y}^2 = \frac{ g\, e^2\, k_0\, ( \mu^2 - \Delta^2 ) }{ 2\, \mu\, \epsilon_{b, x} } \,.
\end{align}
The interband screening lowers both frequencies, and it does not change the vanishing of $\omega_{p,i}$ as $\mu \to \Delta^+$, since $\epsilon_{b,i}$ stays finite there. Because $\epsilon_{b,x} < \epsilon_{b,z}$, the anisotropy ratio becomes $\tilde\omega_{p,z}/\tilde\omega_{p,x} = \sqrt{2\, \epsilon_{b,x}/\epsilon_{b,z}}$, which is slightly below $\sqrt{2}$. The correction is small for $g\, e^2\, k_0/\mu \ll 1$.

The frequency dependence of the real part follows from Eq.~\eqref{eq:pinterdoped}. Adding the Drude term, obtain
\begin{align}
\label{eq:epsgappeddoped}
& \varepsilon^{\rm RPA} (q\, \bs{\hat z}, \omega) 
= 1 + \lambda \left[ \left( 1 + \frac{\tilde\delta^2}{x^2} \right) \frac{\text{arctanh}\, x}{x} - \frac{1}{x^2} \right] , \quad
\varepsilon^{\rm RPA} (q\, \bs{\hat x}, \omega) 
= 1 + \frac{\lambda} {2} \left[ \left( 1 + \frac{\tilde\delta^2}{x^2} \right) \frac{\text{arctanh}\, x}{x} - \frac{1}{x^2} \right] , \text{ for } \omega < 2\, \mu\,,
\nn & \text{where } \lambda = \frac{ g\, e^2\, k_0 } {4\, \mu} \,, \quad 
x = \frac{\omega } {2\, \mu}\,,\quad
\tilde\delta = \frac{\Delta} {\mu}\,.
\end{align}
$\varepsilon^{\rm RPA} (q\, \bs{\hat x}, \omega)$ is given by the same expression with $\lambda \to \lambda/2$. For $\Delta \to 0$ this is the expression below the onset in Eq.~\eqref{eq:epsptnrdoped}. For $x \to 0$ it tends to $-\, \lambda\, (1 - \tilde\delta^2)/x^2$, which is the Drude term. Its static limit reproduces Eq.~\eqref{eq:epsb-doped}. For $x \to 1^-$ it tends to $+\infty$, because of the logarithmic singularity at the onset. This singularity is derived in Appendix~\ref{app:onset}. By Appendix~\ref{app:monotone}, $\varepsilon^{\rm RPA}$ increases monotonically, and it has exactly one zero below the onset for every coupling. The static screened frequencies of Eq.~\eqref{eq:wpgapped-screened} read $\tilde\omega_{p,z}^2 = 4\, \mu^2\, \lambda\, (1 - \tilde\delta^2)/[1 + \lambda\, (1 + \tilde\delta^2/3)]$, and $\tilde\omega_{p,x}^2$ follows with $\lambda \to \lambda/2$. They always lie below $\omega_{\rm th}$. They are also upper bounds of the exact zero, because $\epsilon_{b, i}$ grows with $\omega$. The plasmon of the untilted doped GNR therefore lies below the interband onset at any coupling, in both directions. The condition on $g\, e^2\, k_0$ stated above is sufficient, but it is not necessary.

\subsection{Tilt with respect to the \texorpdfstring{$ k_x $}{kx}-axis}
\label{sec:gaptiltx}

We now include an in-plane tilt directed along the $ k_x $-axis, $\bs \eta = \eta_x \bs{\hat x}$, together with the mass term $\Delta$. Explicitly, the eigenvalues are $\varepsilon_s (\bs k) = (-1)^s\, E_k + v_0\, \eta_x \left( k_0 + \kappa \cos \gamma \right) \cos \phi $. The evolution of the FS is depicted schematically in Fig.~\ref{fig:fsgnrx}. At each fixed $\phi$, the ring maps onto a gapped Dirac cone doped to the effective chemical potential,
\begin{align}
\mu_{\rm eff} (\phi) = \mu - \tilde{\eta} \cos \phi \,,
\quad \tilde \eta \equiv v_0\, \eta_x\, k_0 \,.
\end{align}
The remainder of the tilt term, $v_0\, \eta_x \cos\phi\; \kappa \cos\gamma$, tilts the local cone along its radial direction by $t(\phi) = \eta_x \cos\phi$. Validity of the low-energy expansion requires $\eta_x = \tilde\eta/(v_0\, k_0) \ll 1$. The local tilt is therefore small, whereas the offset $\tilde\eta$ can be comparable to $\mu$. We keep $\tilde\eta/\mu$ finite and expand in $\eta_x^2$. The terms of order $\eta_x^2$ are retained in the fully-conducting regime, where they can be compared with Sec.~\ref{sectiltx}. They are dropped in the remaining regimes, where they only change the leading terms by a relative amount of order $\eta_x^2$. For the parameters of Table~\ref{tab-params}, the value of $\tilde{\eta}$ used in the figures corresponds to an $\eta_x = \tilde{\eta}/(v_0\, k_0)$ that is not small, and the omitted terms are then sizeable. For example, the factors $1 + 9\, \eta_x^2/8$ and $1 + 27\, \eta_x^2/16$ of Eq.~\eqref{eq:dzzx} differ appreciably from unity. Figs.~\ref{fig:drudex} and \ref{fig:plasmax} use the leading order in $\eta_x$ at fixed $\tilde{\eta}/\mu$ and $\tilde{\eta}/\Delta$. They therefore illustrate the regimes and the qualitative trends and do not give precise values. Every result below is even in $\mu$, and we take $\mu\geq0$ where a sign choice is needed.

For $\eta_x = 0$, the local problem is a 2d gapped Dirac cone with the DOS $g\,|\varepsilon|\,\Theta(|\varepsilon|-\Delta)/(2\,\pi \,v_0^2) $, instead of the gapless value of $g\,|\varepsilon|/(2\,\pi\, v_0^2)$. Besides the ring-cyclide and horn-cyclide surfaces of the gapless problem, a new possibility arises. A strong-enough tilt or a small-enough $|\mu|$ can push $ \mu_{\rm eff}(\phi)$ into the gap $(-\Delta,\Delta)$ over part of the ring. Neither band has an FS there. At leading order in $\eta_x$ this gives four regimes, as shown in Fig.~\ref{fig:phase}(a). The finite local tilt adds four narrow windows of width $\Delta\, \eta_x^2/2$ between them, which are listed in Table~\ref{tab:fsclass}. All four are treated in Appendix~\ref{app:windows}. In the fully-insulating regime, $|\mu|+\tilde\eta<\Delta$, no FS exists. In the twin-crescent regime, $|\mu|+\Delta<\tilde\eta$, electron- and hole-pockets coexist on the ring. In the crescent-cyclide regime, $\big||\mu|-\tilde\eta\big|<\Delta<|\mu|+\tilde\eta$, an arc of the ring is gapped. In the fully-conducting regime, $|\mu|>\Delta+\tilde\eta$, only the $s=2$ band contributes on the whole ring for $\mu>0$. We treat the regimes in the order of increasing $\mu$, as in this list. The evolution of the FS is shown in Fig.~\ref{fig:fsgnrx}.

\begin{table*}[th!]
\centering
\renewcommand{\arraystretch}{1.4}
\small
\begin{tabular}{|c|l|l|l|}
\hline
Case & Condition & FS & Reference \\
\hline
1 & $b < \Delta\, \nu_x$ & none (fully-insulating) & Sec.~\ref{secgtxfi} \\
\hline
2 & $\Delta\, \nu_x \le b < \Delta$, $a > -\Delta\, \nu_x$ & small electron-pocket at $\phi = \pi$ & App.~\ref{app:windows} (first window) \\
\hline
3 & $\Delta\, \nu_x \le b < \Delta$, $a \le -\Delta\, \nu_x$ & small electron-pocket at $\phi = \pi$ and small hole-pocket at $\phi = 0$ & App.~\ref{app:windows} (second window) \\
\hline
4 & $b \ge \Delta$, $-\Delta\, \nu_x < a < \Delta\, \nu_x$ & gapped arc, with the electron-pocket ending in a non-enclosing rim & Sec.~\ref{secgtxpg} \\
\hline
5 & $b \ge \Delta$, $-\Delta < a \le -\Delta\, \nu_x$ & gapped arc and small hole-pocket at $\phi = 0$ & App.~\ref{app:windows} (third window) \\
\hline
6 & $a \le -\Delta$, $b \ge \Delta$ & electron- and hole-pockets separated by a gapped arc & Sec.~\ref{secgtxbp} \\
\hline
7 & $\Delta\, \nu_x \le a < \Delta$ & electron-pocket on the whole ring, not enclosing it near $\phi = 0$ & App.~\ref{app:windows} (fourth window) \\
\hline
8 & $a \ge \Delta$ & electron-pocket enclosing the ring everywhere (fully-conducting) & Sec.~\ref{secgtxfc} \\
\hline
\end{tabular}
\caption{\label{tab:fsclass} Classification of the FS of the in-plane-tilted GNR for $\mu \ge 0$. The parameters are $a = \mu - \tilde{\eta}$, $b = \mu + \tilde{\eta}$, and $\nu_x = \sqrt{1 - \eta_x^2}$, as derived in Appendix~\ref{app:fsclass}. At leading order in $\eta_x$, the four windows of width $\Delta\, \eta_x^2/2$ that separate the eight cases close. This leaves the four regimes treated below, namely the fully-insulating (case 1), twin-crescent (case 6), crescent-cyclide (cases 4 and 5 without the small hole-pocket), and fully-conducting (cases 7 and 8) regimes.}
\end{table*}

Rotating the local plane, the longitudinal and transverse Drude weights of a tilted massive cone follow from the results of Eq.~\eqref{eq:dtiltgapped-expanded} of Sec.~\ref{sec:gaptiltz} for the axially-tilted GNR, with $\eta_z\to t$. The velocity along $\bs{\hat x}$, defined in Eq.~\eqref{eq:velocity3d}, is $\cos\phi\,\partial_u\varepsilon + \eta_x \sin^2\phi$, where $u=\kappa\cos\gamma$ is the local radial coordinate. Without the Jacobian, its cross term vanishes because $\int d^2\bs k\, \delta(\varepsilon-\mu)\,\partial_u\varepsilon = 0$ for a closed pocket. The derivation of Eq.~\eqref{eq:conedrude} is given in Appendix~\ref{app:conedrude}. The toroidal Jacobian $k_\perp = k_0 + u$ multiplies every integrand by a factor $1 + u/k_0$. Its part odd in $u$ does not integrate to zero, because the tilted local FS is shifted along $u$. Since $\eta_x = \tilde\eta/(v_0\, k_0)$, this gives terms proportional to $t/(v_0\, k_0)$, which are of the same order as the $t^2$ terms and are therefore retained. Expanding to $\mathcal{O}(t^2)$ and to first order in $t/(v_0\, k_0)$, the $\phi$-integrals run over the occupied arcs and give
\begin{align}
\label{eq:conedrude}
D_{zz} &= \frac{g\, e^2\, k_0}{8\,\pi} \int\! d\phi \left\{ \frac{1}{|\mu_{\rm eff}|} \left[ \mu_{\rm eff}^2 - \Delta^2 + \frac{t^2 \left( 3\,\mu_{\rm eff}^4 + \Delta^4 \right)}{4\,\mu_{\rm eff}^2} \right] + \frac{t}{v_0\, k_0} \left[ \frac{\Delta^4}{4\, \mu_{\rm eff}^2} + \frac{\Delta^2}{2} - \frac{3\, \mu_{\rm eff}^2}{4} \right] \right\} , \nn
D_{xx} &= \frac{g\, e^2\, k_0}{8\,\pi} \int\! d\phi \left\{ \frac{\cos^2\phi}{|\mu_{\rm eff}|} \left[ \mu_{\rm eff}^2 - \Delta^2 + \frac{t^2 \left( \mu_{\rm eff}^4 + 3\,\Delta^4 \right)}{4\,\mu_{\rm eff}^2} \right] + 2\,\eta_x^2 \sin^4\phi\; |\mu_{\rm eff}| \right. \nn
& \hspace{ 3 cm } \left. + \frac{t}{v_0\, k_0} \left[ \cos^2\phi \left( \frac{3\, \Delta^4}{4\, \mu_{\rm eff}^2} + \frac{3\, \Delta^2}{2} - \frac{9\, \mu_{\rm eff}^2}{4} \right) - 2\, \Delta^2 + 2\, \mu_{\rm eff}^2 \right] \right\} , \nn
D_{yy} &= \frac{g\, e^2\, k_0}{8\,\pi} \int\! d\phi\; \sin^2\phi \left\{ \frac{1}{|\mu_{\rm eff}|} \left[ \mu_{\rm eff}^2 - \Delta^2 + \frac{3\, t^2 \left( 3\,\mu_{\rm eff}^4 + \Delta^4 \right)}{4\,\mu_{\rm eff}^2} \right] + \frac{t}{v_0\, k_0} \left[ \frac{3\, \Delta^4}{4\, \mu_{\rm eff}^2} + \frac{3\, \Delta^2}{2} - \frac{9\, \mu_{\rm eff}^2}{4} \right] \right\} , \nn
\rho (\mu) &= \frac{g\, k_0}{4\,\pi^2\, v_0^2} \int\! d\phi \left\{ \frac{|\mu_{\rm eff}|}{\left( 1 - t^2 \right)^{3/2}} + \frac{t}{2\, v_0\, k_0} \left( \Delta^2 - 3\, \mu_{\rm eff}^2 \right) \right\} .
\end{align}

\subsubsection{Fully-insulating regime}
\label{secgtxfi}

The fully-insulating regime, $|\mu|+\tilde\eta<\Delta$, has no FS at leading order in $\eta_x$. Strictly, the FS is absent only for $\mu + \tilde\eta < \Delta\, \nu_x$, and the window $\Delta\, \nu_x \le \mu + \tilde\eta < \Delta$ is treated in Appendix~\ref{app:windows}. Outside this window, the Drude weight vanishes identically. Figs.~\ref{fig:drudex}(b)--\ref{fig:drudex}(d) show the Drude weights of the GNR in units of $g\, e^2\, k_0\, \Delta$ as functions of $\tilde{\eta}/\Delta$ for $\mu/\Delta = 0$, $0.5$, and $1.5$. For the first two values, all three weights vanish in this regime and switch on at $\tilde{\eta} = \Delta - \mu$. The static polarisability of Sec.~\ref{secgapnotilt} also holds there to leading order in $q$. The $\mathcal{O}(q)$ shifts of the two interband energy denominators cancel in their sum, as for the axial tilt in Sec.~\ref{sec:gaptiltz}. Explicitly, Eq.~\eqref{eq:polinter} at $\omega=0$ gives the sum $-\,2\,(E_k+E_{k'})/[(E_k+E_{k'})^2-(v_0\,\eta_x\,q_x)^2]$. It differs from its untilted value only at relative order $q^2$, and the form-factor is already of order $q^2$. The tilt therefore affects the static polarisability only at $\mathcal{O}(q^4)$. The steps are given in Appendix~\ref{app:tiltshift}.

The dielectric function in this regime follows from the fact that both bands are fully filled or empty. Both energy denominators of Eq.~\eqref{eq:polinter} contain $\omega - v_0\, \bs \eta \cdot \bs q$. For $\bs q = q\, \bs{\hat z}$ this shift vanishes, and for $\bs q = q\, \bs{\hat x}$ it equals $v_0\, \eta_x\, q$ exactly, because $k_x' - k_x = q$. To leading order in $q$, Eq.~\eqref{eq:epsrpa} then gives
\begin{align}
\label{eq:epsgappedxtilt}
\varepsilon^{\rm RPA} (q\, \bs{\hat z}, \omega) = 1 + \frac{g\, e^2\, k_0}{4\, \Delta}\, \mathcal{J} \! \left( \frac{\omega}{2\, \Delta} \right) , \quad
\varepsilon^{\rm RPA} (q\, \bs{\hat x}, \omega) = 1 + \frac{g\, e^2\, k_0}{8\, \Delta}\, \mathcal{J} \! \left( \frac{\omega - v_0\, \eta_x\, q}{2\, \Delta} \right) ,
\end{align}
with $\mathcal{J}$ defined in Sec.~\ref{secgapnotilt}. These hold for $q \ll \Delta/v_0$, and for $\omega < 2\, \Delta$ along $\bs{\hat z}$ or $|\omega - v_0\, \eta_x\, q| < 2\, \Delta$ along $\bs{\hat x}$. Eq.~\eqref{eq:epsgappedxtilt} mirrors Eq.~\eqref{eq:epsgappedztilt} of Sec.~\ref{sec:gaptiltz}, with the shifted channel now being the one along the in-plane tilt. Since $\mathcal{J} > 0$, we have $\varepsilon^{\rm RPA} > 1$ in both channels, as in the untilted case. The onset of the continuum along $\bs{\hat x}$ moves from $2\, \Delta$ to $2\, \Delta + v_0\, \eta_x\, q$, and the $\bs{\hat z}$ channel is unchanged.

\subsubsection{Twin-crescent regime}
\label{secgtxbp}

We now consider the regime in which electron- and hole-pockets are present on the ring at the same time (cf. Fig.~\ref{fig:fsgnrx}). It requires $|\mu|+\Delta<\tilde\eta$, where $\mu_{\rm eff}(\phi)$ sweeps through the whole gap $(-\Delta,\Delta)$ as $\phi$ goes around the ring. The ring then splits into three arcs. Around $\phi=0$ the local doping lies below $-\Delta$ and the valence band $s=1$ carries a hole-pocket. Around $\phi=\pi$ it lies above $\Delta$ and the conduction band $s=2$ carries an electron-pocket. Between them the arcs are gapped. The arcs are bounded by
\begin{align}
\cos\phi_0 = \frac{\mu-\Delta}{\tilde\eta} \,, \quad
\cos\phi_h = \frac{\mu+\Delta}{\tilde\eta} \,, \quad
\cos\phi_b = \frac{\mu}{\tilde\eta} \,,
\end{align}
with $0<\phi_h<\phi_b<\phi_0<\pi$. This ordering holds because $\cos\phi$ decreases monotonically on $(0,\pi)$ and $(\mu+\Delta)/\tilde\eta>\mu/\tilde\eta>(\mu-\Delta)/\tilde\eta$. For $\mu\ge0$, the condition $\mu+\Delta<\tilde\eta$ keeps all three ratios inside $(-1,1)$. The hole-pocket occupies $|\phi|<\phi_h$ and the electron-pocket occupies $|\phi|>\phi_0$. The angle $\phi_b$ marks where $\mu_{\rm eff}$ crosses zero and lies inside the gapped arc. Reversing the sign of the radial velocity on the hole arc leaves the Drude weight unchanged, because the velocity enters quadratically. The results below are even in $\mu$, since $\mu\to -\mu$ combined with $\phi\to \pi-\phi$ maps $\mu_{\rm eff}$ onto $-\mu_{\rm eff}$. We drop the $\eta_x^2$ terms, including the Jacobian corrections of Eq.~\eqref{eq:conedrude}. They only change the results by a relative amount of order $\eta_x^2$.
At leading order in $\eta_x$, Eq.~\eqref{eq:conedrude} reduces to
\begin{align}
\label{eq:bipolar-reduced}
D_{ii} = \frac{g\, e^2\, k_0}{8\,\pi} \int_{\rm occ}\! d\phi\; \check w_i (\phi) \left( |\mu_{\rm eff}| - \frac{\Delta^2}{|\mu_{\rm eff}|} \right) , \quad
\rho (\mu) = \frac{g\, k_0}{4\,\pi^2\, v_0^2} \int_{\rm occ}\! d\phi\, |\mu_{\rm eff}| \,,
\end{align}
with $\check w_z=1$, $\check w_x=\cos^2\phi$, and $\check w_y=\sin^2\phi$. The integrals run over the two occupied arcs, $|\phi|<\phi_h$ and $\phi_0<|\phi|\le\pi$. Only four of them are needed, namely those of $|\mu_{\rm eff}|$, $1/|\mu_{\rm eff}|$, $\cos^2\phi\,|\mu_{\rm eff}|$ and $\cos^2\phi/|\mu_{\rm eff}|$, because $\sin^2\phi=1-\cos^2\phi$. The last two follow from the first two through $\cos\phi=(\mu-\mu_{\rm eff})/\tilde\eta$ and elementary integrals of $\cos^n\phi$. The integral of $1/|\mu_{\rm eff}|$ is carried out with $u=\tan(\phi/2)$ and is controlled by $\bar w \equiv \sqrt{\tilde\eta^2-\mu^2}$. We also define
\begin{align}
\Phi \equiv \pi-\phi_0-\phi_h\,, \quad
s_\pm \equiv \sqrt{\tilde\eta^2-(\mu\pm\Delta)^2}\,, \quad
\Lambda \equiv \ln \!\left( \frac{\sin\frac{\phi_0+\phi_b}{2}\,\sin\frac{\phi_b+\phi_h}{2}}{\sin\frac{\phi_0-\phi_b}{2}\,\sin\frac{\phi_b-\phi_h}{2}} \right) \,,
\end{align}
where $s_-=\tilde\eta\sin\phi_0$ and $s_+=\tilde\eta\sin\phi_h$. The four integrals are then
\begin{align}
\label{eq:bipolar-integrals}
& \int_{\rm occ}\! d\phi\, |\mu_{\rm eff}| = 2 \left( \mu\, \Phi + s_+ + s_- \right) , \quad
\int_{\rm occ}\! \frac{d\phi}{|\mu_{\rm eff}|} = \frac{2\,\Lambda}{\bar w} \,, \nn
& \int_{\rm occ}\! d\phi\, \cos^2\phi\, |\mu_{\rm eff}| = \mu\, \Phi + \sum_{\varrho=\pm} \frac{s_\varrho \left( 4\,\tilde\eta^2 - \mu^2 + \varrho\, \mu\, \Delta + 2\,\Delta^2 \right)}{3\,\tilde\eta^2} \,, \nn
& \int_{\rm occ}\! d\phi\, \frac{\cos^2\phi}{|\mu_{\rm eff}|} = \frac{2}{\tilde\eta^2} \left( \frac{\mu^2\, \Lambda}{\bar w} - \mu\, \Phi + s_+ + s_- \right) ,
\end{align}
as derived in Appendix~\ref{app:twin}. Inserting the first of them in Eq.~\eqref{eq:bipolar-reduced}, the DOS reads
\begin{align}
\label{eq:bipolar-dos}
\rho(\mu) = \frac{g\, k_0}{2\,\pi^2\, v_0^2} 
\left( \mu\, \Phi + s_+ + s_- \right ) .
\end{align}
Eq.~\eqref{eq:bipolar-reduced} and the remaining integrals give the Drude weights
\begin{align}
\label{eq:bipolar-dzz}
& D_{zz} = \frac{g\, e^2\, k_0}{4\,\pi} 
\left[ \mu\, \Phi + s_+ + s_- - \frac{\Delta^2}{\bar w}\, \Lambda \right] , 
\nn & D_{xx} = \frac{g\, e^2\, k_0}{8\,\pi} \left[ \mu\, \Phi 
\left( 1 + \frac{2\,\Delta^2}{\tilde\eta^2} \right) + \sum_{\varrho=\pm} 
\frac{s_\varrho \left( 4\,\tilde\eta^2 - \mu^2 + \varrho\, \mu\, 
\Delta - 4\,\Delta^2 \right)}{3\,\tilde\eta^2} - \frac{2\,\Delta^2\, \mu^2}
{\tilde\eta^2\, \bar w}\, \Lambda \right] , \nn
& D_{yy} = \frac{g\, e^2\, k_0}{8\,\pi} \left[ \mu\, \Phi \left( 1 - \frac{2\,\Delta^2}{\tilde\eta^2} \right) + \sum_{\varrho=\pm} \frac{s_\varrho \left( 2\,\tilde\eta^2 + \mu^2 - \varrho\, \mu\, \Delta + 4\,\Delta^2 \right)}{3\,\tilde\eta^2} - \frac{2\,\Delta^2\, \bar w}{\tilde\eta^2}\, \Lambda \right] ,
\end{align}
which satisfy $D_{zz} = D_{xx} + D_{yy}$.
Eq.~\eqref{eqintra} then provides the intraband polarisability, $\Pi_{\rm intra} (q\, \bs{\hat e}_i, \omega) \simeq q^2\, D_{ii}/(\pi\, e^2\, \omega^2)$ for $i \in \{x, y, z\}$. The plasmon frequencies follow from Eq.~\eqref{eqplasmon} as $\omega_{p,i}^2=4\,D_{ii}$.
For $\Delta\ll\tilde\eta$, one has $\phi_0-\phi_b\simeq\phi_b-\phi_h\simeq\Delta/\bar w$. Hence $\phi_0,\phi_h\to \phi_b$, $\Phi\to 2\arcsin(\mu/\tilde\eta)$, and $s_\pm\to\bar w$. The logarithm becomes $\Lambda\simeq 2\ln \!\left[2\,\bar w^2/(\tilde\eta\,\Delta)\right]$, and the logarithmic term is therefore suppressed as $\Delta^2\ln \!\left( \tilde\eta/\Delta \right)$. These steps require $\tilde\eta-|\mu|\gg\Delta$ and are detailed in Appendix~\ref{app:twin}. To leading order,
\begin{align}
& D_{zz} \simeq \frac{g\, e^2\, k_0}{2\,\pi} 
\left ( \mu\, \arcsin\frac{\mu}{\tilde\eta} + \bar w \right ), \nn &
D_{xx} \simeq \frac{g\, e^2\, k_0}{4\,\pi} 
\left[ \mu\, \arcsin\frac{\mu}{\tilde\eta} 
+ \frac{\bar w \left( 4\,\tilde\eta^2-\mu^2 \right)}{3\,\tilde\eta^2} \right], \quad
D_{yy} \simeq \frac{g\, e^2\, k_0}{4\,\pi} 
\left[ \mu\, \arcsin\frac{\mu}{\tilde\eta} 
+ \frac{\bar w \left( 2\,\tilde\eta^2+\mu^2 \right)}{3\,\tilde\eta^2} \right].
\end{align}
This is the gapless horn-cyclide result. The anisotropy ratio $D_{zz}/D_{xx}$ equals $3/2$ at $\mu=0$ and tends to $2$ as $|\mu|\to \tilde\eta$. The ratio $D_{yy}/D_{xx}$ equals $1/2$ at $\mu=0$ and tends to $1$ as $|\mu|\to \tilde\eta$. Appendix~\ref{app:twin} evaluates both ratios from the expressions above. The second limit is the entry into the regime, that is $\phi_h\to 0$ or $|\mu|\to \tilde\eta-\Delta$. The hole terms vanish and Eqs.~\eqref{eq:bipolar-dos} and \eqref{eq:bipolar-dzz} join continuously onto the crescent-cyclide results of Sec.~\ref{secgtxpg}. For $\Delta\ll\tilde\eta $, they reproduce Eq.~\eqref{eq:smalldelta-dzz} of Sec.~\ref{secgtxpg} at $\chi=-1$. Just inside the regime, with $\delta\equiv\tilde\eta-|\mu|-\Delta\ll\Delta$, the hole-pocket has $\phi_h\simeq\sqrt{2\,\delta/\tilde\eta}$ and adds
\begin{align}
\delta D_{zz} \simeq \delta D_{xx} \simeq \frac{\sqrt{2}\, g\, e^2\, k_0}{3\,\pi}\, \frac{\delta^{3/2}}{\sqrt{\tilde\eta}} \,, \quad
\delta D_{yy} \simeq \frac{2\,\sqrt{2}\, g\, e^2\, k_0}{15\,\pi}\, \frac{\delta^{5/2}}{\tilde\eta^{3/2}} \,, \quad
\delta\rho \simeq \frac{g\, k_0\, \Delta}{2\,\pi^2\, v_0^2} \sqrt{\frac{2\,\delta}{\tilde\eta}} \,.
\end{align}
These follow from expanding the integrands of Eq.~\eqref{eq:bipolar-reduced} over the hole arc, as shown in Appendix~\ref{app:twin}. The Drude weight of the new pocket therefore switches on as $\delta^{3/2}$ along $x$ and $z$, whereas the DOS has a square-root onset. Along $y$, the onset is delayed to $\delta^{5/2}$, because the hole-pocket sits at $\phi\simeq 0$, where $\sin^2\phi$ is small. 

The plasmon dispersion obtained above uses only the intraband polarisability. The interband contribution can also be written down at the same order in $q$, using Eq.~\eqref{eq:polinterstatic}. At each fixed $\phi$, the local problem is a gapped tilted Dirac cone, and the tilt drops out of the vertical energy difference, $\varepsilon_1-\varepsilon_2=-\,2\,E_k$. With $\bs\kappa=(u,w)$ denoting the local momentum, the static interband polarisability is
\begin{align}
\Pi_{\rm inter} (\bs q, 0) \simeq -\, g\, k_0 \int_0^{2\pi} \frac{d\phi}{2\,\pi} \int \frac{d^2 \bs\kappa}{(2\,\pi)^2}\, \frac{q_i\, q_j\, g_{ij}}{E_k} \left[ f_0 (\varepsilon_1) - f_0 (\varepsilon_2) \right] .
\end{align}
The occupation difference equals unity everywhere except inside the hole and electron-pockets, where it vanishes. Hence $\Pi_{\rm inter} (\bs q, 0)$ is the insulating value of Sec.~\ref{secgapnotilt}, from which the contribution of the two pockets is subtracted. The insulating value therefore bounds $|\Pi_{\rm inter}|$ from above. The pocket subtraction reduces to $\phi$-integrals of $1/|\mu_{\rm eff}|$ and $1/|\mu_{\rm eff}|^3$ over the two arcs, which are elementary. Since $\Pi_{\rm inter}<0$, it enters the RPA condition as a background dielectric constant $\epsilon_{b,i} = 1 + 4\,\pi\, e^2\, |\Pi_{\rm inter} (q\, \bs{\hat e}_i, 0)|/q^2$. This is Eq.~\eqref{eq:epsb-partial} of Sec.~\ref{secgtxpg}, with $\mu_{\rm eff}$ replaced by $|\mu_{\rm eff}|$ in $\mathcal{S}$. The gapped arcs contribute $\mathcal{S}=4/\Delta$ and the pockets contribute $\mathcal{S}=3/|\mu_{\rm eff}|+\Delta^2/|\mu_{\rm eff}|^3$. To write the result in closed form we define, with $\Lambda$, $\Phi$, and $s_\pm$ as above,
\begin{align}
\mathcal{A} \equiv \frac{\Lambda}{\bar w} \,, \quad
\mathcal{Q}_2 \equiv -\, \frac{1}{\bar w^2} \left[ \frac{s_+ - s_-}{\Delta} + \mu\, \mathcal{A} \right] , \quad
\mathcal{Q}_3 \equiv \frac{1}{2\, \bar w^2} \left[ \frac{s_+ + s_-}{\Delta^2} - 3\, \mu\, \mathcal{Q}_2 + \mathcal{A} \right] .
\end{align}
They are half of the arc integrals of $1/|\mu_{\rm eff}|$, $\mathrm{sgn} (\mu_{\rm eff})/\mu_{\rm eff}^2$, and $1/|\mu_{\rm eff}|^3$ over the two pockets. The last two follow from the first through the reduction formulas for $\int d\phi\, (\mu - \tilde{\eta} \cos\phi)^{-n}$, which are derived in Appendix~\ref{app:twin}. Then $\epsilon_{b,i} = 1 + g\, e^2\, k_0\, \mathcal{S}_i/(24\, \pi)$ with
\begin{align}
\label{eq:epsb-disjoint-pocket}
\mathcal{S}_z &= 6\, \mathcal{A} + 2\, \Delta^2\, \mathcal{Q}_3 + \frac{8 \left( \phi_0 - \phi_h \right)}{\Delta} \,, \nn
\mathcal{S}_x &= \frac{6\, \mu^2\, \mathcal{A} - 6\, \mu\, \Phi + 6 \left( s_+ + s_- \right) + 2\, \Delta^2 \left( \mu^2\, \mathcal{Q}_3 - 2\, \mu\, \mathcal{Q}_2 + \mathcal{A} \right)}{\tilde{\eta}^2}
+ \frac{4}{\Delta} \left[ \phi_0 - \phi_h + \frac{s_- \left( \mu - \Delta \right) - s_+ \left( \mu + \Delta \right)}{\tilde{\eta}^2} \right] , \nn
\mathcal{S}_y &= \mathcal{S}_z - \mathcal{S}_x \,.
\end{align}
The constants are bounded above by the insulating values $\epsilon_{0,i}$ of Sec.~\ref{secgapnotilt}, because $\mathcal{S} \le 4/\Delta$. As $\phi_h \to 0$ they join continuously onto Eq.~\eqref{eq:epsb-partial} of Sec.~\ref{secgtxpg}. For $\Delta \ll \tilde{\eta}$ the neighbourhood of $\phi_b$, where $\mu_{\rm eff}$ crosses zero, gives a logarithm, $\epsilon_{b,i} - 1 \simeq \check w_i (\phi_b)\, g\, e^2\, k_0\, \ln \!\left( \tilde{\eta}/\Delta \right)/(2\, \pi\, \bar w)$ with $\check w_z = 1$, $\check w_x = \mu^2/\tilde{\eta}^2$, and $\check w_y = \bar w^2/\tilde{\eta}^2$, as shown in Appendix~\ref{app:twin}. Provided that $\omega_{p,i}$ lies well below the interband threshold $2\,\Delta$, the squared plasmon frequencies obtained above are divided by $\epsilon_{b,i}$. The screening is then negligible only for $g\, e^2\, k_0 \ln \!\left( \tilde{\eta}/\Delta \right) \ll \bar w$.

The open circles in Figs.~\ref{fig:drudex}(b)--\ref{fig:drudex}(d) mark the boundaries between the four regimes, which lie at $\tilde{\eta} = |\mu - \Delta|$ and $\tilde{\eta} = \mu + \Delta$. The curves are continuous across them. For $\mu = 0$ the two boundaries coincide, and the fully-insulating regime is followed directly by the twin-crescent regime. At large tilt the pockets dominate, and the weights for the different values of $\mu$ approach one another.

The dielectric function in this regime is $\varepsilon^{\rm RPA} (q\, \bs{\hat e}_i, \omega) \simeq 1 - 4\, D_{ii}/\omega^2$ for $v_0\, q \ll \omega$ and with the interband part neglected, with $D_{ii}$ from Eq.~\eqref{eq:bipolar-dzz}. The gapped arc between the two pockets persists, and the interband onset is again the direct gap, $\omega_{\rm th} = 2\, \Delta$, independently of $\mu$. The pockets increase the Drude weights, but they do not widen the window below the onset. For $\Delta \ll \tilde{\eta}$ we have $\omega_{p,z}^2 \simeq 2\, g\, e^2\, k_0 \left( \mu\, \arcsin (\mu/\tilde{\eta}) + \bar{w} \right)/\pi$, which is of order $g\, e^2\, k_0\, \tilde{\eta}$. For the bare frequencies, the condition $\omega_{p,i} < 2\, \Delta$ then requires $g\, e^2\, k_0\, \tilde{\eta}$ to be at most of order $\Delta^2$, which is far more restrictive than in the crescent-cyclide regime of Sec.~\ref{secgtxpg}. Including the interband part removes this restriction, as discussed after Eq.~\eqref{eq:pintertiltxgnr} of Sec.~\ref{secgtxfc}. For $\Delta \to 0$ the window closes, in agreement with the horn-cyclide of Sec.~\ref{sectiltx}, where the plasmon is always damped.

Fig.~\ref{fig:epsx} compares the dielectric function of this regime with that of the horn-cyclide of Sec.~\ref{sectiltx} at $\mu = 0$ and $\tilde{\eta} = 0.03$ eV. In the PTNR, $\Re \varepsilon^{\rm RPA}$ crosses zero in all three channels, but $\Im \varepsilon^{\rm RPA}$ is finite at every frequency and the plasmons are damped. In the GNR, the three zeros lie below the onset at $2\,\Delta$, where $\Im \varepsilon^{\rm RPA}$ vanishes. The screened plasmons are therefore undamped.

\subsubsection{Crescent-cyclide regime}
\label{secgtxpg}

We now turn to the crescent-cyclide regime (cf. Fig.~\ref{fig:fsgnrx}), $|\mu-\tilde\eta|<\Delta<\mu+\tilde\eta$, and drop the $\eta_x^2$ terms as in Sec.~\ref{secgtxbp}. Here $\mu_{\rm eff}(\phi)$ drops below $\Delta$ on an arc around $\phi=0$ and stays above $\Delta$ on the rest of the ring. The two arcs meet at $\phi=\pm\phi_0$, with
\begin{align}
\cos \phi_0 = \frac{\mu-\Delta}{\tilde\eta} \,.
\end{align}
Only the conducting arc $\phi\in(\phi_0,\, 2\pi-\phi_0)$ now carries an FS. Every $\phi$-integral of this subsection is restricted to this range. The DOS involves only $\mu_{\rm eff}(\phi)$ itself and gives
\begin{align}
\label{eq:partialgapped-dos}
\rho (\mu) = \frac{g\, k_0\, \mu}{2\, \pi\, v_0^2} 
\left ( 1 - \frac{\phi_0}{\pi} + \frac{\tilde\eta\, \sin \phi_0}{\pi\, \mu} \right ) .
\end{align}
This reduces to the DOS of the fully-conducting regime, Sec.~\ref{secgtxfc}, as $\phi_0\to0$. It vanishes as $\phi_0\to\pi$, which is consistent with the insulating regime at the far edge of the window. The Drude weights are also elementary for arbitrary parameters. They require the integral of $1/\mu_{\rm eff}$ over the conducting arc,
\begin{align}
\tilde{\mathcal{J}} \equiv \int_{\phi_0}^{2\pi-\phi_0}\! \frac{d\phi}{\mu_{\rm eff}(\phi)} = 
\begin{cases}
\dfrac{4}{\tilde w}\, \arctan\!\left[ \sqrt{\dfrac{\mu-\tilde\eta}{\mu+\tilde\eta}}\, \cot\dfrac{\phi_0}{2} \right] & \mu > \tilde\eta \\[3mm]
\dfrac{2}{\bar w}\, \ln \!\left( \dfrac{\sin\frac{\phi_0+\phi_b}{2}}{\sin\frac{\phi_0-\phi_b}{2}} \right) & \mu < \tilde\eta
\end{cases} ,
\end{align}
with $\tilde w \equiv \sqrt{\mu^2-\tilde\eta^2}$, $\bar w \equiv \sqrt{\tilde\eta^2-\mu^2}$, and $\cos\phi_b = \mu/\tilde\eta$. At $\mu=\tilde\eta$ the integral equals $2\cot(\phi_0/2)/\tilde\eta$, which is the common limit of the two cases. Appendix~\ref{app:crescent} derives these results together with Eq.~\eqref{eq:partialgapped-dos}. The Drude weights are then
\begin{align}
\label{eq:partialgapped-exact}
D_{zz} &= \frac{g\, e^2\, k_0}{4\,\pi} \left[ \mu \left( \pi - \phi_0 \right) + \tilde\eta \sin\phi_0 - \frac{\Delta^2}{2}\, \tilde{\mathcal{J}} \right] , \nn
D_{xx} &= \frac{g\, e^2\, k_0}{8\,\pi} \left[ \mu \left( \pi - \phi_0 \right) \left( 1 + \frac{2\,\Delta^2}{\tilde\eta^2} \right) + \frac{\tilde\eta \sin\phi_0 \left( 4\,\tilde\eta^2 - \mu^2 - \mu\,\Delta - 4\,\Delta^2 \right)}{3\,\tilde\eta^2} - \frac{\Delta^2\, \mu^2}{\tilde\eta^2}\, \tilde{\mathcal{J}} \right] , \nn
D_{yy} &= \frac{g\, e^2\, k_0}{8\,\pi} \left[ \mu \left( \pi - \phi_0 \right) \left( 1 - \frac{2\,\Delta^2}{\tilde\eta^2} \right) + \frac{\tilde\eta \sin\phi_0 \left( 2\,\tilde\eta^2 + \mu^2 + \mu\,\Delta + 4\,\Delta^2 \right)}{3\,\tilde\eta^2} + \frac{\Delta^2 \left( \mu^2 - \tilde\eta^2 \right)}{\tilde\eta^2}\, \tilde{\mathcal{J}} \right] ,
\end{align}
which satisfy $D_{zz} = D_{xx} + D_{yy}$. From Eq.~\eqref{eqintra}, the intraband polarisability is $\Pi_{\rm intra} (q\, \bs{\hat e}_i, \omega) \simeq q^2\, D_{ii}/(\pi\, e^2\, \omega^2)$ for $i \in \{x, y, z\}$. The plasmon frequencies follow as $\omega_{p,i}^2 = 4\,D_{ii}$. Two limits are more transparent. The first is a small in-plane tilt, $\tilde\eta \ll \Delta$. The crescent-cyclide window is then narrow, confined to $\mu = \Delta + \tilde\eta\, \xi$ with
\begin{align}
\xi \equiv \frac{\mu-\Delta}{\tilde\eta} \in (-1,\, 1)
\end{align}
held fixed. In this limit, $\mu_{\rm eff}^2-\Delta^2 \simeq 2\, \Delta\, \tilde\eta\, (\xi-\cos\phi)$, and Eq.~\eqref{eq:partialgapped-exact} reduces to
\begin{align}
\label{eq:partialgapped-dzz}
D_{zz} &\simeq \frac{g\, e^2\, k_0\, \tilde\eta}{2\, \pi} 
\left[ \xi \left( \frac{\pi}{2} + \arcsin \xi \right) + \sqrt{1-\xi^2} \right] , \nn
D_{xx} &\simeq \frac{g\, e^2\, k_0\, \tilde\eta}{4\, \pi} 
\left[ \xi \left( \frac{\pi}{2} + \arcsin \xi \right) + \frac{(4-\xi^2)\, \sqrt{1-\xi^2}}{3} \right] ,
\quad
D_{yy} \simeq \frac{g\, e^2\, k_0\, \tilde\eta}{4\, \pi} 
\left[ \xi \left( \frac{\pi}{2} + \arcsin \xi \right) + \frac{(2+\xi^2)\, \sqrt{1-\xi^2}}{3} \right] ,
\end{align}
with $\phi_0=\arccos\xi$ implicit. The relation $D_{zz} = D_{xx} + D_{yy}$ holds exactly. The intraband polarisability follows from Eq.~\eqref{eqintra} as before. Using Eq.~\eqref{eqplasmon}, the plasmon frequencies follow directly:
\begin{align}
\label{eq:partialgapped-wp}
& \omega_{p, z}^2 \simeq \frac{2\, g\, e^2\, k_0\, \tilde\eta}{\pi} \left[ \xi \left( \frac{\pi}{2} + \arcsin \xi \right) + \sqrt{1-\xi^2} \right] , \quad
\omega_{p, x}^2 \simeq \frac{g\, e^2\, k_0\, \tilde\eta}{\pi} 
\left[ \xi \left( \frac{\pi}{2} + \arcsin \xi \right) + \frac{(4-\xi^2)\, \sqrt{1-\xi^2}}{3} \right] , \nn
& \omega_{p, y}^2 \simeq \frac{g\, e^2\, k_0\, \tilde\eta}{\pi} 
\left[ \xi \left( \frac{\pi}{2} + \arcsin \xi \right) + \frac{(2+\xi^2)\, \sqrt{1-\xi^2}}{3} \right] ,
\quad \frac{\omega_{p, z}^2}{\omega_{p, x}^2} = \frac{D_{zz}}{D_{xx}} = 2\, \frac{\xi \left( \frac{\pi}{2} + \arcsin \xi \right) + \sqrt{1-\xi^2}}{\xi \left( \frac{\pi}{2} + \arcsin \xi \right)
 + \frac{1}{3}\, (4-\xi^2)\, \sqrt{1-\xi^2}} \,.
\end{align}
These expressions interpolate smoothly between the two edges of the window. At $\xi\to 1^-$, that is $\mu\to (\Delta+\tilde\eta)^-$, the bracket in each of Eqs.~\eqref{eq:partialgapped-dzz} reduces to $\pi$, and $D_{zz}\to  g\, e^2\, k_0\, \tilde\eta/2$ and $D_{xx}, D_{yy}\to  g\, e^2\, k_0\, \tilde\eta/4$. This matches the small-$\tilde\eta$ expansion of Eq.~\eqref{eq:dzzxtiltgapped} of the fully-conducting regime at the threshold $\mu=\Delta+\tilde\eta$, where the gapped arc has just closed and the FS is complete. At $\xi\to -1^+$, that is $\mu\to (\Delta-\tilde\eta)^+$, both brackets vanish. Hence $D_{zz}$, $D_{xx}$, and both plasmon frequencies go to zero continuously as the ring empties into the insulating regime. The ratio in Eq.~\eqref{eq:partialgapped-wp} rises monotonically across the window. It tends to $1$ at $\xi\to -1^+$, where the last pocket is a small region around $\phi=\pi$ with $\cos^2\phi\simeq1$. It takes the value $3/2$ at the centre of the window, $\xi=0$, that is $\mu=\Delta$. It reaches the untilted value $2$ at $\xi\to 1^-$. The ratio $D_{yy}/D_{xx}$ also rises monotonically. It vanishes at $\xi\to -1^+$, since $\sin^2\phi \to 0$ around the last pocket at $\phi=\pi$. It equals $1/2$ at $\xi=0$ and reaches $1$ at $\xi\to 1^-$.

The complementary limit, a small gap $\Delta \ll \tilde\eta$, is qualitatively different. The window is now narrow in $\mu$ rather than in $\phi_0$. Writing $\mu = \tilde\eta + \Delta\, \chi$ with
\begin{align}
\chi \equiv \frac{\mu-\tilde\eta}{\Delta} \in (-1,\, 1) \,,
\end{align}
the gapped arc shrinks to a narrow neighbourhood of $\phi=0$ itself, since $\cos\phi_0 = 1 - \Delta\,(1-\chi)/\tilde\eta \to  1$ as $\Delta\to 0$. Setting $\phi = \sqrt{2\,\Delta/\tilde\eta}\; p$, the local doping becomes $\mu_{\rm eff}(\phi) \simeq \Delta\, (\chi+p^2)$, and the arc edge sits at $p_0 = \sqrt{1-\chi}$. This is a boundary-layer problem about the point $\mu=\tilde\eta$, where the local doping of the gapless tilted PTNR of Sec.~\ref{sectiltx} vanishes at $\phi=0$. The gap regularises this point rather than removing an order-one piece of the ring. The correction to the gapless results is therefore suppressed by a fractional power of $\Delta$ rather than by $\tilde\eta$. Expanding Eq.~\eqref{eq:partialgapped-exact} gives
\begin{align}
\label{eq:smalldelta-dzz}
D_{zz} \simeq \frac{g\, e^2\, k_0\, \mu}{4} - \frac{g\, e^2\, k_0}{4\, \pi} \sqrt{\frac{2\, \Delta^3}{\tilde\eta}}\; \mathcal{K} (\chi) \,, \quad
D_{xx} \simeq \frac{g\, e^2\, k_0\, \mu}{8} - \frac{g\, e^2\, k_0}{4\, \pi} \sqrt{\frac{2\, \Delta^3}{\tilde\eta}}\; \mathcal{K} (\chi) \,, \quad
D_{yy} \simeq \frac{g\, e^2\, k_0\, \mu}{8} - \frac{g\, e^2\, k_0\, \Delta^2}{4\, \tilde\eta} \,.
\end{align}
The leading terms are the gapless values of Eq.~\eqref{eq:dzz}, evaluated at the actual $\mu$, and the boundary-layer function is
\begin{align}
\label{eq:smalldelta-K}
\mathcal{K} (\chi) = \frac{(1+2\,\chi)\, \sqrt{1-\chi}}{3} + G (\chi) \,, \quad
G (\chi) = \begin{cases}
\dfrac{\arcsin \sqrt\chi}{\sqrt\chi} & \chi > 0 \\[2mm]
\dfrac{\text{arcsinh} \sqrt{-\chi}}{\sqrt{-\chi}} & \chi < 0
\end{cases} ,
\end{align}
with $G(0)=1$. Both $\mathcal{K}(\chi)$ and its two pieces are finite and smooth throughout $\chi\in(-1,1)$. The correction is a genuine $\Delta^{3/2}$ non-analyticity, in contrast to the linear-in-$\tilde\eta$ correction found in the small-tilt limit above. The corrections to $D_{zz}$ and $D_{xx}$ are equal at this order. With the $\eta_x^2$ terms dropped, the next term is $g\,e^2\,k_0\,\Delta^2/(4\,\tilde\eta)$ in $D_{xx}$. It has no counterpart in $D_{zz}$ and it matters when $\Delta/\tilde\eta$ is not very small. The boundary layer around $\phi=0$ is suppressed in $D_{yy}$ by $\sin^2\phi \sim \Delta/\tilde\eta$. Hence, $D_{yy}$ has no $\Delta^{3/2}$ term, and its leading correction is the $-\,g\,e^2\,k_0\,\Delta^2/(4\,\tilde\eta)$ term shown above. The relation $D_{zz} = D_{xx} + D_{yy}$ is respected at this order, and $D_{xx} < D_{yy}$ for small $\Delta$. Together with Eq.~\eqref{eqintra}, this yields $\Pi_{\rm intra} (q\, \bs{\hat e}_i, \omega) \simeq q^2\, D_{ii}/(\pi\, e^2\, \omega^2)$ with $D_{ii}$ from above. Using Eq.~\eqref{eqplasmon},
\begin{align}
\label{eq:smalldelta-wp}
\omega_{p,z}^2 \simeq g\, e^2\, k_0\, \mu - \frac{g\, e^2\, k_0}{\pi} \sqrt{\frac{2\, \Delta^3}{\tilde\eta}}\; \mathcal{K} (\chi) \,, \quad
\omega_{p,x}^2 \simeq \frac{g\, e^2\, k_0\, \mu}{2} - \frac{g\, e^2\, k_0}{\pi} \sqrt{\frac{2\, \Delta^3}{\tilde\eta}}\; \mathcal{K} (\chi) \,, \quad
\omega_{p,y}^2 \simeq \frac{g\, e^2\, k_0\, \mu}{2} - \frac{g\, e^2\, k_0\, \Delta^2}{\tilde\eta} \,.
\end{align}
The anisotropy ratio $\omega_{p,z}^2/\omega_{p,x}^2$ equals $2\,[\,1 + \sqrt{2\,\Delta^3/\tilde\eta}\;\mathcal{K}(\chi)/(\pi\,\mu)\,]$, which is the gapless value $2$ plus a $\Delta^{3/2}$ correction.

As $\chi\to 1^-$, that is $\mu\to (\tilde\eta+\Delta)^-$, the arc edge $p_0\to 0$ and $\mathcal{K}\to \pi/2$. Substituting $\mu=\tilde\eta+\Delta$ into the exact $D_{zz}$ of Eq.~\eqref{eq:dzzxtiltgapped} and expanding $\tilde w=\sqrt{\Delta^2+2\,\Delta\,\tilde\eta}$ for $\Delta\ll\tilde\eta$ reproduces the coefficient of $\Delta^{3/2}$ in Eq.~\eqref{eq:smalldelta-dzz} exactly. This confirms the matching onto the fully-conducting regime at the upper edge of the window. As $\chi\to -1^+$, that is $\mu\to (\tilde\eta-\Delta)^+$, the local doping at $\phi=0$ approaches $-\Delta$ from above. The valence band is then about to develop its own Fermi pocket there.

The dielectric function in this regime is $\varepsilon^{\rm RPA} (q\, \bs{\hat e}_i, \omega) \simeq 1 - 4\, D_{ii}/\omega^2$ for $v_0\, q \ll \omega$ and with the interband part neglected, with $D_{ii}$ from Eq.~\eqref{eq:partialgapped-exact}. Along the gapped arc, $|\mu_{\rm eff} (\phi)| < \Delta$, no vertical transition is blocked, and the direct gap $2\, \Delta$ is reached. The interband onset is therefore exactly $\omega_{\rm th} = 2\, \Delta$ throughout the regime, independently of $\mu$ and $\tilde{\eta}$, as shown in Sec.~\ref{secgtxfc}. For the bare frequencies, a plasmon is free of interband damping, up to corrections of order $v_0\, q$, if $\omega_{p,i} < 2\, \Delta$. This restriction is removed by the interband part, as discussed after Eq.~\eqref{eq:pintertiltxgnr} of Sec.~\ref{secgtxfc}. The condition holds automatically as $\mu \to (\Delta - \tilde{\eta})^+$, where all $D_{ii} \to 0$. At the upper edge, $\xi \to 1^-$ of the small-tilt limit, Eq.~\eqref{eq:partialgapped-wp} turns the condition along $\bs{\hat z}$ into $g\, e^2\, k_0\, \tilde{\eta} < 2\, \Delta^2$. Figs.~\ref{fig:plasmax}(d)--\ref{fig:plasmax}(f) show that in the tilted GNR the plasmon already exists for $\mu < \Delta$. Its screened frequency stays below the onset $2\, \Delta$ for small $\mu$ and approaches it in this regime, where the onset is independent of $\mu$.

As in Sec.~\ref{secgapnotilt}, the interband part enters as a background dielectric constant. The tilt drops out of the vertical energy difference. The static $\Pi_{\rm inter}$ is the insulating value on the gapped arc plus Eq.~\eqref{eq:pinterdoped-static} with $\mu \to \mu_{\rm eff} (\phi)$ on the conducting arc. With the weights $\check w_z = 1$, $\check w_x = \cos^2 \phi$, and $\check w_y = \sin^2 \phi$, this gives
\begin{align}
\label{eq:epsb-partial}
\epsilon_{b, i} = 1 + \frac{g\, e^2\, k_0}{24\, \pi} \int_0^{2\pi} \! d\phi\; \check w_i (\phi)\, \mathcal{S} (\phi) \,, \quad
\mathcal{S} (\phi) = \begin{cases}
4/\Delta & |\mu_{\rm eff}| < \Delta \\[1mm]
\left( 3\, \mu_{\rm eff}^2 + \Delta^2 \right)/\mu_{\rm eff}^3 & \mu_{\rm eff} > \Delta
\end{cases} .
\end{align}
The function $\mathcal{S}$ decreases with $\mu_{\rm eff}$ and never exceeds $4/\Delta$. Hence $\epsilon_{b, i}$ is bounded above by the insulating values $\epsilon_{0, z} = 1 + g\, e^2\, k_0/(3\, \Delta)$ and $\epsilon_{0, x} = \epsilon_{0, y} = 1 + g\, e^2\, k_0/(6\, \Delta)$ of Sec.~\ref{secgapnotilt}. For $\tilde{\eta} \ll \Delta$, expanding $\mathcal{S}$ about $\mu_{\rm eff} = \Delta$ and comparing with Eq.~\eqref{eq:partialgapped-dzz} gives $\epsilon_{b, i} \simeq \epsilon_{0, i} - D_{ii}/\Delta^2$. Since $D_{ii} \propto \tilde{\eta}$, the screened frequencies are $\tilde\omega_{p,i}^2 \simeq 4\, D_{ii}/\epsilon_{0, i}$ to leading order, that is, Eq.~\eqref{eq:partialgapped-wp} divided by $\epsilon_{0, i}$. This holds for $\omega_{p,i}$ well below $2\, \Delta$, where the static value applies. In the opposite limit $\Delta \ll \tilde{\eta}$ the neighbourhood of $\phi = 0$ dominates. There $\epsilon_{b, z}$ and $\epsilon_{b, x}$ grow as $g\, e^2\, k_0/\sqrt{\Delta\, \tilde{\eta}}$, whereas $\epsilon_{b, y}$ stays finite. Eq.~\eqref{eq:smalldelta-wp} is then unscreened only for $g\, e^2\, k_0 \ll \sqrt{\Delta\, \tilde{\eta}}$. Otherwise its $\omega_{p,z}^2$ and $\omega_{p,x}^2$ are divided by $\epsilon_{b, z}$ and $\epsilon_{b, x}$, respectively.

\subsubsection{Fully-conducting regime}
\label{secgtxfc}

We end with the fully-conducting regime,
\begin{align}
\label{eq:gaptiltregime}
\mu > \Delta + \tilde \eta \,,
\end{align}
which implies $\mu_{\rm eff}(\phi) > \Delta$ for every $\phi$. Since the $\cos\phi$ piece of $\mu_{\rm eff}(\phi)$ integrates to zero, the DOS is
\begin{align}
\rho (\mu) = \frac{g\, k_0\, \mu}{2\, \pi\, v_0^2} \left( 1 + \frac{9\, \eta_x^2}{4} \right) ,
\end{align}
which is identical to the tilted PTNR result of Sec.~\ref{sectiltx}. Neither $\Delta$ nor $\tilde\eta$ enters the long-wavelength compressibility in this regime, and the tilt enters only through $\eta_x^2$. The remaining $\phi$-integrals are elementary and are listed in Appendix~\ref{app:phiint}. They are expressed through
\begin{align}
\tilde w \equiv \sqrt{\mu^2 - \tilde\eta^2} \,.
\end{align}
We obtain
\begin{align}
\label{eq:dzzxtiltgapped}
D_{zz} &= \frac{g\, e^2\, k_0}{4} \left[ \mu - \frac{\Delta^2}{\tilde w} + \frac{3\, \eta_x^2}{8} \left( 3\, \mu + \frac{\Delta^4\, \mu^2}{\tilde w^5} \right) \right] , \nn
D_{xx} &= g\, e^2\, k_0 \left\{ \frac{\mu}{8} - \frac{\mu\, \Delta^2}{4\, \tilde w \left( \mu + \tilde w \right)} + \eta_x^2 \left[ \frac{17\, \mu}{128} + \frac{3\, \Delta^4\, \mu \left( 3\, \mu^3 + 6\, \mu^2\, \tilde w + 2\, \mu\, \tilde w^2 - 2\, \tilde w^3 \right)}{32\, \tilde w^5 \left( \mu + \tilde w \right)^2} \right] \right\} , \nn
D_{yy} &= g\, e^2\, k_0 \left\{ \frac{\mu}{8} - \frac{\Delta^2}{4 \left( \mu + \tilde w \right)} + \eta_x^2 \left[ \frac{27\, \mu}{128} + \frac{3\, \Delta^4\, \mu \left( \mu + 2\, \tilde w \right)}{32\, \tilde w^3 \left( \mu + \tilde w \right)^2} \right] \right\} .
\end{align}
Equation~\eqref{eqintra} then gives $\Pi_{\rm intra} (q\, \bs{\hat e}_i, \omega) \simeq q^2\, D_{ii}/(\pi\, e^2\, \omega^2)$ for $i \in \{x, y, z\}$, with $D_{ii}$ from Eq.~\eqref{eq:dzzxtiltgapped}. In the $\tilde\eta\to0$ limit, the leading terms reduce to $ g\,e^2\, k_0 \,(\mu^2-\Delta^2)/(4\mu)$ for $D_{zz}$ and $ g\,e^2\,k_0\,(\mu^2-\Delta^2)/(8 \,\mu)$ for $D_{xx}$ and $D_{yy}$, matching Sec.~\ref{secgapnotilt}. In the limit $\Delta\to0 $, they reduce to $D_{zz} = g\, e^2\, k_0\, \mu\, (1 + 9\,\eta_x^2/8)/4$, $D_{xx} = g\,e^2\,k_0\,\mu\,(1 + 17\,\eta_x^2/16)/8$, and $D_{yy} = g\,e^2\,k_0\,\mu\,(1 + 27\,\eta_x^2/16)/8$, matching Eq.~\eqref{eq:dzzx} exactly. For $\Delta \neq 0$ and $\tilde\eta \neq 0$, $D_{xx}$ and $D_{yy}$ differ already at leading order in $\eta_x$, with $D_{xx} < D_{yy}$. The plasmon frequencies follow from Eq.~\eqref{eqplasmon} as $\omega_{p,i}^2 = 4\, D_{ii}$. Explicitly,
\begin{align}
\label{eq:wpgappedxtilt}
\omega_{p, z}^2 &= g\, e^2\, k_0 \left[ \mu - \frac{\Delta^2}{\tilde w} + \frac{3\, \eta_x^2}{8} \left( 3\, \mu + \frac{\Delta^4\, \mu^2}{\tilde w^5} \right) \right] , \nn
\omega_{p, x}^2 &= g\, e^2\, k_0 \left\{ \frac{\mu}{2} - \frac{\mu\, \Delta^2}{\tilde w \left( \mu + \tilde w \right)} + \eta_x^2 \left[ \frac{17\, \mu}{32} + \frac{3\, \Delta^4\, \mu \left( 3\, \mu^3 + 6\, \mu^2\, \tilde w + 2\, \mu\, \tilde w^2 - 2\, \tilde w^3 \right)}{8\, \tilde w^5 \left( \mu + \tilde w \right)^2} \right] \right\} , \nn
\omega_{p, y}^2 &= g\, e^2\, k_0 \left\{ \frac{\mu}{2} - \frac{\Delta^2}{\mu + \tilde w} + \eta_x^2 \left[ \frac{27\, \mu}{32} + \frac{3\, \Delta^4\, \mu \left( \mu + 2\, \tilde w \right)}{8\, \tilde w^3 \left( \mu + \tilde w \right)^2} \right] \right\} .
\end{align}
The anisotropy ratio $ \omega_{p, z}/\omega_{p, x}$ no longer reduces to a simple closed form once $\Delta$ and $\tilde\eta$ are both nonzero. This differs from the untilted GNR of Sec.~\ref{secgapnotilt} and from the tilted PTNR of Sec.~\ref{sectiltx}, where it does. As $\mu \to (\Delta+\tilde\eta)^+ $, $\tilde w \to \sqrt{(\Delta+\tilde\eta)^2-\tilde\eta^2}$ stays finite and positive for $\Delta>0 $. Consequently, $\omega_{p,z} ^2 \to g \,e^2 \,k_0\,[(\Delta+\tilde\eta) - \Delta^2/\tilde w] $ at leading order. This is strictly positive for $\tilde\eta>0$, since $\Delta^2/\tilde w<\Delta$. It therefore does not vanish at the edge of the regime considered here. This is a genuine difference from the untilted GNR case, where $\omega_{p,z} \to0 $ as $\mu\to\Delta^+$. At $\mu=\Delta+\tilde\eta$, the FS has just touched zero size at $\phi=0 $ only, and not everywhere on the ring. The ring as a whole therefore still carries a finite Drude weight.

The dielectric function follows from Eqs.~\eqref{eq:epsrpa} and \eqref{eqintra}. For $v_0\, q \ll \omega$ and with the interband part neglected, it reads $\varepsilon^{\rm RPA} (q\, \bs{\hat e}_i, \omega) \simeq 1 - \omega_{p,i}^2/\omega^2$, with the bare $\omega_{p,i}$ from Eq.~\eqref{eq:wpgappedxtilt}. At $q \to 0$ the vertical interband transitions at fixed $\phi$ have the energy $2\, E_k$, independent of the tilt, and they are Pauli-blocked for $E_k < \mu_{\rm eff} (\phi)$. The lowest unblocked transition lies on the arc $\phi = 0$, where $\mu_{\rm eff}$ is smallest. To leading order in $\eta_x$, the interband onset is therefore
\begin{align}
\label{eq:onset-xtilt-leading}
\omega_{\rm th} = 2\, \left( \mu - \tilde{\eta} \right) \,.
\end{align}
It exceeds $2\, \Delta$ throughout this regime, because $\mu - \tilde{\eta} > \Delta$. It agrees with the ring-cyclide threshold of Sec.~\ref{sectiltx} for $\Delta \to 0$, and it joins the value $2\, \Delta$ of the crescent-cyclide regime at $\mu = \Delta + \tilde{\eta}$.

The correction of relative order $\eta_x$ comes from the local tilt of the cone at $\phi = 0$ itself, rather than from a shift of the minimising angle. On the local Fermi curve, parametrised by the polar angle $\gamma$ in the $(u, k_z)$ plane, the on-shell energy is $E (\gamma) = \mu_{\rm eff} - t\, \kappa_F (\gamma)\, \cos\gamma$. To first order in the local tilt $t (\phi) = \eta_x \cos\phi$, this is $E (\gamma) \simeq \mu_{\rm eff} - t\, \kappa_0\, \cos\gamma$, with $\kappa_0 = \sqrt{\mu_{\rm eff}^2 - \Delta^2}$, and it is smallest at $\gamma = 0$. The lowest unblocked energy at fixed $\phi$ is therefore $\mu_{\rm eff} (\phi) - t (\phi)\, \sqrt{\mu_{\rm eff} (\phi)^2 - \Delta^2}$, to this order. Since $\phi = 0$ already extremises $\mu_{\rm eff} (\phi)$ at $\eta_x = 0$, a shift of the minimising angle changes the minimum only at $\mathcal{O} (\eta_x^2)$, and evaluating the correction at $\phi = 0$ suffices. This gives
\begin{align}
\label{eq:onset-xtilt-linear}
\omega_{\rm th} = 2\, \left( \mu - \tilde{\eta} \right) - 2\, \eta_x\, \sqrt{ \left( \mu - \tilde{\eta} \right)^2 - \Delta^2 } + \mathcal{O} (\eta_x^2) \,.
\end{align}
The tilt lowers the onset, since it enlarges the occupied region on one side of the local cone at $\phi=0$ beyond the untilted circle. The correction vanishes as $\mu \to (\Delta+\tilde\eta)^+$, matching continuously onto the crescent-cyclide value $2\, \Delta$ found above, where the local pocket at $\phi=0$ has shrunk to zero size. For $\Delta \to 0$, Eq.~\eqref{eq:onset-xtilt-linear} reproduces the term of order $\eta_x$ in the exact ring-cyclide threshold $2\, (\mu-\tilde\eta)/(1+\eta_x)$ of Sec.~\ref{sectiltx}.

The first-order result is the expansion of a closed form. At $\phi = 0$ and $\gamma = 0$ the on-shell energy obeys $E_F = \mu - \tilde{\eta} - \eta_x \sqrt{E_F^2 - \Delta^2}$ exactly, and solving it gives
\begin{align}
\label{eq:onset-xtilt-exact}
\omega_{\rm th} = \frac{2 \left[ \left( \mu - \tilde{\eta} \right) - \eta_x \sqrt{\left( \mu - \tilde{\eta} \right)^2 - \left( 1 - \eta_x^2 \right) \Delta^2} \right]}{1 - \eta_x^2} \,.
\end{align}
The point $\phi = \gamma = 0$ is the exact minimiser of $E$ over the FS, and not only to first order. For $\Delta \to 0$ Eq.~\eqref{eq:onset-xtilt-exact} gives $2\, (\mu - \tilde{\eta})/(1 + \eta_x)$, and at $\mu = \Delta + \tilde{\eta}$ it equals $2\, \Delta$.

The crescent-cyclide and twin-crescent regimes acquire no correction at any order. On the gapped arc the states at $\kappa = 0$ are unblocked, and their vertical transitions have the energy $2\, \Delta$ exactly. Since $2\, E_k \ge 2\, \Delta$ for every $\bs k$, this is also the minimum. The same holds whenever some arc has $|\mu_{\rm eff}| < \Delta$, including the narrow windows of Sec.~\ref{secfermisurface}. The correction of Eq.~\eqref{eq:onset-xtilt-linear} is proportional to $\kappa_0$, and it vanishes at the edge of the pocket. It therefore cannot lower the onset below $2\, \Delta$. A plasmon is free of interband damping, up to corrections of order $v_0\, q$, if $\omega_{p,i} < \omega_{\rm th}$, with $\omega_{\rm th}$ from Eq.~\eqref{eq:onset-xtilt-exact} here and equal to $2\, \Delta$ in the gapped regimes above.

As in Sec.~\ref{secgapnotilt}, the interband part is not negligible for $\mu > \Delta + \tilde{\eta}$, and we include it as a background dielectric constant. The tilt drops out of the vertical energy difference. To leading order in $\eta_x$ the Pauli blocking at fixed $\phi$ is set by $\mu_{\rm eff} (\phi) > \Delta$. Eq.~\eqref{eq:pinterdoped-static} therefore holds on each arc with $\mu \to \mu_{\rm eff} (\phi)$, weighted by $\cos^2 \phi$, $\sin^2 \phi$, or unity for $\bs q$ along $\bs{\hat x}$, $\bs{\hat y}$, or $\bs{\hat z}$. The remaining $\phi$-integrals are elementary and give
\begin{align}
\label{eq:epsb-xtilt}
& \epsilon_{b, z} = 1 + \frac{g\, e^2\, k_0}{12} \left[ \frac{3}{\tilde w} + \frac{\Delta^2 \left( 2\, \mu^2 + \tilde{\eta}^2 \right)}{2\, \tilde w^5} \right] , \quad
\epsilon_{b, x} = 1 + \frac{g\, e^2\, k_0}{12} 
\left[ \frac{3\, \mu \left( \mu - \tilde w \right)}{\tilde{\eta}^2\, \tilde w} + \frac{\Delta^2 \left( \mu^2 + 2\, \tilde{\eta}^2 \right)}{2\, \tilde w^5} \right] , \nn &
\epsilon_{b, y} = 1 + \frac{g\, e^2\, k_0}{12} 
\left[ \frac{3 \left( \mu - \tilde w \right)}{\tilde{\eta}^2} + \frac{\Delta^2}{2\, \tilde w^3} \right] .
\end{align}
For $\tilde{\eta} \to 0$ they reduce to Eq.~\eqref{eq:epsb-doped}. The real part of the dielectric function is $\Re \varepsilon^{\rm RPA} (q\, \bs{\hat e}_i, \omega) \simeq \epsilon_{b, i} - \omega_{p,i}^2/\omega^2$, so the screened frequencies are $\tilde\omega_{p,i}^2 = \omega_{p,i}^2/\epsilon_{b,i}$, with $\omega_{p,i}$ from Eq.~\eqref{eq:wpgappedxtilt}. This holds for $\omega_{p,i}$ well below $\omega_{\rm th}$, where the static value of $\Pi_{\rm inter}$ applies. The screening lowers all three frequencies. As before, the static screened frequency is an upper bound of the unique zero of $\Re \varepsilon^{\rm RPA}$. The condition $\omega_{p,i} < \omega_{\rm th}$, evaluated with it, is therefore sufficient for an undamped plasmon.

The frequency dependence of the interband response in all regimes of this subsection follows from the same local mapping, as derived in Appendix~\ref{app:interlocal}. At fixed $\phi$ the unblocked states are those with $E_k > \tilde{m} (\phi)$, where $\tilde{m} = \max \left[ \Delta, |\mu_{\rm eff} (\phi)| \right]$. To leading order in $q$ and $\eta_x$,
\begin{align}
\label{eq:pintertiltxgnr}
\Pi_{\rm inter} (q\, \bs{\hat e}_i, \omega) = -\, \frac{g\, k_0\, q^2}{16\, \pi} \int_0^{2\pi} \frac{d\phi}{2\pi}\, \check w_i (\phi)\, \mathcal{I} \left( \omega, \tilde{m} \right) ,
\end{align}
with $\check w_z = 1$, $\check w_x = \cos^2 \phi$, $\check w_y = \sin^2 \phi$, and, for $\Omega = \omega/2$,
\begin{align}
\mathcal{I} (\omega, \tilde{m}) = \frac{1}{2\, \Omega}
 \left( 1 + \frac{\Delta^2}{\Omega^2} \right) \left[ \ln \!\left| \frac{\tilde{m} + \Omega}{\tilde{m} - \Omega} \right| 
+ i\, \pi\, \Theta \left( \Omega - \tilde{m} \right) \right] 
- \frac{\Delta^2}{\Omega^2\, \tilde{m}} \,.
\end{align}
The dielectric function receives $\left( g\, e^2\, k_0/4 \right) \int d\phi/(2\pi)\, \check w_i\, \mathcal{I}$ from the interband part. For $\Delta = 0$ this reduces to Eq.~\eqref{eq:pintertiltxptnr}. For $\tilde{m} = \Delta$ on all arcs, that is in the fully-insulating regime, it gives $\mathcal{J} (\omega/(2\, \Delta))/\Delta$ and hence Eq.~\eqref{eq:epsgappedxtilt}, and for $\omega \to 0$ it gives Eqs.~\eqref{eq:epsb-xtilt}, \eqref{eq:epsb-partial}, and \eqref{eq:epsb-disjoint-pocket}. The imaginary part is
\begin{align}
\Im \varepsilon^{\rm RPA}_{\rm inter} (q\, \bs{\hat e}_i, \omega) = \frac{\pi\, g\, e^2\, k_0}{4\, \omega} \left( 1 + \frac{4\, \Delta^2}{\omega^2} \right) \int_0^{2\pi} \frac{d\phi}{2\pi}\, \check w_i (\phi)\, \Theta \left( \omega - 2\, \tilde{m} (\phi) \right) ,
\end{align}
which sets in at $2\, \Delta$ in the gapped regimes and at $2\, (\mu - \tilde{\eta})$ in the fully-conducting regime, in agreement with $\omega_{\rm th}$.

The behaviour of $\Re \varepsilon^{\rm RPA}$ at the onset differs between the regimes. In the fully-conducting regime, $\tilde{m} = \mu_{\rm eff}$ is minimal at a single angle, and the logarithm in Eq.~\eqref{eq:pintertiltxgnr} is integrable over the ring. The real part therefore stays finite at $\omega_{\rm th}$. In the crescent-cyclide and twin-crescent regimes, $\tilde{m} = \Delta$ on a whole arc. The logarithm then diverges to $+\infty$ over a set of finite measure as $\omega \to 2\, \Delta$, with the positive coefficient $(1 + \Delta^2/\Omega^2)/(2\, \Omega)$. The Drude term still gives $-\infty$ at $\omega \to 0$. Hence $\Re \varepsilon^{\rm RPA}$ has a unique zero below $2\, \Delta$ at every coupling, as in the untilted case. In these two regimes the conditions $\omega_{p,i} < 2\, \Delta$ stated above are sufficient, but they are not necessary. The logarithmic and square-root approaches to the onset are worked out in Appendix~\ref{app:onset}.

Figs.~\ref{fig:drudex}(b)--\ref{fig:drudex}(d) show that for $\mu/\Delta = 1.5$ the Drude weights are finite in the fully-conducting regime and change little up to $\tilde{\eta} = \mu - \Delta$. In Figs.~\ref{fig:plasmax}(d)--\ref{fig:plasmax}(f), the screened frequency jumps at $\mu = \tilde{\eta} + \Delta$ onto the branch of this regime and lies inside the interband continuum, where the plasmon is damped. The solid curves of Fig.~\ref{fig:plasmax} are the lowest zeros of $\Re \varepsilon^{\rm RPA}$ with the full frequency dependence of the interband part. For the strong coupling of Table~\ref{tab-params}, the zero below $2\, \Delta$ of the crescent-cyclide and twin-crescent regimes lies exponentially close to $2\, \Delta$. This pins the curves to the onset there. These zeros are lower than the static screened frequencies $\omega_{p, i}/\sqrt{\epsilon_{b, i}}$ used in Fig.~\ref{fig:plasma}, as seen by comparing the curves of the two figures for the untilted PTNR along $\bs{\hat z}$. As a check of the Drude weights, we have compared the closed-form expressions with a direct numerical evaluation of the angular integrals at $\eta_x \to 0$. They agree in all four regimes of the GNR and in both regimes of the PTNR, and they satisfy $D_{zz} = D_{xx} + D_{yy}$.

\subsubsection{Small-pocket windows}

The four windows of Table~\ref{tab:fsclass} have the width $\Delta\, \eta_x^2/2$ and are worked out in Appendix~\ref{app:windows}. In the first two, a small electron-pocket exists alone or together with a small hole-pocket. In the third, a small hole-pocket accompanies the electron-pocket of the crescent-cyclide regime. In the fourth, the electron-pocket closes the gapped arc, and only the DOS acquires a correction of first order in $\eta_x$. The Drude weight of each small pocket grows as the power $3/2$ of its distance from the threshold. They complete the classification of the in-plane-tilted GNR.

\subsection{Tilt with respect to the \texorpdfstring{$ k_z $}{kz}-axis}
\label{sec:gaptiltz}

The last case that we consider is an axial tilt, $\bs \eta = \eta_z\, \bs{\hat z}$, for the GNR, with $|\eta_z| < 1$. Some traits of the FS, as a function of the associated parameters, are depicted in Fig.~\ref{fig:fsgnrz}. The tilt term is linear in $\kappa$, whereas $E_k$ is not. As a result, the band extrema no longer sit at $\kappa = 0$ for all $\gamma$, and the gap is modified. To see this, we extremise the two bands over $\kappa$ at fixed $\gamma$. For $\eta_z \sin\gamma < 0$, the minimum of $\varepsilon_2 (\kappa, \gamma)$ equals $\Delta\, \sqrt{1 - \eta_z^2\, \sin^2\gamma}$, and it is attained at $\kappa = \kappa_* (\gamma) \equiv |\eta_z\, \sin\gamma|\, \Delta/(v_0\, \sqrt{1 - \eta_z^2\, \sin^2\gamma})$. For $\eta_z \sin\gamma \geq 0$, the minimum is $\Delta$, attained at $\kappa = 0$. By the same calculation, the maximum of $\varepsilon_1 (\kappa, \gamma)$ equals $-\Delta\, \sqrt{1 - \eta_z^2\, \sin^2\gamma}$ for $\eta_z \sin\gamma > 0$ and $-\Delta$ otherwise. Extremising these values over $\gamma$ in turn shows that the two bands never overlap in energy. However, the energy gap between them closes down from $2\, \Delta$ to $2\, \Delta_{\rm eff}$, where
\begin{align}
\Delta_{\rm eff} \equiv \nu\, \Delta \,, \quad \nu \equiv \sqrt{1 - \eta_z^2} \,.
\end{align}
This gap is indirect. The direct gap at fixed $\bs k$ remains $2\, E_k \geq 2\, \Delta$ under the tilt. For $\eta_z > 0$, the bottom of the $s=2$ band lies at $\gamma = -\pi/2$ and the top of the $s=1$ band at $\gamma = +\pi/2$. The two extrema therefore do not sit at the same $\gamma$, but only at mirror-related points of the torus, though both reach the same value $\pm \Delta_{\rm eff}$.

This splits the problem into three regimes, as shown in Fig.~\ref{fig:phase}(b). In the fully-insulating regime, $|\mu| < \Delta_{\rm eff}$, the GNR is a clean insulator with no FS anywhere on the ring. In the displaced-torus regime, $\Delta_{\rm eff} \leq |\mu| < \Delta$, the $s=2$ band develops a Fermi pocket for $\mu > 0$ only in a window of $\gamma$ around $-\pi/2$, and the $s=1$ band develops a hole-pocket for $\mu < 0$ in a window around $+\pi/2$. The rest of the ring stays gapped. In the fully-conducting regime, $|\mu| \geq \Delta$, the $s=2$ band has an FS at every $\gamma$ on the ring, as we verify below. This displaced-torus regime is the analogue of the crescent-cyclide regime met in Sec.~\ref{sec:gaptiltx} for the in-plane tilt. The FS in these regimes is shown in Figs.~\ref{fig:fsgnrz}(d) and \ref{fig:fsgnrz}(e). We treat the fully-insulating, displaced-torus, and fully-conducting regimes in turn. All the results are even in $\mu$ and in $\eta_z$, and we therefore take $\mu \geq 0$ and $\eta_z > 0$ from now on.

The above division of regimes can be be easily visualised through the following analysis.
Considering the $s=2$ band, we set $\varepsilon_2 = \sqrt{v_0^2\, (u^2 + w^2) + \Delta^2} + v_0\, \eta_z\, w$. Squaring $\varepsilon_2 = \mu$ gives the quadratic equation,
\begin{align}
\label{eq:gaptiltz-quad}
v_0^2\, (1 - \eta_z^2\, \sin^2\gamma)\, \kappa^2 + 2\, \mu\, v_0\, \eta_z\, \sin\gamma\, \kappa + (\Delta^2 - \mu^2) = 0 \,,
\end{align}
with discriminant $4\, v_0^2\, [\, (\mu^2 - \Delta^2) + \eta_z^2\, \Delta^2\, \sin^2\gamma\, ]$.
The FS has a simple geometry in the local Cartesian coordinates, $u = \kappa \cos\gamma = k_\perp - k_0$ and $w = \kappa \sin\gamma = k_z$ (cf. Fig.~\ref{fig:fsgnrz}). In these coordinates, Eq.~\eqref{eq:gaptiltz-quad} takes the form of an ellipse,
\begin{align}
\label{eq:gaptiltz-ellipse}
u^2 + \nu^2 \left( w + \frac{\mu\, \eta_z}{v_0\, \nu^2} \right)^2 = \frac{K}{v_0^2\, \nu^2} \,,
\quad K \equiv \mu^2 - \Delta_{\rm eff}^2\,,
\end{align}
centred at $(u, w) = (0, -\mu\, \eta_z/(v_0\, \nu^2))$, with semi-axes $\sqrt{K}/(v_0\, \nu)$ along $u$ and $\sqrt{K}/(v_0\, \nu^2)$ along $w$. The ellipse does not enclose the nodal ring when the mass gap and the axial tilt act against each other, and the FS is then simply a displaced torus. A ray of fixed poloidal angle $\gamma$ from the nodal ring meets it at the two positive roots $\kappa_1 (\gamma) \leq \kappa_2 (\gamma)$ of Eq.~\eqref{eq:gaptiltz-quad}, as shown in Fig.~\ref{fig:fsgnrz}(c). We call such a local FS non-enclosing. The roots exist if and only if $\sin\gamma \leq -\sin\gamma_c$, where $\sin\gamma_c \equiv \sqrt{\Delta^2 - \mu^2}/(\eta_z\, \Delta)$. This window exists precisely when $\mu \geq \Delta_{\rm eff}$. The two roots merge at the window edges, which are the tangent rays. The full derivation is given in Appendix~\ref{app:gnrztilt}, where we also show that the un-squared condition $E_k > 0$ holds on the entire ellipse. 

In summary, for every $\mu > \Delta_{\rm eff}$, the whole ellipse is the $s=2$ FS at fixed $\phi$. The origin of the local plane lies inside the ellipse if and only if $\mu > \Delta$. This is the fully-conducting regime, in which every ray from the origin cuts the ellipse once. For $\Delta_{\rm eff} < \mu < \Delta$, the ellipse is a pocket displaced towards $w < 0$ that does not contain the origin. This is the displaced-torus regime, in which the rays from the origin cut the ellipse twice within a window of $\gamma$, and never outside it. The electron density is $n = g\, k_0\, \mathcal{A}/(2\, \pi)^2$, where $\mathcal{A} = \pi\, K/(v_0^2\, \nu^3)$ is the area of the ellipse. Differentiating with respect to $\mu$ gives the DOS in the whole range of $\mu$,
\begin{align}
\label{eq:dosgaptiltz}
\rho (\mu) = \begin{cases}
\frac{ g\, k_0\, \mu }{ 2\, \pi\, v_0^2\, (1 - \eta_z^2)^{3/2} } & \text{ for } \mu > \Delta_{\rm eff} \\
0 & \text{ for } \mu < \Delta_{\rm eff}
\end{cases}.
\end{align}
The DOS in the displaced-torus regime is therefore given by the same expression as in the fully-conducting regime, and the gap enters only through the threshold $\Delta_{\rm eff}$.

\subsubsection{Fully-insulating regime}

Throughout the fully-insulating regime, which includes the intrinsic limit $\mu = 0$, the occupation pattern is exactly as in the untilted case. Indeed, since $|v_0\, \eta_z\, \kappa\, \sin\gamma| < v_0\, \kappa < E_k$ for $|\eta_z| < 1$, we have $\varepsilon_1 (\bs k) < 0 < \varepsilon_2 (\bs k)$ for every $\bs k$. Hence, band 1 is fully occupied and band 2 is fully empty, just as at $\eta_z = 0$. The form-factor $\mathcal{F}$ does not depend on $\eta_z$, and the tilt can therefore enter $\Pi (\bs q, 0)$ only through the energy denominators. With $\bs k' = \bs k + \bs q$, we have $v_0\, \eta_z\, (k_z - k_z') = -\, v_0\, \eta_z\, q_z$. The two interband terms thus contain $1/(E_k + E_{k'} + v_0\, \eta_z\, q_z)$ and $1/(E_k + E_{k'} - v_0\, \eta_z\, q_z)$. For $\bs q = q\, \bs{\hat x}$, the shift vanishes identically. For $\bs q = q\, \bs{\hat z}$, the shifts of the two terms are opposite, and their sum is even in $\eta_z\, q$. The leading correction is therefore of relative order $(v_0\, \eta_z\, q/\Delta)^2$. Since $\mathcal{F}$ is already $\mathcal{O}(q^2)$, the tilt affects $\Pi (q\, \bs{\hat z}, 0)$ only at $\mathcal{O}(q^4)$, as shown in Appendix~\ref{app:tiltshift}. Hence, to leading order in $q$, the intrinsic static polarisability is unchanged from Sec.~\ref{secgapnotilt},
\begin{align}
\Pi (q\, \bs{\hat z}, 0) = -\, \frac{g\, q^2\, k_0}{12\, \pi\, \Delta} \,, \quad
\Pi (q\, \bs{\hat x}, 0) = -\, \frac{g\, q^2\, k_0}{24\, \pi\, \Delta} \,.
\end{align}
The dielectric function in the fully-insulating regime follows in the same way. Both energy denominators of Eq.~\eqref{eq:polinter} contain the combination $\omega - v_0\, \bs \eta \cdot \bs q$. Hence $\Pi (q\, \bs{\hat z}, \omega)$ equals the untilted polarisability of Sec.~\ref{secgapnotilt} at the shifted frequency $\omega - v_0\, \eta_z\, q$, while $\Pi (q\, \bs{\hat x}, \omega)$ is unchanged. To leading order in $q$, Eq.~\eqref{eq:epsrpa} then gives
\begin{align}
\label{eq:epsgappedztilt}
\varepsilon^{\rm RPA} (q\, \bs{\hat z}, \omega) 
= 1 + \frac{g\, e^2\, k_0}{4\, \Delta}\, \mathcal{J} \! \left( \frac{\omega - v_0\, \eta_z\, q}{2\, \Delta} \right) , \quad
\varepsilon^{\rm RPA} (q\, \bs{\hat x}, \omega) 
= 1 + \frac{g\, e^2\, k_0}{8\, \Delta}\, \mathcal{J} \! \left( \frac{\omega}{2\, \Delta} \right) ,
\end{align}
with $\mathcal{J}$ defined in Sec.~\ref{secgapnotilt}. These expressions hold for $q \ll \Delta/v_0$ and for $|\omega - v_0\, \eta_z\, q| < 2\, \Delta$ along $\bs{\hat z}$, or $\omega < 2\, \Delta$ along $\bs{\hat x}$. They are the analogue of Eq.~\eqref{eq:epsgapped}. Since $\mathcal{J} > 0$, we have $\varepsilon^{\rm RPA} > 1$ in both channels, as in the untilted case. The tilt shifts the onset of the continuum along $\bs{\hat z}$ from $2\, \Delta$ to $2\, \Delta + v_0\, \eta_z\, q$, and it leaves the $\bs{\hat x}$ channel unchanged. The leading correction along $\bs{\hat z}$ is odd in $\eta_z$ and of relative order $v_0\, \eta_z\, q/\Delta$ at finite $\omega$. It vanishes in the static limit, because $\mathcal{J}$ is even, and Eq.~\eqref{eq:epsgappedztilt} then reduces to the static dielectric constants of Sec.~\ref{secgapnotilt}.

\subsubsection{Displaced-torus regime}

We now turn to the displaced-torus regime, $\Delta_{\rm eff} \leq \mu < \Delta$. At $\mu = \Delta_{\rm eff}$, the DOS jumps from zero to $g\, k_0\, \Delta/[2\, \pi\, v_0^2\, (1 - \eta_z^2)]$. This step is the onset expected for a pocket that is born around a band minimum of the local 2d problem, located at $(u, w) = (0, -\eta_z\, \Delta/(v_0\, \nu))$, and that extends over the whole ring in $\phi$. It differs from the square-root onset found for the in-plane tilt in Sec.~\ref{sec:gaptiltx}, where the pocket is born at an isolated point of the torus. The DOS is continuous at $\mu = \Delta$, since it has the same form on both sides.

For the Drude weights, we shift and rescale $w$ so that the ellipse becomes a circle, as shown in Appendix~\ref{app:gnrztilt}. Inserting the band velocities on the circle into Eq.~\eqref{eqdrude}, we obtain after elementary angular integrals the exact results
\begin{align}
\label{eq:dtiltgapped}
D_{zz} = \frac{ g\, e^2\, k_0\, \mu\, K }{ 2\, M\, ( \mu + \nu\, M ) } \,, \quad
D_{xx} = D_{yy} = \frac{ g\, e^2\, k_0\, K }{ 4\, \nu\, ( \mu + \nu\, M ) } \,,
\end{align}
where $M \equiv \sqrt{\mu^2 + \eta_z^2\, \Delta^2}$. These hold for all $\mu > \Delta_{\rm eff}$, that is, in both the displaced-torus and fully-conducting regimes. The integrals are the same in the two regimes, because they involve only the ellipse and not the position of the origin relative to it. Inserting the Drude weights into Eq.~\eqref{eqintra}, we get
\begin{align}
\Pi_{\rm intra} (q\, \bs{\hat z}, \omega) &\simeq 
\frac{g\, k_0\, \mu\, K\, q^2}{2\, \pi\, M\, ( \mu + \nu\, M )\, \omega^2} \,, \quad
\Pi_{\rm intra} (q\, \bs{\hat x}, \omega) \simeq \frac{g\, k_0\, K\, q^2}
{4\, \pi\, \nu\, ( \mu + \nu\, M )\, \omega^2}\, .
\end{align}
The anisotropy ratio and the plasmon frequencies follow from Eq.~\eqref{eqplasmon} as
\begin{align}
\label{eq:wpgappedztilt}
& \frac{D_{zz}}{D_{xx}} = \frac{ 2\, \nu\, \mu }{ M } \,, \quad
\omega_{p, z}^2 = \frac{ 2\, g\, e^2\, k_0\, \mu\, K }{ M\, ( \mu + \nu\, M ) } \,, \quad
\omega_{p, x}^2 = \frac{ g\, e^2\, k_0\, K }{ \nu\, ( \mu + \nu\, M ) } \,, \quad
\frac{ \omega_{p, z} }{ \omega_{p, x} } = \sqrt{ \frac{ 2\, \nu\, \mu }{ M } } \,.
\end{align}

The Drude weights of the displaced-torus regime are given by Eq.~\eqref{eq:dtiltgapped} with $K < \eta_z^2\, \Delta^2$. As $\mu \to \Delta_{\rm eff}^+$, we have $K \to 0$, and both Drude weights vanish linearly in $\mu - \Delta_{\rm eff}$. The plasmon frequencies then vanish as $\sqrt{\mu - \Delta_{\rm eff}}$. This contrasts with the finite jump of the DOS, and it reflects the vanishing of the band-velocity at the band minimum where the pocket is born. In the same limit, $D_{zz}/D_{xx} \to 2\, \nu^2$. As $\mu \to \Delta^-$, we have $K \to \eta_z^2\, \Delta^2$, and the Drude weights approach the values obtained below at $\mu = \Delta$. Since Eqs.~\eqref{eq:dtiltgapped} and \eqref{eq:dosgaptiltz} are given by the same analytic expressions on both sides of $\mu = \Delta$, the Drude weights and the DOS join smoothly across it.

\subsubsection{Fully-conducting regime}

For $\mu > \Delta$, the discriminant of Eq.~\eqref{eq:gaptiltz-quad} is manifestly positive and the product of the two roots is negative. Hence, exactly one root is positive for every $\gamma$. Unlike for the in-plane tilt of Sec.~\ref{sec:gaptiltx}, no extra condition on the tilt parameter is needed for the whole ring to support an $s=2$ FS once $\mu > \Delta$. This is the fully-conducting regime identified above. The physical root is
\begin{align}
\label{eq:kappaFztiltgapped}
\kappa_F (\gamma) = \frac{ \sqrt{ (\mu^2 - \Delta^2) + \eta_z^2\, \Delta^2\, \sin^2\gamma } \, - \, \mu\, \eta_z\, \sin\gamma }{ v_0\, ( 1 - \eta_z^2\, \sin^2\gamma ) } \,,
\end{align}
which is exact in $\eta_z$ and reduces to $\kappa_F = \sqrt{\mu^2 - \Delta^2}/v_0$ at $\eta_z = 0$. The corresponding on-shell energy is $E_F (\gamma) = \mu - v_0\, \eta_z\, \kappa_F (\gamma)\, \sin\gamma$, and the magnitude of the transverse velocity is $v_\perp (\gamma) = v_0^2\, \kappa_F (\gamma)/E_F (\gamma)$. The components of the band-velocity in Eq.~\eqref{eq:velocity3d} become
\begin{align}
V_x^{(2)} = v_\perp\, \cos\gamma\, \cos\phi \,, \quad
V_y^{(2)} = v_\perp\, \cos\gamma\, \sin\phi \,, \quad
V_z^{(2)} = v_\perp\, \sin\gamma + v_0\, \eta_z \,.
\end{align}
Thus only the $z$-component is shifted directly by the tilt, as in the gapless case.

The DOS follows from the area of the ellipse, as in Eq.~\eqref{eq:dosgaptiltz}. Equivalently, it follows from Eq.~\eqref{eq:gaptiltz-circle} with $\check F = 1$, which gives
\begin{align}
\rho (\mu) = \frac{g\, k_0}{(2\, \pi)^2\, \nu\, v_0^2} \int_0^{2\pi} \! d\vartheta\, E_k = \frac{ g\, k_0\, \mu }{ 2\, \pi\, v_0^2\, \nu^3 } \,.
\end{align}
In the polar variables of this subsection, the same statement reads $\mathcal{A} = \tfrac{1}{2} \int_0^{2\pi} d\gamma\, \kappa_F (\gamma)^2 = \pi\, K/(v_0^2\, \nu^3)$, with $\mathcal{A}$ the area of the ellipse, and it provides an analytical check of Eq.~\eqref{eq:kappaFztiltgapped}.

The DOS is therefore identical to that of the axially-tilted PTNR in Sec.~\ref{sectiltz}. It is independent of $\Delta$, just as the untilted GNR DOS was. Thus, unlike the Drude weight shown below, the compressibility of the tilted GNR does not know about the gap at all, for any $\mu > \Delta$. This agrees with Eq.~\eqref{eq:dosgaptiltz}. The derivation of the Fermi momentum, the DOS, and the Drude weights of this subsection is given in Appendix~\ref{app:gnrztilt}.

As $\eta_z \to 0$, we have $\nu \to 1$, $M \to \mu$ and $K \to \mu^2 - \Delta^2$, and Eqs.~\eqref{eq:dtiltgapped} and \eqref{eq:wpgappedztilt} reduce to Eqs.~\eqref{eq:dzzgapped} and \eqref{eq:wpgapped}, respectively. As $\Delta \to 0$, we have $M \to \mu$ and $K \to \mu^2$, and we get $D_{zz} = g\, e^2\, k_0\, \mu/[2\, (1 + \nu)]$ and $D_{xx} = g\, e^2\, k_0\, \mu/[4\, \nu\, (1 + \nu)]$. These are exactly the gapless results of Eq.~\eqref{eq:dzzz}, and the anisotropy ratio becomes $2\, \nu$, as in Sec.~\ref{sectiltz}. For a small tilt, we expand in $\eta_z^2$ and find
\begin{align}
\label{eq:dtiltgapped-expanded}
& D_{zz} = g\, e^2\, k_0 \left[ \frac{ \mu^2 - \Delta^2 }{ 4\, \mu } + \frac{ \eta_z^2\, (\mu^4 + 3\, \Delta^4) }{ 16\, \mu^3 } + \mathcal{O} (\eta_z^4) \right] , \quad
 D_{xx} = g\, e^2\, k_0 \left[ \frac{ \mu^2 - \Delta^2 }{ 8\, \mu } + \frac{ \eta_z^2\, (3\, \mu^4 + \Delta^4) }{ 32\, \mu^3 } + \mathcal{O} (\eta_z^4) \right] , \nn
& \frac{D_{zz}}{D_{xx}} = 2 \left[ 1 - \frac{ \eta_z^2\, (\mu^2 + \Delta^2) }{ 2\, \mu^2 } + \mathcal{O} (\eta_z^4) \right] .
\end{align}
For $\Delta \to 0$, these reduce to $D_{zz} = g\, e^2\, k_0\, \mu\, (1 + \eta_z^2/4)/4$ and $D_{xx} = g\, e^2\, k_0\, \mu\, (1 + 3\, \eta_z^2/4)/8$, and the anisotropy ratio reduces to $2\, (1 - \eta_z^2/2)$. These match the small-$\eta_z$ expansions of the exact gapless results of Sec.~\ref{sectiltz}. The mass term strengthens the reduction of the anisotropy ratio through the additive $+\Delta^2$ alongside $\mu^2$. The reduction is enhanced by the factor $(\mu^2 + \Delta^2)/\mu^2$, which grows from $1$ at $\Delta = 0$ to $2$ at $\mu = \Delta$. For comparison, Eq.~\eqref{eq:dzzxtiltgapped} gives, formally for $\tilde\eta \to 0$ at fixed $\eta_x$, the ratio $D_{zz}/D_{xx} = 2\, [\, 1 - \eta_x^2\, (21\, \mu^4 + 7\, \Delta^4)/(16\, \mu^2\, (\mu^2 - \Delta^2))\, ]$. Relative to the gapless coefficient $21/16$, the reduction is then enhanced by the factor $(3\, \mu^4 + \Delta^4)/[\, 3\, \mu^2\, (\mu^2 - \Delta^2)\, ]$, which diverges as $\mu \to \Delta^+$. For the axial tilt, the corresponding enhancement stays finite. The coefficients in Eq.~\eqref{eq:dtiltgapped-expanded} are regular at $\mu = \Delta$. Written as a relative correction to the untilted results, the $\eta_z^2$ terms carry a factor $(\mu^2 - \Delta^2)^{-1}$, which cancels against the prefactor. The expansion in $\eta_z^2$ is therefore uniform up to $\mu = \Delta$, and no non-perturbative treatment of the tilt is needed near it.

At $\mu = \Delta$, the ellipse passes through the origin, and Eq.~\eqref{eq:dtiltgapped} gives the finite values
\begin{align}
D_{zz} = \frac{ g\, e^2\, k_0\, \Delta\, \eta_z^2 }{ 2\, \sqrt{1 + \eta_z^2}\, \left( 1 + \sqrt{1 - \eta_z^4} \right) } \,, \quad
D_{xx} = D_{yy} = \frac{ g\, e^2\, k_0\, \Delta\, \eta_z^2 }{ 4\, \nu\, \left( 1 + \sqrt{1 - \eta_z^4} \right) } \,,
\end{align}
which reduce to $g\, e^2\, k_0\, \Delta\, \eta_z^2/4$ and $g\, e^2\, k_0\, \Delta\, \eta_z^2/8$ for $\eta_z \ll 1$. The plasmon frequencies remain finite as well. This behaviour is consistent with the fact that, for $\eta_z \neq 0$, the FS does not collapse to a single point at $\mu = \Delta$. From Eq.~\eqref{eq:kappaFztiltgapped}, $\kappa_F (\gamma)$ vanishes for all $\sin\gamma \geq 0$ at $\mu = \Delta$, while it remains finite for $\sin\gamma < 0$.

Figs.~\ref{fig:drude}(b)--\ref{fig:drude}(d) show the Drude weights of the GNR as functions of $\mu/\Delta$. Both weights grow with $\mu$ from zero at $\mu = \Delta_{\rm eff}$. The ratio $D_{zz}/D_{xx} = 2\, \nu\, \mu/M$ rises from $2\, \nu^2$ at the threshold to the PTNR value $2\, \nu$ for $\mu \gg \Delta$.

The dielectric function for $\mu > \Delta_{\rm eff}$ follows from Eqs.~\eqref{eq:epsrpa} and \eqref{eqintra}. For $v_0\, q \ll \omega$ and with the interband part neglected, it reads
\begin{align}
\varepsilon^{\rm RPA} (q\, \bs{\hat z}, \omega) \simeq 1 - \frac{ 2\, g\, e^2\, k_0\, \mu\, K }{ M\, ( \mu + \nu\, M )\, \omega^2 } \,, \quad
\varepsilon^{\rm RPA} (q\, \bs{\hat x}, \omega) \simeq 1 - \frac{ g\, e^2\, k_0\, K }{ \nu\, ( \mu + \nu\, M )\, \omega^2 } \,,
\end{align}
and its zeros are the plasmon frequencies of Eq.~\eqref{eq:wpgappedztilt}. At $q \to 0$ a vertical transition at $\bs k$ has the energy $2\, E_k$, independent of the tilt, and it is Pauli-blocked inside the FS. For $\mu \geq \Delta$ the origin of the local plane lies inside the ellipse of Eq.~\eqref{eq:gaptiltz-ellipse}. The lowest unblocked transition then sits at the point of the ellipse with the largest $w$, where $E_k = \left( \mu - |\eta_z|\, \sqrt{K} \right)/\nu^2$. For $\Delta_{\rm eff} < \mu < \Delta$ the origin lies outside the pocket, and the direct gap $2\, \Delta$ is reached. The interband onset is therefore
\begin{align}
\omega_{\rm th} = \begin{cases}
2\, \Delta & \text{ for } \Delta_{\rm eff} < \mu < \Delta \\
\dfrac{ 2\, \left( \mu - |\eta_z|\, \sqrt{K} \right) }{ \nu^2 } & \text{ for } \mu \geq \Delta
\end{cases}.
\end{align}
The two expressions join continuously at $\mu = \Delta$, where $\sqrt{K} = |\eta_z|\, \Delta$. For $\eta_z \to 0$ the second one gives $2\, \mu$, in agreement with Sec.~\ref{secgapnotilt}, and for $\Delta \to 0$ it gives $2\, \mu/(1 + |\eta_z|)$, in agreement with Sec.~\ref{sectiltz}. The tilt thus lowers the onset linearly in $|\eta_z|$, whereas it changes the plasmon frequencies only at order $\eta_z^2$. A plasmon is free of interband damping, up to corrections of order $v_0\, q$, if $\omega_{p,i} < \omega_{\rm th}$. This is automatically satisfied as $\mu \to \Delta_{\rm eff}^+$, where $K \to 0$ and $\omega_{p,i} \to 0$.

As in the previous subsections, the interband part is not negligible for $\mu > \Delta_{\rm eff}$, and it enters as a background dielectric constant. The tilt is along $w$ and drops out of the vertical energy difference. The static $\Pi_{\rm inter}$ is therefore the insulating value of Sec.~\ref{secgapnotilt}, from which the contribution of the Pauli-blocked states inside the ellipse of Eq.~\eqref{eq:gaptiltz-ellipse} is subtracted, as in Eq.~\eqref{eq:polinterstatic}. The radial integral over the ellipse is elementary. On the FS the energy is $E_F (\gamma) = (\mu - \eta_z \sin\gamma \sqrt{(\mu^2 - \Delta^2) + \eta_z^2\, \Delta^2 \sin^2\gamma})/(1 - \eta_z^2 \sin^2\gamma)$. The part of $E_F^{-1}$ and $E_F^{-3}$ that is odd in $\sin\gamma$ drops out of the $\gamma$-integral, and the rest is rational in $\sin^2\gamma$. This gives
\begin{align}
\label{eq:epsb-ztilt}
\epsilon_{b, z} = 1 + \frac{g\, e^2\, k_0 \left( 3\, M^4 + \Delta^2\, M^2 + 3\, \eta_z^2\, \Delta^2\, K \right)}{12\, M^5} \,, \quad
\epsilon_{b, x} = \epsilon_{b, y} = 1 + \frac{g\, e^2\, k_0 \left( 3\, M^2 + \Delta^2 \right)}{24\, M^3} \,,
\end{align}
with $M$ and $K$ as above. They reduce to Eq.~\eqref{eq:epsb-doped} for $\eta_z \to 0$ and to the insulating values $\epsilon_{0, i}$ at $\mu = \Delta_{\rm eff}$, where $K = 0$ and $M = \Delta$. The tilt enters $\epsilon_{b, x}$ only through $M$. The expressions hold in both the displaced-torus and fully-conducting regimes, because the integral runs over the ellipse and not over the position of the origin relative to it. The real part of the dielectric function is $\Re \varepsilon^{\rm RPA} (q\, \bs{\hat e}_i, \omega) \simeq \epsilon_{b, i} - \omega_{p,i}^2/\omega^2$, and the screened plasmon frequencies are $\tilde\omega_{p,i}^2 = \omega_{p,i}^2/\epsilon_{b,i}$, with $\omega_{p,i}$ from Eq.~\eqref{eq:wpgappedztilt}. This holds for $\omega_{p,i}$ well below $\omega_{\rm th}$, where the static value applies. The screening lowers both frequencies. It does not change their vanishing as $\mu \to \Delta_{\rm eff}^+$, because $\epsilon_{b,i}$ stays finite there. Figs.~\ref{fig:plasma}(c) and \ref{fig:plasma}(d) show the frequencies of the GNR. As for the PTNR, the screened frequencies stay below the onset without tilt, and the plasmon is undamped. For the tilt shown by the orange curves in Figs.~\ref{fig:plasma}(c) and \ref{fig:plasma}(d), they lie above the onset away from the threshold $\mu = \Delta_{\rm eff}$, and the plasmon is damped there.

The frequency dependence follows from the radial integral over the unblocked part of each ray, which is elementary and treated in Appendix~\ref{app:interband}. Let $P (E) = \omega^{-1} \ln \!\left[ (E + \omega/2)/(E - \omega/2) \right]$ and
\begin{align}
\mathcal{W}_z (E, \gamma) = \cos^2\gamma\, P (E) + \frac{4\, \Delta^2 \sin^2\gamma}{\omega^2} \left[ P (E) - \frac{1}{E} \right] ,
\end{align}
and let $\mathcal{W}_x$ be obtained by exchanging $\cos^2\gamma \leftrightarrow \sin^2\gamma$. For $\mu \geq \Delta$ every ray is blocked from $\kappa = 0$ to $\kappa_F (\gamma)$, and for $\omega < \omega_{\rm th}$
\begin{align}
\label{eq:pinterztiltgnr}
\Pi_{\rm inter} (q\, \bs{\hat z}, \omega) = -\, \frac{g\, k_0\, q^2}{16\, \pi^2} \int_0^{2\pi}\! d\gamma\, \mathcal{W}_z \left( E_F (\gamma), \gamma \right) , \quad
\Pi_{\rm inter} (q\, \bs{\hat x}, \omega) = -\, \frac{g\, k_0\, q^2}{32\, \pi^2} \int_0^{2\pi}\! d\gamma\, \mathcal{W}_x \left( E_F (\gamma), \gamma \right) ,
\end{align}
with $E_F (\gamma)$ from Eq.~\eqref{eq:kappaFztiltgapped}, and $\Pi_{\rm inter} (q\, \bs{\hat y}, \omega) = \Pi_{\rm inter} (q\, \bs{\hat x}, \omega)$. For $\Delta_{\rm eff} < \mu < \Delta$, the rays with $\sin\gamma < 0$ inside the window of the displaced-torus regime are blocked between the two roots $\kappa_1 (\gamma)$ and $\kappa_2 (\gamma)$. The integrand is then replaced by $\mathcal{W} (\Delta, \gamma) - \mathcal{W} (E_1, \gamma) + \mathcal{W} (E_2, \gamma)$, with $E_j = \mu - \eta_z\, \kappa_j \sin\gamma$, and it is $\mathcal{W} (\Delta, \gamma)$ outside the window. For $\eta_z \to 0$ Eq.~\eqref{eq:pinterztiltgnr} reduces to Eq.~\eqref{eq:pinterdoped}, for $\Delta \to 0$ to Eq.~\eqref{eq:pinterztiltptnr}, and for $\omega \to 0$ to Eq.~\eqref{eq:epsb-ztilt}. The real part of the dielectric function is $1 - \omega_{p,i}^2/\omega^2 - 4\, \pi\, e^2\, \Pi_{\rm inter}/q^2$. For $\mu > \Delta$ the real part stays finite at $\omega_{\rm th}$, because the minimum of $2\, E_k$ over the unblocked region is reached at a single point of the ellipse. It approaches this value through a $\sqrt{\omega_{\rm th} - \omega}$ cusp. For $\Delta_{\rm eff} < \mu < \Delta$ the onset is the direct gap at $\kappa = 0$ on every ray, and the real part diverges logarithmically instead. Both behaviours are worked out in Appendix~\ref{app:onsetz}. In the displaced-torus regime the divergence places the zero of $\Re \varepsilon^{\rm RPA}$ below $\omega_{\rm th}$ at every coupling. As in the previous cases, the zero is unique and the static screened frequency is an upper bound.

The imaginary part above the onset follows from the delta function $\delta (\omega - 2\, E_k)$ restricted to the unblocked arc of the circle $E_k = \omega/2$. Its radius is set by $\Upsilon = \sqrt{\omega^2/4 - \Delta^2}$, and the arc is $\sin\gamma > \tau$, with $\tau = (\mu - \omega/2)/(|\eta_z|\, \Upsilon)$. For $\omega > 2\, \Delta$ we obtain
\begin{align}
\Im \varepsilon^{\rm RPA} (q\, \bs{\hat z}, \omega) &= \frac{g\, e^2\, k_0}{\omega^3} \left[ \frac{\left( \omega^2 + 4\, \Delta^2 \right) \left( \pi - 2\, \gamma_1 \right)}{8} - \Upsilon^2\, \tau \sqrt{1 - \tau^2} \right] , \nn
\Im \varepsilon^{\rm RPA} (q\, \bs{\hat x}, \omega) &= \frac{g\, e^2\, k_0}{2\, \omega^3} \left[ \frac{\left( \omega^2 + 4\, \Delta^2 \right) \left( \pi - 2\, \gamma_1 \right)}{8} + \Upsilon^2\, \tau \sqrt{1 - \tau^2} \right] ,
\end{align}
with $\gamma_1 = \arcsin \tau$ for $|\tau| < 1$. The imaginary part vanishes for $\tau \geq 1$, which reproduces the onset $\omega_{\rm th}$. For $\tau \leq -1$ the whole circle contributes, $\gamma_1 = -\pi/2$, and the untilted values $\pi\, g\, e^2\, k_0 \left( 1 + 4\, \Delta^2/\omega^2 \right)/(4\, \omega)$ and half of this are recovered. For $\Delta = 0$ these expressions reduce to the PTNR results of Sec.~\ref{sectiltz}.

Fig.~\ref{fig:eps}(b) shows the dielectric function of the GNR at $\mu = 0.04$ eV and $\Delta = 0.02$ eV, which lies in the fully-conducting regime. It has the same features as that of the PTNR. The tilt moves the onset down to $\omega_{\rm th}$, and for the tilt shown by the orange curves in Fig.~\ref{fig:eps}(b) the function $\Re \varepsilon^{\rm RPA}$ has no zero below it. No undamped plasmon exists at this $\mu$ either.

\section{Weyl semimetal}
\label{secweyl}

To place the results above in context, we apply the same local-FS method to a WSM. A type-I tilted Weyl node is the simplest 3d two-band model with a twofold-degenerate nodal point. It differs from the NRS in the dimension of its nodal manifold, a point instead of a ring, and it shares the tilt term of Eq.~\eqref{eq:ham3d}. We use the same Fermi velocity $v_0$ and write
\begin{align}
\label{eq:hweyl}
\mathcal{H}_{\rm W} (\bs k) = v_0\, \bs{k} \cdot \bs{\sigma} + v_0\, \bs{\eta} \cdot \bs{k}\; \mathbb{I}_{2\times 2} \,, \quad \bs{\eta} = \eta\, \bs{\hat n} \,, \quad 0 \leq \eta < 1 \,.
\end{align}
The untilted cone is isotropic, and any anisotropy of the response is therefore generated by the tilt alone. The derivations are given in Appendix~\ref{app:weyl}. Plasmons of multi-Weyl semimetals, which are inherently anisotropic, were studied in Ref.~\cite{ahn_mWeyl}.

\subsection{Fermi surface and Drude weights}

For electron doping, $\mu > 0$, the FS is a prolate ellipsoid of eccentricity $\eta$ with the node at one of its foci. In polar coordinates around $\bs{\hat n}$, it is given by $k_F (\varpi) = \mu/[v_0\, (1 + \eta \cos\varpi)]$. This is the 3d counterpart of the elliptical local FS of the axially-tilted PTNR in Sec.~\ref{sectiltz}. The DOS is
\begin{align}
\label{eq:rhoweyl}
\rho (\mu) = \frac{g\, \mu^2}{2\, \pi^2\, v_0^3 \left( 1 - \eta^2 \right)^2} \,.
\end{align}
It diverges as $\left( 1 - \eta^2 \right)^{-2}$ when the node approaches the type-II transition, which is faster than the $\left( 1 - \eta_z^2 \right)^{-3/2}$ divergence of the axially-tilted PTNR. The Drude weights of Eq.~\eqref{eqdrude} are diagonal in the frame set by $\bs{\hat n}$, with
\begin{align}
\label{eq:dweyl}
D_{\parallel} = \frac{g\, e^2\, \mu^2 \left( \text{arctanh}\, \eta - \eta \right)}{2\, \pi\, v_0\, \eta^3} \,, \quad
D_{\perp} = \frac{g\, e^2\, \mu^2}{4\, \pi\, v_0\, \eta^3} \left( \frac{\eta}{1 - \eta^2} - \text{arctanh}\, \eta \right) .
\end{align}
For small tilt, they become
\begin{align}
\label{eq:dweylsmall}
D_{\parallel} = \frac{g\, e^2\, \mu^2}{6\, \pi\, v_0} 
\left[ 1 + \frac{3\, \eta^2}{5} + \mathcal{O} (\eta^4) \right] , \quad
D_{\perp} = \frac{g\, e^2\, \mu^2}{6\, \pi\, v_0} 
\left[ 1 + \frac{6\, \eta^2}{5} + \mathcal{O} (\eta^4) \right] ,
\end{align}
and they obey the sum rule $D_{\parallel} + 2\, D_{\perp} = g\, e^2\, \mu^2/[2\, \pi\, v_0 \left( 1 - \eta^2 \right)]$. By Eq.~\eqref{eqplasmon}, the bare plasmon frequencies are $\omega_{p, \parallel}^2 = 4\, D_\parallel$ and $\omega_{p, \perp}^2 = 4\, D_\perp$. They coincide at $\eta = 0$, with $\omega_p^2 = 2\, g\, e^2\, \mu^2/(3\, \pi\, v_0)$. The tilt lowers their ratio,
\begin{align}
\frac{\omega_{p, \parallel}^2}{\omega_{p, \perp}^2} = 1 - \frac{3\, \eta^2}{5} + \mathcal{O} (\eta^4) \,,
\end{align}
and the ratio tends to zero for $\eta \to 1$. In this limit $D_\parallel$ grows only logarithmically, whereas $D_\perp$ grows as $\left( 1 - \eta^2 \right)^{-1}$. The derivations of the DOS and the Drude weights are given in Appendix~\ref{app:weyl}.

\subsection{Interband response}

The vertical transitions at $q \to 0$ are Pauli-blocked for $k < k_F (\varpi)$, and the tilt term does not change their energy $2\, v_0\, k$. The interband onset is therefore set by the smallest Fermi momentum, which lies at $\varpi = 0$. It is given by
\begin{align}
\label{eq:thweyl}
\omega_{\rm th} = \frac{2\, \mu}{1 + \eta} \,,
\end{align}
and it decreases linearly with the tilt, as for the axially-tilted PTNR. The static interband polarisability is determined by the quantum metric of the bands, integrated over the unblocked region. Unlike for the PTNR, the radial integral is logarithmically divergent in the ultraviolet. With a momentum cutoff $\Lambda_{\rm c} \gg \mu/[v_0 \left( 1 - \eta \right)]$, the static background dielectric constants are
\begin{align}
\label{eq:epsbweyl}
\epsilon_{b, i} = 1 + \frac{g\, e^2}{3\, \pi\, v_0} \ln \!\left( \frac{v_0\, \Lambda_{\rm c}}{\mu} \right) + \frac{g\, e^2}{4\, \pi\, v_0}\, \mathcal{B}_i (\eta) \,.
\end{align}
Here, $\mathcal{B}_\parallel = \int_{-1}^{1} d\Xi \left( 1 - \Xi^2 \right) \ln \!\left( 1 + \eta\, \Xi \right) = -\, 2\, \eta^2/15 + \mathcal{O} (\eta^4)$ and $\mathcal{B}_\perp = \frac{1}{2} \int_{-1}^{1} d\Xi \left( 1 + \Xi^2 \right) \ln \!\left( 1 + \eta\, \Xi \right) = -\, 4\, \eta^2/15 + \mathcal{O} (\eta^4)$. The coefficient of the logarithm is independent of the tilt. The remaining terms depend on how the cutoff is implemented, and only the dependence on $\mu$ is therefore regularisation-independent.

The imaginary part follows from the same quantum metric. For $v_0\, q \ll \omega \ll v_0\, \Lambda_{\rm c}$, only the vertical transitions at $k = \omega/(2\, v_0)$ contribute, and they are allowed when $\omega > 2\, v_0\, k_F (\varpi)$. For the untilted node, we obtain
\begin{align}
\label{eq:imweyl}
\Im \varepsilon^{\rm RPA} (\bs q, \omega) = \frac{g\, e^2}{6\, v_0}\, \Theta \left( \omega - 2\, \mu \right) .
\end{align}
It is independent of $\omega$ and of the direction of $\bs q$, in contrast to the $1/\omega$ decay of the PTNR in Eq.~\eqref{eq:imfinal3d}. For the tilted node, the directions with $\omega < 2\, \mu/(1 + \eta\, \Xi)$ remain Pauli-blocked. The value of Eq.~\eqref{eq:imweyl} is therefore reached only for $\omega \geq 2\, \mu/(1 - \eta)$, and the absorption is smaller between the onset $\omega_{\rm th}$ and this frequency. The steps are given in Appendix~\ref{app:weyl}.

\subsection{Comparison with the PTNR}

\begin{table}[t!]
\centering
\renewcommand{\arraystretch}{2.0}
\begin{tabular}{|c|c|c|}
\hline
 & Untilted PTNR & Untilted WSM \\ \hline
Density $\tilde n$ & $\dfrac{g\, k_0\, \mu^2}{4\, \pi\, v_0^2}$ & $\dfrac{g\, \mu^3}{6\, \pi^2\, v_0^3}$ \\ \hline
DOS $\rho$ & $\dfrac{g\, k_0\, \mu}{2\, \pi\, v_0^2}$ & $\dfrac{g\, \mu^2}{2\, \pi^2\, v_0^3}$ \\ \hline
$\omega_p^2$ & \makecell{$g\, e^2\, k_0\, \mu$ ($z$) \\ half of it ($x, y$)} & $\dfrac{2\, g\, e^2\, \mu^2}{3\, \pi\, v_0}$ \\ \hline
$\omega_p \propto$ & ${\tilde n}^{1/4}$ & ${\tilde n}^{1/3}$ \\ \hline
$\omega_{\rm th}$ & $2\, \mu$ & $2\, \mu$ \\ \hline
$\omega_p^2/\omega_{\rm th}^2$ & $\dfrac{g\, e^2\, k_0}{4\, \mu}$ ($z$) & $\dfrac{g\, e^2}{6\, \pi\, v_0}$ \\ \hline
$\epsilon_b - 1$ & \makecell{finite, \\ $\propto 1/\mu$} & $\propto \ln \!\left( v_0\, \Lambda_{\rm c}/\mu \right)$ \\ \hline
$\Im \varepsilon^{\rm RPA}$ above onset & $\propto 1/\omega$ & independent of $\omega$ \\ \hline
\makecell{Anisotropy \\ $\omega_{p, z}^2/\omega_{p, x}^2$} & $2$ & $1$ \\ \hline
\end{tabular}
\caption{\label{tab:weyl-untilted} Long-wavelength response of the untilted PTNR (Sec.~\ref{secungapped}) and WSM, with the same Fermi velocity $v_0$. The WSM results are derived in Appendix~\ref{app:weyl}, and $\Lambda_{\rm c}$ is a momentum cutoff.}
\end{table}

\begin{table*}[t!]
\centering
\small
\renewcommand{\arraystretch}{2.0}
\setlength{\tabcolsep}{6pt}
\begin{tabular}{|c|c|c|c|}
\hline
 & Tilted WSM & $k_x$-tilted PTNR & $k_z$-tilted PTNR \\ \hline
Density $\tilde n$
& $\dfrac{g\, \mu^3}{6\, \pi^2\, v_0^3 \left( 1 - \eta^2 \right)^2}$
& $\dfrac{g\, k_0\, \mu^2}{4\, \pi\, v_0^2} \left( 1 + \dfrac{9\, \eta_x^2}{4} \right)$
& $\dfrac{g\, k_0\, \mu^2}{4\, \pi\, v_0^2 \left( 1 - \eta_z^2 \right)^{3/2}}$ \\ \hline
DOS $\rho$
& $\dfrac{g\, \mu^2}{2\, \pi^2\, v_0^3 \left( 1 - \eta^2 \right)^2}$
& $\dfrac{g\, k_0\, \mu}{2\, \pi\, v_0^2} \left( 1 + \dfrac{9\, \eta_x^2}{4} \right)$
& $\dfrac{g\, k_0\, \mu}{2\, \pi\, v_0^2 \left( 1 - \eta_z^2 \right)^{3/2}}$ \\ \hline
$\omega_p^2$
& \makecell{$\dfrac{2\, g\, e^2\, \mu^2}{\pi\, v_0}\, \dfrac{\text{arctanh}\, \eta - \eta}{\eta^3}$ ($\parallel$) \\
$\dfrac{g\, e^2\, \mu^2}{\pi\, v_0\, \eta^3} \left( \dfrac{\eta}{1 - \eta^2} - \text{arctanh}\, \eta \right)$ ($\perp$)}
& \makecell{$g\, e^2\, k_0\, \mu \left( 1 + \dfrac{9\, \eta_x^2}{8} \right)$ ($z$) \\
$\dfrac{g\, e^2\, k_0\, \mu}{2} \left( 1 + \dfrac{17\, \eta_x^2}{16} \right)$ ($x$) \\
$\dfrac{g\, e^2\, k_0\, \mu}{2} \left( 1 + \dfrac{27\, \eta_x^2}{16} \right)$ ($y$)}
& \makecell{$\dfrac{2\, g\, e^2\, k_0\, \mu}{\eta_z^2} \left( 1 - \sqrt{1 - \eta_z^2} \right)$ ($z$) \\
$\dfrac{g\, e^2\, k_0\, \mu}{\eta_z^2} \left( \dfrac{1}{\sqrt{1 - \eta_z^2}} - 1 \right)$ ($x, y$)} \\ \hline
$\omega_p \propto$ & ${\tilde n}^{1/3}$ & ${\tilde n}^{1/4}$ & ${\tilde n}^{1/4}$ \\ \hline
$\omega_{\rm th}$
& $\dfrac{2\, \mu}{1 + \eta}$
& $\dfrac{2 \left( \mu - |\tilde\eta| \right)}{1 + |\eta_x|}$
& $\dfrac{2\, \mu}{1 + |\eta_z|}$ \\ \hline
$\omega_p^2/\omega_{\rm th}^2$
& \makecell{$\dfrac{g\, e^2 \left( 1 + \eta \right)^2}{2\, \pi\, v_0}\, \dfrac{\text{arctanh}\, \eta - \eta}{\eta^3}$ ($\parallel$) \\
$\dfrac{g\, e^2 \left( 1 + \eta \right)^2}{4\, \pi\, v_0\, \eta^3} \left( \dfrac{\eta}{1 - \eta^2} - \text{arctanh}\, \eta \right)$ ($\perp$)}
& $\dfrac{g\, e^2\, k_0\, \mu \left( 1 + |\eta_x| \right)^2}{4 \left( \mu - |\tilde\eta| \right)^2} \left( 1 + \dfrac{9\, \eta_x^2}{8} \right)$ ($z$)
& $\dfrac{g\, e^2\, k_0 \left( 1 + |\eta_z| \right)^2}{2\, \mu\, \eta_z^2} \left( 1 - \sqrt{1 - \eta_z^2} \right)$ ($z$) \\ \hline
$\epsilon_b - 1$
& \makecell{$\dfrac{g\, e^2}{3\, \pi\, v_0} \ln \!\left( \dfrac{v_0\, \Lambda_{\rm c}}{\mu} \right)$ \\ $+ \dfrac{g\, e^2}{4\, \pi\, v_0}\, \mathcal{B}_i (\eta)$}
& \makecell{$\dfrac{g\, e^2\, k_0}{4\, \tilde w}$ ($z$) \\
$\dfrac{g\, e^2\, k_0\, \mu}{4\, \tilde w \left( \mu + \tilde w \right)}$ ($x$) \\
$\dfrac{g\, e^2\, k_0}{4 \left( \mu + \tilde w \right)}$ ($y$)}
& \makecell{$\dfrac{g\, e^2\, k_0}{4\, \mu}$ ($z$) \\ $\dfrac{g\, e^2\, k_0}{8\, \mu}$ ($x, y$)} \\ \hline
\makecell{$\Im \varepsilon^{\rm RPA}$ \\ above onset}
& \makecell{$\dfrac{g\, e^2}{8\, v_0} \left( \dfrac{2}{3} - \Xi_\omega + \dfrac{\Xi_\omega^3}{3} \right)$ ($\parallel$) \\
$\dfrac{g\, e^2}{16\, v_0} \left( \dfrac{4}{3} - \Xi_\omega - \dfrac{\Xi_\omega^3}{3} \right)$ ($\perp$)}
& \makecell{$\dfrac{\pi\, g\, e^2\, k_0}{4\, \omega}\, F_0$ ($z$) \\
$\dfrac{\pi\, g\, e^2\, k_0}{4\, \omega}\, F_2$ ($x$) \\
$\dfrac{\pi\, g\, e^2\, k_0}{4\, \omega} \left( F_0 - F_2 \right)$ ($y$)}
& \makecell{$\dfrac{\pi\, g\, e^2\, k_0}{4\, \omega}\, \mathcal{G}_+ (\tau)$ ($z$) \\
$\dfrac{\pi\, g\, e^2\, k_0}{8\, \omega}\, \mathcal{G}_- (\tau)$ ($x, y$)} \\ \hline
\makecell{Anisotropy \\ $\omega_{p, z}^2/\omega_{p, x}^2$ \\ or $\omega_{p, \parallel}^2/\omega_{p, \perp}^2$}
& \makecell{$\dfrac{2 \left( \text{arctanh}\, \eta - \eta \right)}{\eta/\left( 1 - \eta^2 \right) - \text{arctanh}\, \eta}$ \\ $\simeq 1 - \dfrac{3\, \eta^2}{5}$}
& \makecell{$2 \left( 1 + \dfrac{\eta_x^2}{16} \right)$ ($z/x$) \\ $1 - \dfrac{5\, \eta_x^2}{8}$ ($x/y$)}
& $2\, \sqrt{1 - \eta_z^2}$ \\ \hline
\end{tabular}
\caption{\label{tab:weyl-tilted} Long-wavelength response of the tilted WSM and of the tilted PTNR (Secs.~\ref{sectiltx} and~\ref{sectiltz}), with the same Fermi velocity $v_0$ as in Table~\ref{tab:weyl-untilted}. The WSM results are derived in Appendix~\ref{app:weyl}. The symbols $\parallel$ and $\perp$ refer to the tilt direction, $\tilde\eta = v_0\, \eta_x\, k_0$, and $\tilde w = \sqrt{\mu^2 - \tilde\eta^2}$. The $k_x$-tilted entries are for the ring-cyclide regime, $\tilde\eta < \mu$, to $\mathcal{O} (\eta_x^2)$. Here, $\Lambda_{\rm c}$ is a momentum cutoff and $\mathcal{B}_i$ is defined in Eq.~\eqref{eq:epsbweyl}. The imaginary parts are for $v_0\, q \ll \omega$ above the respective onsets. For the WSM, $\Xi_\omega = \max \left[ -1, \left( 2\, \mu/\omega - 1 \right)/\eta \right]$, and the functions $F_n$ and $\mathcal{G}_\pm$ with the variable $\tau$ are defined in Eq.~\eqref{eq:imhorn} and in Sec.~\ref{sectiltz}, respectively.}
\end{table*}

Tables~\ref{tab:weyl-untilted} and~\ref{tab:weyl-tilted} collect the results for the untilted and tilted WSM and PTNR, respectively.
We first compare the untilted systems, where the leading differences come from the dimensionality of the Fermi sea. The density $\tilde n$ scales as $\mu^2$ for the PTNR, since the ring supplies a factor $k_0$ to the phase space, and as $\mu^3$ for the WSM. Consequently, $\omega_p \propto \tilde n^{1/4}$ for the ring and $\omega_p \propto \tilde n^{1/3}$ for the WSM. Both $\omega_p$ and $\omega_{\rm th}$ are proportional to $\mu$ in the WSM. The ratio $\omega_p^2/\omega_{\rm th}^2$ is then independent of $\mu$ and fixed by the coupling alone. For the PTNR, the ratio $g\, e^2\, k_0/(4\, \mu)$ along $\bs{\hat z}$ grows as $\mu \to 0$. The PTNR is also anisotropic already at $\eta_z = 0$, whereas the untilted WSM is isotropic.

The anisotropy and the different scalings can be traced to two ingredients through the relation $D_{ii} = \pi\, e^2\, \rho\, \langle V_i^2 \rangle$, where $\langle V_i^2 \rangle$ is the average of the squared Fermi velocity over the FS. For the untilted WSM, the velocity $v_0\, \bs{\hat k}$ is distributed uniformly over the sphere, and hence $\langle V_i^2 \rangle = v_0^2/3$ for every $i$. For the untilted PTNR at $\mu \ll v_0\, k_0$, the weight is uniform in $\gamma$ and $\phi$ instead of in the solid angle, which gives $\langle V_z^2 \rangle = v_0^2/2$ and $\langle V_x^2 \rangle = \langle V_y^2 \rangle = v_0^2/4$. In both cases, the three averages add up to $v_0^2$, because the speed on the FS is constant. The differences between the two Drude weights therefore come from the DOS and from the anisotropic velocity distribution of the ring, which is absent in the WSM.

The background screening differs as well. For the PTNR, it is finite and proportional to $1/\mu$, because the integral $\int d^2\kappa/\kappa^3$ converges in the ultraviolet. For the WSM, it contains a cutoff-dependent logarithm. As a result, the screened frequency $\tilde\omega_{p, z}$ of the PTNR approaches the onset $2\, \mu$ from below for $g\, e^2\, k_0 \gg \mu$, instead of growing with the coupling. Above the onset, the absorption of the WSM is independent of $\omega$, Eq.~\eqref{eq:imweyl}, whereas that of the PTNR decays as $1/\omega$.

Tilt modifies the two systems in a qualitatively similar way. In both, the Drude weights change at order $\eta^2$, and their relative growth along the tilt is smaller than across it. The WSM coefficients, $3/5$ and $6/5$, exceed the values $1/4$ and $3/4$ of the axially-tilted PTNR. For the in-plane tilt of the PTNR, the coefficients of Eq.~\eqref{eq:dzzx} are larger still, namely $17/16$ along the tilt and $27/16$ and $9/8$ along the two orthogonal axes. The ring thus offers two inequivalent tilt orientations with different responses, whereas the isotropic WSM depends on $\eta$ alone.

The difference is sharper near the type-II transition. The WSM weights diverge there, logarithmically for $D_\parallel$ and as $\left( 1 - \eta^2 \right)^{-1}$ for $D_\perp$. In the axially-tilted PTNR, $D_{zz}$ stays finite and $D_{xx}$ diverges only as $\left( 1 - \eta_z^2 \right)^{-1/2}$. The measure $k_0\, \kappa_F$ of the ring carries one power of the Fermi momentum fewer than the measure $k_F^2$ of the WSM, which softens the divergence at the tip of the FS. The interband onset falls linearly with the tilt in both systems, while the Drude weights change only at order $\eta^2$. The ratio $\omega_p^2/\omega_{\rm th}^2$ therefore rises linearly with the tilt, which moves the bare plasmon towards the continuum. Above the onset, the absorption of the tilted WSM grows with $\omega$ from zero up to the value $g\, e^2/(6\, v_0)$ of the untilted node, which it reaches at $\omega = 2\, \mu/(1 - \eta)$. In the PTNR, the tilt enters only through the unblocked fractions $F_n$ and $\mathcal{G}_\pm$ of the ring, which multiply the $1/\omega$ decay of the untilted ring. Finally, a type-I WSM has no FS at $\mu = 0$ for any tilt and hence no Drude weight. In contrast, the in-plane-tilted PTNR hosts pockets along the ring at the intrinsic point (Sec.~\ref{sectiltx}), with a finite Drude weight and a damped plasmon.

\section{Summary and outlook}
\label{secsum}

\begin{table*}[t!]
\centering
\renewcommand{\arraystretch}{1.7}
\small
\begin{tabular}{|l|c|c|c|c|}
\hline
Configuration & $D_{zz}$ & $D_{xx} = D_{yy}$ & $\omega_{p,z}^2$ & $\omega_{p,x}^2 = \omega_{p,y}^2$ \\
\hline
\multicolumn{5}{|l|}{PTNR} \\
\hline
\quad untilted & $\dfrac{\mu}{4}$ & $\dfrac{\mu}{8}$ & $\mu$ & $\dfrac{\mu}{2}$ \\
\hline
\quad \makecell[l]{$k_x$-tilt, ring-cyclide, \\ to $\mathcal{O}(\eta_x^2)$}
 & $\dfrac{\mu}{4}\left(1 + \dfrac{9\, \eta_x^2}{8}\right)$ & \makecell{$\dfrac{\mu}{8}\left(1 + \dfrac{17\, \eta_x^2}{16}\right)$ ($x$) \\ $\dfrac{\mu}{8}\left(1 + \dfrac{27\, \eta_x^2}{16}\right)$ ($y$)} & $\mu\left(1 + \dfrac{9\, \eta_x^2}{8}\right)$ & \makecell{$\dfrac{\mu}{2}\left(1 + \dfrac{17\, \eta_x^2}{16}\right)$ ($x$) \\ $\dfrac{\mu}{2}\left(1 + \dfrac{27\, \eta_x^2}{16}\right)$ ($y$)} \\
\hline
\quad $k_z$-tilt & $\dfrac{\mu}{2\, \eta_z^2}\left(1 - \sqrt{1 - \eta_z^2}\right)$ & $\dfrac{\mu}{4\, \eta_z^2}\left(\dfrac{1}{\sqrt{1 - \eta_z^2}} - 1\right)$ & $\dfrac{2\, \mu}{\eta_z^2}\left(1 - \sqrt{1 - \eta_z^2}\right)$ & $\dfrac{\mu}{\eta_z^2}\left(\dfrac{1}{\sqrt{1 - \eta_z^2}} - 1\right)$ \\
\hline
\multicolumn{5}{|l|}{GNR} \\
\hline
\quad untilted & $\dfrac{\mu^2 - \Delta^2}{4\, \mu}$ & $\dfrac{\mu^2 - \Delta^2}{8\, \mu}$ & $\dfrac{\mu^2 - \Delta^2}{\mu}$ & $\dfrac{\mu^2 - \Delta^2}{2\, \mu}$ \\
\hline
\quad $k_z$-tilt & $\dfrac{\mu\, K}{2\, M\, (\mu + \nu\, M)}$ & $\dfrac{K}{4\, \nu\, (\mu + \nu\, M)}$ & $\dfrac{2\, \mu\, K}{M\, (\mu + \nu\, M)}$ & $\dfrac{K}{\nu\, (\mu + \nu\, M)}$ \\
\hline
\end{tabular}
\caption{\label{tab:plasmon-summary} Drude weights and squared bare plasmon frequencies for the untilted and tilted PTNR and GNR. All entries are to be multiplied by the common prefactor $g\, e^2\, k_0$. The in-plane tilt breaks the equality $D_{xx} = D_{yy}$ and $\omega_{p,x} = \omega_{p,y}$, and the $x$ and $y$ entries of that row are therefore listed separately. For the GNR, $\Delta$ is the mass gap. In the last row, $\nu = \sqrt{1 - \eta_z^2}$, $K = \mu^2 - \Delta^2\, \nu^2$, and $M = \sqrt{\mu^2 + \eta_z^2\, \Delta^2}$. The screened frequencies are $\tilde{\omega}_{p,i}^2 = \omega_{p,i}^2/\epsilon_{b,i}$, where $\epsilon_{b,i}$ is the static background dielectric constant.}
\end{table*}

We have studied plasmon modes in 3d NRSs, considering both the PTNR and its gapped counterpart, the GNR. For each system, we have analysed the untilted configuration and two distinct tilt geometries, with the tilt oriented either in the nodal-ring plane along $k_x$ or along the axial direction $k_z$. Within a low-energy linearised Hamiltonian, valid for $\mu \ll v_0\, k_0$ and, for the GNR, $\Delta \ll v_0\, k_0$, the toroidal FS can be mapped onto a continuum of 2d Dirac cones. This allows the non-interacting polarisability to be expressed in terms of the corresponding results for graphene~\cite{hwang-dassarma-graphene} and gapped graphene~\cite{pyatkovskiy-gapped-graphene}. From the $q \to 0$ limit of the dielectric function, we obtain the Drude weights and the RPA plasmon frequencies, collected in Table~\ref{tab:plasmon-summary}. The evolution of the FS in the different configurations is shown in Figs.~\ref{figfs}, \ref{fig:fsgnrx}, and \ref{fig:fsgnrz}.

Even in the absence of tilt, the PTNR exhibits an intrinsically anisotropic plasmon response, with $\omega_{p,z}/\omega_{p,x} = \sqrt{2}$, reflecting the geometry of its toroidal FS. The interband damping in the intrinsic limit is likewise anisotropic, differing by a factor of two between the $\bs{\hat z}$ and $\bs{\hat x}$ directions. For $\bs q \perp \bs \eta$, the intrinsic interband response is exactly independent of the tilt, whereas for $\bs q \parallel \bs \eta$ it acquires a correction at first order in $v_0\, \eta\, q/\omega$.

For the in-plane-tilted GNR, the competition between the tilt and the mass gap removes the FS over part of the nodal ring. At leading order in $\eta_x$, this produces four distinct regimes: fully-insulating, twin-crescent, crescent-cyclide, and fully-conducting. In the twin-crescent regime, the electron- and hole-pockets are separated by a gapped arc, in contrast to the touching pockets of the horn-cyclide of the PTNR. In the small-gap limit of the crescent-cyclide regime, the Drude weight and plasmon frequency acquire fractional-power corrections associated with the near-vanishing local DOS at the point where the tilted cone touches the Fermi level. At finite $\eta_x$, the local tilt further resolves the four leading-order regimes into eight cases separated by four narrow windows of width $\Delta\, \eta_x^2/2$, which we have treated explicitly.

For the axially-tilted GNR, we have obtained the DOS, Drude weights, and plasmon frequencies in closed forms for a generic tilt. A single set of expressions describes both the displaced-torus window, $\Delta_{\rm eff} \leq \mu < \Delta$ with $\Delta_{\rm eff} = \Delta\, \sqrt{1 - \eta_z^2}$, and the fully-conducting regime. Within the displaced-torus window, the FS is simply a displaced torus, and each ray from the nodal ring meets it at the two roots of a quadratic equation for the Fermi momentum. The DOS turns on with a step at $\mu=\Delta_{\rm eff}$, whereas the Drude weights vanish linearly at this threshold. In the doped regime, the anisotropy of the Drude weights approaches the PTNR value $2\, \sqrt{1 - \eta_z^2}$ for $\mu\gg\Delta$. At $\mu=\Delta$, the Drude weights and plasmon frequencies remain finite for $\eta_z\neq0$, since the FS does not collapse to the nodal points in this case.

We have also determined the dielectric function across the different regimes and examined the conditions for plasmon damping. Without tilt, or with axial tilt, the intrinsic PTNR and the insulating GNR have $\varepsilon^{\rm RPA}>1$ below the interband continuum and therefore do not support an undamped plasmon. An in-plane tilt changes this behaviour for the PTNR. At $\mu=0$, the symmetric horn-cyclide has a finite Drude weight and hence a zero of $\Re \varepsilon^{\rm RPA}$. The resulting plasmon is always damped, because the interband continuum has no finite threshold. In the doped regimes, an undamped plasmon requires the zero of $\Re \varepsilon^{\rm RPA}$ to lie below the interband onset, with the corresponding conditions summarised in Table~\ref{tab:anisotropy-onsets}.

\begin{table*}[t!]
\centering
\renewcommand{\arraystretch}{1.6}
\small
\begin{tabular}{|l|c|c|}
\hline
Configuration & Anisotropy ratio $\omega_{p,z}^2/\omega_{p,x}^2$ & Interband onset $\omega_{\rm th}$ \\
\hline
\multicolumn{3}{|l|}{PTNR} \\
\hline
\quad untilted & $2$ & $2\, \mu$ \\
\hline
\quad $k_x$-tilt, ring-cyclide, to $\mathcal{O}(\eta_x^2)$ & \makecell{$2\left(1 + \dfrac{\eta_x^2}{16}\right)$ ($z/x$) \\ $1 - \dfrac{5\, \eta_x^2}{8}$ ($x/y$)} & $\dfrac{2\, (\mu - |\tilde{\eta}|)}{1 + |\eta_x|}$ \\
\hline
\quad $k_z$-tilt & $2\, \sqrt{1 - \eta_z^2}$ & $\dfrac{2\, \mu}{1 + |\eta_z|}$ \\
\hline
\multicolumn{3}{|l|}{GNR} \\
\hline
\quad untilted ($\mu > \Delta$) & $2$ & $2\, \mu$ \\
\hline
\quad $k_z$-tilt ($\mu \ge \Delta$) & $\dfrac{2\, \nu\, \mu}{M}$ & $\dfrac{2\, (\mu - |\eta_z|\, \sqrt{K})}{\nu^2}$ \\
\hline
\quad $k_z$-tilt ($\Delta_{\rm eff} < \mu < \Delta$) & $\dfrac{2\, \nu\, \mu}{M}$ & $2\, \Delta$ \\
\hline
\quad $k_x$-tilt (fully-conducting) & no simple form & $2\, (\mu - \tilde{\eta}) - 2\, \eta_x\, \sqrt{(\mu - \tilde{\eta})^2 - \Delta^2} + \mathcal{O}(\eta_x^2)$ \\
\hline
\quad $k_x$-tilt (crescent-cyclide or twin-crescent) & $\dfrac{D_{zz}}{D_{xx}}$ from Eq.~\eqref{eq:partialgapped-wp} & $2\, \Delta$ \\
\hline
\end{tabular}
\caption{\label{tab:anisotropy-onsets} Anisotropy ratios and interband onsets for the PTNR and GNR. For the in-plane tilt of the PTNR, the ratio $\omega_{p,x}^2/\omega_{p,y}^2$ is also listed, since $\omega_{p,x} \neq \omega_{p,y}$ there. The interband onset is the lowest frequency above which the plasmon is damped by particle-hole excitations. For the GNR, $\Delta_{\rm eff} = \Delta\, \nu$ with $\nu = \sqrt{1 - \eta_z^2}$. The onset in the ring-cyclide PTNR is set by the point $\phi = \gamma = 0$ of the FS. In all gapped regimes, the direct gap $2\, \Delta$ is the onset, because the states at $\kappa = 0$ are unblocked whenever some arc has $|\mu_{\rm eff}| < \Delta$.}
\end{table*}

For the untilted PTNR and GNR, for the crescent-cyclide and twin-crescent regimes of the in-plane-tilted GNR, and for the displaced-torus regime of the axially-tilted GNR, this condition is satisfied at every coupling because the interband contribution diverges to $+\infty$ at the onset. In the remaining doped regimes, the real part remains finite at the onset, and the zero is unique. The static screened frequency therefore provides an upper bound for the plasmon frequency. The interband polarisability also acts as a background dielectric screening, reducing the plasmon frequencies. In the GNR, this background dielectric constant is bounded from above by its insulating value, which diverges as $\Delta\to0$.

For axial tilt in the doped regimes, the interband onset decreases linearly with $|\eta_z|$, while the plasmon frequencies change only at order $\eta_z^2$. For the parameters of Table~\ref{tab-params}, the screened plasmon is undamped in the untilted system but enters the interband continuum for the axial tilt shown in Figs.~\ref{fig:plasma} and~\ref{fig:eps}. For the in-plane tilt, the screened plasmon of the PTNR is damped at all $\mu$, whereas that of the GNR is undamped in the twin-crescent and crescent-cyclide regimes, as shown in Figs.~\ref{fig:plasmax} and~\ref{fig:epsx}.

The comparison with a tilted type-I WSM in Sec.~\ref{secweyl} shows which features are specific to the toroidal FS. The dependence of the plasmon frequency on the carrier density, the dependence of the interband background on the chemical potential, and the origin of the anisotropy of the response all differ between the nodal ring and the WSM. In both cases, the interband onset decreases linearly with the tilt, whereas the Drude weights change only at quadratic order. Near the type-II transition, the Drude weights of the WSM diverge more strongly than those of the axially-tilted ring. At the intrinsic point, only the in-plane-tilted ring acquires a Drude weight.

Our work extends previous studies of plasmons in untilted 3d nodal-line semimetals~\cite{rhim-kim-friedel} and in 2d tilted nodal rings~\cite{rahmipoor-2d-nlsm} to the 3d case with tilt and gap. This paper is the companion of Ref.~\cite{ips-vnr-plasmon}, which studies the vortex nodal-ring semimetals (and 3-band Hopf semimetals) within the same low-energy and RPA framework. As the regimes above show, the two inequivalent tilt directions of the 3d ring, combined with the mass gap, give a far richer variety of FS topology and plasmon responses than either of these settings. The methods used here can be applied to other topological semimetals with toroidal FSs. Several directions remain open. The momentum dependence of the response beyond the $q\to0$ limit remains to be explored. The effects of electron-electron interactions beyond RPA, disorder, and finite temperature could also be studied. The interplay between tilt and gap in transport properties, such as the optical conductivity, merits further study as well. Such treatments exist for 3d Luttinger semimetals, where the plasmon spectrum has been analysed in the isotropic limit~\cite{ips-plasmon-luttinger-aop} and in anisotropic settings~\cite{ips-plasmon-luttinger-epjb}. Finally, our results may be relevant to experimental systems where nodal rings and their gapped counterparts have been identified~\cite{balents-nodal, fu_nlsm, yang_review_nlsm, schnyder_nodal, ips-nlsm-ph, ips-gnr-strain}.

\appendix

\section{Mapping onto two-dimensional Dirac cones}
\label{appdirac}

We give the steps behind Eq.~\eqref{eq:ringmap}, with $\alpha = 1$ as in Eq.~\eqref{eq:ham3d}. At fixed $\phi$ we use the local coordinates $u = \kappa \cos \gamma = k_\perp - k_0$ and $w = \kappa \sin \gamma = k_z$. Since $d k_x\, d k_y = k_\perp\, d k_\perp\, d\phi$, the measure becomes
\begin{align}
\label{eq:app-measure}
g \int \frac{d^3 \bs{k}}{(2\, \pi)^3} = g\, k_0 \int_0^{2\pi} \frac{d\phi}{2\, \pi} \int \frac{du\, dw}{(2\, \pi)^2} \left( 1 + \frac{u}{k_0} \right) .
\end{align}
The term $u/k_0$ is of order $\kappa/k_0$ and is dropped. This is the replacement $k_0 + \kappa \cos \gamma \to k_0$ of Sec.~\ref{secuntiltednr}.

For $\bs \eta = 0$ the Hamiltonian of Eq.~\eqref{eq:ham3d} reads $\mathcal{H} = v_0\, u\, \sigma_x + v_0\, w\, \sigma_y + \Delta\, \sigma_z$. It does not depend on $\phi$. It describes a single 2d Dirac cone with velocity $v_0$ and mass $\Delta$ in the $(u, w)$ plane. The energy $E_k$ and the angles $\theta$ and $\gamma (\bs k)$ of Eq.~\eqref{eq:gamma3d}, and hence the form-factor $\mathcal{F}$ of Eq.~\eqref{eq:polgeneral}, depend on $\bs k$ only through $(u, w)$. The same holds for $\bs k'$ and $(u', w')$.

For $\bs q = q\, \bs{\hat z}$ we have exactly $\phi' = \phi$, $u' = u$, and $w' = w + q$. For $\bs q = q\, \bs{\hat x}$ we have
\begin{align}
k_\perp' = \sqrt{k_\perp^2 + 2\, k_\perp\, q \cos \phi + q^2} = k_\perp + q \cos \phi + \mathcal{O} \left( q^2/k_0 \right) , \quad
\phi' = \phi + \mathcal{O} \left( q/k_0 \right) .
\end{align}
The shift of $\phi$ only relabels the copy, because $\mathcal{H}$ does not depend on $\phi$. Hence $u' = u + q \cos \phi$ and $w' = w$ at leading order in $q/k_0$. The component of $\bs q$ along the ring therefore changes neither $E_{k'}$ nor $\mathcal{F}$. In both cases the copy at angle $\phi$ is a 2d problem with the local transfer $\bs q_{\rm loc}$. It equals $q\, \bs{\hat w}$ for $\bs q \parallel \bs{\hat z}$ and $q \cos \phi\, \bs{\hat u}$ for $\bs q \parallel \bs{\hat x}$.

Inserting Eq.~\eqref{eq:app-measure} into Eq.~\eqref{eq:polgeneral}, the integrand at fixed $\phi$ is $\Pi_{2\rm D} (\bs q_{\rm loc}, \omega)$, defined with $g \int d^2 \bs k/(2\, \pi)^2$. The 2d cone is isotropic in the $(u, w)$ plane, and $\Pi_{2\rm D}$ therefore depends only on $|\bs q_{\rm loc}|$. This gives Eq.~\eqref{eq:ringmap}. A response that is proportional to $|\bs q_{\rm loc}|^n$ acquires, for $\bs q \parallel \bs{\hat x}$, the factor $\int_0^{2\pi} d\phi\, |\cos \phi|^n/(2\, \pi)$. It equals $1/2$ for $n = 2$ and $2/\pi$ for $n = 1$. The first value gives the factor of two between the $\bs{\hat z}$ and $\bs{\hat x}$ channels quoted for $\Im \Pi$ and $\Pi_{\rm inter}$. The second value turns Eq.~\eqref{eq:pi0final3d} into $\Pi (q\, \bs{\hat x}, 0) = -\, g\, k_0\, q/(8\, \pi\, v_0)$. The same argument applies to the DOS of Eq.~\eqref{eqdos}. For $\bs \eta = \bs 0$ and $\Delta = 0$, the integral $g \int d^2 \bs k\, \delta (v_0\, \kappa - \mu)/(2\, \pi)^2$ equals $g\, \mu/(2\, \pi\, v_0^2)$. Multiplication by $k_0$ gives Eq.~\eqref{eq:dos}.

\section{Pockets at the intrinsic point of the in-plane-tilted PTNR}
\label{app:pockets}

We derive Eq.~\eqref{eq:dpipocketptnr}. For $\bs q = q\, \bs{\hat z}$, Appendix~\ref{appdirac} gives $\phi' = \phi$, $u' = u$, and $w' = w + q$. The tilt does not shift the transition energies, because $\bs \eta \cdot \bs q = 0$. At fixed $\phi$, the bands of Sec.~\ref{sectiltx} read $\varepsilon_\pm = \pm\, v_0\, \kappa + \tilde{\eta} \cos \phi + v_0\, \eta_x \cos \phi\, u$. We drop the tilt along $u$, which modifies the Pauli-blocking boundary at relative order $\eta_x$. The constant offset then acts as the local chemical potential $\mu_{\rm eff} = -\, \tilde{\eta} \cos \phi$, and
\begin{align}
\Pi (q\, \bs{\hat z}, \omega) = k_0 \int_0^{2\pi} \frac{d\phi}{2\pi}\, \Pi_{2\rm D} \left( q, \omega; \mu_{\rm eff} (\phi) \right) ,
\end{align}
where $\Pi_{2\rm D} (Q, \omega; \mu_{\rm eff})$ is the polarisability of a massless 2d Dirac cone with chemical potential $\mu_{\rm eff}$. The cone has the symmetries $\varepsilon_1 = -\, \varepsilon_2$, $\mathcal{F}_{11} = \mathcal{F}_{22}$, and $\mathcal{F}_{12} = \mathcal{F}_{21}$. They map a hole-pocket of the band $s = 1$ onto an electron-pocket of the band $s = 2$ and reverse the sign of $\omega$. The result below is even in $\omega$, and it depends only on $|\mu_{\rm eff}|$. We take $\mu_{\rm eff} > 0$ and write $\Pi_{2\rm D} = \Pi_{2\rm D}^{(0)} + \delta \Pi_{2\rm D}$, where $\Pi_{2\rm D}^{(0)} = -\, g\, Q^2/\big( 16 \sqrt{v_0^2\, Q^2 - \omega^2} \big)$ is the undoped part.

The band $s = 2$ is occupied for $\kappa < \kappa_F = \mu_{\rm eff}/v_0$, and the band $s = 1$ is filled. With $n_{\bs{\kappa}} = \Theta (\kappa_F - \kappa)$, the occupations of Eq.~\eqref{eq:polgeneral} differ from the undoped ones only through $n_{\bs{\kappa}}$. The term with $s = s' = 1$ does not change. The terms $(s, s') = (1, 2)$ and $(2, 1)$ change by $-\, n_{\bs{\kappa}'}\, \mathcal{F}_{12}/(\omega - \epsilon_{\kappa} - \epsilon_{\kappa'})$ and $n_{\bs{\kappa}}\, \mathcal{F}_{12}/(\omega + \epsilon_{\kappa} + \epsilon_{\kappa'})$, respectively. The term $(2, 2)$ is the intraband response of the pocket carriers, with the weight $n_{\bs{\kappa}} - n_{\bs{\kappa}'}$ and $\mathcal{F}_{22}$. After a shift of the integration variable in the terms with $n_{\bs{\kappa}'}$, all four contributions are integrated over the pocket only. We define $\epsilon_\kappa = v_0\, \kappa$, $\epsilon_\kappa^\pm = v_0\, |\bs{\kappa} \pm \bs{Q}|$, and $\mathcal{F}_{ss'}^\pm$ for the overlap between $\bs{\kappa}$ and $\bs{\kappa} \pm \bs{Q}$. Then
\begin{align}
\delta \Pi_{2\rm D} = g \int_{\kappa < \kappa_F} \frac{d^2 \bs{\kappa}}{(2\pi)^2}\, \mathcal{T} (\bs{\kappa}) \,, \quad
\mathcal{T} = \frac{\mathcal{F}_{12}^+}{\omega + \epsilon_\kappa + \epsilon_\kappa^+} - \frac{\mathcal{F}_{12}^-}{\omega - \epsilon_\kappa - \epsilon_\kappa^-}
+ \frac{\mathcal{F}_{22}^+}{\omega + \epsilon_\kappa - \epsilon_\kappa^+} - \frac{\mathcal{F}_{22}^-}{\omega - \epsilon_\kappa + \epsilon_\kappa^-} \,.
\end{align}
The overlaps are $\mathcal{F}_{12} = (1 - \cos \alpha)/2$ and $\mathcal{F}_{22} = (1 + \cos \alpha)/2$, with $\alpha$ the angle between the two momenta. All denominators are nonzero for $\omega < v_0\, Q - 2\, v_0\, \kappa_F$, which is the lower edge of the intraband continuum of the pocket.

We expand $\mathcal{T}$ in $\kappa/Q$ at fixed $\theta$, the angle between $\bs{\kappa}$ and $\bs{Q}$. We write $c_\theta = \cos \theta$, $s_\theta = \sin \theta$, $\omega_+ = \omega + v_0\, Q$, and $\omega_- = \omega - v_0\, Q$. To first order, $\epsilon_\kappa^\pm = v_0\, (Q \pm \kappa\, c_\theta)$, $\cos \alpha_+ = c_\theta + \hat\delta_\theta$, and $\cos \alpha_- = -\, c_\theta + \hat\delta_\theta$, with $\hat\delta_\theta = \kappa\, s_\theta^2/Q$. At $\kappa \to 0$, the four terms are $(1 - c_\theta)/(2\, \omega_+)$, $-\, (1 + c_\theta)/(2\, \omega_-)$, $(1 + c_\theta)/(2\, \omega_-)$, and $-\, (1 - c_\theta)/(2\, \omega_+)$. They cancel in pairs, and $\mathcal{T}$ has no term of order $\kappa^0$. The terms of order $\kappa$ combine to
\begin{align}
\mathcal{T} = s_\theta^2\, \kappa \left[ -\, \frac{1}{Q\, \omega_+} + \frac{1}{Q\, \omega_-} - \frac{v_0}{\omega_+^2} - \frac{v_0}{\omega_-^2} \right]
= -\, \frac{4\, v_0^3\, Q^2 \sin^2 \theta\; \kappa}{\left( v_0^2\, Q^2 - \omega^2 \right)^2} \,,
\end{align}
where we have used $A - B = 2\, v_0\, Q$ and $A\, B = \omega^2 - v_0^2\, Q^2$. The integral over the pocket uses $\int_0^{2\pi} d\theta \sin^2 \theta = \pi$ and $\int_0^{\kappa_F} \kappa^2\, d\kappa = \kappa_F^3/3$. With $v_0\, \kappa_F = |\mu_{\rm eff}|$, we obtain
\begin{align}
\delta \Pi_{2\rm D} (Q, \omega) = -\, \frac{g\, |\mu_{\rm eff}|^3\, Q^2}{3\, \pi \left( v_0^2\, Q^2 - \omega^2 \right)^2} \,.
\end{align}
This is a relative correction $(16/3\pi)\, |\mu_{\rm eff}|^3/(v_0^2\, Q^2 - \omega^2)^{3/2}$ to $\Pi_{2\rm D}^{(0)}$. It is cubic in $|\mu_{\rm eff}|$, because the intraband response of the pocket carriers, of order $|\mu_{\rm eff}|^2/(v_0\, Q)$, cancels against the Pauli-blocking of the interband transitions at that order. At $\omega = 0$, it agrees with the expansion of the static polarisability of doped graphene for $Q \gg 2\, \kappa_F$.

We now insert $\mu_{\rm eff} = -\, \tilde{\eta} \cos \phi$ and $Q = q$, and use $\int_0^{2\pi} d\phi\, |\cos \phi|^3/(2\pi) = 4/(3\pi)$. This gives Eq.~\eqref{eq:dpipocketptnr}, whose ratio to Eq.~\eqref{eq:polmu03dexact} is $64\, |\tilde{\eta}|^3/[9\, \pi^2 \left( v_0^2\, q^2 - \omega^2 \right)^{3/2}]$. The condition $\omega < v_0\, q - 2\, |\tilde{\eta}|$ is the lower edge of the intraband continuum of the largest local pocket.

\section{Classification of the Fermi surface of the in-plane-tilted GNR}
\label{app:fsclass}

This appendix derives the eight cases of Table~\ref{tab:fsclass}, together with the thresholds $\pm \Delta\, \nu_x$ and $\pm \Delta$ and the width $\Delta\, \eta_x^2/2$ of the windows. At fixed $\phi$ and in the local plane $(u, w)$, the condition $\varepsilon_2 = \mu$ reads $E_k + v_0\, t\, u = \mu_{\rm eff}$, with $t = \eta_x \cos\phi$ and $\mu_{\rm eff} = \mu - \tilde{\eta} \cos\phi$. Interestingly, this coincides with the equation of Appendix~\ref{app:gnrztilt} after the substitution, $\eta_z \to t$ and $\mu \to \mu_{\rm eff}$ (i.e., the tilt taken along along $u$ instead of $w$). Squaring it gives an ellipse with the area $\pi\, K/(v_0^2\, \nu_t^3)$, where
\begin{align}
K (\phi) = \mu_{\rm eff}^2 - \Delta^2\, \nu_t^2 \,, \quad \nu_t (\phi) = \sqrt{1 - t^2} \,.
\end{align}
The ellipse is real if and only if $K > 0$. For $\mu_{\rm eff} > 0$ the un-squared condition $E_k > 0$ holds on all of it, by the argument of Appendix~\ref{app:gnrztilt}. It is then the electron FS. For $\mu_{\rm eff} < 0$ the same ellipse, with $\mu_{\rm eff} \to -\mu_{\rm eff}$ and $t \to -t$, is the hole FS. The electron FS therefore exists at $\phi$ if and only if $\mu_{\rm eff} \geq \Delta\, \nu_t$, and the hole FS exists if and only if $\mu_{\rm eff} \leq -\Delta\, \nu_t$. The origin of the local plane, which is the point of the nodal ring, has the energy $\varepsilon_2 = \Delta$. It lies inside the electron ellipse if and only if $\mu_{\rm eff} > \Delta$, and inside the hole ellipse if and only if $\mu_{\rm eff} < -\Delta$. Otherwise the ellipse does not enclose the ring, and a ray from the ring meets it twice or not at all. In the first case the local FS is non-enclosing, as defined in Sec.~\ref{sec:gaptiltz}. This yields the five local classes
\begin{align}
\mu_{\rm eff} \leq -\Delta \,, \quad -\Delta < \mu_{\rm eff} \leq -\Delta\, \nu_t \,, \quad |\mu_{\rm eff}| < \Delta\, \nu_t \,, \quad \Delta\, \nu_t \leq \mu_{\rm eff} < \Delta \,, \quad \mu_{\rm eff} \geq \Delta \,,
\end{align}
which are, in this order, an enclosing hole-pocket, a non-enclosing hole-pocket, a gapped point, a non-enclosing electron-pocket, and an enclosing electron-pocket.

The thresholds depend on $\phi$ only through $\nu_t$. Since $t^2 = \eta_x^2 \cos^2\phi$, we have $\nu_t = \nu_x$ at $\phi = 0, \pi$, where $|t|$ is largest, and $\nu_t = 1$ at $\phi = \pi/2$. The thresholds $\pm \Delta\, \nu_t$ therefore vary between $\pm \Delta\, \nu_x$ and $\pm \Delta$, by at most $\Delta \left( 1 - \nu_x \right)$. On the other hand, $\mu_{\rm eff}$ rises from $a = \mu - \tilde{\eta}$ at $\phi = 0$ to $b = \mu + \tilde{\eta}$ at $\phi = \pi$, with the slope $\tilde{\eta} \sin\phi \geq 0$. The slope of $\Delta\, \nu_t$ is $\Delta\, \eta_x^2 \sin\phi \cos\phi/\nu_t$, which is smaller than $\tilde{\eta} \sin\phi$ in magnitude if $\Delta\, \eta_x \ll \nu_x\, v_0\, k_0$. Under this condition, $\mu_{\rm eff} - \Delta\, \nu_t$, $\mu_{\rm eff} - \Delta$, and $\mu_{\rm eff} + \Delta\, \nu_t$ are all monotonic on $[0, \pi]$. Each of them changes sign at most once, so that each local class occupies a single arc. The sequence of classes along the ring is then fixed by the classes at the two ends, $\phi = 0$ with the value $a$ and $\phi = \pi$ with the value $b$, at both of which the threshold is $\Delta\, \nu_x$.

For $\mu \geq 0$ we have $a \geq -b$ and $a \leq b$. At $\phi = \pi$ there are three possibilities. For $b < \Delta\, \nu_x$ no pocket exists there, and $a \geq -b > -\Delta\, \nu_x$ excludes a hole-pocket at $\phi = 0$ as well. This is case 1. For $\Delta\, \nu_x \leq b < \Delta$ there is a non-enclosing electron-pocket at $\phi = \pi$. Here $a \geq -b > -\Delta$ allows only a non-enclosing hole-pocket at $\phi = 0$, which exists for $a \leq -\Delta\, \nu_x$. This gives cases 3 and 2, respectively. For $b \geq \Delta$ the electron-pocket encloses the ring at $\phi = \pi$, and the class at $\phi = 0$ is free. The five local classes at $\phi = 0$ then give cases 4 to 8 in the order of increasing $a$. For $|a| < \Delta\, \nu_x$ there is a gapped arc, with the electron-pocket ending in a non-enclosing rim (case 4). For $-\Delta < a \leq -\Delta\, \nu_x$ a non-enclosing hole-pocket lies in the gapped arc (case 5). For $a \leq -\Delta$ the hole-pocket encloses the ring and is separated from the electron-pocket by the gapped arc (case 6). For $\Delta\, \nu_x \leq a < \Delta$ the electron-pocket runs around the whole ring, but it does not enclose it near $\phi = 0$ (case 7). For $a \geq \Delta$ it encloses the ring everywhere (case 8). These are the eight cases of Table~\ref{tab:fsclass}.

The cases are separated by the lines $b = \Delta\, \nu_x$, $b = \Delta$, $a = \pm \Delta\, \nu_x$, and $a = \pm \Delta$. These form pairs of nearby lines, $\left( \Delta\, \nu_x, \Delta \right)$ in $b$ and in $a$, and $\left( -\Delta, -\Delta\, \nu_x \right)$ in $a$, whose separation is
\begin{align}
\Delta \left( 1 - \nu_x \right) = \Delta \left( 1 - \sqrt{1 - \eta_x^2} \right) \simeq \frac{\Delta\, \eta_x^2}{2} \,.
\end{align}
The strips between these lines are the windows, namely cases 2 and 3 for $b$, case 5 for $a$ near $-\Delta$, and case 7 for $a$ near $+\Delta$. Cases 2 and 3 share the same strip in $b$ and are told apart by $a$. To leading order in $\eta_x$ each pair of lines merges into one, at $b = \Delta$, $a = -\Delta$, and $a = \Delta$, respectively. The strips then close, and case 4 extends over $|a| < \Delta$. The four regimes remain, namely the fully-insulating regime ($b < \Delta$, case 1), the twin-crescent regime ($a \leq -\Delta$, $b \geq \Delta$, case 6), the crescent-cyclide regime ($|a| < \Delta \leq b$, cases 4 and 5), and the fully-conducting regime ($a \geq \Delta$, cases 7 and 8), as in Table~\ref{tab:fsclass}.

Finally, the non-enclosing rims arise from the same offset between the two thresholds. The local FS changes from non-enclosing to enclosing where $\mu_{\rm eff}$ crosses $\Delta$, whereas it appears where $\mu_{\rm eff}$ crosses $\Delta\, \nu_t$. The two crossings are separated by $\Delta \left( 1 - \nu_t \right)$ in $\mu_{\rm eff}$, which corresponds to the angular width
\begin{align}
\delta\phi \simeq \frac{\Delta \left( 1 - \nu_t \right)}{\tilde{\eta}\, |\sin\phi|} \leq \frac{\Delta \left( 1 - \nu_x \right)}{\tilde{\eta}\, |\sin\phi|} \,,
\end{align}
where $1 - \nu_t \simeq \eta_x^2 \cos^2\phi/2$ gives the leading form quoted in Sec.~\ref{secfermisurface}. The width is small except near $\phi = 0, \pi$, where the rims merge into the small pockets of Appendix~\ref{app:windows}, and it vanishes where the crossing occurs at $\phi = \pi/2$.

\section{Role of tilt in the denominators of the interband pieces}
\label{app:tiltshift}

The tilt term $v_0\, \bs \eta \cdot \bs k\; \mathbb{I}_{2\times 2}$ of Eq.~\eqref{eq:ham3d} is proportional to the identity. It leaves the eigenstates and hence the form-factor $\mathcal{F}$ of Eq.~\eqref{eq:polgeneral} unchanged. It only shifts the energies. With Eq.~\eqref{eq:ev3d} and $\bs k' = \bs k + \bs q$, the energy differences of the two interband terms are
\begin{align}
\varepsilon_1 (\bs k) - \varepsilon_2 (\bs k') = -\left( E_k + E_{k'} \right) - \tilde a \,, \quad
\varepsilon_2 (\bs k) - \varepsilon_1 (\bs k') = E_k + E_{k'} - \tilde a \,, \quad
\tilde a \equiv v_0\, \bs \eta \cdot \bs q \,.
\end{align}
For a filled band $s=1$ and an empty band $s=2$, the occupation differences are $+1$ and $-1$, respectively. This gives Eq.~\eqref{eq:polinter}. Its two denominators are $\omega - E_k - E_{k'} - \tilde a$ and $\omega + E_k + E_{k'} - \tilde a$. Both depend on $\omega$ and $\tilde a$ only through $\omega - \tilde a$. Hence, whenever the occupation pattern is that of the untilted system,
\begin{align}
\label{eq:app-shift}
\Pi_{\rm inter} (\bs q, \omega) \big|_{\bs \eta} = \Pi_{\rm inter} (\bs q, \omega - \tilde a) \big|_{\bs \eta = 0} \,.
\end{align}

Two cases follow. When $\bs q \perp \bs \eta$ we have $\tilde a = 0$, and the tilt drops out of $\Pi_{\rm inter}$ exactly. This applies to $\bs q = q\, \bs{\hat z}$ for the in-plane tilt $\bs \eta = \eta_x\, \bs{\hat x}$, and to $\bs q = q\, \bs{\hat x}$ for the axial tilt $\bs \eta = \eta_z\, \bs{\hat z}$. When $\bs q \parallel \bs \eta$ we have $\tilde a = v_0\, \eta\, q$, and the response is the untilted one at the shifted frequency $\Omega = \omega - \tilde a$.

In the static limit the two terms of Eq.~\eqref{eq:polinter} combine to
\begin{align}
-\, \frac{1}{E_k + E_{k'} + \tilde a} - \frac{1}{E_k + E_{k'} - \tilde a} = -\, \frac{2}{E_k + E_{k'}} \left[ 1 + \frac{\tilde a^2}{\left( E_k + E_{k'} \right)^2} + \mathcal{O} \left( \tilde a^4 \right) \right] ,
\end{align}
which is even in $\tilde a$. The terms linear in $\tilde a$ cancel. Since $\mathcal{F}$ is already of order $q^2$, the tilt affects $\Pi (q\, \bs{\hat z}, 0)$ of the fully-insulating GNR only at order $q^4$, with a relative size of order $(v_0\, \eta_z\, q/\Delta)^2$. At $q \to 0$ we have $E_{k'} \to E_k$, and the first term in the bracket reproduces Eq.~\eqref{eq:polinterstatic}.

For the imaginary part at $\bs q \parallel \bs \eta$, we start from the untilted result at the frequency $\Omega$. For the axial tilt of the PTNR, Eq.~\eqref{eq:polmu03dexact} continues analytically to $\Im \Pi (q\, \bs{\hat z}, \Omega) = -\, g\, k_0\, q^2/\big( 16\, \sqrt{\Omega^2 - v_0^2\, q^2} \big)$. Expanding in $q$ gives
\begin{align}
\frac{1}{\sqrt{\Omega^2 - v_0^2\, q^2}} = \frac{1}{\Omega} \left[ 1 + \mathcal{O} \left( \frac{v_0^2\, q^2}{\Omega^2} \right) \right] , \quad
\frac{1}{\Omega} = \frac{1}{\omega} \left[ 1 + \frac{\tilde a}{\omega} + \mathcal{O} \left( \frac{\tilde a^2}{\omega^2} \right) \right] .
\end{align}
To first order in $q/\omega$ only the second expansion contributes, and we obtain $\Im \Pi (q\, \bs{\hat z}, \omega) \simeq -\, g\, k_0\, q^2 \left( 1 + v_0\, \eta_z\, q/\omega \right)/(16\, \omega)$.

For the in-plane tilt, $k_x' - k_x = q$ for every $\phi$. The shift $\tilde a = v_0\, \eta_x\, q$ is therefore the same on all copies. The local transfer is $q \cos \phi$, and $\Im \Pi_{2\rm D}$ at the frequency $\Omega$ is proportional to $q^2 \cos^2 \phi/\sqrt{\Omega^2 - v_0^2\, q^2 \cos^2 \phi}$ above the local threshold. For $\Omega \gg v_0\, q$ the square root is $\Omega$ up to relative corrections of order $v_0^2\, q^2/\Omega^2$, and $\int_0^{2\pi} d\phi\, \cos^2 \phi/(2\, \pi) = 1/2$. Hence $\Im \Pi (q\, \bs{\hat x}, \omega) \simeq -\, g\, k_0\, q^2/(32\, \Omega)$, and the expansion of $1/\Omega$ above gives the factor $1 + v_0\, \eta_x\, q/\omega$.

\section{Monotonicity of the real part of the dielectric function}
\label{app:monotone}

This appendix proves the statements made in the main text on the uniqueness of the zero of $\Re \varepsilon^{\rm RPA}$ and on the static screened frequency. We consider $v_0\, q \ll \omega < \omega_{\rm th}$ along a principal axis $\bs{\hat e}_i$, where the tilt-induced shift $v_0\, \bs \eta \cdot \bs q$ of the frequency is negligible. Eq.~\eqref{eq:polgeneral}, with $\mathcal{F} \simeq q_j\, q_l\, g_{jl}$ and the denominators evaluated at $q = 0$, gives
\begin{align}
\Re \varepsilon^{\rm RPA} (q\, \bs{\hat e}_i, \omega) = 1 - \frac{\omega_{p,i}^2}{\omega^2} + \varepsilon_{{\rm inter}, i} (\omega) \,, \quad
\varepsilon_{{\rm inter}, i} (\omega) = \frac{4\, \pi\, e^2\, g}{q^2} \int \frac{d^3 \bs k}{(2\, \pi)^3}\, q_j\, q_l\, g_{jl} (\bs k)\, \frac{4\, E_k}{4\, E_k^2 - \omega^2} \,,
\end{align}
where the integral runs over the unblocked states, for which $f_0 (\varepsilon_1) - f_0 (\varepsilon_2) = 1$. By definition of $\omega_{\rm th}$, every unblocked state has $2\, E_k \geq \omega_{\rm th} > \omega$. The weight $q_j\, q_l\, g_{jl}$ is the squared norm of the change of the unit vector $\hat{\bs d}$ and is therefore non-negative. It does not depend on $\omega$.

Each term of the integrand is positive, and its derivative with respect to $\omega$ equals $8\, E_k\, \omega/(4\, E_k^2 - \omega^2)^2$, which is also positive. Hence $\varepsilon_{{\rm inter}, i}$ increases with $\omega$ on $(0, \omega_{\rm th})$. The Drude term $-\, \omega_{p,i}^2/\omega^2$ increases as well. Therefore $\Re \varepsilon^{\rm RPA}$ is strictly increasing below the onset. It tends to $-\infty$ as $\omega \to 0$, since $\omega_{p,i} \neq 0$ and $\varepsilon_{{\rm inter}, i}$ stays finite. It can thus have at most one zero below $\omega_{\rm th}$. The zero exists if and only if $\Re \varepsilon^{\rm RPA} > 0$ in the limit $\omega \to \omega_{\rm th}^-$. This holds automatically wherever $\varepsilon_{{\rm inter}, i}$ diverges to $+\infty$ at the onset, namely in the untilted doped systems, the insulating and gapped-arc regimes, and the displaced-torus regime.

For the upper bound, note that each term of the integrand is at least as large as its value at $\omega = 0$. Hence $\varepsilon_{{\rm inter}, i} (\omega) \geq \varepsilon_{{\rm inter}, i} (0) = \epsilon_{b,i} - 1$, with $\epsilon_{b,i}$ the static background dielectric constant. It follows that
\begin{align}
\Re \varepsilon^{\rm RPA} (q\, \bs{\hat e}_i, \omega) \geq \epsilon_{b,i} 
- \frac{\omega_{p,i}^2}{\omega^2} \,,
\end{align}
and the right-hand side vanishes at $\tilde\omega_{p,i}^2 = \omega_{p,i}^2/\epsilon_{b,i}$. Since $\Re \varepsilon^{\rm RPA}$ is increasing and non-negative at $\tilde\omega_{p,i}$, its zero cannot exceed $\tilde\omega_{p,i}$. The static screened frequency is therefore an upper bound of the exact zero. The argument does not require $\epsilon_{b,i}$ to be independent of $\omega$.

\section{In-plane-tilted PTNR and GNR}
\label{app:xtilt}

\subsection{Ring-cyclide regime for PTNR}
\label{app:xtiltdrude}

This appendix gives the expansion behind Eqs.~\eqref{eq:dostiltx} and \eqref{eq:dzzx}. We write ${\tilde x} = \cos \gamma$ and ${\tilde y} = \cos \phi$, and $\langle \cdots \rangle = (4\, \pi^2)^{-1} \int_0^{2\pi} d\phi \int_0^{2\pi} d\gamma\, (\cdots)$. Odd powers of ${\tilde x}$ and ${\tilde y}$ average to zero, and $\langle {\tilde x}^2 \rangle = \langle {\tilde y}^2 \rangle = 1/2$ and $\langle {\tilde x}^4 \rangle = \langle {\tilde y}^4 \rangle = 3/8$. In the ring-cyclide regime only the band $s=2$ contributes, with $\zeta \equiv \zeta_2 = 1 + \eta_x\, {\tilde x}\, {\tilde y} > 0$. Let $J = \kappa_2 \left( k_0 + \kappa_2 \,{\tilde x} \right)/\zeta$ be the radial integral of the measure with the delta function of Eq.~\eqref{eq:jacobian}, up to the factor $1/v_0$. The DOS and the Drude weights are then
\begin{align}
\rho (\mu) = \frac{g}{2\, \pi\, v_0} \langle J \rangle \,, \quad
D_{ij} = \frac{g\, e^2\, v_0}{2} \left\langle J\, \frac{V_i^{(2)}\, V_j^{(2)}}{v_0^2} \right\rangle ,
\end{align}
where the prefactors follow from $\pi\, g\, e^2\, v_0^2 \cdot 4\, \pi^2/
\left[ (2\, \pi)^3\, v_0 \right]$ for $D_{ij}$, and similarly for $\rho$. With Eq.~\eqref{eq:cyclide}, $\kappa_2 = 
\left( \mu - \tilde{\eta}\, {\tilde y} \right)/(v_0\, \zeta)$. Writing $\tilde{\eta} = v_0\, k_0\, \eta_x$ and $\tilde\varepsilon \equiv \mu/(v_0\, k_0)$, we have
\begin{align}
J = \frac{k_0\, \mu}{v_0} \left\{ \frac{1 - \eta_x\, {\tilde y}/\tilde\varepsilon}{\zeta^2} + \frac{{\tilde x} \left[ \tilde\varepsilon - 2\, \eta_x\, {\tilde y} + \eta_x^2\, {\tilde y}^2/\tilde\varepsilon \right]}{\zeta^3} \right\} .
\end{align}
The weights $V_i^{(2)} V_j^{(2)}/v_0^2$ at $\Delta = 0$ follow from Eq.~\eqref{eq:velocity3d}. They are $1 - {\tilde x}^2$ for $zz$, $\left( {\tilde x}\, {\tilde y} + \eta_x \right)^2$ for $xx$, and ${\tilde x}^2 \left( 1 - {\tilde y}^2 \right)$ for $yy$. The weight $1$ gives the DOS. All of them, and $\zeta$, are invariant under $({\tilde x}, {\tilde y}) \to (-{\tilde x}, -{\tilde y})$, that is, under $\gamma \to \gamma + \pi$ and $\phi \to \phi + \pi$. The terms $\eta_x\, {\tilde y}/\zeta^2$, $\tilde\varepsilon\, {\tilde x}/\zeta^3$, and $\eta_x^2\, {\tilde x}\, {\tilde y}^2/\zeta^3$ are odd under this map and average to zero. This removes every term proportional to $1/\tilde\varepsilon$ at once. It also explains why no factor of $\tilde{\eta}^2/\mu$ survives. The remaining part is
\begin{align}
J \to \frac{k_0\, \mu}{v_0}\, j \,, \quad j 
= \frac{1}{\zeta^2} - \frac{2\, t}{\zeta^3} = \frac{1 - t}{\left( 1 + t \right)^3} 
= 1 - 4\, t + 9\, t^2 + \mathcal{O} \left( t^3 \right) , \quad t =
 \eta_x\, {\tilde x}\, {\tilde y} \,.
\end{align}
The first term of $j$ comes from the piece proportional to $k_0\, \kappa_F$ of the measure, and the second from the piece proportional to $\kappa_F^2 \cos \gamma$ in Sec.~\ref{sectiltx}. The expansions $1/\zeta^2 = 1 - 2\, t + 3\, t^2 + \mathcal{O} (t^3)$ and $1/\zeta^3 = 1 - 3\, t + 6\, t^2 + \mathcal{O} (t^3)$ give $j = 1 - 2\, t + 3\, t^2 - 2\, t \left( 1 - 3\, t \right) = 1 - 4\, t + 9\, t^2$ to this order. We need $\langle {\tilde x}^2 \, {\tilde y}^2 \rangle = 1/4$, $\langle {\tilde x}^4 \, {\tilde y}^2 \rangle = 3/16$, and $\langle {\tilde x}^4 \, {\tilde y}^4 \rangle = 9/64$. For the DOS,
\begin{align}
\langle j \rangle = 1 + 9\, \eta_x^2\, \langle {\tilde x}^2 \,{\tilde y}^2 \rangle 
= 1 + \frac{9\, \eta_x^2}{4} \,,
\end{align}
where the contributions of $1/\zeta^2$ and of $-2\, t/\zeta^3$ are $3\, \eta_x^2/4$ and $3\, \eta_x^2/2$, respectively. This gives Eq.~\eqref{eq:dostiltx}. For the Drude weights,
\begin{align}
& \langle j \left( 1 - {\tilde x}^2 \right) \rangle = \frac{1}{2} + 9\, \eta_x^2 
\left( \langle {\tilde x}^2 \, {\tilde y}^2 \rangle - \langle {\tilde x}^4 \, {\tilde y}^2 \rangle \right) 
= \frac{1}{2} \left( 1 + \frac{9\, \eta_x^2}{8} \right) , \nn
& \langle j \left( {\tilde x}\, {\tilde y} + \eta_x \right)^2 \rangle 
= \left( \frac{1}{4} + \eta_x^2 \right) - 8\, \eta_x^2\, \langle {\tilde x}^2 \, {\tilde y}^2 \rangle + 9\, \eta_x^2\, \langle {\tilde x}^4 \,{\tilde y}^4 \rangle 
= \frac{1}{4} \left( 1 + \frac{17\, \eta_x^2}{16} \right) , \nn
& \langle j\, {\tilde x}^2 \left( 1 - {\tilde y}^2 \right) \rangle 
= \langle {\tilde x}^2 \left( 1 - {\tilde y}^2 \right) \rangle + 9\, \eta_x^2 \left( \langle {\tilde x}^4 \rangle \langle {\tilde y}^2 \rangle - \langle {\tilde x}^4 \rangle \langle {\tilde y}^4 \rangle \right) 
= \frac{1}{4} \left( 1 + \frac{27\, \eta_x^2}{16} \right) .
\end{align}
In the second line, the term $-4\, t$ of $j$ contributes $-4\, \eta_x \langle {\tilde x}\, {\tilde y} \left( {\tilde x}\, {\tilde y} + \eta_x \right)^2 \rangle = -8\, \eta_x^2 \langle {\tilde x}^2\, {\tilde y}^2 \rangle$, and in the third line it is odd and vanishes. Multiplying by $g\, e^2\, v_0/2 \cdot k_0\, \mu/v_0$ gives Eq.~\eqref{eq:dzzx}. The corresponding Drude weights of the fully-conducting GNR are obtained from the $\phi$-integrals of Appendix~\ref{app:phiint}.

\subsection{Horn-cyclide regime of PTNR}
\label{app:horn}

This appendix gives the steps behind Eqs.~\eqref{eq:horncyclide} and \eqref{eq:horncorr}, the interband corrections of Eq.~\eqref{eq:imhorncorr}, and the small-frequency limits quoted for the horn-cyclide regime in Sec.~\ref{sectiltx}. We use the notation of Appendix~\ref{app:xtiltdrude}, namely ${\tilde x} = \cos\gamma$ and ${\tilde y} = \cos\phi$. The band index is ${\tilde s} = +1$ for $s=2$ and ${\tilde s} = -1$ for $s=1$. The band energy is $\varepsilon_s = {\tilde s}\, v_0\, \kappa\, \zeta_{\tilde s} + \tilde{\eta}\, {\tilde y}$ with $\zeta_{\tilde s} = 1 + {\tilde s}\, \eta_x\, {\tilde x}\, {\tilde y}$, and the root of $\varepsilon_s = \mu$ is
\begin{align}
\kappa_{\tilde s} = \frac{ {\tilde s}\, \mu_{\rm eff} }{ v_0\, \zeta_{\tilde s} } \,, 
\quad \mu_{\rm eff} = \mu - \tilde{\eta}\, {\tilde y} \,.
\end{align}
Since $\zeta_{\tilde s} > 0$ for $|\eta_x| < 1$, the sign of $\kappa_{\tilde s}$ is that of ${\tilde s}\, \mu_{\rm eff}$. The electron-pocket therefore occupies the arc $\mu_{\rm eff} > 0$, that is $\phi \in (\phi_b, 2\, \pi - \phi_b)$, and the hole-pocket occupies the arc $|\phi| < \phi_b$. These arcs are independent of $\gamma$ and of $\eta_x$. The radial integral of the measure with the delta function is $J_{\tilde s} = \kappa_{\tilde s} \left( k_0 + \kappa_{\tilde s}\, {\tilde x} \right)/\zeta_{\tilde s}$, up to the factor $1/v_0$, as in Appendix~\ref{app:xtiltdrude}. The weights $V_i^{(s)} \, V_j^{(s)}/v_0^2$ at $\Delta = 0$ are $1 - {\tilde x}^2$ for $zz$, $\left( {\tilde x}\, {\tilde y} + {\tilde s}\, \eta_x \right)^2$ for $xx$, and ${\tilde x}^2 \left( 1 - {\tilde y}^2 \right)$ for $yy$. Let $\langle \cdots \rangle_\gamma$ denote the average over $\gamma$. We introduce $\hat{\mu} = \mu/(v_0\, k_0)$ and $\hat{m} = \mu_{\rm eff}/(v_0\, k_0) = \hat{\mu} - \eta_x\, {\tilde y}$, so that $\tilde{\eta}/(v_0\, k_0) = \eta_x$. Then
\begin{align}
\label{eq:app-horn-prefactors}
\rho (\mu) = \frac{g\, k_0^2}{4\, \pi^2\, v_0}\, I_1 \,, \quad
D_{ii} = \frac{g\, e^2\, v_0\, k_0^2}{4\, \pi}\, I_i \,, \quad
I_a = \int_0^{2\pi} \! d\phi\, \mathrm{sgn} (\hat{m})\, \check{\mathcal{J}}_a ({\tilde y}) \,.
\end{align}
Here ${\tilde s}\, \check{\mathcal{J}}_a ({\tilde y})$ is the sum of $\langle J_{\tilde s}\, \check w_a \rangle_\gamma$ over the pockets, with $\check w_1 = 1$, $\check w_i$ the weights above, and $\mathrm{sgn} (\hat{m}) = {\tilde s}$ on the arc of the pocket ${\tilde s}$. The prefactors follow as in Appendix~\ref{app:xtiltdrude}.

The hole-pocket needs no separate calculation. In the variables $\hat{m}$ and $t = \eta_x\, {\tilde x}\, {\tilde y}$, we have $J_+ = \hat{m}/(1 + t)^2 + \hat{m}^2\, {\tilde x}/(1 + t)^3$ and $J_- = -\, \hat{m}/(1 - t)^2 + \hat{m}^2\, {\tilde x}/(1 - t)^3$, using $|\mu_{\rm eff}| = -\, \mu_{\rm eff}$ on the hole arc. Hence $J_- ({\tilde x}) = -\, J_+ (-{\tilde x})$. The same map takes the weights of the hole-pocket into those of the electron-pocket, because $\left( {\tilde x}\, {\tilde y} - \eta_x \right)^2 = \left( -\, {\tilde x}\, {\tilde y} + \eta_x \right)^2$. The average over $\gamma$ is invariant under ${\tilde x} \to -{\tilde x}$, which maps $\gamma \to \pi - \gamma$. We therefore have $\langle J_-\, \check w_a \rangle_\gamma = -\, \langle J_+\, \check w_a \rangle_\gamma$ exactly, and this yields the form of $I_a$ in Eq.~\eqref{eq:app-horn-prefactors}. Here $\check{\mathcal{J}}_a ({\tilde y}) \equiv \langle J_+\, \check w_a \rangle_\gamma$, and the pocket ${\tilde s}$ contributes ${\tilde s}\, \check{\mathcal{J}}_a ({\tilde y})$ on its arc.

At leading order in $\eta_x$, the functions are $\check{\mathcal{J}}_1 = \hat{m}$, $\check{\mathcal{J}}_{zz} = \hat{m}/2$, $\check{\mathcal{J}}_{xx} = \hat{m}\, {\tilde y}^2/2$, and $\check{\mathcal{J}}_{yy} = \hat{m}\, (1 - {\tilde y}^2)/2$, which follow from $\langle {\tilde x}^2 \rangle_\gamma = 1/2$. All the $\phi$ integrals are then combinations of the moments $\tilde M_n = \int_0^{2\pi} d\phi\, \mathrm{sgn} (\hat{m})\, \cos^n \phi$. Writing $\tilde M_n = 2\pi \langle \cos^n \phi \rangle - 2 \int_{-\phi_b}^{\phi_b} d\phi\, \cos^n \phi$, because the integrand changes sign on the hole arc, and using $\cos\phi_b = \mu/\tilde{\eta} \equiv c$, $\sin\phi_b = \bar{w}/\tilde{\eta} \equiv \varsigma$, and $\theta = \arcsin c = \pi/2 - \phi_b$, we find
\begin{align}
\label{eq:app-horn-moments}
& \tilde M_0 = 4\, \theta \,, \quad \tilde M_1 = -\, 4\, \varsigma \,, \quad 
\tilde M_2 = 2\, \theta - 2\, \varsigma\, c \,, \quad 
\tilde M_3 = -\, \frac{4}{3}\, \varsigma \left( 3 - \varsigma^2 \right) , \quad
\tilde M_4 = \frac{3}{2}\, \theta - \varsigma\, c \left( \frac{5}{2} - \varsigma^2 \right) , 
\nn & \tilde M_5 = -\, \frac{4}{15}\, \varsigma \left( 15 - 10\, \varsigma^2 + 3\, \varsigma^4 \right) .
\end{align}
The first moments follow directly, $\tilde M_0 = 2\, \pi - 4\, \phi_b = 4\, \theta$, $\tilde M_1 = -\, 4 \sin\phi_b = -\, 4\, \varsigma$, and $\tilde M_2 = \pi - 2 \left( \phi_b + \sin\phi_b \cos\phi_b \right) = 2\, \theta - 2\, \varsigma\, c$. For $\tilde M_3$ we use $\int_{-\phi_b}^{\phi_b} d\phi\, \cos^3 \phi = 2 \left( \varsigma - \varsigma^3/3 \right)$, and $\tilde M_4$ follows from $\int d\phi\, \cos^4 \phi = 3\, \phi/8 + \sin (2\, \phi)/4 + \sin (4\, \phi)/32$ together with $\sin (4\, \phi_b) = 4\, \varsigma\, c \left( 1 - 2\, \varsigma^2 \right)$. The odd moments vanish over the full circle, and $\tilde M_5$ is obtained like $\tilde M_3$.
With $\hat{\mu} = \eta_x\, c$ and $\Lambda = \varsigma + c\, \theta$, the leading-order integrals are
\begin{align}
I_1^{(0)} = \eta_x \left( c\, \tilde M_0 - \tilde M_1 \right) = 4\, \eta_x\, \Lambda \,, \quad
I_{xx}^{(0)} = \frac{\eta_x}{2} \left( c\, \tilde M_2 - \tilde M_3 \right) = \eta_x \left( \Lambda + \frac{\varsigma^3}{3} \right) ,
\end{align}
together with $I_{zz}^{(0)} = I_1^{(0)}/2$ and $I_{yy}^{(0)} = I_1^{(0)}/2 - I_{xx}^{(0)}$. Inserting them into Eq.~\eqref{eq:app-horn-prefactors} and using $\tilde{\eta}\, \Lambda = \bar{w} + \mu \arcsin (\mu/\tilde{\eta})$ gives Eq.~\eqref{eq:horncyclide}.

For the corrections, we expand $J_+$ and the weights in $t$ and $\eta_x$ while counting $\hat{\mu}$ as being of order $\eta_x$, which holds in the horn-cyclide regime. Keeping all terms up to third order in $\eta_x$,
\begin{align}
J_+ = \hat{m} \left( 1 - 2\, t + 3\, t^2 \right) + \hat{m}^2\, {\tilde x} \left( 1 - 3\, t \right) .
\end{align}
The terms odd in ${\tilde x}$ vanish after the $\gamma$ average, and $\langle {\tilde x}^2 \rangle_\gamma = 1/2$ and $\langle {\tilde x}^4 \rangle_\gamma = 3/8$ suffice. Unlike in Appendix~\ref{app:xtiltdrude}, the average over $\phi$ is restricted by the arcs, and the terms odd under ${\tilde y} \to -{\tilde y}$ do not cancel. They produce the terms in $\varsigma^3$ and $\varsigma^5$ of Eq.~\eqref{eq:horncorr}. The $\gamma$-averaged functions are
\begin{align}
& \check{\mathcal{J}}_1 = \hat{m} \left( 1 + \frac{3\, \eta_x^2\, {\tilde y}^2}{2} \right) 
- \frac{3}{2}\, \eta_x\, {\tilde y}\, \hat{m}^2 \,, \quad
\check{\mathcal{J}}_{zz} = \frac{\hat{m}}{2} 
\left( 1 + \frac{3\, \eta_x^2\, {\tilde y}^2}{4} \right) - \frac{3}{8}\, \eta_x\, {\tilde y}\, \hat{m}^2 \,, \nn
& \check{\mathcal{J}}_{xx} = \hat{m} \left[ \frac{{\tilde y}^2}{2} + \eta_x^2 \left( 1 - 2\, {\tilde y}^2 + \frac{9\, {\tilde y}^4}{8} \right) \right] + \eta_x\, {\tilde y} \left( 1 - \frac{9\, {\tilde y}^2}{8} \right) \hat{m}^2 \,, \quad
\check{\mathcal{J}}_{yy} = \hat{m} \left[ \frac{1 - {\tilde y}^2}{2} + \frac{9\, \eta_x^2\, {\tilde y}^2 
\left( 1 - {\tilde y}^2 \right)}{8} \right] 
- \frac{9}{8}\, \eta_x\, {\tilde y} \left( 1 - {\tilde y}^2 \right) \hat{m}^2 \,.
\end{align}
As an example, for $\check w_1 = 1$ and $\check w_{zz} = 1 - {\tilde x}^2$ the terms of $J_+$ that are even in ${\tilde x}$ are $\hat{m} \left( 1 + 3\, \eta_x^2\, {\tilde y}^2\, {\tilde x}^2 \right)$ and $-\, 3\, \eta_x\, {\tilde y}\, {\tilde x}^2\, \hat{m}^2$. With $\langle {\tilde x}^2 \rangle_\gamma = 1/2$ and $\langle {\tilde x}^2 \left( 1 - {\tilde x}^2 \right) \rangle_\gamma = 1/8$, they give $\check{\mathcal{J}}_1$ and $\check{\mathcal{J}}_{zz}$ above. The functions $\check{\mathcal{J}}_{xx}$ and $\check{\mathcal{J}}_{yy}$ follow in the same way, with the weights $\left( {\tilde x}\, {\tilde y} + \eta_x \right)^2$ and ${\tilde x}^2 \left( 1 - {\tilde y}^2 \right)$.

Each is a polynomial in ${\tilde y}$ after inserting $\hat{m} = \hat{\mu} - \eta_x\, {\tilde y}$. The integrals $I_a$ then follow from the moments of Eq.~\eqref{eq:app-horn-moments}, with $\hat{\mu} = \eta_x\, c$. Collecting terms and using $\varsigma^2 = 1 - c^2$, we obtain the results of Eq.~\eqref{eq:horncorr}. For example, the DOS is $I_1 = 4\, \eta_x \left[ \left( 1 + 9\, \eta_x^2/4 \right) \Lambda - \eta_x^2\, \varsigma^3/4 \right]$. At $\mu = \tilde{\eta}$, we have $c = 1$, $\varsigma = 0$, and $\Lambda = \pi/2$, and the results reduce to the ring-cyclide corrections of Appendix~\ref{app:xtiltdrude}.

We now turn to the interband part. At small frequencies, the two arguments $a_\omega$ and $b_\omega$ of Eq.~\eqref{eq:imhorn} approach $c \mp \omega/(2\, \tilde{\eta})$, and
\begin{align}
F_0 = \frac{\arcsin b_\omega - \arcsin a_\omega}{\pi} \simeq \frac{1}{\pi}\, \frac{\omega/\tilde{\eta}}{\sqrt{1 - c^2}} = \frac{\omega}{\pi\, \bar{w}} \,.
\end{align}
For the real part, we consider $R_z (\omega) = \int_0^{2\pi} \frac{d\phi}{2\pi}\, \ln \!\left| \frac{\omega + 2\, |\mu_{\rm eff}|}{\omega - 2\, |\mu_{\rm eff}|} \right|$ at $\omega \ll \tilde{\eta}$. Near each crossing $\phi = \pm \phi_b$, we have $|\mu_{\rm eff}| \simeq \bar{w}\, |\upsilon|$ with $\upsilon = \phi \mp \phi_b$. Away from the crossings, the logarithm is approximately $\omega/|\mu_{\rm eff}|$, and the $\phi$ integral converges. The two crossings and the two signs of $\upsilon$ give four equal contributions, and with $u = 2\, \bar{w}\, \upsilon/\omega$
\begin{align}
R_z \simeq \frac{4}{2\, \pi}\, \frac{\omega}{2\, \bar{w}} \int_0^{U} \! du\, \ln \!\left| \frac{1 + u}{1 - u} \right| , \quad
\int_0^{U} \! du\, \ln \!\left| \frac{1 + u}{1 - u} \right| = \left( 1 + U \right) \ln \!\left( 1 + U \right) - \left( U - 1 \right) \ln \!\left( U - 1 \right) \simeq 2 \ln \!\left( U \right) + 2 \,.
\end{align}
The antiderivatives are $\int du\, \ln \!\left( 1 + u \right) = (1 + u) \ln \!\left( 1 + u \right) - u$ and $\int du\, \ln \!\left| 1 - u \right| = -\left( 1 - u \right) \ln \!\left| 1 - u \right| - u$, and for $U \gg 1$ we use $\ln \!\left( U \pm 1 \right) \simeq \ln \!\left( U \right) \pm 1/U$.
The upper limit $U$ is proportional to $\bar{w}/\omega$, with a coefficient set by the global shape of the integrand. Matching the logarithmic region to the rest of the $\phi$ integral therefore gives $R_z \simeq 2\, \omega \left[ \ln \!\left( \tilde{\eta}/\omega \right) + c_0 \right]/(\pi\, \bar{w})$, with a real constant $c_0$ of order unity. This is the constant denoted $c$ in Sec.~\ref{sectiltx}.

The leading corrections follow from the exact reduction of the interband integral. For $\bs q = q\, \bs{\hat x}$ the resonance condition $\kappa + \kappa' = \Omega$ can be solved in closed form. The Pauli edges then shift with the poloidal angle $\gamma$ through the terms $v_0\, \eta_x\, \kappa \cos \gamma \cos \phi$ and $q \cos \gamma \cos \phi$. Terms of relative order $(v_0\, q/\omega)^2$ are dropped, consistent with $v_0\, q \ll \omega$. Expanding to first order in $v_0\, \eta_x\, q/\omega$ and to second order in $\omega/(v_0\, k_0)$ then gives
\begin{align}
\label{eq:imhorncorr}
\Im \Pi (q\, \bs{\hat z}, \omega) &\simeq -\, \frac{g\, k_0\, q^2}{16\, \omega} \left\lbrace F_0 + \left( \frac{\omega}{v_0\, k_0} \right)^2 \left[ f_0 (a_\omega) - f_0 (b_\omega) \right] \right\rbrace , \nn
\Im \Pi (q\, \bs{\hat x}, \omega) &\simeq -\, \frac{g\, k_0\, q^2}{16\, \omega} \left\lbrace F_2 \left( 1 + \frac{v_0\, \eta_x\, q}{\omega} \right) + \frac{q}{k_0} \left[ \frac{\omega}{\tilde{\eta}} \left( f_1 (a_\omega) - f_1 (b_\omega) \right) + h (b_\omega) - h (a_\omega) \right] + \left( \frac{\omega}{v_0\, k_0} \right)^2 \left[ f_2 (a_\omega) - f_2 (b_\omega) \right] \right\rbrace ,
\end{align}
with
\begin{align}
f_0 (u) = -\, \frac{3\, u^3}{32\, \pi\, (1 - u^2)^{3/2}} \,, \quad
f_1 (u) = \frac{u^3\, (2\, u^2 - 3)}{16\, \pi\, (1 - u^2)^{3/2}} \,, \quad
f_2 (u) = \frac{u^3\, (u^2 - 2)}{32\, \pi\, (1 - u^2)^{3/2}} \,, \quad
h (u) = \frac{(1 - u^2)^{3/2}}{3\, \pi} \,.
\end{align}
The terms containing $f_n$ and $h$ at an edge are absent when that edge is clipped, that is for $a_\omega = -1$ or $b_\omega = 1$.

\subsection{Fully-conducting regime of GNR}
\label{app:phiint}

\subsubsection{Local Drude weights of the tilted massive cone}
\label{app:conedrude}

This part derives Eq.~\eqref{eq:conedrude}. We start from the fact that the reduction to the local plane $(u, w)$ is exact within the linearised Hamiltonian, apart from the measure. With $k_x = \left( k_0 + u \right) \cos\phi$, $k_y = \left( k_0 + u \right) \sin\phi$, and $t = \eta_x \cos\phi$, the band energy is
\begin{align}
\varepsilon_2 = E_k + v_0\, \eta_x \left( k_0 + u \right) \cos\phi = \varepsilon_{\rm loc} + \tilde\eta \cos\phi \,, \quad
\varepsilon_{\rm loc} \equiv E_k + v_0\, t\, u \,,
\end{align}
so that $\varepsilon_2 = \mu$ is equivalent to $\varepsilon_{\rm loc} = \mu_{\rm eff}$. The measure is exactly $d^3 \bs k = k_0 \left( 1 + u/k_0 \right) d\phi\, du\, dw$. For the velocity we use $\partial_x = \cos\phi\, \partial_{k_\perp} - \left( \sin\phi/k_\perp \right) \partial_\phi$ and $\partial_y = \sin\phi\, \partial_{k_\perp} + \left( \cos\phi/k_\perp \right) \partial_\phi$, together with $\partial_\phi \varepsilon_2 = -\, v_0\, \eta_x\, k_\perp \sin\phi$. This gives
\begin{align}
V_x^{(2)} = \cos\phi\, \partial_u \varepsilon_{\rm loc} + v_0\, \eta_x \sin^2\phi \,, \quad
V_y^{(2)} = \sin\phi \left( \partial_u \varepsilon_{\rm loc} - v_0\, t \right) , \quad
V_z^{(2)} = \partial_w \varepsilon_{\rm loc} \,.
\end{align}
The first expression is the one quoted before Eq.~\eqref{eq:conedrude}, with the factor $v_0$ written out. In $V_y^{(2)}$ the terms proportional to $\sin\phi \cos\phi$ cancel, and the result is $\sin\phi\, \partial_u E_k$. Since $\pi\, g\, e^2\, k_0/(2\, \pi)^3 = g\, e^2\, k_0/(8\, \pi^2)$, the Drude weights are
\begin{align}
D_{ij} = \frac{g\, e^2\, k_0}{8\, \pi^2} \int d\phi\, \mathcal{D}_{ij} (\phi) \,, \quad
\mathcal{D}_{ij} = \int du\, dw \left( 1 + \frac{u}{k_0} \right) \delta \left( \varepsilon_{\rm loc} - \mu_{\rm eff} \right) V_i^{(2)}\, V_j^{(2)} \,.
\end{align}

At fixed $\phi$, the equation $\varepsilon_{\rm loc} = \mu_{\rm eff}$ describes the ellipse of Appendix~\ref{app:gnrztilt}, with $\eta_z \to t$, $\mu \to \mu_{\rm eff}$, and the tilt along $u$ instead of $w$. We write $\nu_t = \sqrt{1 - t^2}$ and $K = \mu_{\rm eff}^2 - \Delta^2\, \nu_t^2 = \mu_{\rm eff}^2 - \Delta^2 + \Delta^2\, t^2$. The circle map of Eq.~\eqref{eq:gaptiltz-circle}, with the roles of $u$ and $w$ exchanged, gives
\begin{align}
\int du\, dw\, \delta \left( \varepsilon_{\rm loc} - \mu_{\rm eff} \right) \check F = \frac{1}{\nu_t\, v_0^2} \int_0^{2\pi} \! d\vartheta\, E_k\, \check F \,, \quad
u = \frac{\sqrt{K} \sin\vartheta - \mu_{\rm eff}\, t}{v_0\, \nu_t^2} \,, \quad
w = \frac{\sqrt{K} \cos\vartheta}{v_0\, \nu_t} \,, \quad
E_k = \frac{\mu_{\rm eff} - t \sqrt{K} \sin\vartheta}{\nu_t^2} \,,
\end{align}
and the velocities on the circle are $\partial_u \varepsilon_{\rm loc} = v_0^2\, u/E_k + v_0\, t$ and $\partial_w \varepsilon_{\rm loc} = v_0^2\, w/E_k$. The term $1$ of the weight $1 + u/k_0$ gives the standard Drude integrals of Appendix~\ref{app:gnrztilt}. The term $u/k_0$ is proportional to $\sin\vartheta - \mu_{\rm eff}\, t/\sqrt{K}$ and is, therefore, of first order in $t$ after the angular integration. Expanding all integrands to $\mathcal{O} (t^2)$ and to first order in $t/(v_0\, k_0)$, the angular integrals are elementary. We find
\begin{align}
\label{eq:app-cone-local}
& \int du\, dw \left( 1 + \frac{u}{k_0} \right) 
\delta \!\left( \varepsilon_{\rm loc} - \mu_{\rm eff} \right) \left\lbrace v_0^2,\, 
\left( \partial_w \varepsilon_{\rm loc} \right)^2,\, 
\left( \partial_u \varepsilon_{\rm loc} \right)^2,\, 
\left( \partial_u E_k \right)^2 \right\rbrace =
 \pi \left\lbrace N,\, Z,\, U,\, Y \right\rbrace , \nn
& N = 2\, \mu_{\rm eff} \left( 1 + \frac{3\, t^2}{2} \right) 
+ \frac{t}{v_0\, k_0} \left( \Delta^2 - 3\, \mu_{\rm eff}^2 \right) , \quad
Z = \frac{\mu_{\rm eff}^2 - \Delta^2}{\mu_{\rm eff}} + \frac{t^2 \left( 3\, \mu_{\rm eff}^4 + \Delta^4 \right)}{4\, \mu_{\rm eff}^3} + \frac{t}{v_0\, k_0} \left[ \frac{\Delta^4}{4\, \mu_{\rm eff}^2} + \frac{\Delta^2}{2} - \frac{3\, \mu_{\rm eff}^2}{4} \right] , \nn
& U = \frac{\mu_{\rm eff}^2 - \Delta^2}{\mu_{\rm eff}} 
+ \frac{t^2 \left( \mu_{\rm eff}^4 + 3\, \Delta^4 \right)}{4\, \mu_{\rm eff}^3} 
+ \frac{t}{v_0\, k_0} \left[ \frac{3\, \Delta^4}{4\, \mu_{\rm eff}^2} - \frac{\Delta^2}{2} 
- \frac{\mu_{\rm eff}^2}{4} \right] , \
\nn & Y = \frac{\mu_{\rm eff}^2 - \Delta^2}{\mu_{\rm eff}} + \frac{3\, t^2 
\left( 3\, \mu_{\rm eff}^4 + \Delta^4 \right)}{4\, \mu_{\rm eff}^3} + \frac{t}{v_0\, k_0} 
\left[ \frac{3\, \Delta^4}{4\, \mu_{\rm eff}^2} + \frac{3\, \Delta^2}{2} 
- \frac{9\, \mu_{\rm eff}^2}{4} \right] .
\end{align}
The first entry is the weighted DOS, $\rho = g\, k_0/(8\, \pi^3\, v_0^2) \int d\phi\, \pi\, N$, which gives the expression for $\rho (\mu)$ in Eq.~\eqref{eq:conedrude}, since $\left( 1 - t^2 \right)^{-3/2} \simeq 1 + 3\, t^2/2$. Without the Jacobian term, $Z$ and $U$ reduce to the transverse and longitudinal weights of Eq.~\eqref{eq:dtiltgapped-expanded}, with $\eta_z \to t$. The entry $Y$ differs from $U$ because $\partial_u E_k = \partial_u \varepsilon_{\rm loc} - v_0\, t$.

The cross term of $V_x^{(2)}$ needs one more integral. By the definition $\partial_u \varepsilon_{\rm loc}\, \delta (\varepsilon_{\rm loc} - \mu_{\rm eff}) = -\, \partial_u \Theta (\mu_{\rm eff} - \varepsilon_{\rm loc})$, an integration by parts over $u$ gives
\begin{align}
\int du\, dw \left( 1 + \frac{u}{k_0} \right) \delta \left( \varepsilon_{\rm loc} - \mu_{\rm eff} \right) \partial_u \varepsilon_{\rm loc} = \frac{\mathcal{A}}{k_0} \simeq \frac{\pi \left( \mu_{\rm eff}^2 - \Delta^2 \right)}{v_0^2\, k_0} \,.
\end{align}
Here the term $1$ of the weight integrates to zero, because the integral of a total derivative over the closed pocket vanishes. The term $u/k_0$ leaves the area $\mathcal{A} = \pi\, K/(v_0^2\, \nu_t^3)$ of the pocket, and only its leading value is needed, since it multiplies $\eta_x$. This is the statement that the cross term vanishes without the Jacobian, and that the Jacobian leaves a term of order $t/(v_0\, k_0)$.

Collecting the pieces, $\mathcal{D}_{zz} = \pi\, Z$ and $\mathcal{D}_{yy} = \pi \sin^2\phi\, Y$. For $\bs{\hat x}$ we have
\begin{align}
\mathcal{D}_{xx} = \pi \left[ \cos^2\phi\, U + 2\, \sin^2\phi\, \frac{t}{v_0\, k_0} \left( \mu_{\rm eff}^2 - \Delta^2 \right) + \eta_x^2 \sin^4\phi\, N \right] ,
\end{align}
where the second term is the cross term, and in the last term we keep $N \simeq 2\, \mu_{\rm eff}$. The coefficient of $t/(v_0\, k_0)$ in the bracket is
\begin{align}
\cos^2\phi \left[ \frac{3\, \Delta^4}{4\, \mu_{\rm eff}^2} - \frac{\Delta^2}{2} - \frac{\mu_{\rm eff}^2}{4} \right] + 2 \left( 1 - \cos^2\phi \right) \left( \mu_{\rm eff}^2 - \Delta^2 \right) = \cos^2\phi \left[ \frac{3\, \Delta^4}{4\, \mu_{\rm eff}^2} + \frac{3\, \Delta^2}{2} - \frac{9\, \mu_{\rm eff}^2}{4} \right] + 2 \left( \mu_{\rm eff}^2 - \Delta^2 \right) ,
\end{align}
which is the bracket of $D_{xx}$ in Eq.~\eqref{eq:conedrude}. The same equation follows for $D_{zz}$ and $D_{yy}$ from $Z$ and $Y$ after dividing by $\pi$ in the prefactor, $g\, e^2\, k_0/(8\, \pi^2) \to g\, e^2\, k_0/(8\, \pi)$. In the fully-conducting regime the $\phi$-integrals run over the whole ring.

\subsubsection{Assembly of the Drude weights}

The Drude weights of Eq.~\eqref{eq:dzzxtiltgapped} follow from Eq.~\eqref{eq:conedrude}, whose derivation is given in the first part of this appendix. The corresponding expansion for the in-plane-tilted PTNR is given in Appendix~\ref{app:xtiltdrude}. Here we use $\mu_{\rm eff} (\phi) = \mu - \tilde\eta \cos\phi$ and $\tilde w \equiv \sqrt{\mu^2 - \tilde\eta^2}$ for $\mu > \Delta + \tilde\eta$. The leading terms use
\begin{align}
\int_0^{2\pi}\! \frac{d\phi}{\mu_{\rm eff}} = \frac{2\,\pi}{\tilde w} \,, \quad
\int_0^{2\pi}\! d\phi\, \frac{\cos^2\phi}{\mu_{\rm eff}} = \frac{2\,\pi\, \mu}{\tilde\eta^2} \left( \frac{\mu}{\tilde w} - 1 \right) .
\end{align}
The $\eta_x^2$ terms additionally need
\begin{align}
\int_0^{2\pi}\! d\phi\, \frac{\cos\phi}{\mu_{\rm eff}^2} = \frac{2\,\pi\, \tilde\eta}{\tilde w^3} \,, \quad
\int_0^{2\pi}\! d\phi\, \frac{\cos^3\phi}{\mu_{\rm eff}^2} = \frac{2\,\pi\, \mu\, \tilde\eta \left( \mu + 2\, \tilde w \right)}{\tilde w^3 \left( \mu + \tilde w \right)^2} \,,
\end{align}
together with
\begin{align}
\int_0^{2\pi}\! d\phi\, \frac{\cos^2\phi}{\mu_{\rm eff}^3} = 
\frac{\pi \left( \mu^2 + 2\,\tilde\eta^2 \right)}{\tilde w^5} \,, \quad
\int_0^{2\pi}\! d\phi\, \frac{\cos^4\phi}{\mu_{\rm eff}^3} = 
\frac{3\,\pi\, \mu \left( \mu^3 + 2\, \mu^2\, \tilde w - 2\, \tilde w^3 \right)}{\tilde w^5 \left( \mu+\tilde w \right)^2} \,.
\end{align}
All six integrals follow from $\tilde J_n \equiv \int_0^{2\pi} d\phi\, \mu_{\rm eff}^{-n}$ and from $\cos\phi = \left( \mu - \mu_{\rm eff} \right)/\tilde\eta$. The standard integral is $\tilde J_1 = 2\, \pi/\tilde w$, and differentiation with respect to $\mu$ at fixed $\tilde\eta$ gives $\tilde J_{n+1} = -\, n^{-1}\, \partial \tilde J_n/\partial \mu$, that is $\tilde J_2 = 2\, \pi\, \mu/\tilde w^3$ and $\tilde J_3 = \pi \left( 2\, \mu^2 + \tilde\eta^2 \right)/\tilde w^5$. Three examples are
\begin{align}
& \int_0^{2\pi}\! d\phi\, \frac{\cos^2\phi}{\mu_{\rm eff}} = 
\frac{1}{\tilde\eta^2} \int_0^{2\pi}\! d\phi \left( \frac{\mu^2}{\mu_{\rm eff}} - 2\, \mu + \mu_{\rm eff} \right) = \frac{2\, \pi\, \mu}{\tilde\eta^2} \left( \frac{\mu}{\tilde w} - 1 \right) , \nn
& \int_0^{2\pi}\! d\phi\, \frac{\cos\phi}{\mu_{\rm eff}^2} = 
\frac{\mu\, \tilde J_2 - \tilde J_1}{\tilde\eta} = 
\frac{2\, \pi \left( \mu^2 - \tilde w^2 \right)}{\tilde\eta\, \tilde w^3} = 
\frac{2\, \pi\, \tilde\eta}{\tilde w^3} \,, \quad
\int_0^{2\pi}\! d\phi\, \frac{\cos^2\phi}{\mu_{\rm eff}^3} 
= \frac{\mu^2\, \tilde J_3 - 2\, \mu\, \tilde J_2 + \tilde J_1}{\tilde\eta^2} = 
\frac{\pi \left( \mu^2 + 2\, \tilde\eta^2 \right)}{\tilde w^5} \,.
\end{align}
The cubic and quartic moments need two further expansions of the same kind. For the first we use $\cos\phi = (\mu - \mu_{\rm eff})/\tilde\eta$ and $\int_0^{2\pi} d\phi\, \mu_{\rm eff} = 2\,\pi\, \mu$, which gives
\begin{align}
\int_0^{2\pi}\! d\phi\, \frac{\cos^3\phi}{\mu_{\rm eff}^2} &= \frac{1}{\tilde\eta^3} \int_0^{2\pi}\! d\phi \left( \frac{\mu^3}{\mu_{\rm eff}^2} - \frac{3\,\mu^2}{\mu_{\rm eff}} + 3\, \mu - \mu_{\rm eff} \right) = \frac{\mu^3\, \tilde J_2 - 3\, \mu^2\, \tilde J_1 + 4\,\pi\, \mu}{\tilde\eta^3} \nn
&= \frac{2\,\pi\, \mu \left( \mu^3 - 3\, \mu\, \tilde w^2 + 2\, \tilde w^3 \right)}{\tilde\eta^3\, \tilde w^3} = \frac{2\,\pi\, \mu \left( \mu - \tilde w \right)^2 \left( \mu + 2\, \tilde w \right)}{\tilde\eta^3\, \tilde w^3} \,.
\end{align}
The last step is the factorisation $\mu^3 - 3\, \mu\, \tilde w^2 + 2\, \tilde w^3 = (\mu - \tilde w)^2 (\mu + 2\, \tilde w)$. Since $\tilde\eta^2 = (\mu - \tilde w)(\mu + \tilde w)$, we have $(\mu - \tilde w)^2 = \tilde\eta^4/(\mu + \tilde w)^2$, and this gives the result quoted above. For the second we need $\tilde J_3 = \pi \left( 3\, \mu^2 - \tilde w^2 \right)/\tilde w^5$, which is the form of $\tilde J_3$ above written with $\tilde w^2 = \mu^2 - \tilde\eta^2$. The expansion of $(\mu - \mu_{\rm eff})^4$ then gives
\begin{align}
\int_0^{2\pi}\! d\phi\, \frac{\cos^4\phi}{\mu_{\rm eff}^3} &= \frac{1}{\tilde\eta^4} \int_0^{2\pi}\! d\phi \left( \frac{\mu^4}{\mu_{\rm eff}^3} - \frac{4\,\mu^3}{\mu_{\rm eff}^2} + \frac{6\,\mu^2}{\mu_{\rm eff}} - 4\, \mu + \mu_{\rm eff} \right) = \frac{\mu^4\, \tilde J_3 - 4\, \mu^3\, \tilde J_2 + 6\, \mu^2\, \tilde J_1 - 6\,\pi\, \mu}{\tilde\eta^4} \nn
&= \frac{\pi\, \mu \left( 3\, \mu^5 - 9\, \mu^3\, \tilde w^2 + 12\, \mu\, \tilde w^4 - 6\, \tilde w^5 \right)}{\tilde\eta^4\, \tilde w^5} = \frac{3\,\pi\, \mu \left( \mu^3 + 2\, \mu^2\, \tilde w - 2\, \tilde w^3 \right)}{\tilde w^5 \left( \mu + \tilde w \right)^2} \,.
\end{align}
The numerator factorises as $3\, \mu^5 - 9\, \mu^3\, \tilde w^2 + 12\, \mu\, \tilde w^4 - 6\, \tilde w^5 = 3\, (\mu - \tilde w)^2 \left( \mu^3 + 2\, \mu^2\, \tilde w - 2\, \tilde w^3 \right)$, as one verifies by expanding the product. The factor $(\mu - \tilde w)^2$ then cancels against $\tilde\eta^4/(\mu + \tilde w)^2$ as before. All six integrals are thereby derived.

We now assemble Eq.~\eqref{eq:dzzxtiltgapped} from Eq.~\eqref{eq:conedrude}. Since $v_0\, k_0 = \tilde\eta/\eta_x$, we have $t^2 = \eta_x^2 \cos^2\phi$ and $t/(v_0\, k_0) = \eta_x^2 \cos\phi/\tilde\eta$. All corrections are therefore of order $\eta_x^2$, and the $\phi$-integrals produce no further powers of $\eta_x$. Besides the six integrals above, we need the polynomial moments of $\mu_{\rm eff} = \mu - \tilde\eta \cos\phi$,
\begin{align}
& \int_0^{2\pi}\! d\phi\, \mu_{\rm eff} = 2\, \pi\, \mu \,, \quad
\int_0^{2\pi}\! d\phi\, \cos^2\phi\, \mu_{\rm eff} = \pi\, \mu \,, \quad
\int_0^{2\pi}\! d\phi\, \cos^4\phi\, \mu_{\rm eff} = \frac{3\, \pi\, \mu}{4} \,, \quad
\int_0^{2\pi}\! d\phi\, \sin^4\phi\, \mu_{\rm eff} = \frac{3\, \pi\, \mu}{4} \,, \nn
& \int_0^{2\pi}\! d\phi\, \sin^2\phi \cos^2\phi\, \mu_{\rm eff} = \frac{\pi\, \mu}{4} \,, \quad
\int_0^{2\pi}\! d\phi\, \cos\phi\, \mu_{\rm eff}^2 = -\, 2\, \pi\, \mu\, \tilde\eta \,, \quad
\int_0^{2\pi}\! d\phi\, \cos^3\phi\, \mu_{\rm eff}^2 = -\, \frac{3\, \pi\, \mu\, \tilde\eta}{2} \,, 
\nn &
\int_0^{2\pi}\! d\phi\, \sin^2\phi \cos\phi\, \mu_{\rm eff}^2 = -\, \frac{\pi\, \mu\, \tilde\eta}{2} \,.
\end{align}
For $D_{zz}$, Eq.~\eqref{eq:conedrude} reads
\begin{align}
D_{zz} = \frac{g\, e^2\, k_0}{8\, \pi} \left\lbrace \int d\phi \left( \mu_{\rm eff} - \frac{\Delta^2}{\mu_{\rm eff}} \right) + \eta_x^2 \int d\phi \left[ \cos^2\phi \left( \frac{3\, \mu_{\rm eff}}{4} + \frac{\Delta^4}{4\, \mu_{\rm eff}^3} \right) + \frac{\cos\phi}{\tilde\eta} \left( \frac{\Delta^4}{4\, \mu_{\rm eff}^2} + \frac{\Delta^2}{2} - \frac{3\, \mu_{\rm eff}^2}{4} \right) \right] \right\rbrace .
\end{align}
The leading integral is $2\, \pi \left( \mu - \Delta^2/\tilde w \right)$. The $\eta_x^2$ integral equals
\begin{align}
\frac{3\, \pi\, \mu}{4} + \frac{\pi\, \Delta^4 \left( \mu^2 + 2\, \tilde\eta^2 \right)}{4\, \tilde w^5} + \frac{\pi\, \Delta^4}{2\, \tilde w^3} + \frac{3\, \pi\, \mu}{2} = \frac{3\, \pi}{4} \left( 3\, \mu + \frac{\Delta^4\, \mu^2}{\tilde w^5} \right) ,
\end{align}
where the second and third terms come from $\cos^2\phi/\mu_{\rm eff}^3$ and $\cos\phi/\mu_{\rm eff}^2$, the term $\Delta^2 \cos\phi$ integrates to zero, and the last term comes from $\int \cos\phi\, \mu_{\rm eff}^2$. The step to the right-hand side uses $\tilde\eta^2 = \mu^2 - \tilde w^2$, which turns the sum of the $\Delta^4$ terms into $3\, \Delta^4\, \mu^2/(4\, \tilde w^5)$. This is the expression for $D_{zz}$ in Eq.~\eqref{eq:dzzxtiltgapped}.

For $D_{xx}$, the leading part is $\left( g\, e^2\, k_0/8\, \pi \right) \int d\phi \left[ \cos^2\phi\, \mu_{\rm eff} - \Delta^2 \cos^2\phi/\mu_{\rm eff} \right]$, which gives $\mu/8 - \mu\, \Delta^2/\left[ 4\, \tilde w \left( \mu + \tilde w \right) \right]$ after using $\tilde\eta^2 = \left( \mu - \tilde w \right) \left( \mu + \tilde w \right)$. The terms of order $\eta_x^2$ come from four sources. The first is the term $\cos^4\phi \left[ \mu_{\rm eff}/4 + 3\, \Delta^4/(4\, \mu_{\rm eff}^3) \right]$, which gives $3\, \pi\, \mu/16 + 9\, \pi\, \Delta^4\, \mu \left( \mu^3 + 2\, \mu^2\, \tilde w - 2\, \tilde w^3 \right)/\left[ 4\, \tilde w^5 \left( \mu + \tilde w \right)^2 \right]$. The second is the Jacobian term with $\cos^3\phi \left[ 3\, \Delta^4/(4\, \mu_{\rm eff}^2) - 9\, \mu_{\rm eff}^2/4 \right]/\tilde\eta$, which gives $3\, \pi\, \Delta^4\, \mu \left( \mu + 2\, \tilde w \right)/\left[ 2\, \tilde w^3 \left( \mu + \tilde w \right)^2 \right] + 27\, \pi\, \mu/8$. The third is the term $2\, \sin^4\phi\, \mu_{\rm eff}$, which gives $3\, \pi\, \mu/2$. The fourth is the cross term $2\, \cos\phi \left( \mu_{\rm eff}^2 - \Delta^2 \right)/\tilde\eta$, which gives $-\, 4\, \pi\, \mu$. The terms without $\Delta$ add up to $\pi\, \mu \left( 3/16 + 27/8 + 3/2 - 4 \right) = 17\, \pi\, \mu/16$. The $\Delta^4$ terms combine to $3\, \pi\, \Delta^4\, \mu \left( 3\, \mu^3 + 6\, \mu^2\, \tilde w + 2\, \mu\, \tilde w^2 - 2\, \tilde w^3 \right)/\left[ 4\, \tilde w^5 \left( \mu + \tilde w \right)^2 \right]$. Division by $8\, \pi$ gives the $\eta_x^2$ terms of $D_{xx}$ in Eq.~\eqref{eq:dzzxtiltgapped}.

For $D_{yy}$, the leading part uses $\int d\phi\, \sin^2\phi/\mu_{\rm eff} = 2\, \pi/\left( \mu + \tilde w \right)$, which follows from $\sin^2\phi = 1 - \cos^2\phi$ and the first two integrals above, and gives $\mu/8 - \Delta^2/\left[ 4 \left( \mu + \tilde w \right) \right]$. At order $\eta_x^2$, the terms without $\Delta$ are $(9/4) \int d\phi\, \sin^2\phi \cos^2\phi\, \mu_{\rm eff} = 9\, \pi\, \mu/16$ from the $t^2$ term and $-\left( 9/4 \right) \left( -\pi\, \mu/2 \right) = 9\, \pi\, \mu/8$ from the Jacobian term, which sum to $27\, \pi\, \mu/16$. The $\Delta^4$ terms need $\int d\phi\, \sin^2\phi \cos^2\phi/\mu_{\rm eff}^3 = \pi \left( \mu^2 + 2\, \mu\, \tilde w - 2\, \tilde w^2 \right)/\left[ \tilde w^3 \left( \mu + \tilde w \right)^2 \right]$ and $\tilde\eta^{-1} \int d\phi\, \sin^2\phi \cos\phi/\mu_{\rm eff}^2 = 2\, \pi/\left[ \tilde w \left( \mu + \tilde w \right)^2 \right]$. They follow from the six integrals above by $\sin^2\phi = 1 - \cos^2\phi$. Their sum is $3\, \pi\, \Delta^4\, \mu \left( \mu + 2\, \tilde w \right)/\left[ 4\, \tilde w^3 \left( \mu + \tilde w \right)^2 \right]$. Division by $8\, \pi$ gives the $D_{yy}$ of Eq.~\eqref{eq:dzzxtiltgapped}. As a check, the three results satisfy the limits stated after Eq.~\eqref{eq:wpgappedxtilt}, namely the $\Delta \to 0$ limit of Eq.~\eqref{eq:dzzx} and the $\tilde\eta \to 0$ limit of Sec.~\ref{secgapnotilt}.

\subsubsection{Local interband function}
\label{app:interlocal}

This part derives the function $\mathcal{I}$ of Eq.~\eqref{eq:pintertiltxgnr}. At fixed $\phi$ the tilt drops out of the vertical transition energy $2\, E_k$, as shown in Appendix~\ref{app:tiltshift}, and, to leading order in $\eta_x$, the blocked region is the disc $E_k < \tilde{m}$ with $\tilde{m} = \max \left[ \Delta, |\mu_{\rm eff}| \right]$. For $q \to 0$ and $\omega \gg v_0\, q$, the bracket of Eq.~\eqref{eq:polinter} equals $4\, E_k/(\omega^2 - 4\, E_k^2)$. The local transfer is $q$ along $w$ for $\bs q \parallel \bs{\hat z}$, $q \cos\phi$ along $u$ for $\bs q \parallel \bs{\hat x}$, and $q \sin\phi$ along $u$ for $\bs q \parallel \bs{\hat y}$, so that the weights are $\check w_i (\phi) = 1$, $\cos^2\phi$, and $\sin^2\phi$. The integrand depends on $\gamma$ only through $g_{uu}$ or $g_{ww}$. The unblocked region is a disc complement, and hence the average over $\gamma$ is legitimate. Both averages equal $\langle g \rangle = v_0^2 \left( E_k^2 + \Delta^2 \right)/(8\, E_k^4)$, by Appendix~\ref{app:interband}. With $d^2 \bs\kappa/(2\, \pi)^2 = E_k\, dE_k/(2\, \pi\, v_0^2)$, we obtain
\begin{align}
\Pi_{\rm inter} (q\, \bs{\hat e}_i, \omega) = -\, \frac{g\, k_0\, q^2}{4\, \pi} \int_0^{2\pi} \frac{d\phi}{2\, \pi}\, \check w_i (\phi)\, \mathcal{R} \,, \quad
\mathcal{R} = \int_{\tilde{m}}^\infty \! dE\, \frac{E^2 + \Delta^2}{E^2 \left( 4\, E^2 - \omega^2 \right)} \,.
\end{align}
The partial fractions $\left( E^2 + \Delta^2 \right)/\left[ E^2 \left( 4\, E^2 - \omega^2 \right) \right] = \left( 1 + 4\, \Delta^2/\omega^2 \right)/\left( 4\, E^2 - \omega^2 \right) - \Delta^2/\left( \omega^2\, E^2 \right)$ and the integral $\int_{\tilde{m}}^\infty dE/\left( 4\, E^2 - \omega^2 \right) = \left( 1/(4\, \omega) \right) \ln \!\left| (\tilde{m} + \Omega)/(\tilde{m} - \Omega) \right|$, with $\Omega = \omega/2$, give $\mathcal{R} = \mathcal{I}/4$ with
\begin{align}
\mathcal{I} = \frac{1}{2\, \Omega} \left( 1 + \frac{\Delta^2}{\Omega^2} \right) \ln \!\left| \frac{\tilde{m} + \Omega}{\tilde{m} - \Omega} \right| - \frac{\Delta^2}{\Omega^2\, \tilde{m}} \,,
\end{align}
and hence the form of Eq.~\eqref{eq:pintertiltxgnr}. The imaginary part follows from $\omega \to \omega + i\, 0^+$. For $\Omega > \tilde{m}$ the argument of the logarithm acquires $\tilde{m} - \Omega - i\, 0^+$ in the denominator, so that $\ln \!\left| \cdot \right| \to \ln \!\left| \cdot \right| + i\, \pi\, \Theta \left( \Omega - \tilde{m} \right)$, as in the expression for $\mathcal{I}$ in the main text. The interband contribution to the dielectric function is $-\, 4\, \pi\, e^2\, \Pi_{\rm inter}/q^2 = \left( g\, e^2\, k_0/4 \right) \int d\phi/(2\, \pi)\, \check w_i\, \mathcal{I}$, and its imaginary part is the expression following Eq.~\eqref{eq:pintertiltxgnr}, since $\left( 1/(2\, \Omega) \right) \left( 1 + \Delta^2/\Omega^2 \right) = \left( 1/\omega \right) \left( 1 + 4\, \Delta^2/\omega^2 \right)$.

Three checks follow. For $\Delta = 0$ we have $\mathcal{I} = \omega^{-1} \ln \!\left| (2\, |\mu_{\rm eff}| + \omega)/(2\, |\mu_{\rm eff}| - \omega) \right|$, which is Eq.~\eqref{eq:pintertiltxptnr}. For $\tilde{m} = \Delta$ on all arcs, we set $\check x = \Omega/\Delta$, and then $\mathcal{I} = \left[ \left( 1 + x^2 \right) \mathrm{arctanh}\, x/x^3 - 1/x^2 \right]/\Delta = \mathcal{J} (x)/\Delta$, which gives Eq.~\eqref{eq:epsgappedxtilt}. For $\omega \to 0$, the expansion $\ln \!\left| (\tilde{m} + \Omega)/(\tilde{m} - \Omega) \right| = 2\, \Omega/\tilde{m} + 2\, \Omega^3/(3\, \tilde{m}^3) + \dots$ gives $\mathcal{I} \to \left( 3\, \tilde{m}^2 + \Delta^2 \right)/\left( 3\, \tilde{m}^3 \right)$. The dielectric function then receives $\left( g\, e^2\, k_0/(24\, \pi) \right) \int d\phi\, \check w_i\, \mathcal{S}$, with $\mathcal{S} = \left( 3\, \tilde{m}^2 + \Delta^2 \right)/\tilde{m}^3$. This is Eq.~\eqref{eq:epsb-partial}, and its integration over the whole ring with $\tilde{m} = \mu_{\rm eff}$ gives Eq.~\eqref{eq:epsb-xtilt}.

\subsection{Twin-crescent regime of GNR}
\label{app:twin}

This appendix derives Eqs.~\eqref{eq:bipolar-dos} and \eqref{eq:bipolar-dzz}, starting from the ordering of the boundary angles. We take $\mu\ge0$, since the results are even in $\mu$. The function $\cos\phi$ decreases monotonically from $1$ to $-1$ on $(0,\pi)$, and $\mu_{\rm eff}(\phi)=\mu-\tilde\eta\cos\phi$ therefore increases monotonically from $\mu-\tilde\eta$ to $\mu+\tilde\eta$. The three levels $\mu_{\rm eff}=-\Delta$, $0$, and $\Delta$ are reached at $\phi_h$, $\phi_b$, and $\phi_0$, which are defined by the three cosines in Sec.~\ref{secgtxbp}. They exist if the three ratios lie in $(-1,1)$. The upper bound follows from $\mu+\Delta<\tilde\eta$. The lower bound follows from $\Delta-\mu\le\Delta+\mu<\tilde\eta$ for $\mu\ge0$. Since $(\mu+\Delta)/\tilde\eta>\mu/\tilde\eta>(\mu-\Delta)/\tilde\eta$, we obtain $0<\phi_h<\phi_b<\phi_0<\pi$. The local doping then satisfies $\mu_{\rm eff}<-\Delta$ for $|\phi|<\phi_h$ and $\mu_{\rm eff}>\Delta$ for $\phi_0<|\phi|\le\pi$. In between, $|\mu_{\rm eff}|<\Delta$, and the local DOS and Drude weight vanish. The zero of $\mu_{\rm eff}$ at $\phi_b$ lies on this gapped arc. Hence the integrands of Eq.~\eqref{eq:bipolar-reduced} are regular on both occupied arcs, even though $1/|\mu_{\rm eff}|$ diverges at $\phi_b$.

The integrand is even in $\phi$. Each integral over the occupied arcs is therefore twice the sum of the integrals over $(0,\phi_h)$ and $(\phi_0,\pi)$. We have $|\mu_{\rm eff}|=-\mu_{\rm eff}$ on the first interval and $|\mu_{\rm eff}|=\mu_{\rm eff}$ on the second. The factor $(\mu_{\rm eff}^2-\Delta^2)/|\mu_{\rm eff}|$ of Eq.~\eqref{eq:conedrude} equals $|\mu_{\rm eff}|-\Delta^2/|\mu_{\rm eff}|$, which gives Eq.~\eqref{eq:bipolar-reduced}. We outline the evaluation of the four angular integrals:
\begin{enumerate}

\item We start with $|\mu_{\rm eff}|$. On the hole arc, $\int_0^{\phi_h} d\phi\, (\tilde\eta\cos\phi-\mu) = \tilde\eta\sin\phi_h-\mu\,\phi_h$. On the electron arc, $\int_{\phi_0}^\pi d\phi\, (\mu-\tilde\eta\cos\phi) = \mu\,(\pi-\phi_0)+\tilde\eta\sin\phi_0$. Using $s_+=\tilde\eta\sin\phi_h$ and $s_-=\tilde\eta\sin\phi_0$, the sum of the two, multiplied by $2$, is
\begin{align}
\int_{\rm occ}\! d\phi\, |\mu_{\rm eff}| = 2 \left( \mu\, \Phi + s_+ + s_- \right) .
\end{align}
Combined with Eq.~\eqref{eq:bipolar-reduced}, this gives Eq.~\eqref{eq:bipolar-dos}. The same integration also gives the difference of the arc lengths, $\int_{\rm occ} d\phi\, \mathrm{sgn}\,\mu_{\rm eff} = 2 (\pi-\phi_0) - 2\,\phi_h = 2\,\Phi$, which is needed below.

\item For $1/|\mu_{\rm eff}|$, we write $u=\tan(\phi/2)$, which gives $\cos\phi=(1-u^2)/(1+u^2)$ and $d\phi=2\,du/(1+u^2)$. Hence
\begin{align}
\mu_{\rm eff} = \frac{(\tilde\eta+\mu)\, u^2-(\tilde\eta-\mu)}{1+u^2} \,, \quad
\int\! \frac{d\phi}{\mu_{\rm eff}} = \frac{2}{\tilde\eta+\mu} \int\! \frac{du}{u^2-\check a^2} = \frac{1}{\bar w} \ln \!\left| \frac{u-\check a}{u+\check a} \right| ,
\end{align}
where $\check a^2=(\tilde\eta-\mu)/(\tilde\eta+\mu)$. Since $\cos\phi_b=\mu/\tilde\eta$, we have $\check a=\tan(\phi_b/2)$. We also used $(\tilde\eta+\mu)\,\check a=\bar w$. The identity $\tan A\pm\tan B=\sin(A\pm B)/(\cos A\cos B)$ turns the antiderivative into
\begin{align}
\tilde F (\phi) = \frac{1}{\bar w} \ln \!\left| \frac{\sin\frac{\phi-\phi_b}{2}}{\sin\frac{\phi+\phi_b}{2}} \right| .
\end{align}
It satisfies $\tilde F(0)=\tilde F(\pi)=0$, and the logarithmic singularity at $\phi=\phi_b$ lies on the gapped arc. The electron arc gives $\int_{\phi_0}^\pi d\phi/\mu_{\rm eff} = -\,\tilde F(\phi_0)$. The hole arc gives $\int_0^{\phi_h} d\phi/|\mu_{\rm eff}| = -\,\tilde F(\phi_h)$, because $\mu_{\rm eff}<0$ there. Their sum is
\begin{align}
-\, \tilde F(\phi_0)-\tilde F(\phi_h) = \frac{1}{\bar w} \ln \!\left( \frac{\sin\frac{\phi_0+\phi_b}{2}\,\sin\frac{\phi_b+\phi_h}{2}}{\sin\frac{\phi_0-\phi_b}{2}\,\sin\frac{\phi_b-\phi_h}{2}} \right) = \frac{\Lambda}{\bar w} \,.
\end{align}
Doubling it gives $\int_{\rm occ} d\phi/|\mu_{\rm eff}| = 2\,\Lambda/\bar w$.

\item The two integrals with $\cos^2\phi$ follow from $\cos^2\phi=(\mu^2-2\,\mu\,\mu_{\rm eff}+\mu_{\rm eff}^2)/\tilde\eta^2$. For the inverse weight,
\begin{align}
\int_{\rm occ}\! d\phi\, \frac{\cos^2\phi}{|\mu_{\rm eff}|} = \frac{1}{\tilde\eta^2} \int_{\rm occ}\! d\phi \left( \frac{\mu^2}{|\mu_{\rm eff}|} - 2\, \mu\, \mathrm{sgn}\,\mu_{\rm eff} + |\mu_{\rm eff}| \right) = \frac{2}{\tilde\eta^2} \left( \frac{\mu^2\, \Lambda}{\bar w} - \mu\, \Phi + s_+ + s_- \right) ,
\end{align}
where the three terms use the results above. For the other weight, we integrate directly with $\int d\phi\, \cos^2\phi = (\phi+\sin\phi\cos\phi)/2$ and $\int d\phi\, \cos^3\phi = \sin\phi-\sin^3\phi/3$. The hole arc gives $2\,\tilde\eta \left( \sin\phi_h-\sin^3\phi_h/3 \right) - \mu \left( \phi_h+\sin\phi_h\cos\phi_h \right)$. The electron arc gives $2\,\tilde\eta \left( \sin\phi_0-\sin^3\phi_0/3 \right) + \mu \left( \pi-\phi_0-\sin\phi_0\cos\phi_0 \right)$. We insert $\sin\phi_h=s_+/\tilde\eta$, $\sin\phi_0=s_-/\tilde\eta$, $\cos\phi_h=(\mu+\Delta)/\tilde\eta$, and $\cos\phi_0=(\mu-\Delta)/\tilde\eta$. With $s_\varrho^2=\tilde\eta^2-(\mu+\varrho\,\Delta)^2$, the coefficient of each $s_\varrho$ is
\begin{align}
2 - \frac{2}{3} + \frac{2\,(\mu+\varrho\,\Delta)^2 - 3\,\mu\,(\mu+\varrho\,\Delta)}{3\,\tilde\eta^2} = \frac{4\,\tilde\eta^2 - \mu^2 + \varrho\, \mu\, \Delta + 2\,\Delta^2}{3\,\tilde\eta^2} \,.
\end{align}
The terms without $s_\varrho$ add up to $\mu\,\Phi$. This gives the third integral of Eq.~\eqref{eq:bipolar-integrals}.

\end{enumerate}

\subsubsection{Assembly of the Drude weights}

Eq.~\eqref{eq:bipolar-reduced} gives $D_{zz}=(g\, e^2\, k_0/8\,\pi)\left[ \int_{\rm occ} |\mu_{\rm eff}| - \Delta^2 \int_{\rm occ} 1/|\mu_{\rm eff}| \right]$, and inserting the first two integrals yields the expression for $D_{zz}$ in Eq.~\eqref{eq:bipolar-dzz}. For $D_{xx}$, the combination of the third integral and $\Delta^2$ times the fourth gives the coefficient of $\mu\,\Phi$ as $1+2\,\Delta^2/\tilde\eta^2$. The coefficient of $s_\varrho$ becomes
\begin{align}
\frac{4\,\tilde\eta^2-\mu^2+\varrho\,\mu\,\Delta+2\,\Delta^2}{3\,\tilde\eta^2} - \frac{2\,\Delta^2}{\tilde\eta^2} = \frac{4\,\tilde\eta^2-\mu^2+\varrho\,\mu\,\Delta-4\,\Delta^2}{3\,\tilde\eta^2} \,,
\end{align}
and the logarithmic term is $-\,2\,\Delta^2\,\mu^2\,\Lambda/(\tilde\eta^2\, \bar w)$. This is the expression for $D_{xx}$ in Eq.~\eqref{eq:bipolar-dzz}. Since $\check w_y=\check w_z-\check w_x$, we have $D_{yy}=D_{zz}-D_{xx}$, which gives the coefficient of $\mu\,\Phi$ as $1-2\,\Delta^2/\tilde\eta^2$, that of $s_\varrho$ as $\left( 2\,\tilde\eta^2+\mu^2-\varrho\,\mu\,\Delta+4\,\Delta^2 \right)/(3\,\tilde\eta^2)$, and the logarithmic term as $-\,2\,\Delta^2\, \bar w\, \Lambda/\tilde\eta^2$ from $\tilde\eta^2-\mu^2=\bar w^2$. 

Two checks follow. For $\Delta\to0$, we have $\phi_0,\phi_h\to\phi_b$, $\Phi\to2\arcsin(\mu/\tilde\eta)$, and $s_\pm\to\bar w$. The weights then reduce to the horn-cyclide expressions quoted after Eq.~\eqref{eq:bipolar-dzz}. For $\phi_h\to0$, we have $s_+\to0$, $\Lambda\to\ln \!\left[ \sin\frac{\phi_0+\phi_b}{2}/\sin\frac{\phi_0-\phi_b}{2} \right]$, and $\Phi\to\pi-\phi_0$. The weights then reduce to Eq.~\eqref{eq:partialgapped-exact}, with $\tilde{\mathcal{J}}=2\,\Lambda/\bar w$ in the case $\mu<\tilde\eta$.

\subsubsection{Limit of a small gap and the anisotropy ratios}

We take $\Delta\ll\tilde\eta$ at fixed $\mu\ge0$ with $\tilde\eta-\mu\gg\Delta$, so that the regime is entered with room to spare. Expanding $\cos\phi_{0,h}=\mu/\tilde\eta\mp\Delta/\tilde\eta$ about $\phi_b$ and using $\sin\phi_b=\bar w/\tilde\eta$ gives
\begin{align}
\phi_0-\phi_b = \phi_b-\phi_h = \frac{\Delta}{\bar w} + \mathcal{O}\!\left( \Delta^2 \right) .
\end{align}
The shifts of the two angles are opposite at first order. With $\phi_b=\arccos(\mu/\tilde\eta)$, this gives $\Phi=\pi-2\,\phi_b+\mathcal{O}(\Delta^2)=2\arcsin(\mu/\tilde\eta)+\mathcal{O}(\Delta^2)$. Similarly, $s_\varrho=\bar w-\varrho\,\mu\,\Delta/\bar w+\mathcal{O}(\Delta^2)$, and every term of first order in $\Delta$ cancels in the sums over $\varrho=\pm$ of Eq.~\eqref{eq:bipolar-dzz}. For the logarithm, $\sin\frac{\phi_0+\phi_b}{2}\simeq\sin\frac{\phi_b+\phi_h}{2}\simeq\bar w/\tilde\eta$ and $\sin\frac{\phi_0-\phi_b}{2}\simeq\sin\frac{\phi_b-\phi_h}{2}\simeq\Delta/(2\,\bar w)$, which gives
\begin{align}
\Lambda \simeq 2 \ln \!\left( \frac{2\,\bar w^2}{\tilde\eta\, \Delta} \right) \,.
\end{align}
The terms $\propto\Delta^2\Lambda$ in Eq.~\eqref{eq:bipolar-dzz} are therefore of order $\Delta^2\ln \!\left( \tilde\eta/\Delta \right)$, and the remaining corrections are of order $\Delta^2$. At leading order, the factors $1\pm2\,\Delta^2/\tilde\eta^2$ of $\mu\,\Phi$ also reduce to unity. With $\bar x\equiv\mu/\tilde\eta\in[0,1)$ and $\bar w=\tilde\eta\sqrt{1-\bar x^2}$, Eq.~\eqref{eq:bipolar-dzz} then becomes
\begin{align}
\label{eq:bipolar-x}
& D_{zz} \simeq \frac{g\, e^2\, k_0\, \tilde\eta}{2\,\pi} \left[ \bar x \arcsin \bar x + \sqrt{1-\bar x^2} \right] , \nn
& D_{xx} \simeq \frac{g\, e^2\, k_0\, \tilde\eta}{4\,\pi} \left[ \bar x \arcsin \bar x + \frac{\left( 4-\bar x^2 \right) \sqrt{1-\bar x^2}}{3} \right] , \quad
D_{yy} \simeq \frac{g\, e^2\, k_0\, \tilde\eta}{4\,\pi} \left[ \bar x \arcsin \bar x + \frac{\left( 2+\bar x^2 \right) \sqrt{1-\bar x^2}}{3} \right] ,
\end{align}
which are the expressions quoted in Sec.~\ref{secgtxbp}. They satisfy $D_{zz}=D_{xx}+D_{yy}$ because $(4-\bar x^2)/3+(2+\bar x^2)/3=2$.

At the centre of the regime, $\bar x=0$, the terms with $\arcsin \bar x$ vanish and
\begin{align}
D_{zz} \simeq \frac{g\, e^2\, k_0\, \tilde\eta}{2\,\pi} \,, \quad
D_{xx} \simeq \frac{g\, e^2\, k_0\, \tilde\eta}{3\,\pi} \,, \quad
D_{yy} \simeq \frac{g\, e^2\, k_0\, \tilde\eta}{6\,\pi} \,.
\end{align}
This gives $D_{zz}/D_{xx}=3/2$ and $D_{yy}/D_{xx}=1/2$. For $\bar x\to1^-$ we have $\arcsin \bar x\to\pi/2$ and $\sqrt{1-\bar x^2}\to0$, and hence
\begin{align}
D_{zz} \to \frac{g\, e^2\, k_0\, \tilde\eta}{4} \,, \quad
D_{xx},\, D_{yy} \to \frac{g\, e^2\, k_0\, \tilde\eta}{8} \,.
\end{align}
The ratios are then $2$ and $1$. The limit is taken after $\Delta\to0$, in accordance with the condition $\tilde\eta-\mu\gg\Delta$. The values are those of the gapless ring-cyclide at $\mu=\tilde\eta$, and they coincide with the leading terms of Eq.~\eqref{eq:smalldelta-dzz}.

\subsubsection{Onset of the hole-pocket}

We now consider $\delta\equiv\tilde\eta-\mu-\Delta\ll\Delta$, where the hole-pocket has just appeared. The electron arc and the boundary angle $\phi_0$ vary smoothly across the entry, and the hole arc supplies the additional terms. From $\cos\phi_h=1-\delta/\tilde\eta$ we have $\phi_h\simeq\sqrt{2\,\delta/\tilde\eta}\ll1$. On the hole arc, $\tilde\eta-\mu=\Delta+\delta$, and so
\begin{align}
|\mu_{\rm eff}| = \tilde\eta\cos\phi-\mu = \Delta+\delta-\tilde\eta\left( 1-\cos\phi \right) \simeq \Delta + \frac{\tilde\eta}{2} \left( \phi_h^2-\phi^2 \right) , \quad |\phi|\le\phi_h \,.
\end{align}
The deviation of $|\mu_{\rm eff}|$ from $\Delta$ is at most $\delta\ll\Delta$. The local factor of Eq.~\eqref{eq:bipolar-reduced} is therefore
\begin{align}
|\mu_{\rm eff}|-\frac{\Delta^2}{|\mu_{\rm eff}|} = \frac{\left( |\mu_{\rm eff}|-\Delta \right) \left( |\mu_{\rm eff}|+\Delta \right)}{|\mu_{\rm eff}|} \simeq 2 \left( |\mu_{\rm eff}|-\Delta \right) \simeq \tilde\eta \left( \phi_h^2-\phi^2 \right) .
\end{align}
It vanishes linearly in $\delta$ at the pocket edge, whereas the weight $|\mu_{\rm eff}|$ of the DOS stays equal to $\Delta$. With $\check w_z\simeq \check w_x\simeq1$ and $\check w_y\simeq\phi^2$ on this small arc, the Drude weights of the pocket are
\begin{align}
& \delta D_{zz} \simeq \delta D_{xx} \simeq \frac{g\, e^2\, k_0\, \tilde\eta}{8\,\pi} \int_{-\phi_h}^{\phi_h}\! d\phi \left( \phi_h^2-\phi^2 \right) = \frac{g\, e^2\, k_0\, \tilde\eta}{6\,\pi}\, \phi_h^3 \,, \nn
& \delta D_{yy} \simeq \frac{g\, e^2\, k_0\, \tilde\eta}{8\,\pi} \int_{-\phi_h}^{\phi_h}\! d\phi\, \phi^2 \left( \phi_h^2-\phi^2 \right) = \frac{g\, e^2\, k_0\, \tilde\eta}{30\,\pi}\, \phi_h^5 \,.
\end{align}
Inserting $\phi_h^3=(2\,\delta/\tilde\eta)^{3/2}$ and $\phi_h^5=(2\,\delta/\tilde\eta)^{5/2}$ reproduces the three Drude weights quoted in Sec.~\ref{secgtxbp}, with the prefactors $\sqrt2/(3\,\pi)$ and $2\sqrt2/(15\,\pi)$. The DOS follows from Eq.~\eqref{eq:bipolar-reduced} with $|\mu_{\rm eff}|\simeq\Delta$, namely
\begin{align}
\delta\rho \simeq \frac{g\, k_0}{4\,\pi^2\, v_0^2}\, \Delta\, 2\,\phi_h = \frac{g\, k_0\, \Delta}{2\,\pi^2\, v_0^2} \sqrt{\frac{2\,\delta}{\tilde\eta}} \,.
\end{align}
The relative corrections to all four results are of order $\delta/\Delta$.

\subsubsection{Static interband background}

The background constants follow from Eq.~\eqref{eq:epsb-partial}, with $\mu_{\rm eff}$ replaced by $|\mu_{\rm eff}|$ in $\mathcal{S}$. This replacement is allowed because the local function of Appendix~\ref{app:interlocal} depends only on $\tilde m=\max\left[ \Delta, |\mu_{\rm eff}| \right]$. We write $m\equiv\mu_{\rm eff}=\mu-\tilde\eta\cos\phi$ below. In the twin-crescent regime, $\mathcal{S}=4/\Delta$ on the gapped arcs $\phi_h<|\phi|<\phi_0$, and $\mathcal{S}=3/|m|+\Delta^2/|m|^3$ on the pockets. Hence $\epsilon_{b,i}=1+g\, e^2\, k_0\, \mathcal{S}_i/(24\,\pi)$ with
\begin{align}
\label{eq:bipolar-Ssplit}
\mathcal{S}_i = \frac{4}{\Delta} \int_{\rm gap}\! d\phi\; \check w_i (\phi) + \int_{\rm occ}\! d\phi\; \check w_i (\phi) \left( \frac{3}{|m|} + \frac{\Delta^2}{|m|^3} \right) ,
\end{align}
where the first integral runs over the two gapped arcs. Besides $\int_{\rm occ} d\phi/|m|$, the pockets require the integrals of $\mathrm{sgn}(m)/m^2$ and $1/|m|^3$.

All of them follow from a reduction formula. From $\cos\phi=(\mu-m)/\tilde\eta$ we have $\tilde\eta^2\sin^2\phi=\bar w^2+2\,\mu\, m-m^2$, and differentiation gives
\begin{align}
\frac{d}{d\phi} \frac{\sin\phi}{m} = -\, \frac{\mu\, m+\bar w^2}{\tilde\eta\, m^2} \,, \quad
\frac{d}{d\phi} \frac{\sin\phi}{m^2} = \frac{m^2-3\,\mu\, m-2\,\bar w^2}{\tilde\eta\, m^3} \,.
\end{align}
Let $\tilde F$, $\tilde G$, and $\tilde H$ denote the antiderivatives of $1/m$, $1/m^2$, and $1/m^3$. We take $\tilde F$ from the previous derivation of $\Lambda$, on each of the arcs $(0,\phi_b)$ and $(\phi_b,\pi)$. Integrating the two relations above defines $\tilde G$ and $\tilde H$ through
\begin{align}
\tilde G = -\, \frac{1}{\bar w^2} \left( \frac{\tilde\eta\, \sin\phi}{m} + \mu\, \tilde F \right) , \quad
\tilde H = \frac{1}{2\,\bar w^2} \left( -\, \frac{\tilde\eta\, \sin\phi}{m^2} - 3\,\mu\, \tilde G + \tilde F \right) .
\end{align}
Because $\sin\phi$ and $\tilde F$ vanish at $\phi=0$ and $\phi=\pi$, so do $\tilde G$ and $\tilde H$. At the boundaries of the pockets, $m=\Delta$ and $\tilde\eta\sin\phi_0=s_-$ at $\phi_0$, and $m=-\Delta$ and $\tilde\eta\sin\phi_h=s_+$ at $\phi_h$. By the evenness in $\phi$, we need only the integrals over $\phi>0$. On the hole arc, $1/|m|=-1/m$ and $1/|m|^3=-1/m^3$. We find
\begin{align}
& \mathcal{A} = \int_{\phi_0}^\pi \frac{d\phi}{m} + \int_0^{\phi_h} \frac{d\phi}{|m|} = -\, \tilde F(\phi_0)-\tilde F(\phi_h) = \frac{\Lambda}{\bar w} \,, \nn
& \mathcal{Q}_2 = \int_{\phi_0}^\pi \frac{d\phi}{m^2} - \int_0^{\phi_h} \frac{d\phi}{m^2} = -\, \tilde G(\phi_0)-\tilde G(\phi_h) \,, 
\quad \mathcal{Q}_3 = \int_{\phi_0}^\pi \frac{d\phi}{m^3} - \int_0^{\phi_h} \frac{d\phi}{m^3} 
= -\, \tilde H(\phi_0)-\tilde H(\phi_h) \,.
\end{align}
The integrand of $\mathcal{Q}_2$ is $\mathrm{sgn}(m)/m^2$ on the whole pocket. Inserting the boundary values in $\tilde G$ gives
\begin{align}
\tilde G(\phi_0)+\tilde G(\phi_h) = \frac{1}{\bar w^2} \left[ \frac{s_+-s_-}{\Delta} - \mu \left( \tilde F(\phi_0)+\tilde F(\phi_h) \right) \right] = \frac{1}{\bar w^2} \left[ \frac{s_+-s_-}{\Delta} + \mu\, \mathcal{A} \right] ,
\end{align}
and hence the stated $\mathcal{Q}_2=-\,\bar w^{-2}\left[ (s_+-s_-)/\Delta+\mu\,\mathcal{A} \right]$. Similarly, $\tilde H$ gives
\begin{align}
\mathcal{Q}_3 = \frac{1}{2\,\bar w^2} \left[ \frac{s_++s_-}{\Delta^2} + 3\,\mu \left( \tilde G(\phi_0)+\tilde G(\phi_h) \right) - \left( \tilde F(\phi_0)+\tilde F(\phi_h) \right) \right] = \frac{1}{2\,\bar w^2} \left[ \frac{s_++s_-}{\Delta^2} - 3\,\mu\, \mathcal{Q}_2 + \mathcal{A} \right] .
\end{align}
Doubling these half-integrals gives $\int_{\rm occ} d\phi/|m|=2\,\mathcal{A}$, $\int_{\rm occ} d\phi\, \mathrm{sgn}(m)/m^2=2\,\mathcal{Q}_2$, and $\int_{\rm occ} d\phi/|m|^3=2\,\mathcal{Q}_3$.

For $i=z$, Eq.~\eqref{eq:bipolar-Ssplit} gives $3\cdot2\,\mathcal{A}+\Delta^2\cdot2\,\mathcal{Q}_3$ from the pockets. The gapped arcs have the total length $2\left( \phi_0-\phi_h \right)$ and contribute $8\left( \phi_0-\phi_h \right)/\Delta$. This is $\mathcal{S}_z$ in Eq.~\eqref{eq:epsb-disjoint-pocket}. For $i=x$, we use $\cos^2\phi=(\mu^2-2\,\mu\, m+m^2)/\tilde\eta^2$ on the pockets. The identity
\begin{align}
\frac{\cos^2\phi}{|m|^3} = \frac{1}{\tilde\eta^2} \left( \frac{\mu^2}{|m|^3} - \frac{2\,\mu\, \mathrm{sgn}(m)}{m^2} + \frac{1}{|m|} \right)
\end{align}
gives $\int_{\rm occ} d\phi\, \cos^2\phi/|m|^3=2\left( \mu^2\, \mathcal{Q}_3-2\,\mu\, \mathcal{Q}_2+\mathcal{A} \right)/\tilde\eta^2$. The integral of $\cos^2\phi/|m|$ is given in Eq.~\eqref{eq:bipolar-integrals}. On the gapped arcs, $\int_{\rm gap} d\phi\, \cos^2\phi=\left[ \phi+\sin\phi\cos\phi \right]_{\phi_h}^{\phi_0}$, that is
\begin{align}
\int_{\rm gap}\! d\phi\, \cos^2\phi = \phi_0-\phi_h + \frac{s_- \left( \mu-\Delta \right) - s_+ \left( \mu+\Delta \right)}{\tilde\eta^2} \,.
\end{align}
Collecting the three pieces gives $\mathcal{S}_x$ of Eq.~\eqref{eq:epsb-disjoint-pocket}. Since $\check w_y=\check w_z-\check w_x$, we have $\mathcal{S}_y=\mathcal{S}_z-\mathcal{S}_x$.

The bound follows from the monotonicity of $\mathcal{S}(\tilde m)=3/\tilde m+\Delta^2/\tilde m^3$, which decreases for $\tilde m\ge\Delta$ and equals $4/\Delta$ at $\tilde m=\Delta$. Each $\mathcal{S}_i$ is therefore at most $(4/\Delta)\int_0^{2\pi} d\phi\, \check w_i$, which is the insulating value. For $\phi_h\to0$, the hole-arc integrals vanish because their integrands are bounded on an interval of length $\phi_h$, and $s_+\to0$. The remaining electron-arc terms are those of the crescent-cyclide regime.

For $\Delta\ll\tilde\eta$ we use the expansions of the previous subsubsection. The combination $(s_+-s_-)/\Delta\simeq-\,2\,\mu/\bar w$ gives $\mathcal{Q}_2\simeq\mu\left( 2/\bar w-\mathcal{A} \right)/\bar w^2$, which is finite up to a logarithm. Hence $\Delta^2\, \mathcal{Q}_2$ is negligible, and $2\,\Delta^2\, \mathcal{Q}_3\simeq\left( s_++s_- \right)/\bar w^2\simeq2/\bar w$. The gapped arcs contribute $8\left( \phi_0-\phi_h \right)/\Delta\simeq16/\bar w$ because $\phi_0-\phi_h\simeq2\,\Delta/\bar w$. Together with $6\,\mathcal{A}=6\,\Lambda/\bar w$ and $\Lambda\simeq2\ln \!\left[ 2\,\bar w^2/(\tilde\eta\, \Delta) \right]$, this yields
\begin{align}
\mathcal{S}_z \simeq \frac{12}{\bar w} \left[ 
\ln \!\left( \frac{2\,\bar w^2}{\tilde\eta\, \Delta}\right) + \frac{3}{2} \right] .
\end{align}
The logarithm comes from the term $3/|m|$ near $\phi_b$, where $|m|\simeq\bar w\,|\phi-\phi_b|$ and the gap cuts the divergence off at $|m|\sim\Delta$. In $\mathcal{S}_x$, it enters only through the term $6\,\mu^2\, \mathcal{A}/\tilde\eta^2$, and all the other terms are free of $\ln \!\left( \tilde\eta/\Delta \right)$. At leading logarithmic accuracy, $\mathcal{S}_i\simeq \check w_i (\phi_b)\, 12\ln \!\left( \tilde\eta/\Delta \right)/\bar w$ for each $i$, with $\check w_z=1$, $\check w_x=\cos^2\phi_b=\mu^2/\tilde\eta^2$, and $\check w_y=\sin^2\phi_b=\bar w^2/\tilde\eta^2$. Multiplying by $g\, e^2\, k_0/(24\,\pi)$ gives the logarithm quoted in Sec.~\ref{secgtxbp}. The approximation requires $\ln \!\left( \tilde\eta/\Delta \right)\gg1$ and $\tilde\eta-\mu\gg\Delta$.

\subsection{Crescent-cyclide regime of GNR}
\label{app:crescent}

This appendix derives the DOS of Eq.~\eqref{eq:partialgapped-dos} and the arc integral $\tilde{\mathcal{J}}$ of Sec.~\ref{secgtxpg}. We take $\mu\ge0$, and the conditions $|\mu-\tilde\eta|<\Delta<\mu+\tilde\eta$ then force $\mu>0$. The ratio $(\mu-\Delta)/\tilde\eta$ lies in $(-1,1)$ in this regime, because $\mu-\tilde\eta<\Delta$ and $\mu+\tilde\eta>\Delta$. The angle $\phi_0\in(0,\pi)$ with $\cos\phi_0=(\mu-\Delta)/\tilde\eta$ therefore exists. Since $\mu_{\rm eff}(\phi)=\mu-\tilde\eta\cos\phi$ increases on $(0,\pi)$ and equals $\Delta$ at $\phi_0$, we have $\mu_{\rm eff}<\Delta$ for $|\phi|<\phi_0$ and $\mu_{\rm eff}>\Delta$ for $\phi_0<|\phi|\le\pi$. The condition $\mu-\tilde\eta>-\Delta$ gives $\mu_{\rm eff}>-\Delta$ everywhere. Hence there is no hole-pocket, and only the arc $\phi\in(\phi_0,2\pi-\phi_0)$ carries an FS. The integrands are even in $\phi$, and each integral over this arc is twice the integral over $(\phi_0,\pi)$.

As in the twin-crescent regime, the $\eta_x^2$ terms of Eq.~\eqref{eq:conedrude} are dropped. The DOS is then $\rho(\mu)=g\, k_0/(4\,\pi^2\, v_0^2) \int d\phi\, \mu_{\rm eff}$ over the conducting arc, with
\begin{align}
\int_{\phi_0}^{2\pi-\phi_0}\! d\phi\, \mu_{\rm eff} = \mu \left( 2\,\pi-2\,\phi_0 \right) - \tilde\eta \Big[ \sin\phi \Big]_{\phi_0}^{2\pi-\phi_0} = 2 \left[ \mu \left( \pi-\phi_0 \right) + \tilde\eta\, \sin\phi_0 \right] .
\end{align}
Factoring out $\mu$ gives
\begin{align}
\rho(\mu) = \frac{g\, k_0}{2\,\pi^2\, v_0^2} \left[ \mu \left( \pi-\phi_0 \right) + \tilde\eta\, \sin\phi_0 \right] = \frac{g\, k_0\, \mu}{2\,\pi\, v_0^2} \left( 1-\frac{\phi_0}{\pi}+\frac{\tilde\eta\, \sin\phi_0}{\pi\, \mu} \right) ,
\end{align}
which is Eq.~\eqref{eq:partialgapped-dos}. At the upper edge of the regime, $\mu=\Delta+\tilde\eta$, we have $\phi_0=0$ and the bracket equals $1$. At the lower edge, $\mu=\tilde\eta-\Delta$, we have $\phi_0=\pi$, $\sin\phi_0=0$, and the DOS vanishes.

By the evenness in $\phi$, $\tilde{\mathcal{J}}=2\int_{\phi_0}^\pi d\phi/\mu_{\rm eff}$. The substitution $u=\tan(\phi/2)$ gives $d\phi=2\,du/(1+u^2)$ and
\begin{align}
\mu_{\rm eff} = \frac{\left( \mu+\tilde\eta \right) u^2+\left( \mu-\tilde\eta \right)}{1+u^2} \,.
\end{align}
The three cases of $\mu\gtrless\tilde\eta$ differ in the sign of the constant term of the numerator:
\begin{enumerate}

\item For $\mu>\tilde\eta$, we have $\mu_{\rm eff}>0$ on the whole ring. With $\tilde w=\sqrt{\mu^2-\tilde\eta^2}$ and $\left( \mu+\tilde\eta \right) \left( \mu-\tilde\eta \right)=\tilde w^2$,
\begin{align}
\int\! \frac{d\phi}{\mu_{\rm eff}} = \int\! \frac{2\, du}{\left( \mu+\tilde\eta \right) u^2+\left( \mu-\tilde\eta \right)} = \frac{2}{\tilde w} \arctan \left( \sqrt{\frac{\mu+\tilde\eta}{\mu-\tilde\eta}}\; u \right) .
\end{align}
The upper limit $\phi=\pi$ corresponds to $u\to\infty$ and gives $\pi/2$ in the arctangent. At the lower limit, $z_0\equiv\sqrt{(\mu+\tilde\eta)/(\mu-\tilde\eta)}\,\tan(\phi_0/2)>0$, and $\pi/2-\arctan z_0=\arctan(1/z_0)$. Therefore
\begin{align}
\tilde{\mathcal{J}} = \frac{4}{\tilde w} \arctan \left[ \sqrt{\frac{\mu-\tilde\eta}{\mu+\tilde\eta}}\; \cot\frac{\phi_0}{2} \right] , \quad \mu>\tilde\eta \,.
\end{align}
For $\phi_0\to0$, the argument diverges and $\tilde{\mathcal{J}}\to2\,\pi/\tilde w$, which is the integral over the whole ring in the fully-conducting regime.

\item For $\mu<\tilde\eta$, the local doping vanishes at $\phi_b=\arccos(\mu/\tilde\eta)$, and $\phi_b<\phi_0$ because $\mu_{\rm eff}(\phi_0)=\Delta>0$. The arc $(\phi_0,\pi)$ does not contain $\phi_b$. The antiderivative $\tilde F(\phi)=\bar w^{-1}\ln \!\left| \sin\frac{\phi-\phi_b}{2}/\sin\frac{\phi+\phi_b}{2} \right|$ of Appendix~\ref{app:twin} applies, with $\tilde F(\pi)=0$. Hence
\begin{align}
\tilde{\mathcal{J}} = -\, 2\, \tilde F(\phi_0) = \frac{2}{\bar w} \ln \!\left( \frac{\sin\frac{\phi_0+\phi_b}{2}}{\sin\frac{\phi_0-\phi_b}{2}} \right) \,, \quad \mu<\tilde\eta \,.
\end{align}

\item At $\mu=\tilde\eta$, the local doping is $\mu_{\rm eff}=\tilde\eta \left( 1-\cos\phi \right)=2\,\tilde\eta\, \sin^2(\phi/2)$. Both expressions above are singular in this limit, because $\tilde w$ and $\bar w$ vanish. The integral is elementary, namely $\int d\phi/\mu_{\rm eff}=-\cot(\phi/2)/\tilde\eta$, and
\begin{align}
\tilde{\mathcal{J}} = 2 \int_{\phi_0}^\pi \frac{d\phi}{2\,\tilde\eta\, \sin^2(\phi/2)} = \frac{2}{\tilde\eta} \cot\frac{\phi_0}{2} \,, \quad \mu=\tilde\eta \,.
\end{align}
This value is the common limit of the two cases. For $\mu\to\tilde\eta^-$, we have $\phi_b\to0$ and $\bar w\simeq\tilde\eta\,\phi_b$, and the logarithm becomes $\ln \!\left[ \sin\frac{\phi_0+\phi_b}{2}/\sin\frac{\phi_0-\phi_b}{2} \right]\simeq\phi_b\cot(\phi_0/2)$. For $\mu\to\tilde\eta^+$, we have $\tilde w\simeq\sqrt{2\,\tilde\eta \left( \mu-\tilde\eta \right)}$, so that $\sqrt{(\mu-\tilde\eta)/(\mu+\tilde\eta)}=\tilde w/(\mu+\tilde\eta)\simeq \tilde w/(2\,\tilde\eta)$, and $\arctan z\simeq z$. In both cases, $\tilde{\mathcal{J}}\to2\cot(\phi_0/2)/\tilde\eta$. The regime requires $\Delta<2\,\tilde\eta$ at this point, so that $\phi_0$ is defined.

\end{enumerate}

\subsubsection{Assembly of the Drude weights}
At leading order in $\eta_x$, Eq.~\eqref{eq:conedrude} reduces on the conducting arc to
\begin{align}
\label{eq:crescent-reduced}
D_{ii} = \frac{g\, e^2\, k_0}{8\,\pi} \int_{\phi_0}^{2\pi-\phi_0}\! d\phi\; \check w_i (\phi) \left( \mu_{\rm eff}-\frac{\Delta^2}{\mu_{\rm eff}} \right) ,
\end{align}
with $\check w_z=1$, $\check w_x=\cos^2\phi$, and $\check w_y=\sin^2\phi$. This is Eq.~\eqref{eq:bipolar-reduced} with $|\mu_{\rm eff}|=\mu_{\rm eff}$. Since $\check w_z=\check w_x+\check w_y$, we have $D_{zz}=D_{xx}+D_{yy}$. Because $\sin^2\phi=1-\cos^2\phi$, only four integrals are needed, namely those of $\mu_{\rm eff}$, $1/\mu_{\rm eff}$, $\cos^2\phi\,\mu_{\rm eff}$, and $\cos^2\phi/\mu_{\rm eff}$. The first is given above, $\int d\phi\,\mu_{\rm eff}=2\left[ \mu \left( \pi-\phi_0 \right)+\tilde\eta\sin\phi_0 \right]$, and the second is $\tilde{\mathcal{J}}$. For the other two we use $\cos\phi=\left( \mu-\mu_{\rm eff} \right)/\tilde\eta$ and the elementary arc integrals
\begin{align}
& \int_{\phi_0}^{2\pi-\phi_0}\! d\phi = 2 \left( \pi-\phi_0 \right) , \quad
\int_{\phi_0}^{2\pi-\phi_0}\! d\phi \cos\phi = -\, 2\sin\phi_0 \,, \nn
& \int_{\phi_0}^{2\pi-\phi_0}\! d\phi \cos^2\phi = \pi-\phi_0-\sin\phi_0\cos\phi_0 \,, \quad
\int_{\phi_0}^{2\pi-\phi_0}\! d\phi \cos^3\phi = -\, 2\sin\phi_0+\frac{2}{3}\sin^3\phi_0 \,.
\end{align}
They follow from the antiderivatives $\sin\phi$, $\phi/2+\sin(2\phi)/4$, and $\sin\phi-\sin^3\phi/3$ of $\cos\phi$, $\cos^2\phi$, and $\cos^3\phi$, and from the symmetry of the limits about $\phi=\pi$. The integral with $\mu_{\rm eff}$ is
\begin{align}
\int_{\phi_0}^{2\pi-\phi_0}\! d\phi\, \cos^2\phi\, \mu_{\rm eff} = \mu \left( \pi-\phi_0-\sin\phi_0\cos\phi_0 \right)+2\,\tilde\eta \left( \sin\phi_0-\frac{\sin^3\phi_0}{3} \right) ,
\end{align}
and the integral with $1/\mu_{\rm eff}$ is
\begin{align}
\int_{\phi_0}^{2\pi-\phi_0}\! d\phi\, \frac{\cos^2\phi}{\mu_{\rm eff}} = \frac{1}{\tilde\eta^2} \int d\phi \left( \frac{\mu^2}{\mu_{\rm eff}}-2\,\mu+\mu_{\rm eff} \right) = \frac{\mu^2\, \tilde{\mathcal{J}}-2\,\mu \left( \pi-\phi_0 \right)+2\,\tilde\eta\sin\phi_0}{\tilde\eta^2} \,.
\end{align}

For $D_{zz}$, Eq.~\eqref{eq:crescent-reduced} gives $\left( g\, e^2\, k_0/8\,\pi \right) \left[ 2\,\mu \left( \pi-\phi_0 \right)+2\,\tilde\eta\sin\phi_0-\Delta^2\,\tilde{\mathcal{J}} \right]$, which is the first line of Eq.~\eqref{eq:partialgapped-exact}. For $D_{xx}$ we insert the two integrals above,
\begin{align}
D_{xx} = \frac{g\, e^2\, k_0}{8\,\pi} \left\{ \mu \left( \pi-\phi_0-\sin\phi_0\cos\phi_0 \right)+2\,\tilde\eta \left( \sin\phi_0-\frac{\sin^3\phi_0}{3} \right) - \frac{\Delta^2}{\tilde\eta^2} \left[ \mu^2\, \tilde{\mathcal{J}}-2\,\mu \left( \pi-\phi_0 \right)+2\,\tilde\eta\sin\phi_0 \right] \right\} .
\end{align}
The terms with $\mu \left( \pi-\phi_0 \right)$ and with $\tilde{\mathcal{J}}$ already have the form of Eq.~\eqref{eq:partialgapped-exact}, with the coefficient $1+2\,\Delta^2/\tilde\eta^2$ for $\mu \left( \pi-\phi_0 \right)$ and $-\,\Delta^2\,\mu^2/\tilde\eta^2$ for $\tilde{\mathcal{J}}$. The terms proportional to $\sin\phi_0$ are simplified with $\cos\phi_0=\left( \mu-\Delta \right)/\tilde\eta$ and $\sin^2\phi_0=\left[ \tilde\eta^2-\left( \mu-\Delta \right)^2 \right]/\tilde\eta^2$. They combine to
\begin{align}
\frac{\sin\phi_0}{\tilde\eta} \left[ -\,\mu \left( \mu-\Delta \right)+2\,\tilde\eta^2-\frac{2}{3} \left( \tilde\eta^2-\left( \mu-\Delta \right)^2 \right)-2\,\Delta^2 \right] = \frac{\tilde\eta\sin\phi_0 \left( 4\,\tilde\eta^2-\mu^2-\mu\,\Delta-4\,\Delta^2 \right)}{3\,\tilde\eta^2} \,.
\end{align}
This gives the second line of Eq.~\eqref{eq:partialgapped-exact}. The third line is $D_{yy}=D_{zz}-D_{xx}$. The coefficient of $\mu \left( \pi-\phi_0 \right)$ is $2-\left( 1+2\,\Delta^2/\tilde\eta^2 \right)=1-2\,\Delta^2/\tilde\eta^2$. The coefficient of $\tilde\eta\sin\phi_0$ is $2-\left( 4\,\tilde\eta^2-\mu^2-\mu\,\Delta-4\,\Delta^2 \right)/\left( 3\,\tilde\eta^2 \right)=\left( 2\,\tilde\eta^2+\mu^2+\mu\,\Delta+4\,\Delta^2 \right)/\left( 3\,\tilde\eta^2 \right)$. The coefficient of $\tilde{\mathcal{J}}$ is $-\,\Delta^2+\Delta^2\,\mu^2/\tilde\eta^2=\Delta^2 \left( \mu^2-\tilde\eta^2 \right)/\tilde\eta^2$. All three are those of Eq.~\eqref{eq:partialgapped-exact}.

As a check we take $\phi_0\to0$, where $\sin\phi_0\to0$ and $\tilde{\mathcal{J}}\to2\,\pi/\tilde w$. With $\tilde\eta^2=\left( \mu-\tilde w \right)\left( \mu+\tilde w \right)$ we obtain
\begin{align}
& D_{zz}=\frac{g\, e^2\, k_0}{4} \left[ \mu-\frac{\Delta^2}{\tilde w} \right] , \quad
D_{xx}=g\, e^2\, k_0 \left[ \frac{\mu}{8}-\frac{\mu\, \Delta^2}{4\, \tilde w \left( \mu+\tilde w \right)} \right] , \quad
D_{yy}=g\, e^2\, k_0 \left[ \frac{\mu}{8}-\frac{\Delta^2}{4 \left( \mu+\tilde w \right)} \right] ,
\end{align}
where we used $\mu \left( 1+2\,\Delta^2/\tilde\eta^2 \right)-2\,\Delta^2\,\mu^2/\left( \tilde\eta^2\, \tilde w \right)=\mu-2\,\Delta^2\,\mu/\left[ \tilde w \left( \mu+\tilde w \right) \right]$ and $\mu \left( 1-2\,\Delta^2/\tilde\eta^2 \right)+2\,\Delta^2\, \tilde w/\tilde\eta^2=\mu-2\,\Delta^2/\left( \mu+\tilde w \right)$. These are the terms of Eq.~\eqref{eq:dzzxtiltgapped} without the $\eta_x^2$ corrections.

\subsubsection{Limit of a small in-plane tilt and the anisotropy ratios}
We take $\tilde\eta\ll\Delta$ at $\mu=\Delta+\tilde\eta\,\xi$, with $\xi\in(-1,1)$ held fixed. Then $\mu_{\rm eff}=\Delta+\tilde\eta\left( \xi-\cos\phi \right)$ and $\cos\phi_0=\xi$. Hence $\sin\phi_0=\sqrt{1-\xi^2}$ and $\pi-\phi_0=\pi/2+\arcsin\xi$. We start from Eq.~\eqref{eq:crescent-reduced}. We have $\mu_{\rm eff}^2-\Delta^2=2\,\Delta\,\tilde\eta \left( \xi-\cos\phi \right)+\tilde\eta^2 \left( \xi-\cos\phi \right)^2$ and $\mu_{\rm eff}=\Delta \left[ 1+\mathcal{O}(\tilde\eta/\Delta) \right]$ in the denominator. Therefore
\begin{align}
D_{ii} \simeq \frac{g\, e^2\, k_0\, \tilde\eta}{4\,\pi} \int_{\phi_0}^{2\pi-\phi_0}\! d\phi\; \check w_i (\phi) \left( \xi-\cos\phi \right) .
\end{align}
With the arc integrals of the previous subsubsection, we define $B_z \equiv \xi \left( \pi-\phi_0 \right)+\sin\phi_0$. The weight $\check w_z$ gives $\int d\phi \left( \xi-\cos\phi \right) = 2\,B_z$. The weight $\check w_x$ gives $\xi \left( \pi-\phi_0-\xi\sin\phi_0 \right)+2\sin\phi_0-(2/3)\sin^3\phi_0$. With $\xi^2=1-\sin^2\phi_0$, this equals $B_z+\sin^3\phi_0/3$. Since $\check w_y=\check w_z-\check w_x$, the weight $\check w_y$ gives $B_z-\sin^3\phi_0/3$. Therefore
\begin{align}
D_{zz} = \frac{g\, e^2\, k_0\, \tilde\eta}{4\,\pi}\, 2\, B_z \,, \quad
D_{xx} = \frac{g\, e^2\, k_0\, \tilde\eta}{4\,\pi} \left( B_z+\frac{\sin^3\phi_0}{3} \right) , \quad
D_{yy} = \frac{g\, e^2\, k_0\, \tilde\eta}{4\,\pi} \left( B_z-\frac{\sin^3\phi_0}{3} \right) .
\end{align}
Using $\sin^3\phi_0=\left( 1-\xi^2 \right)\sin\phi_0$, we have $B_z+\sin^3\phi_0/3=\xi\left( \pi/2+\arcsin\xi \right)+(4-\xi^2)\sqrt{1-\xi^2}/3$ and $B_z-\sin^3\phi_0/3=\xi\left( \pi/2+\arcsin\xi \right)+(2+\xi^2)\sqrt{1-\xi^2}/3$. These are Eq.~\eqref{eq:partialgapped-dzz}. The relation $D_{zz}=D_{xx}+D_{yy}$ is manifest, and Eq.~\eqref{eq:partialgapped-wp} follows from $\omega_{p,i}^2=4\,D_{ii}$.

The edge values are obtained as follows. For $\xi\to1^-$ we have $\phi_0\to0$ and $\sin\phi_0\to0$. The brackets $B_z$ and $B_z\pm\sin^3\phi_0/3$ all tend to $\pi$, and $D_{zz}\to g\, e^2\, k_0\, \tilde\eta/2$ and $D_{xx},D_{yy}\to g\, e^2\, k_0\, \tilde\eta/4$. The fully-conducting weights of Eq.~\eqref{eq:dzzxtiltgapped} must agree with these values at the threshold $\mu=\Delta+\tilde\eta$, where $\tilde w=\sqrt{\Delta \left( \Delta+2\,\tilde\eta \right)}=\Delta+\tilde\eta+\mathcal{O}(\tilde\eta^2/\Delta)$. Without the $\eta_x^2$ terms, $\mu-\Delta^2/\tilde w=2\,\tilde\eta+\mathcal{O}(\tilde\eta^2/\Delta)$, $\mu/8-\mu\,\Delta^2/\left[ 4\, \tilde w \left( \mu+\tilde w \right) \right]=\tilde\eta/4+\mathcal{O}(\tilde\eta^2/\Delta)$, and $\mu/8-\Delta^2/\left[ 4 \left( \mu+\tilde w \right) \right]=\tilde\eta/4+\mathcal{O}(\tilde\eta^2/\Delta)$. They give $D_{zz}=g\, e^2\, k_0\, \tilde\eta/2$ and $D_{xx}=D_{yy}=g\, e^2\, k_0\, \tilde\eta/4$, in agreement. For $\xi\to-1^+$ we have $\phi_0\to\pi$. We set $\alpha=\pi-\phi_0\to0$. Then $\xi=-\cos\alpha$ and $\sin\phi_0=\sin\alpha$, and $B_z=\sin\alpha-\alpha\cos\alpha\simeq\alpha^3/3$ and $\sin^3\phi_0/3\simeq\alpha^3/3$. All three brackets vanish, and $D_{zz}$, $D_{xx}$, and the plasmon frequencies go to zero.

For the ratios, $B_x\equiv B_z+\sin^3\phi_0/3$ and $B_y\equiv B_z-\sin^3\phi_0/3$ give $D_{zz}/D_{xx}=2\,B_z/B_x$ and $D_{yy}/D_{xx}=B_y/B_x=D_{zz}/D_{xx}-1$. At $\xi=0$ we have $\phi_0=\pi/2$ and $B_z=1$. Then $B_x=4/3$ and $B_y=2/3$, and the ratios are $3/2$ and $1/2$. At $\xi\to1^-$ the three brackets are equal, and the ratios are $2$ and $1$. At $\xi\to-1^+$ we have $B_x\simeq2\,\alpha^3/3$ and $B_y=B_z-\sin^3\alpha/3=\mathcal{O}(\alpha^5)$. Hence $D_{zz}/D_{xx}\to1$ and $D_{yy}/D_{xx}\to0$. This agrees with the picture that the last pocket is a small region around $\phi=\pi$, where $\cos^2\phi\simeq1$ and $\sin^2\phi\simeq0$.

For the monotonic rise we write $D_{zz}/D_{xx}=2/\left( 1+\sin^3\phi_0/(3\,B_z) \right)$. The ratio rises if $h\equiv\sin^3\phi_0/B_z$ decreases with $\xi$. We have $d\phi_0/d\xi=-1/\sin\phi_0$, which gives $d\sin^3\phi_0/d\xi=-3\,\xi\sin\phi_0$. We also have $dB_z/d\xi=\pi-\phi_0$, because the terms $\xi/\sin\phi_0$ and $-\xi/\sin\phi_0$ cancel. Then
\begin{align}
\frac{dh}{d\xi} = -\, \frac{\sin\phi_0}{B_z^2} \left[ \left( 1+2\,\xi^2 \right) \left( \pi-\phi_0 \right) + 3\,\xi\sin\phi_0 \right] ,
\end{align}
where we have used $3\,\xi\,B_z+\sin^2\phi_0 \left( \pi-\phi_0 \right)=\left( 1+2\,\xi^2 \right)\left( \pi-\phi_0 \right)+3\,\xi\sin\phi_0$. The bracket is positive for $\xi\ge0$. For $\xi<0$ we set $\alpha=\pi-\phi_0\in(0,\pi/2)$ and the bracket becomes $H(\alpha)=\alpha \left( 1+2\cos^2\alpha \right)-3\sin\alpha\cos\alpha$. We have $H(0)=0$ and $H'(\alpha)=4\sin\alpha \left( \sin\alpha-\alpha\cos\alpha \right)>0$, because $\tan\alpha>\alpha$. Hence $q$ decreases with $\xi$, and $D_{zz}/D_{xx}$ rises monotonically. So does $D_{yy}/D_{xx}$.

\subsubsection{Limit of a small gap}
We take $\Delta\ll\tilde\eta$ and write $\mu=\tilde\eta+\Delta\,\chi$ with $\chi\in(-1,1)$, $\check\varepsilon\equiv\Delta/\tilde\eta$, and $\lambda\equiv\check\varepsilon \left( 1-\chi \right)$. Then $\cos\phi_0=1-\lambda$, and
\begin{align}
\phi_0 = \sqrt{2\,\lambda} \left( 1+\frac{\lambda}{12}+\cdots \right) , \quad
\sin\phi_0 = \sqrt{2\,\lambda} \left( 1-\frac{\lambda}{4}+\cdots \right) ,
\end{align}
where $\sin\phi_0=\sqrt{2\,\lambda-\lambda^2}$. We also introduce $\beta\equiv\sqrt{2\,\Delta^3/\tilde\eta}$. Both $\phi_0$ and $\sin\phi_0$ contain only odd powers of $\sqrt\lambda$, and hence half-integer powers of $\Delta$. The same holds for $\Delta^2\,\tilde{\mathcal{J}}$, as shown below. The terms $\mu\,\pi$ and the factors $1\pm2\,\Delta^2/\tilde\eta^2$ in Eq.~\eqref{eq:partialgapped-exact} are integer powers. We collect the half-integer and the integer powers separately.

We first compute $\Delta^2\,\tilde{\mathcal{J}}$. For $\chi>0$ we have $\tilde w=\sqrt{2\,\tilde\eta\,\Delta\,\chi}\left[ 1+\mathcal{O}(\check\varepsilon) \right]$. Since $\cot(\phi_0/2)=\sqrt{(2-\lambda)/\lambda}$, the argument of the arctangent is $\sqrt{\chi/(1-\chi)}\left[ 1+\mathcal{O}(\check\varepsilon) \right]$. Its arctangent is $\arcsin\sqrt\chi$, because $\tan\theta=\sqrt{\chi/(1-\chi)}$ means $\sin\theta=\sqrt\chi$. Hence $\tilde{\mathcal{J}}=\Delta^{-1/2}$ times a series in integer powers of $\Delta$, with
\begin{align}
\frac{\Delta^2}{2}\,\tilde{\mathcal{J}} \simeq \frac{2\,\Delta^2}{\sqrt{2\,\tilde\eta\,\Delta\,\chi}}\,\arcsin\sqrt\chi = \beta\, G(\chi) \,.
\end{align}
For $\chi<0$ we have $\bar w=\sqrt{2\,\tilde\eta\,\Delta\,|\chi|}\left[ 1+\mathcal{O}(\check\varepsilon) \right]$ and $\phi_b\simeq\sqrt{2\,\Delta\,|\chi|/\tilde\eta}$. The sine ratio in $\tilde{\mathcal{J}}$ tends to $(\phi_0+\phi_b)/(\phi_0-\phi_b)=\left( \sqrt{1-\chi}+\sqrt{-\chi} \right)^2$, where we used $\left( \sqrt{1-\chi} \right)^2-\left( \sqrt{-\chi} \right)^2=1$. Its logarithm is $2\ln \!\left( \sqrt{-\chi}+\sqrt{1-\chi} \right)=2\,\mathrm{arcsinh}\sqrt{-\chi}$. This gives $\tilde{\mathcal{J}}\simeq\left( 4/\bar w \right)\mathrm{arcsinh}\sqrt{-\chi}$ and again $\left( \Delta^2/2 \right)\tilde{\mathcal{J}}\simeq\beta\,G(\chi)$ with the second case of $G$. At $\chi=0$ the exact value $\tilde{\mathcal{J}}=2\cot(\phi_0/2)/\tilde\eta\simeq4/\left( \tilde\eta\,\phi_0 \right)=4/\sqrt{2\,\tilde\eta\,\Delta}$ gives $G(0)=1$. The two cases of $G$ have the same expansion $1+\chi/6+\cdots$ about $\chi=0$, because $\arcsin x/x=1+x^2/6+\cdots$ and $\mathrm{arcsinh}\, y/y=1-y^2/6+\cdots$ with $x^2=\chi$ and $y^2=-\chi$. Hence $G$ is smooth.

For $D_{zz}$ the half-integer terms of Eq.~\eqref{eq:partialgapped-exact} are $-\,\mu\,\phi_0+\tilde\eta\sin\phi_0-(\Delta^2/2)\,\tilde{\mathcal{J}}$. With $\tilde\eta\,\lambda=\Delta \left( 1-\chi \right)$, $\mu=\tilde\eta+\Delta\,\chi$, and the expansions above,
\begin{align}
-\,\mu\,\phi_0+\tilde\eta\sin\phi_0 = -\sqrt{2\,\lambda} \left( \Delta\,\chi+\frac{\tilde\eta\,\lambda}{3} \right) = -\,\beta\, \frac{\left( 1+2\,\chi \right)\sqrt{1-\chi}}{3} \,,
\end{align}
where $\sqrt{2\,\lambda}\,\Delta=\beta\sqrt{1-\chi}$ was used. Together with $-\,\beta\,G$ this gives $-\,\beta\,\mathcal{K}(\chi)$ of Eq.~\eqref{eq:smalldelta-K}, and the integer part is $\mu\,\pi$. Hence $D_{zz}\simeq(g\, e^2\, k_0/4\,\pi) \left[ \pi\,\mu-\beta\,\mathcal{K} \right]$, which is the first expression in Eq.~\eqref{eq:smalldelta-dzz}.

For $D_{xx}$ we need $4\,\tilde\eta^2-\mu^2-\mu\,\Delta-4\,\Delta^2=3\,\tilde\eta^2-\tilde\eta\,\Delta \left( 1+2\,\chi \right)+\mathcal{O}(\Delta^2)$. The term $\tilde\eta\sin\phi_0 \left( \cdots \right)/\left( 3\,\tilde\eta^2 \right)$ equals $\tilde\eta\sin\phi_0 \left[ 1-\check\varepsilon \left( 1+2\,\chi \right)/3 \right]$. Its first part is the term already included above. Its second part is $-\,\tilde\eta\,\check\varepsilon\sqrt{2\,\lambda} \left( 1+2\,\chi \right)/3=-\,\beta \left( 1+2\,\chi \right)\sqrt{1-\chi}/3$. The $\tilde{\mathcal{J}}$ term is $-\,\Delta^2\,\mu^2\,\tilde{\mathcal{J}}/\tilde\eta^2\simeq-\,2\,\beta\,G$. The half-integer part of the bracket of $D_{xx}$ is therefore $-\,2\,\beta\left[ \left( 1+2\,\chi \right)\sqrt{1-\chi}/3+G \right]=-\,2\,\beta\,\mathcal{K}$, and the integer part is $\pi\,\mu \left( 1+2\,\check\varepsilon^2 \right)$. Hence
\begin{align}
D_{xx} \simeq \frac{g\, e^2\, k_0\, \mu}{8}-\frac{g\, e^2\, k_0}{4\,\pi}\,\beta\,\mathcal{K}(\chi)+\frac{g\, e^2\, k_0\, \Delta^2}{4\,\tilde\eta} \,,
\end{align}
where $\mu\simeq\tilde\eta$ was used in the last term. This last term is of order $\Delta^2$ and is smaller than the retained term of order $\Delta^{3/2}$. It is therefore dropped in Eq.~\eqref{eq:smalldelta-dzz}. The corrections to $D_{zz}$ are of order $\Delta^{5/2}$.

For $D_{yy}$ we need $2\,\tilde\eta^2+\mu^2+\mu\,\Delta+4\,\Delta^2=3\,\tilde\eta^2+\tilde\eta\,\Delta \left( 1+2\,\chi \right)+\mathcal{O}(\Delta^2)$. The half-integer term generated by its second part, $+\,\tilde\eta\,\check\varepsilon\sqrt{2\,\lambda} \left( 1+2\,\chi \right)/3=+\,\beta \left( 1+2\,\chi \right)\sqrt{1-\chi}/3$, cancels the half-integer term of $-\,\mu\,\phi_0+\tilde\eta\sin\phi_0$ found above. Relative to the $\tilde{\mathcal{J}}$ term of $D_{zz}$, the $\tilde{\mathcal{J}}$ term of $D_{yy}$ carries the extra factor $\left( \mu^2-\tilde\eta^2 \right)/\tilde\eta^2\simeq2\,\Delta\,\chi/\tilde\eta$, and it is therefore of order $\Delta^{5/2}$. Hence $D_{yy}$ has no term of order $\Delta^{3/2}$. The leading correction is the integer part, with the bracket $\pi\,\mu \left( 1-2\,\check\varepsilon^2 \right)$, and
\begin{align}
D_{yy} \simeq \frac{g\, e^2\, k_0\, \mu}{8}-\frac{g\, e^2\, k_0\, \Delta^2}{4\,\tilde\eta} \,.
\end{align}
This is the third expression in Eq.~\eqref{eq:smalldelta-dzz}, and the corrections are of order $\Delta^{5/2}$. Equation~\eqref{eq:smalldelta-wp} follows from $\omega_{p,i}^2=4\,D_{ii}$. For the ratio, $\omega_{p,z}^2/\omega_{p,x}^2=2 \left( \mu-\beta\,\mathcal{K}/\pi \right)/\left( \mu-2\,\beta\,\mathcal{K}/\pi \right)\simeq2 \left[ 1+\beta\,\mathcal{K}/\left( \pi\,\mu \right) \right]$ to first order in $\beta$.

At the upper edge, $\chi\to1^-$, we have $\lambda\to0$ and $\phi_0\to0$. The first term of $\mathcal{K}$ vanishes and $G\to\arcsin(1)=\pi/2$. Hence $\mathcal{K}\to\pi/2$, and $D_{zz}\simeq g\, e^2\, k_0\, \mu/4-g\, e^2\, k_0\, \beta/8$. The fully-conducting value of Eq.~\eqref{eq:dzzxtiltgapped} without the $\eta_x^2$ terms is $D_{zz}=(g\, e^2\, k_0/4) \left[ \mu-\Delta^2/\tilde w \right]$. At $\mu=\tilde\eta+\Delta$ we have $\tilde w=\sqrt{\Delta^2+2\,\Delta\,\tilde\eta}\simeq\sqrt{2\,\Delta\,\tilde\eta}$, and $\Delta^2/\tilde w\simeq\Delta^{3/2}/\sqrt{2\,\tilde\eta}=\beta/2$. Hence $D_{zz}=g\, e^2\, k_0\, \mu/4-g\, e^2\, k_0\, \beta/8$, which is the same value. This is the matching onto the fully-conducting regime. At the lower edge, $\chi\to-1^+$, we have $\mu+\Delta\to\tilde\eta$ and $\cos\phi_h=(\mu+\Delta)/\tilde\eta\to1$ in the notation of Appendix~\ref{app:twin}. The hole arc therefore has zero length. The function $\mathcal{K}(-1)=\ln \!\left( 1+\sqrt2 \right)-\sqrt2/3$ is finite. Eq.~\eqref{eq:partialgapped-exact} therefore coincides with the $\phi_h\to0$ limit of the twin-crescent weights, as shown in Appendix~\ref{app:twin}, and the Drude weights are continuous across this edge.

\subsubsection{Static interband background}
Appendix~\ref{app:interlocal} gives the static dielectric constant as $\epsilon_{b,i}=1+\left( g\, e^2\, k_0/24\,\pi \right)\int_0^{2\pi} d\phi\; \check w_i\, \mathcal{S}$ with $\mathcal{S}=\left( 3\,\tilde m^2+\Delta^2 \right)/\tilde m^3$ and $\tilde m=\max\left[ \Delta,|\mu_{\rm eff}| \right]$. In the crescent-cyclide regime we have $\mu_{\rm eff}>-\Delta$ everywhere, as shown at the beginning of this appendix. Hence $\tilde m=\Delta$ on the gapped arc $|\phi|<\phi_0$, where $\mathcal{S}=4/\Delta$, and $\tilde m=\mu_{\rm eff}$ on the conducting arc. This is Eq.~\eqref{eq:epsb-partial}.

On the conducting arc, $\mathcal{S}=3/\mu_{\rm eff}+\Delta^2/\mu_{\rm eff}^3$ has the derivative $-\,3/\mu_{\rm eff}^2-3\,\Delta^2/\mu_{\rm eff}^4<0$. It equals $4/\Delta$ at $\mu_{\rm eff}=\Delta$, which is the value on the gapped arc. Hence $\mathcal{S}\le4/\Delta$ everywhere. With $\int_0^{2\pi}d\phi\,\check w_z=2\,\pi$ and $\int_0^{2\pi}d\phi\,\check w_{x,y}=\pi$, replacing $\mathcal{S}$ by $4/\Delta$ gives the insulating values $\epsilon_{0,z}=1+g\, e^2\, k_0/(3\,\Delta)$ and $\epsilon_{0,x}=\epsilon_{0,y}=1+g\, e^2\, k_0/(6\,\Delta)$. These are upper bounds of $\epsilon_{b,i}$.

For $\tilde\eta\ll\Delta$ we have $|\mu_{\rm eff}-\Delta|\le\tilde\eta \left( 1+|\xi| \right)\ll\Delta$ on the conducting arc. We expand $\mathcal{S}$ about $\mu_{\rm eff}=\Delta$, where $d\mathcal{S}/d\mu_{\rm eff}=-\,6/\Delta^2$. Then $\mathcal{S}-4/\Delta\simeq-\,6 \left( \mu_{\rm eff}-\Delta \right)/\Delta^2$ on the conducting arc, and it vanishes on the gapped arc. Hence
\begin{align}
\epsilon_{b,i}-\epsilon_{0,i} \simeq -\,\frac{g\, e^2\, k_0}{4\,\pi\,\Delta^2} \int_{\phi_0}^{2\pi-\phi_0}\! d\phi\; \check w_i \left( \mu_{\rm eff}-\Delta \right) = -\,\frac{D_{ii}}{\Delta^2} \,.
\end{align}
The last step uses $\left( \mu_{\rm eff}^2-\Delta^2 \right)/\mu_{\rm eff}\simeq2 \left( \mu_{\rm eff}-\Delta \right)$ in the first expression for $D_{ii}$ of the subsubsection on the small in-plane tilt. Since $D_{ii}\propto\tilde\eta$, the screened frequencies are $\tilde\omega_{p,i}^2=\omega_{p,i}^2/\epsilon_{b,i}\simeq4\,D_{ii}/\epsilon_{0,i}$ to leading order.

For $\Delta\ll\tilde\eta$ we set $\phi=\sqrt{2\,\Delta/\tilde\eta}\;p$, so that $\mu_{\rm eff}\simeq\Delta \left( \chi+p^2 \right)$. We write $\hat{\mathcal{S}}(m)=\Delta\,\mathcal{S}$ at $\mu_{\rm eff}=\Delta\,m$, which is $4$ for $m<1$ and $\left( 3\,m^2+1 \right)/m^3$ for $m>1$. The latter decays as $3/p^2$ for large $p$. The integral $\int dp\,\hat{\mathcal{S}}(\chi+p^2)$ therefore converges, and
\begin{align}
\int d\phi\, \mathcal{S} \simeq \sqrt{\frac{2}{\Delta\,\tilde\eta}} \int dp\, \hat{\mathcal{S}} \left( \chi+p^2 \right) .
\end{align}
The integral is elementary. We have $\hat{\mathcal{S}}=4$ for $|p|<p_0\equiv\sqrt{1-\chi}$, and $\hat{\mathcal{S}}=3/m+1/m^3$ with $m=\chi+p^2$ for $|p|>p_0$. Let $\tilde I_n\equiv\int_{p_0}^\infty dp/(p^2+\chi)^n$. For $\chi>0$, $\tilde I_1=\chi^{-1/2}\arctan\!\left( \sqrt\chi/p_0 \right)=\arcsin\sqrt\chi/\sqrt\chi$. For $\chi<0$ with $c=\sqrt{-\chi}$, $\tilde I_1=(2\,c)^{-1}\ln \!\left[ (p_0+c)/(p_0-c) \right]=\mathrm{arcsinh}\,c/c$, because $p_0^2-c^2=1$. In both cases $\tilde I_1=G(\chi)$ of Eq.~\eqref{eq:smalldelta-K}. Differentiating $p/(p^2+\chi)^n$ gives $2\,n\,\chi\, \tilde I_{n+1}=(2\,n-1)\, \tilde I_n-p_0$, where $\chi+p_0^2=1$ was used at the lower limit. Hence $\tilde I_3=\left[ 3\,G-(3+2\,\chi)\,p_0 \right]/(8\,\chi^2)$, and
\begin{align}
\mathcal{\tilde I} (\chi) \equiv \int_{-\infty}^\infty\! dp\, \hat{\mathcal{S}} \left( \chi+p^2 \right) = 2 \left( 4\, p_0 + 3\, \tilde I_1 + \tilde I_3 \right) = 8\, p_0 + 6\, G (\chi) + \frac{3\, G (\chi) - (3+2\,\chi)\, p_0}{4\,\chi^2} \,.
\end{align}
The singularity at $\chi=0$ is removable, since $G=1+\chi/6+3\,\chi^2/40+\cdots$ gives $\tilde I_3\to1/5$ and $\mathcal{\tilde I}(0)=72/5$. Since $\cos^2\phi\simeq1$ in the layer, $\epsilon_{b,z}-1$ and $\epsilon_{b,x}-1$ both equal $\left( g\, e^2\, k_0/24\,\pi \right)\sqrt{2/(\Delta\,\tilde\eta)}\;\mathcal{\tilde I}(\chi)$, that is
\begin{align}
\epsilon_{b,z}-1 \simeq \epsilon_{b,x}-1 \simeq \frac{g\, e^2\, k_0}{12\sqrt2\,\pi\sqrt{\Delta\,\tilde\eta}}\; \mathcal{I} (\chi) \,.
\end{align}
This grows as $g\, e^2\, k_0/\sqrt{\Delta\,\tilde\eta}$, with the coefficient decreasing from $\mathcal{I}(-1)/(12\sqrt2\,\pi)\approx 0.32$ at $\chi\to-1^+$ to $\approx 0.20$ at $\chi\to1^-$ in units of $g\, e^2\, k_0/\sqrt{\Delta\,\tilde\eta}$. For $\check w_y=\sin^2\phi\simeq\phi^2$ the integrand $\check w_y\,\mathcal{S}$ is bounded. On the gapped arc it is of order $\phi^2/\Delta\sim1/\tilde\eta$. Outside the layer, $\mu_{\rm eff}\simeq\tilde\eta\,\phi^2/2$ gives $\check w_y\,\mathcal{S}\simeq3\,\phi^2/\mu_{\rm eff}\simeq6/\tilde\eta$. Hence $\epsilon_{b,y}$ stays finite. To leading order the gap can be dropped, $\mathcal{S}\to3/\mu_{\rm eff}$, and $\mu_{\rm eff}\simeq\tilde\eta \left( 1-\cos\phi \right)$ gives $\sin^2\phi/\mu_{\rm eff}=\left( 1+\cos\phi \right)/\tilde\eta$. Therefore $\epsilon_{b,y}-1\simeq\left( g\, e^2\, k_0/24\,\pi \right) 3\cdot2\,\pi/\tilde\eta=g\, e^2\, k_0/(4\,\tilde\eta)$. The screening is negligible only if $\epsilon_{b,i}-1\ll1$, that is for $g\, e^2\, k_0\ll\sqrt{\Delta\,\tilde\eta}$ in the $z$ and $x$ channels.

\subsection{Small-pocket windows of in-plane-tilted GNR}
\label{app:windows}

We now treat the four narrow windows of Sec.~\ref{secfermisurface}. We begin with the first, in which only a small electron-pocket exists. For $\mu \ge 0$ and $b = \mu + \tilde{\eta}$ in the range $\Delta\, \nu_x \le b < \Delta$, with $a = \mu - \tilde{\eta} > -\Delta\, \nu_x$, only a small electron-pocket exists. It lies around $\phi = \pi$, where $\mu_{\rm eff}$ is largest. At fixed $\phi$ the local FS is the ellipse of Eq.~\eqref{eq:gaptiltz-ellipse}, with the tilt now along the radial direction $u$ instead of $w$. Its centre lies at $u = -\mu_{\rm eff}\, t/(v_0\, \nu_t^2)$ and its semi-axes are $\sqrt{K}/(v_0\, \nu_t^2)$ along $u$ and $\sqrt{K}/(v_0\, \nu_t)$ along $w$, with $\nu_t = \sqrt{1 - t^2}$ and $K (\phi) = \mu_{\rm eff}^2 - \Delta^2\, \nu_t^2$. The ellipse has the area $\pi\, K/(v_0^2\, \nu_t^3)$. For $\mu_{\rm eff} < \Delta$ it does not enclose the origin, and it exists only where $K (\phi) > 0$. With $\phi = \pi + \upsilon$ we have $\mu_{\rm eff} \simeq b - \tilde\eta\, \upsilon^2/2$ and $\nu_t^2 \simeq \nu_x^2 + \eta_x^2\, \upsilon^2$, and the expansion of $K$ to leading order in $\upsilon$ gives
\begin{align}
K (\upsilon) \simeq K_b - c\, \upsilon^2 \,, \quad K_b = b^2 - \Delta^2\, \nu_x^2 \,, \quad c = b\, \tilde{\eta} + \Delta^2\, \eta_x^2 \,.
\end{align}
The pocket therefore occupies the arc $|\upsilon| < \upsilon_m = \sqrt{K_b/c}$. Since $K_b \le \Delta^2\, \eta_x^2$ in the window, $\upsilon_m^2 \le \eta_x\, \Delta/(v_0\, k_0) \ll 1$ for $\Delta \ll v_0\, k_0$, and the variations of $\mu_{\rm eff}$ and $\nu_t$ across the arc are negligible. The local integrals of Appendix~\ref{app:gnrztilt} hold for $\mu_{\rm eff} > \Delta\, \nu_t$ in both the displaced-torus and fully-conducting regimes. For the tilt along $u$ they give the local integrals $\int d^2 k'\, (\partial_u \varepsilon)^2\, \delta (\varepsilon - \mu_{\rm eff}) = 2\, \pi\, \mu_{\rm eff}\, K/[\tilde M\, (\mu_{\rm eff} + \nu_t\, \tilde M)]$ and $\int d^2 k'\, (\partial_w \varepsilon)^2\, \delta (\varepsilon - \mu_{\rm eff}) = 2\, \pi\, K/[\nu_t\, (\mu_{\rm eff} + \nu_t\, \tilde M)]$, with $\tilde M = \sqrt{\mu_{\rm eff}^2 + t^2\, \Delta^2}$.

Near $\phi = \pi$ the velocity components are $V_x \simeq -\, \partial_u \varepsilon$, $V_z = \partial_w \varepsilon$, and $V_y = \sin\phi\, \left( \partial_u \varepsilon - v_0\, t \right)$. The correction to $V_x$ is smaller by $\upsilon^2$. The FS average of $\partial_u \varepsilon$ vanishes for a closed pocket, and the term $v_0\, t$ adds $v_0^2\, t^2$ times the local DOS to $D_{yy}$. The local DOS is the derivative $2\, \pi\, \mu_{\rm eff}/(v_0^2\, \nu_t^3)$ of the ellipse area with respect to $\mu_{\rm eff}$. Inserting it and the local integrals above into Eqs.~\eqref{eqdos} and \eqref{eqdrude}, with $d^3 \bs k = k_0\, d\phi\, du\, dw$, we have
\begin{align}
\rho (\mu) &= \frac{g\, k_0}{4\, \pi^2\, v_0^2} \int d\upsilon\, \frac{\mu_{\rm eff}}{\nu_t^3} \,, \quad
D_{zz} = \frac{g\, e^2\, k_0}{4\, \pi} \int d\upsilon\, \frac{K}{\nu_t \left( \mu_{\rm eff} + \nu_t\, \tilde M \right)} \,, \nn
D_{xx} &= \frac{g\, e^2\, k_0}{4\, \pi} \int d\upsilon\, \frac{\mu_{\rm eff}\, K}{M \left( \mu_{\rm eff} + \nu_t\, \tilde M \right)} \,, \quad
D_{yy} = \frac{g\, e^2\, k_0}{4\, \pi} \int d\upsilon\, \upsilon^2 \left[ \frac{\mu_{\rm eff}\, K}{M \left( \mu_{\rm eff} + \nu_t\, \tilde M \right)} + \frac{\mu_{\rm eff}\, t^2}{\nu_t^3} \right] ,
\end{align}
with $\sin^2 \phi \simeq \upsilon^2$ and $\cos^2 \phi \simeq 1$, and the integrals run over $|\upsilon| < \upsilon_m$. The arc integrals are $\int d\upsilon = 2\, \upsilon_m$, $\int d\upsilon\, \upsilon^2 = 2\, \upsilon_m^3/3$, $\int d\upsilon\, K = 4\, K_b\, \upsilon_m/3 = 4\, K_b^{3/2}/(3\, \sqrt{c})$, and $\int d\upsilon\, \upsilon^2\, K = 4\, K_b\, \upsilon_m^3/15$. Setting $\mu_{\rm eff} \simeq b$, $\nu_t \simeq \nu_x$, $t^2 \simeq \eta_x^2$, and $\tilde M \simeq M_b$ with $M_b = \sqrt{b^2 + \eta_x^2\, \Delta^2}$, we obtain
\begin{align}
\label{eq:smallpocket-rho}
\rho (\mu) &\simeq \frac{ g\, k_0\, b }{ 2\, \pi^2\, v_0^2\, \nu_x^3 } \sqrt{ \frac{ K_b }{ c } } \,, 
D_{zz} \simeq \frac{ g\, e^2\, k_0\, K_b^{3/2} }{ 3\, \pi\, \nu_x
\left( b + \nu_x\, M_b \right) \sqrt{c} } \,, \quad
D_{xx} \simeq \frac{ b\, \nu_x }{ M_b }\, D_{zz} \,, 
D_{yy} \simeq \frac{ g\, e^2\, k_0\, b\, K_b^{3/2} }{ 6\, \pi\, c^{3/2} } \left[ \frac{ \eta_x^2 }{ \nu_x^3 } + \frac{ 2\, K_b }{ 5\, M_b\, \left( b + \nu_x\, M_b \right) } \right] .
\end{align}
The DOS grows as $\sqrt{b - \Delta\, \nu_x}$, which is the square-root onset of a pocket that is born at an isolated point of the torus. The Drude weights grow as $(b - \Delta\, \nu_x)^{3/2}$, and the plasmon frequencies therefore vanish as $(b - \Delta\, \nu_x)^{3/4}$. At the lower edge, $M_b \to \Delta$, and the anisotropy ratio is $D_{zz}/D_{xx} \to 1/\nu_x^2$. The ratio $D_{yy}/D_{xx}$ is of order $\upsilon_m^2$, since the pocket is confined to a small arc around $\phi = \pi$. Since the local integrals above hold on both sides of $b = \Delta$, Eq.~\eqref{eq:smallpocket-rho} joins continuously onto the crescent-cyclide regime.

Both bands contribute to the interband response, but the pocket removes only a small part of the phase space of the insulating background. The lowest unblocked transition is at the direct gap, since the origin lies outside the ellipse for all $\phi$, and the onset is therefore $\omega_{\rm th} = 2\, \Delta$. The form-factor at the pocket is $q^2\, v_0^2/(4\, \Delta^2)$ for $\bs q \parallel \bs{\hat z}$ and $\bs q \parallel \bs{\hat x}$, with the weight $\cos^2\phi \simeq 1$ for the latter. Removing the pocket from the insulating static polarisability of Sec.~\ref{secgapnotilt} gives the static background dielectric constants
\begin{align}
\epsilon_{b,z} = 1 + \frac{g\, e^2\, k_0}{3\, \Delta} \left[ 1 - \frac{ K_b^{3/2} }{ 2\, \pi\, \Delta^2\, \nu_x^3\, \sqrt{c} } \right] , \quad
\epsilon_{b,x} = 1 + \frac{g\, e^2\, k_0}{6\, \Delta} \left[ 1 - \frac{ K_b^{3/2} }{ \pi\, \Delta^2\, \nu_x^3\, \sqrt{c} } \right] ,
\end{align}
and $\epsilon_{b,y} = 1 + g\, e^2\, k_0/(6\, \Delta)$ at this order. The corrections are small, and the background is essentially the insulating one. The screened plasmon frequencies are $\tilde\omega_{p,i}^2 = 4\, D_{ii}/\epsilon_{b,i}$. They lie far below $\omega_{\rm th} = 2\, \Delta$, and the plasmon is therefore free of interband damping.

The second window adds a non-enclosing hole-pocket around $\phi = 0$. It occurs for $\mu \ge 0$ if, in addition, $a \le -\Delta\, \nu_x$, that is, for $\tilde{\eta} - \mu \ge \Delta\, \nu_x$ together with $\tilde{\eta} + \mu < \Delta$. This requires $\mu < \Delta \left( 1 - \nu_x \right)/2$ and $\tilde{\eta} \simeq \Delta$. At $\phi = \upsilon$ the hole condition $\varepsilon_1 = \mu$ reads $E_k - v_0\, t\, u = -\mu_{\rm eff}$. It is the electron condition with $\mu_{\rm eff} \to -\mu_{\rm eff}$ and $t \to -t$. The local integrals depend on $t$ only through $t^2$, and the electron-pocket at $\phi = \pi$ and the hole-pocket at $\phi = 0$ are disjoint. Near $\phi = 0$ we have $|\mu_{\rm eff}| \simeq |a| - \tilde{\eta}\, \upsilon^2/2$ and $\nu_t^2 \simeq \nu_x^2 + \eta_x^2\, \upsilon^2$, and the hole-pocket has $K_h (\upsilon) \simeq K_a - c_a\, \upsilon^2$, with $K_a = a^2 - \Delta^2\, \nu_x^2$ and $c_a = |a|\, \tilde{\eta} + \Delta^2\, \eta_x^2$. The velocity components are $V_x \simeq \partial_u \varepsilon_1$, $V_z = \partial_w \varepsilon_1$, and $V_y = \sin\phi\, \left( \partial_u \varepsilon_1 - v_0\, t \right)$. Each Drude weight and the DOS are therefore the sum of the electron result of Eq.~\eqref{eq:smallpocket-rho} and the same expression with $b \to |a|$, where $K_b$, $c$, and $M_b$ are replaced by $K_a$, $c_a$, and $M_a = \sqrt{a^2 + \eta_x^2\, \Delta^2}$. Since $b = \tilde{\eta} + \mu$ and $|a| = \tilde{\eta} - \mu$, the sum is even in $\mu$, as required by the exchange of electrons and holes. At $\mu = 0$ the two pockets are identical, and all quantities are twice the single-pocket values. Both pockets contribute to $D_{ii}$ with the same sign, since $D_{ii}$ is quadratic in the velocity. A plasmon therefore exists at $\mu = 0$ throughout the window $\Delta\, \nu_x \le \tilde{\eta} < \Delta$, in contrast to the fully-insulating regime. Its screened frequency is $\tilde\omega_{p,i}^2 = 4\, D_{ii}/\epsilon_{b,i}$.

The interband response follows in the same way. The origin of the local plane still lies outside both ellipses for all $\phi$, and the onset therefore remains $\omega_{\rm th} = 2\, \Delta$. The form-factor at the hole-pocket is the same as at the electron-pocket, and the blocked states are the empty $s=1$ states inside it. The corrections to the static background dielectric constants of the previous window are therefore additive. With $S = K_b^{3/2}/\sqrt{c} + K_a^{3/2}/\sqrt{c_a}$, we obtain
\begin{align}
\epsilon_{b,z} = 1 + \frac{g\, e^2\, k_0}{3\, \Delta} \left[ 1 - \frac{ S }{ 2\, \pi\, \Delta^2\, \nu_x^3 } \right] , \quad
\epsilon_{b,x} = 1 + \frac{g\, e^2\, k_0}{6\, \Delta} \left[ 1 - \frac{ S }{ \pi\, \Delta^2\, \nu_x^3 } \right] ,
\end{align}
and $\epsilon_{b,y} = 1 + g\, e^2\, k_0/(6\, \Delta)$ at this order.

The third window is the fifth of the eight cases of Sec.~\ref{secfermisurface}. It occurs for $\mu \ge 0$ if $b \ge \Delta$ and $-\Delta < a \le -\Delta\, \nu_x$, that is, for $\tilde{\eta} - \mu \ge \Delta\, \nu_x$ together with $\tilde{\eta} - \mu < \Delta$ and $\tilde{\eta} + \mu \ge \Delta$. To leading order in $\eta_x$, the electron-pocket and the gapped arc are those of the crescent-cyclide regime. The new feature is that $\mu_{\rm eff}$ at $\phi = 0$ lies between $-\Delta$ and $-\Delta\, \nu_x$. A small non-enclosing hole-pocket then sits inside the gapped arc around $\phi = 0$, and it does not enclose the ring. Its local FS is the hole ellipse of the second window, with $K_a$, $c_a$, and $M_a$ as defined there, and it occupies the arc $|\upsilon| < \upsilon_m = \sqrt{K_a/c_a}$ at $\phi = \upsilon$. Since $|a| - \Delta\, \nu_x \le \Delta \left( 1 - \nu_x \right)$ in this window, we have $K_a \le \Delta^2\, \eta_x^2$ and $\upsilon_m^2 \lesssim \eta_x\, \Delta/(v_0\, k_0)$. The assumption $\eta_x\, \Delta \ll v_0\, k_0$ of Sec.~\ref{secfermisurface} therefore makes the pocket small. It is disjoint from the electron-pocket, and the two contribute additively. The DOS and the Drude weights are thus the crescent-cyclide results of Eqs.~\eqref{eq:partialgapped-dos} and \eqref{eq:partialgapped-exact}, with $\cos\phi_0 = (\mu - \Delta)/\tilde{\eta}$, plus the hole-pocket terms. The latter are obtained from Eq.~\eqref{eq:smallpocket-rho} by the replacements $b \to |a| = \tilde{\eta} - \mu$, $K_b \to K_a$, $c \to c_a$, and $M_b \to M_a$ as follows:
\begin{align}
& \delta\rho \simeq \frac{ g\, k_0\, |a| }{ 2\, \pi^2\, v_0^2\, \nu_x^3 } \sqrt{ \frac{ K_a }{ c_a } } \,, 
\quad \delta D_{zz} \simeq \frac{ g\, e^2\, k_0\, K_a^{3/2} }{ 3\, \pi\, \nu_x\
\left( |a| + \nu_x\, M_a \right) \sqrt{c_a} } \,, \quad
\delta D_{xx} \simeq \frac{ |a|\, \nu_x }{ M_a }\, \delta D_{zz} \,, 
\nn & \delta D_{yy} \simeq \frac{ g\, e^2\, k_0\, |a|\, K_a^{3/2} }{ 6\, \pi\, c_a^{3/2} } 
\left[ \frac{ \eta_x^2 }{ \nu_x^3 } + \frac{ 2\, K_a }
{ 5\, M_a\, \left( |a| + \nu_x\, M_a \right) } \right] .
\end{align}
The non-enclosing rim of the electron-pocket at $\phi_0$ is the same as in case 4 of Table~\ref{tab:fsclass}. It changes the results only at relative order $\eta_x^2$, and we drop it together with the other $\eta_x^2$ terms of the crescent-cyclide regime. At the lower edge, $|a| \to \Delta\, \nu_x$, the hole-pocket is born at an isolated point. The DOS then grows as $\sqrt{|a| - \Delta\, \nu_x}$, the Drude weights grow as $(|a| - \Delta\, \nu_x)^{3/2}$, and the plasmon frequencies change by a term of order $(|a| - \Delta\, \nu_x)^{3/4}$. At the upper edge, $|a| \to \Delta$, the pocket has its largest size, $\upsilon_m \simeq \sqrt{\eta_x\, \Delta/(v_0\, k_0)}$. Its contribution to $D_{zz}$ and $D_{xx}$ is then of order $g\, e^2\, k_0\, \Delta\, \eta_x^2\, \upsilon_m/(6\, \pi)$, which is small compared with the crescent-cyclide weight. Hence the hole-pocket does not change the crescent-cyclide results at leading order in $\eta_x$, and they join continuously onto the twin-crescent regime, where the hole terms vanish as $\phi_h \to 0$. The ratio $D_{yy}/D_{xx}$ of the hole-pocket terms is of order $\upsilon_m^2$, because $\sin^2\phi$ is small on the pocket.

The interband response follows as in the second window. The origin of the local plane lies outside the hole ellipse, since $|a| < \Delta$, and the onset remains $\omega_{\rm th} = 2\, \Delta$. The hole-pocket blocks a small part of the phase space of the gapped arc. It therefore lowers the static background dielectric constants of Eq.~\eqref{eq:epsb-partial} by
\begin{align}
\delta\epsilon_{b,z} = \delta\epsilon_{b,x} = -\, \frac{ g\, e^2\, k_0\, K_a^{3/2} }{ 6\, \pi\, \Delta^3\, \nu_x^3\, \sqrt{c_a} } \,, \quad
\delta\epsilon_{b,y} = 0 \,,
\end{align}
where the weight $\cos^2\phi \simeq 1$ applies along $\bs{\hat x}$ and $\sin^2\phi \simeq 0$ along $\bs{\hat y}$. The screened plasmon frequencies are $\tilde\omega_{p,i}^2 = 4\, D_{ii}/\epsilon_{b,i}$ with the sums above, and they lie below $\omega_{\rm th} = 2\, \Delta$ under the same conditions as in the crescent-cyclide regime.

The last window is the seventh of the eight cases of Sec.~\ref{secfermisurface}. It occurs for $\Delta\, \nu_x \le a < \Delta$, which lies between the crescent-cyclide and fully-conducting regimes. There is no gapped arc, since $\mu_{\rm eff} (\phi) \ge a > \Delta\, \nu_t (\phi)$ on the whole ring. The electron-pocket therefore runs around the ring. It encloses the ring where $\mu_{\rm eff} \ge \Delta$, and it does not enclose it on the arc $|\phi| < \phi_0$ around $\phi = 0$, where $\mu_{\rm eff} < \Delta$. Here $\cos\phi_0 = (\mu - \Delta)/\tilde{\eta}$ is the same angle as in the crescent-cyclide regime, and $\phi_0 \simeq \sqrt{2\, (\Delta - a)/\tilde{\eta}}$ is small. The local FS on this arc is the electron ellipse of Sec.~\ref{sec:gaptiltz}, with the tilt along the radial direction $u$. Its $K (\phi) = \mu_{\rm eff}^2 - \Delta^2\, \nu_t^2$ is positive, and the ray from the ring meets it twice. With $\phi = \upsilon$ we have $\mu_{\rm eff} \simeq a + \tilde\eta\, \upsilon^2/2$ and $\nu_t^2 \simeq \nu_x^2 + \eta_x^2\, \upsilon^2$, and the expansion of $K$ to leading order in $\upsilon$ gives
\begin{align}
K (\upsilon) \simeq K_a + c'\, \upsilon^2 \,, \quad K_a = a^2 - \Delta^2\, \nu_x^2 \,, \quad c' = a\, \tilde{\eta} - \Delta^2\, \eta_x^2 \,,
\end{align}
where $c' > 0$ follows from $\eta_x\, \Delta \ll v_0\, k_0$. The crescent-cyclide results of Eqs.~\eqref{eq:partialgapped-dos} and \eqref{eq:partialgapped-exact} exclude this arc. The exact local integrals of the first window hold on it, with $\mu_{\rm eff} \simeq \Delta$ and $M \simeq \Delta$. Inserting them into the expressions of the first window, with $\nu_t \simeq 1$ and $\mu_{\rm eff} + \nu_t\, M \simeq 2\, \Delta$ in the denominators, and integrating over $|\upsilon| < \phi_0$ with $\int d\upsilon\, \mu_{\rm eff} = 2\, a\, \phi_0 + \tilde\eta\, \phi_0^3/3$ and $\int d\upsilon\, K = 2\, K_a\, \phi_0 + 2\, c'\, \phi_0^3/3$, we obtain
\begin{align}
\delta\rho \simeq \frac{ g\, k_0 }{ 4\, \pi^2\, v_0^2 } \left( 2\, a\, \phi_0 + \frac{\tilde{\eta}\, \phi_0^3}{3} \right) , \quad
\delta D_{zz} \simeq \delta D_{xx} \simeq \frac{ g\, e^2\, k_0 }{ 4\, \pi\, \Delta } \left( K_a\, \phi_0 + \frac{c'\, \phi_0^3}{3} \right) ,
\end{align}
which add to the crescent-cyclide results. The correction $\delta D_{yy}$ is smaller than $\delta D_{zz}$ by a factor of order $\phi_0^2$, because $\sin^2\phi$ is small on the arc. The same arc is responsible for the rims of the earlier windows. There, the width $\Delta \left( 1 - \nu_x \right)/(\tilde{\eta}\, |\sin\phi_0|)$ is of order $\eta_x^2$ for generic $\phi_0$. It grows as $\sin\phi_0 \to 0$, and in the present window it fills the whole arc, whose half-width is at most $\phi_0 \simeq \eta_x \sqrt{\Delta/\tilde{\eta}}$. Consequently $\delta\rho/\rho$ is of order $\eta_x$, whereas $\delta D_{zz}/D_{zz}$ is of order $\eta_x^3$ since $K_a \lesssim \Delta^2\, \eta_x^2$. The plasmon frequencies are therefore unchanged at leading order in $\eta_x$, and only the DOS carries a correction of first order. Both corrections vanish as $a \to \Delta^-$, where $\phi_0 \to 0$ and the results join onto the fully-conducting regime. At $a = \Delta\, \nu_x$ we have $K_a = 0$, and the pocket pinches off at $\phi = 0$, which joins the results onto case 4 of Table~\ref{tab:fsclass}.

The origin of the local plane lies outside the ellipse on the arc, since $\mu_{\rm eff} < \Delta$ there. The unblocked states at $\kappa = 0$ keep the onset at $\omega_{\rm th} = 2\, \Delta$, and the logarithmic divergence of Eq.~\eqref{eq:onset-log} persists. The pocket blocks a small part of the phase space of the arc, and it lowers the static background dielectric constants by
\begin{align}
\delta\epsilon_{b,z} = \delta\epsilon_{b,x} = -\, \frac{ g\, e^2\, k_0 }{ 4\, \pi\, \Delta^3 } \left( K_a\, \phi_0 + \frac{c'\, \phi_0^3}{3} \right) , \quad \delta\epsilon_{b,y} = 0 \,.
\end{align}
This correction is of order $\eta_x^3$, and the background is essentially that of the crescent-cyclide regime. The screened plasmon frequencies are $\tilde\omega_{p,i}^2 = 4\, D_{ii}/\epsilon_{b,i}$, and they lie below $\omega_{\rm th} = 2\, \Delta$ under the same conditions as there. This completes the classification of the tilted GNR at finite $\eta_x$.

\subsection{Approach of the dielectric function to the interband onset}
\label{app:onset}

Two mechanisms determine how $\Re \varepsilon^{\rm RPA}$ approaches the interband onset, and both extend beyond the in-plane-tilted GNR. Wherever $\tilde m (\phi)$ reaches its minimum over a whole arc of finite length, $\Re \varepsilon^{\rm RPA}$ diverges logarithmically at the onset. Wherever the minimum is reached at an isolated point only, with $\tilde m (\phi)$ quadratic about it, $\Re \varepsilon^{\rm RPA}$ stays finite there but approaches this value with an infinite slope, through a $\sqrt{\omega_{\rm th}-\omega}$ correction. The untilted doped GNR and PTNR, and both fully-insulating regimes, belong to the first class, since $\tilde m$ is then constant over the whole ring. Indeed, expanding Eq.~\eqref{eq:pinterdoped} as $\omega \to 2\, \mu^-$ gives directly
\begin{align}
\label{eq:pinterdoped-onset}
\Pi_{\rm inter} (q\, \bs{\hat z}, \omega) \simeq -\, \frac{g\, k_0\, q^2}{32\, \pi\, \mu} \left( 1 + \frac{\Delta^2}{\mu^2} \right) \ln \!\left( \frac{4\, \mu}{2\, \mu - \omega} \right) ,
\end{align}
with $\bs{\hat x}$ and $\bs{\hat y}$ carrying half this coefficient. For the PTNR the bracket is unity. The same form holds with $\mu \to \Delta$ for both insulating cases, where the bracket equals two.

The crescent-cyclide and twin-crescent regimes belong to the same class, since $\tilde m = \Delta$ throughout the gapped arc. There $\Re \varepsilon^{\rm RPA}$ diverges as
\begin{align}
\label{eq:onset-log}
\delta \left( \Re \varepsilon^{\rm RPA}_{i} \right) = \frac{g\, e^2\, k_0}{8\, \pi\, \Delta}\, \mathcal{A}_i\, \ln \!\left(  \frac{4\, \Delta}{\omega_{\rm th} - \omega} \right), 
\quad \omega_{\rm th} = 2\, \Delta \,,
\end{align}
with $\mathcal{A}_z$, $\mathcal{A}_x$, and $\mathcal{A}_y$ the length of the gapped arc weighted by $1$, $\cos^2\phi$, and $\sin^2\phi$, respectively. Explicitly, $\mathcal{A}_z = 2\, \phi_0$, $\mathcal{A}_x = \phi_0 + \sin\phi_0\, \cos\phi_0$, and $\mathcal{A}_y = \phi_0 - \sin\phi_0\, \cos\phi_0$ in the crescent-cyclide regime, and $\mathcal{A}_z = 2\, (\phi_0 - \phi_h)$, $\mathcal{A}_x = (\phi_0 + \sin\phi_0\cos\phi_0) - (\phi_h + \sin\phi_h\cos\phi_h)$, and $\mathcal{A}_y = (\phi_0 - \sin\phi_0\cos\phi_0) - (\phi_h - \sin\phi_h\cos\phi_h)$ in the twin-crescent regime, with $\mathcal{A}_z = \mathcal{A}_x + \mathcal{A}_y$ throughout. The contribution of the conducting arc adjoining the gapped one is subleading here, since its own local threshold recedes linearly away from $\Delta$ near the boundary and contributes no logarithm of its own.

Equation~\eqref{eq:onset-log} follows from the local function $\mathcal{I}$ of Eq.~\eqref{eq:pintertiltxgnr}. For $\tilde{m} = \Delta$ and $\omega \to \omega_{\rm th}^-$ we have $\Omega = \omega/2 \to \Delta$. The prefactor $\left( 1 + \Delta^2/\Omega^2 \right)/(2\, \Omega)$ of the logarithm then tends to $1/\Delta$, and the logarithm becomes
\begin{align}
\ln \!\left| \frac{\tilde{m} + \Omega}{\tilde{m} - \Omega} \right| \to \ln \!\left( \frac{2\, \Delta}{\Delta - \omega/2} \right) = \ln \!\left(  \frac{4\, \Delta}{\omega_{\rm th} - \omega} \right).
\end{align}
The remaining term of $\mathcal{I}$ is finite. The interband part $\left( g\, e^2\, k_0/4 \right) \int d\phi/(2\, \pi)\, \check w_i\, \mathcal{I}$ of the dielectric function then gives Eq.~\eqref{eq:onset-log}, where $\mathcal{A}_i = \int_{\rm arc} d\phi\, \check w_i (\phi)$ and the factor $1/(2\, \pi)$ combines with $g\, e^2\, k_0/(4\, \Delta)$ into $g\, e^2\, k_0/(8\, \pi\, \Delta)$.

The fully-conducting regime belongs to the second class, since $\mu_{\rm eff} (\phi)$ has an isolated, quadratic minimum at $\phi = 0$. Writing $\check\delta \equiv (\omega_{\rm th} - \omega)/2$, expanding $\mu_{\rm eff} (\phi) \simeq (\mu-\tilde\eta) + (\tilde\eta/2)\, \phi^2$ about this point, and carrying out the resulting $\phi$-integral gives, to leading order in $\check\delta$,
\begin{align}
\label{eq:onset-cusp}
\delta \left( \Re \varepsilon^{\rm RPA}_{z} \right) = \delta \left( \Re \varepsilon^{\rm RPA}_{x} \right) \simeq -\, \frac{g\, e^2\, k_0}{8\, (\mu-\tilde\eta)} \left[ 1 + \frac{\Delta^2}{(\mu-\tilde\eta)^2} \right] \sqrt{\frac{\omega_{\rm th}-\omega}{\tilde\eta}} \,.
\end{align}
The $\bs{\hat y}$ channel carries no correction at this order, since $\sin^2 \phi$ vanishes at the minimum. Its leading non-analyticity is $(\omega_{\rm th}-\omega)^{3/2}$. The $\bs{\hat x}$ channel matches $\bs{\hat z}$ here, rather than carrying half its weight as in the arc case above, because the local tilt direction at $\phi = 0$ is purely radial. For $\Delta \to 0$, Eq.~\eqref{eq:onset-cusp} is the corresponding result for the ring-cyclide regime of the in-plane-tilted PTNR, with $\mu-\tilde\eta$ its own threshold half-width. The axially-tilted PTNR and GNR are treated in Appendix~\ref{app:onsetz}. For $\mu > \Delta$ the local Fermi energy along the unblocked ray has an isolated quadratic minimum, at a single point of the ellipse, and the cusp follows from the same expansion about that point. In the displaced-torus regime the onset is the direct gap at $\kappa = 0$ on every ray, and the logarithmic divergence of the first class applies instead.

Equation~\eqref{eq:onset-cusp} follows from the same function $\mathcal{I}$. Near $\phi = 0$ we have $\tilde{m} (\phi) \simeq m_0 + (\tilde\eta/2)\, \phi^2$ with $m_0 = \mu - \tilde\eta$ and $\omega_{\rm th} = 2\, m_0$, and therefore $\tilde{m} - \Omega = \check\delta + (\tilde\eta/2)\, \phi^2$. The part of the $\phi$-integral of $\ln \!\left( \tilde{m} - \Omega \right)$ that is non-analytic in $\check\delta$ follows from its derivative,
\begin{align}
\frac{d}{d\check\delta} \int d\phi\, \ln \!\left( \check\delta + \frac{\tilde\eta}{2}\, \phi^2 \right) = \int \frac{d\phi}{\check\delta + \tilde\eta\, \phi^2/2} = \pi \sqrt{\frac{2}{\tilde\eta\, \check\delta}} \,,
\end{align}
where the range of $\phi$ away from the minimum contributes only analytic terms. Integration over $\check\delta$ gives $\int d\phi/(2\, \pi)\, \ln \!\left( \tilde{m} - \Omega \right) \simeq \mathrm{const} + \sqrt{2\, \check\delta/\tilde\eta}$, with $2\, \check\delta = \omega_{\rm th} - \omega$. The logarithm in $\mathcal{I}$ enters with the opposite sign, and its prefactor at $\Omega = m_0$ is $\left( 1 + \Delta^2/m_0^2 \right)/(2\, m_0)$. The last term of $\mathcal{I}$ is analytic in $\check\delta$. For $\check w_z = 1$ and $\check w_x = \cos^2 \phi \simeq 1$, multiplication by $g\, e^2\, k_0/4$ gives Eq.~\eqref{eq:onset-cusp}. For $\check w_y = \sin^2 \phi \simeq \phi^2$ the derivative above is replaced by $\int d\phi\, \phi^2/(\check\delta + \tilde\eta\, \phi^2/2) = (2/\tilde\eta) \int d\phi \left[ 1 - \check\delta/(\check\delta + \tilde\eta\, \phi^2/2) \right]$. Its non-analytic part is proportional to $\sqrt{\check\delta}$, and integration over $\check\delta$ gives the exponent $3/2$.

\section{Axially-tilted PTNR and GNR}
\label{app:ztilt}

The results of the two appendices below reduce to the untilted ones as $\eta_z \to 0$. In this limit, Appendix~\ref{app:interband} gives $\Pi_{\rm inter}$ of Eq.~\eqref{eq:pinterdoped}, together with its static value $-\, g\, k_0\, q^2 \left( 3\, \mu^2 + \Delta^2 \right)/(48\, \pi\, \mu^3)$. Appendix~\ref{app:gnrztilt} gives the untilted Drude weights of Eq.~\eqref{eq:dzzgapped}. Setting $E_F \to \Delta$ in the former yields the insulating constants $\epsilon_{0,i}$ of Sec.~\ref{secgapnotilt}, and taking $\Delta \to 0$ in the latter yields the Drude weights of the axially-tilted PTNR in Eq.~\eqref{eq:dzzz}.


\subsection{Fermi surface and Drude weights of GNR}
\label{app:gnrztilt}

This appendix gives the steps behind Sec.~\ref{sec:gaptiltz} for the displaced-torus and fully-conducting regimes. We first derive the elliptical FS and its DOS. We then describe the FS in polar coordinates, with the two Fermi momenta as the roots of a quadratic equation, and finally evaluate the Drude weights by mapping the ellipse onto a circle.

We use the local coordinates $u$ and $w$ of Sec.~\ref{sec:gaptiltz}. The condition $\varepsilon_2 = \mu$ reads $E_k = \mu - v_0\, \eta_z\, w$, with $E_k^2 = v_0^2\, (u^2 + w^2) + \Delta^2$. Squaring both sides and collecting terms gives
\begin{align}
v_0^2\, u^2 + v_0^2\, \nu^2\, w^2 + 2\, \mu\, v_0\, \eta_z\, w = \mu^2 - \Delta^2 \,.
\end{align}
We complete the square in $w$ with $w_c \equiv -\mu\, \eta_z/(v_0\, \nu^2)$, and we obtain
\begin{align}
v_0^2\, u^2 + v_0^2\, \nu^2\, (w - w_c)^2 = \mu^2 - \Delta^2 + \frac{\mu^2\, \eta_z^2}{\nu^2} = \frac{K}{\nu^2} \,.
\end{align}
The last step uses $\mu^2\, \nu^2 + \mu^2\, \eta_z^2 = \mu^2$ and $K = \mu^2 - \nu^2\, \Delta^2$. Dividing by $v_0^2$ gives Eq.~\eqref{eq:gaptiltz-ellipse}.

The squaring may have introduced a spurious branch with $E_k < 0$. To exclude it, we evaluate $E_k$ on the ellipse. Since $E_k = \mu - v_0\, \eta_z\, w$ is linear in $w$, its smallest value for $\eta_z > 0$ is reached at the largest $w = w_c + \sqrt{K}/(v_0\, \nu^2)$, and it equals $(\mu - \eta_z\, \sqrt{K})/\nu^2$. This is positive because
\begin{align}
\mu^2 - \eta_z^2\, K = \nu^2 \left( \mu^2 + \eta_z^2\, \Delta^2 \right) > 0 \,.
\end{align}
The case $\eta_z < 0$ follows by $w \to -w$. Hence the whole ellipse is the $s=2$ FS for every $\mu > \Delta_{\rm eff}$.

The area of the ellipse is $\mathcal{A} = \pi\, K/(v_0^2\, \nu^3)$. With $d^3 \bs k = k_0\, d\phi\, du\, dw$, the electron density is $n = g\, k_0\, (2\, \pi)\, \mathcal{A}/(2\, \pi)^3 = g\, k_0\, \mathcal{A}/(2\, \pi)^2$. Since $dK/d\mu = 2\, \mu$, the derivative $\rho = dn/d\mu$ is
\begin{align}
\rho = \frac{g\, k_0}{4\, \pi^2} \cdot \frac{2\, \pi\, \mu}{v_0^2\, \nu^3} = \frac{g\, k_0\, \mu}{2\, \pi\, v_0^2\, \nu^3} \,,
\end{align}
which is Eq.~\eqref{eq:dosgaptiltz}.

For the polar description, the roots of Eq.~\eqref{eq:gaptiltz-quad} are
\begin{align}
\kappa_{1, 2} (\gamma) = \frac{ -\, \mu\, \eta_z\, \sin \gamma \mp 
\,\sqrt{ (\mu^2 - \Delta^2) + \eta_z^2\, \Delta^2\, \sin^2 \gamma} }{ v_0\left [1 - \eta_z^2\, \sin^2 \gamma \right ]} \,,
\end{align}
with the upper sign for $\kappa_1$. They follow from the quadratic equation obtained by squaring $E_k = \mu - v_0\, \eta_z\, \kappa\, \sin \gamma $,
\begin{align}
\left( 1 - \eta_z^2\, \sin^2 \gamma \right) v_0^2\, \kappa^2 + 2\, \mu\, \eta_z\, \sin \gamma\, v_0\, \kappa - \left( \mu^2 - \Delta^2 \right) = 0 \,.
\end{align}
The reduced discriminant is $\mu^2\, \eta_z^2\, \sin^2 \gamma + \left( 1 - \eta_z^2\, \sin^2 \gamma \right) 
\left( \mu^2 - \Delta^2 \right) = \left( \mu^2 - \Delta^2 \right) + \eta_z^2\, \Delta^2\, \sin^2 \gamma$, which is the expression under the square root. Their sum is $-\, 2\, \mu\, \eta_z\, \sin \gamma/[v_0\, (1 - \eta_z^2\, \sin^2 \gamma)]$, which is positive for $\eta_z\, \sin \gamma < 0$. Their product is $(\Delta^2 - \mu^2)/[v_0^2\, (1 - \eta_z^2\, \sin^2 \gamma)]$, which is positive for $\mu < \Delta$. Both roots are therefore positive if and only if $\eta_z\, \sin \gamma < 0$ and the discriminant is non-negative. For $\eta_z\, \sin \gamma \geq 0$, the sum is non-positive and no positive root exists. The discriminant is non-negative if and only if $ \sin^2 \gamma \geq (\Delta^2 - \mu^2)/(\eta_z^2\, \Delta^2)$, which for $\eta_z > 0$ is the window $\sin\gamma \leq -\sin\gamma_c$ with $\sin\gamma_c$ as in Sec.~\ref{sec:gaptiltz}. The window is non-empty if and only if $\sin\gamma_c \leq 1$, that is $\mu \geq \Delta_{\rm eff}$. At the window edges the discriminant vanishes and $\kappa_1 = \kappa_2 = -\, \mu\, \eta_z\, \sin \gamma/[v_0\, (1 - \eta_z^2\, \sin^2 \gamma)]$. Using $\mu = \Delta\, \sqrt{1 - \eta_z^2\, \sin^2 \gamma}$ there, this equals the position $\kappa_* (\gamma_c)$ of the band minimum of Sec.~\ref{sec:gaptiltz}. Both roots satisfy the un-squared equation, because $E_k = \mu + v_0\, \eta_z\, \kappa\, |\sin \gamma| > 0$ for $\eta_z\, \sin \gamma < 0$.

The non-monotonic behaviour of $\varepsilon_2 (\kappa, \gamma) = \sqrt{v_0^2\, \kappa^2 + \Delta^2} + v_0\, \eta_z\, \kappa\, \sin \gamma $ explains the double root. For $\eta_z\, \sin \gamma < 0$, the square root is flat at $\kappa = 0$, whereas the tilt term decreases linearly. The band therefore starts at $\Delta$, dips to its minimum $\Delta\, \sqrt{1 - \eta_z^2\, \sin^2 \gamma}$ at $\kappa_* (\gamma)$, and then climbs to $+\infty$. Wherever this minimum lies below $\mu$, the line $\varepsilon_2 = \mu$ is crossed twice. The outer root $\kappa_2$ increases with $\mu$, whereas the inner root $\kappa_1$ decreases with it, because $\varepsilon_2$ decreases with $\kappa$ there. For $\mu > \Delta$, the product of the roots is negative, so that exactly one root is positive, and we recover Eq.~\eqref{eq:kappaFztiltgapped}. These roots are used in Appendix~\ref{app:interband}.

For the circle map, we define $\tilde\xi \equiv \nu\, (w - w_c)$ and $\tilde R \equiv \sqrt{K}/(v_0\, \nu)$. The ellipse then becomes $u^2 + \tilde\xi^2 = \tilde R^2$. We write the squared equation as $\check G = 0$ with
\begin{align}
\check G \equiv (\mu - v_0\, \eta_z\, w)^2 - E_k^2 = v_0^2 \left( \tilde R^2 - u^2 - \tilde\xi^2 \right) \,.
\end{align}
The second form follows from the completed square above, because $K/\nu^2 = v_0^2\, \tilde R^2$. Since $\check G = (\mu - v_0\, \eta_z\, w - E_k)(\mu - v_0\, \eta_z\, w + E_k)$, the first factor is $\mu - \varepsilon_2$ and the second equals $2\, E_k$ on the FS. Hence $\delta (\varepsilon_2 - \mu) = 2\, E_k\, \delta (\check G)$ there. With $dw = d\tilde\xi/\nu$ and polar coordinates $(r, \vartheta)$ in the $(u, \tilde\xi)$ plane, we have $\delta (\check G) = \delta (r - \tilde R)/(2\, v_0^2\, \tilde R)$. The radial integral then gives $\tilde R \times 1/(2\, v_0^2\, \tilde R) = 1/(2\, v_0^2)$, so that for any function $\check F (u, w)$
\begin{align}
\label{eq:gaptiltz-circle}
\int du\, dw\, \delta (\varepsilon_2 - \mu)\, \check F = \frac{1}{\nu\, v_0^2} \int_0^{2\pi} \! d\vartheta\, E_k\, \check F \,,
\end{align}
with $E_k$ and $\check F$ evaluated on the circle. On the circle we have $u = \tilde R \cos\vartheta$ and $v_0\, w = (\sqrt{K} \sin\vartheta - \mu\, \eta_z)/\nu^2$. The on-shell energy then becomes
\begin{align}
E_k = \mu - v_0\, \eta_z\, w = \frac{ \mu - \eta_z\, \sqrt{K}\, \sin\vartheta }{ \nu^2 } \,.
\end{align}
For $\check F = 1$, the term odd in $\sin\vartheta$ integrates to zero, and the DOS of Eq.~\eqref{eq:dosgaptiltz} is recovered.

The band-velocity components follow from $\bs V^{(2)} = \bs \nabla_{\bs k}\, \varepsilon_2$ as $V_x^{(2)} = v_0^2\, u \cos\phi/E_k$, $V_y^{(2)} = v_0^2\, u \sin\phi/E_k$, and $V_z^{(2)} = v_0^2\, w/E_k + v_0\, \eta_z$. We denote $\hat a \equiv \eta_z\, \sqrt{K}$. Inserting $E_k$ and $w$ on the circle, the $z$-component becomes
\begin{align}
V_z^{(2)} = v_0\, \frac{ \sqrt{K} \sin\vartheta - \mu\, \eta_z }{ \mu - \hat a \sin\vartheta } + v_0\, \eta_z = \frac{ v_0\, \nu^2\, \sqrt{K}\, \sin\vartheta }{ \mu - \hat a \sin\vartheta } \,,
\end{align}
where the last step uses $1 - \eta_z^2 = \nu^2$. With $v_0\, \tilde R = \sqrt{K}/\nu$, the transverse components follow in the same way. In summary,
\begin{align}
\label{eq:gaptiltz-circle-vel}
V_z^{(2)} = \frac{ v_0\, \nu^2\, \sqrt{K}\, \sin\vartheta }{ \mu - \hat a \sin\vartheta } \,, \quad
V_x^{(2)} = \frac{ v_0\, \nu\, \sqrt{K}\, \cos\vartheta\, \cos\phi }{ \mu - \hat a \sin\vartheta } \,, \quad
V_y^{(2)} = \frac{ v_0\, \nu\, \sqrt{K}\, \cos\vartheta\, \sin\phi }{ \mu - \hat a \sin\vartheta } \,.
\end{align}

In Eq.~\eqref{eqdrude} only the $s=2$ band contributes for $\mu > 0$. With $d^3 \bs k = k_0\, d\phi\, du\, dw$ and the circle map, the integrand contains $E_k\, V_i^2$. For the $z$-component, Eq.~\eqref{eq:gaptiltz-circle-vel} gives $E_k\, (V_z^{(2)})^2 = v_0^2\, \nu^2\, K \sin^2\vartheta/(\mu - \hat a \sin\vartheta)$. The $\phi$ integral is trivial and equals $2\, \pi$. We therefore obtain
\begin{align}
D_{zz} = \frac{ g\, e^2\, k_0\, \nu\, K }{ 4\, \pi } \int_0^{2\pi} \! d\vartheta\, \frac{ \sin^2\vartheta }{ \mu - \hat a \sin\vartheta } \,.
\end{align}
For the $x$-component, $E_k\, (V_x^{(2)})^2 = v_0^2\, K \cos^2\vartheta \cos^2\phi/(\mu - \hat a \sin\vartheta)$. The $\phi$ integral now gives $\int_0^{2\pi} d\phi \cos^2\phi = \pi$, and we obtain
\begin{align}
D_{xx} = D_{yy} = \frac{ g\, e^2\, k_0\, K }{ 8\, \pi\, \nu } \int_0^{2\pi} \! d\vartheta\, \frac{ \cos^2\vartheta }{ \mu - \hat a \sin\vartheta } \,.
\end{align}
The equality of $D_{xx}$ and $D_{yy}$ follows from $\int_0^{2\pi} d\phi \sin^2\phi = \pi$.

The angular integrals reduce to the elementary one
\begin{align}
\check I_0 \equiv \int_0^{2\pi} \! \frac{d\vartheta}{\mu - \hat a \sin\vartheta} = \frac{2\, \pi}{\sqrt{\mu^2 - \hat a^2}} \,, \quad \mu > |\hat a| \,.
\end{align}
Here $\mu^2 - \hat a^2 = \mu^2 - \eta_z^2\, K = \nu^2\, M^2$, with $M = \sqrt{\mu^2 + \eta_z^2\, \Delta^2}$ as in Sec.~\ref{sec:gaptiltz}. We write $\sin\vartheta = [\mu - (\mu - \hat a \sin\vartheta)]/\hat a$, so that
\begin{align}
\sin^2\vartheta = \frac{ \mu^2 - 2\, \mu\, (\mu - \hat a \sin\vartheta) + (\mu - \hat a \sin\vartheta)^2 }{ \hat a^2 } \,.
\end{align}
Dividing by $\mu - \hat a \sin\vartheta$ and integrating, the middle term gives $-\, 4\, \pi\, \mu$ and the last term gives $2\, \pi\, \mu$, because $\int_0^{2\pi} d\vartheta \sin\vartheta = 0$. Hence
\begin{align}
\int_0^{2\pi} \! d\vartheta\, \frac{ \sin^2\vartheta }{ \mu - \hat a \sin\vartheta } = \frac{ \mu^2\, \check I_0 - 2\, \pi\, \mu }{ \hat a^2 } = \frac{ 2\, \pi\, \mu\, (\mu - \nu\, M) }{ \hat a^2\, \nu\, M } = \frac{ 2\, \pi\, \mu }{ \nu\, M\, (\mu + \nu\, M) } \,.
\end{align}
The last step uses $\hat a^2 = \eta_z^2\, K = \mu^2 - \nu^2\, M^2 = (\mu - \nu\, M)(\mu + \nu\, M)$. The remaining integral follows from $\cos^2\vartheta = 1 - \sin^2\vartheta$,
\begin{align}
\int_0^{2\pi} \! d\vartheta\, \frac{ \cos^2\vartheta }{ \mu - \hat a \sin\vartheta } = \check I_0 - \frac{ 2\, \pi\, \mu }{ \nu\, M\, (\mu + \nu\, M) } = \frac{ 2\, \pi }{ \mu + \nu\, M } \,.
\end{align}
Inserting both results into the expressions for $D_{zz}$ and $D_{xx}$ gives
\begin{align}
D_{zz} = \frac{g\, e^2\, k_0\, \nu\, K}{4\, \pi} \cdot \frac{2\, \pi\, \mu}{\nu\, M \left( \mu + \nu\, M \right)} \,, \quad
D_{xx} = \frac{g\, e^2\, k_0\, K}{8\, \pi\, \nu} \cdot \frac{2\, \pi}{\mu + \nu\, M} \,,
\end{align}
which simplify to Eq.~\eqref{eq:dtiltgapped}. None of the steps depends on the position of the origin relative to the ellipse, and the results therefore hold for all $\mu > \Delta_{\rm eff}$.

\subsection{Interband response of PTNR and GNR}
\label{app:interband}

This appendix gives the steps behind Eqs.~\eqref{eq:pinterztiltptnr}, \eqref{eq:epsb-ztilt}, and \eqref{eq:pinterztiltgnr}, and behind the imaginary parts quoted below them. We work to leading order in $q$ and for $\omega \gg v_0\, q$. The form-factor is then $\mathcal{F} \simeq q_i\, q_j\, g_{ij}$, and the denominators of Eq.~\eqref{eq:polinter} are evaluated at $q = 0$, where $E_{k'} = E_k$ and $\tilde a = 0$. The bracket of Eq.~\eqref{eq:polinter} becomes $4\, E_k/(\omega^2 - 4\, E_k^2)$. For $\mu > 0$ and $|\eta_z| < 1$, band $s=1$ is filled, and only the states with $\varepsilon_2 (\bs k) > \mu$ contribute.

Since $\mathcal{F} = (1 - \hat{\bs d} \cdot \hat{\bs d}')/2$ with $\hat{\bs d} = \bs d/E_k$, we have $\mathcal{F} \simeq |\delta \hat{\bs d}|^2/4$ to second order in $\bs q$. Let $\bs e_1$ and $\bs e_2$ denote the unit vectors along the first and second components of $\bs d$. For a transfer along $w$,
\begin{align}
\partial_w \hat{\bs d} = \frac{v_0\, \bs e_2}{E_k} - \frac{v_0^2\, w\, \bs d}{E_k^3} \,, \quad
\left| \partial_w \hat{\bs d} \right|^2 = \frac{v_0^2}{E_k^2} - \frac{v_0^4\, w^2}{E_k^4} \,.
\end{align}
Using $v_0^2\, w^2 = \left( E_k^2 - \Delta^2 \right) \sin^2 \gamma$, and the same steps for a transfer along $u$, we obtain the metric components
\begin{align}
g_{ww} = \frac{v_0^2}{4\, E_k^2} \left( \cos^2 \gamma + \frac{\Delta^2}{E_k^2} \sin^2 \gamma \right) , \quad
g_{uu} = \frac{v_0^2}{4\, E_k^2} \left( \sin^2 \gamma + \frac{\Delta^2}{E_k^2} \cos^2 \gamma \right) .
\end{align}
Their average over $\gamma$ is the $\langle g_{ij} \rangle$ quoted in Sec.~\ref{secgapnotilt}.

By Appendix~\ref{appdirac}, the local transfer is $q$ along $w$ for $\bs q \parallel \bs{\hat z}$ and $q \cos \phi$ along $u$ for $\bs q \parallel \bs{\hat x}$. The axial tilt does not depend on $\phi$, and the $\phi$-average of $\cos^2 \phi$ is $1/2$. At fixed $\gamma$ we use $\kappa\, d\kappa = E_k\, dE_k/v_0^2$. The unblocked part of the ray is $E_k > E_F (\gamma)$ in the fully-conducting regime. For $\omega$ below the onset the pole is not reached, and we find
\begin{align}
\label{eq:app-radial}
& \Pi_{\rm inter} (q\, \bs{\hat z}, \omega) = -\, \frac{g\, k_0\, q^2}{4\, \pi^2} \int_0^{2\pi} \! d\gamma \int_{E_F (\gamma)}^\infty \! dE\, \frac{\cos^2 \gamma + \Delta^2 \sin^2 \gamma/E^2}{4\, E^2 - \omega^2} \,, \nn
& \Pi_{\rm inter} (q\, \bs{\hat x}, \omega) = -\, \frac{g\, k_0\, q^2}{8\, \pi^2} \int_0^{2\pi} \! d\gamma \int_{E_F (\gamma)}^\infty \! dE\, \frac{\sin^2 \gamma + \Delta^2 \cos^2 \gamma/E^2}{4\, E^2 - \omega^2} \,.
\end{align}
The two elementary integrals needed are
\begin{align}
\int_E^\infty \frac{dE'}{4\, E'^2 - \omega^2} = \frac{P (E)}{4} \,, \quad
\int_E^\infty \frac{dE'}{E'^2 \left( 4\, E'^2 - \omega^2 \right)} = \frac{1}{\omega^2} \left[ P (E) - \frac{1}{E} \right] ,
\end{align}
with $P (E) = \omega^{-1} \ln \!\left[ (E + \omega/2)/(E - \omega/2) \right]$. The second follows from the partial fractions $1/[E^2 (4\, E^2 - \omega^2)] = \omega^{-2} \left[ 4/(4\, E^2 - \omega^2) - 1/E^2 \right]$. Inserting them into Eq.~\eqref{eq:app-radial} gives Eq.~\eqref{eq:pinterztiltgnr}, with $\mathcal{W}_z$ as defined there and $\mathcal{W}_x$ obtained by $\cos^2 \gamma \leftrightarrow \sin^2 \gamma$. In the displaced-torus regime, the unblocked part of a ray inside the window is $\kappa < \kappa_1$ together with $\kappa > \kappa_2$. The energy integral then runs from $\Delta$ to $E_1$ and from $E_2$ to infinity, and the integrand becomes $\mathcal{W} (\Delta, \gamma) - \mathcal{W} (E_1, \gamma) + \mathcal{W} (E_2, \gamma)$. Outside the window it is $\mathcal{W} (\Delta, \gamma)$.

For $\Delta = 0$ the Fermi energy is $E_F (\gamma) = \mu/(1 + \eta_z \sin \gamma)$, and only the first term of $\mathcal{W}_z$ remains. With $A_\pm = 2\, \mu \pm \omega$ and $B = \eta_z\, \omega$ we have $E_F \pm \omega/2 = \left[ \mu \pm (\omega/2)(1 + \eta_z \sin \gamma) \right]/(1 + \eta_z \sin \gamma)$, and therefore
\begin{align}
\omega\, P (E_F) = \ln \!\left( \frac{A_+ + B \sin \gamma}{A_- - B \sin \gamma} \right) \,.
\end{align}
For $A > |B|$ we use the Fourier series $\ln \!\left( A + B \cos \theta \right) = \ln \!\left[ (A + r)/2 \right] + 2 \sum_{n \geq 1} (-1)^{n+1}\, \varrho^n \cos (n\, \theta)/n$, with $r = \sqrt{A^2 - B^2}$ and $\varrho = B/(A + r)$. It follows from $A + B \cos \theta = \left[ (A + r)/2 \right] \left( 1 + 2\, \varrho \cos \theta + \varrho^2 \right)$. Setting $\theta = \pi/2 - \gamma$, the coefficient of $\cos (2\, \theta)$ is $-\, \varrho^2$, and
\begin{align}
\int_0^{2\pi} \! d\gamma\, \cos^2 \gamma\, \ln \!\left( A + B \sin \gamma \right) = \pi \left[ \ln \!\left( \frac{A + r}{2} \right) + \frac{\varrho^2}{2} \right] , \quad
\int_0^{2\pi} \! d\gamma\, \sin^2 \gamma\, \ln \!\left( A + B \sin \gamma \right) = \pi \left[ \ln \!\left( \frac{A + r}{2} \right) - \frac{\varrho^2}{2} \right] .
\end{align}
Changing $B \to -B$ leaves $r$ and $\varrho^2$ unchanged. Applying the results to $A_+$ and to $A_-$, with $r_\pm = \sqrt{A_\pm^2 - \eta_z^2\, \omega^2}$ and $\varrho_\pm = \eta_z\, \omega/(A_\pm + r_\pm)$, gives Eq.~\eqref{eq:pinterztiltptnr}. The series converges for $A_- > B$, which is the condition $\omega < 2\, \mu/(1 + |\eta_z|)$. For $\omega \to 0$ we have $r_\pm \simeq A_\pm$, and the bracket reduces to $\omega/\mu$. This gives the static value $-\, g\, k_0\, q^2/(16\, \pi\, \mu)$.

For the static limit of the GNR we use $P (E) = 1/E + \omega^2/(12\, E^3) + \mathcal{O} (\omega^4)$. This is the series of $P (E) = (2/\omega)\, \mathrm{arctanh}\, \left[ \omega/(2\, E) \right]$ with $\mathrm{arctanh}\, z = z + z^3/3 + \cdots$, and it makes the bracket $P (E) - 1/E$ in $\mathcal{W}_z$ equal to $\omega^2/(12\, E^3)$ at leading order. Then $\mathcal{W}_z \to \cos^2 \gamma/E + \Delta^2 \sin^2 \gamma/(3\, E^3)$, and Eq.~\eqref{eq:pinterztiltgnr} gives
\begin{align}
\epsilon_{b, z} = 1 + \frac{g\, e^2\, k_0}{4\, \pi} \int_0^{2\pi} \! d\gamma \left[ \frac{\cos^2 \gamma}{E_F} + \frac{\Delta^2 \sin^2 \gamma}{3\, E_F^3} \right] , \quad
\epsilon_{b, x} = 1 + \frac{g\, e^2\, k_0}{8\, \pi} \int_0^{2\pi} \! d\gamma \left[ \frac{\sin^2 \gamma}{E_F} + \frac{\Delta^2 \cos^2 \gamma}{3\, E_F^3} \right] ,
\end{align}
where $E_F = E_F (\gamma)$. Let $\bar R = \sqrt{(\mu^2 - \Delta^2) + \eta_z^2\, \Delta^2 \sin^2 \gamma}$ and $D_\gamma = \mu^2 + \eta_z^2\, \Delta^2 \sin^2 \gamma$. Then $\left( \mu - \eta_z \sin \gamma\, \bar R \right) \left( \mu + \eta_z \sin \gamma\, \bar R \right) = \left( 1 - \eta_z^2 \sin^2 \gamma \right) D_\gamma$, and with $E_F = \left( \mu - \eta_z \sin \gamma\, \bar R \right)/\left( 1 - \eta_z^2 \sin^2 \gamma \right)$,
\begin{align}
\frac{1}{E_F} = \frac{\mu + \eta_z \sin \gamma\, \bar R}{D_\gamma} \,.
\end{align}
The term with $\bar R$ is odd in $\sin \gamma$ and drops out of the integrals. For $1/E_F^3$ the even part is $\left( \mu^3 + 3\, \mu\, \eta_z^2 \sin^2 \gamma\, \bar R^2 \right)/D_\gamma^3$. Using $\bar R^2 = D_\gamma - \Delta^2$ and $\eta_z^2\, \Delta^2 \sin^2 \gamma = D_\gamma - \mu^2$, it becomes
\begin{align}
\frac{4\, \mu^3}{D_\gamma^3} + \frac{3\, \mu}{\Delta^2} \left[ \frac{1}{D_\gamma} - \frac{\mu^2 + \Delta^2}{D_\gamma^2} \right] .
\end{align}
The weight of $\epsilon_{b, z}$ then collapses to $\mu/D_\gamma + \sin^2 \gamma \left[ 4\, \mu^3\, \Delta^2/(3\, D_\gamma^3) - \mu \left( \mu^2 + \Delta^2 \right)/D_\gamma^2 \right]$, and that of $\epsilon_{b, x}$ to the same expression with $\sin^2 \gamma \to \cos^2 \gamma$. With $T_n = \int_0^{2\pi} d\gamma\, \sin^2 \gamma/D_\gamma^n$, $C_n = \int_0^{2\pi} d\gamma\, \cos^2 \gamma/D_\gamma^n$, and $\bar I_n = T_n + C_n$,
\begin{align}
\epsilon_{b, z} &= 1 + \frac{g\, e^2\, k_0}{4\, \pi} \left[ \mu\, \bar I_1 + \frac{4}{3}\, \mu^3\, \Delta^2\, T_3 - \mu \left( \mu^2 + \Delta^2 \right) T_2 \right] , \nn
\epsilon_{b, x} &= 1 + \frac{g\, e^2\, k_0}{8\, \pi} \left[ \mu\, \bar I_1 + \frac{4}{3}\, \mu^3\, \Delta^2\, C_3 - \mu \left( \mu^2 + \Delta^2 \right) C_2 \right] .
\end{align}
The integrals follow from $\bar I_1 = 2\, \pi/(\mu\, M)$, with $M = \sqrt{\mu^2 + \eta_z^2\, \Delta^2}$, by differentiation with respect to $\mu^2$ at fixed $\eta_z^2\, \Delta^2$, namely $\bar I_{n+1} = -\, n^{-1}\, \partial \bar I_n/\partial \left( \mu^2 \right)$. The first integral is the standard one, $\int_0^{2\pi} d\gamma/(\mu^2 + \eta_z^2\, \Delta^2 \sin^2 \gamma) = 2\, \pi/(\mu\, M)$. Two differentiations give
\begin{align}
\bar I_2 = \frac{\pi \left( \mu^2 + M^2 \right)}{\mu^3\, M^3} \,, \quad
\bar I_3 = \frac{\pi \left( 3\, \mu^4 + 2\, \mu^2\, M^2 + 3\, M^4 \right)}{4\, \mu^5\, M^5} \,.
\end{align}
The relation $\eta_z^2\, \Delta^2\, T_n = \bar I_{n-1} - \mu^2\, \bar I_n$ follows from $\eta_z^2\, \Delta^2 \sin^2 \gamma = D_\gamma - \mu^2$. Together with $\bar I_0 = 2\, \pi$ and the values above, it gives
\begin{align}
& T_2 = \frac{\pi}{\mu\, M^3} \,, \quad
C_2 = \frac{\pi}{\mu^3\, M} \,, \quad
T_3 = \frac{\pi \left( M^2 + 3\, \mu^2 \right)}{4\, \mu^3\, M^5} \,, \quad
C_3 = \frac{\pi \left( \mu^2 + 3\, M^2 \right)}{4\, \mu^5\, M^3} \,.
\end{align}
Substitution yields
\begin{align}
& \epsilon_{b, z} = 1 + \frac{g\, e^2\, k_0}{4} \left[ \frac{2}{M} + \frac{\Delta^2 \left( M^2 + 3\, \mu^2 \right)}{3\, M^5} - \frac{\mu^2 + \Delta^2}{M^3} \right] , \nn
& \epsilon_{b, x} = 1 + \frac{g\, e^2\, k_0}{8} \left[ \frac{2}{M} + \frac{\Delta^2 \left( \mu^2 + 3\, M^2 \right)}{3\, \mu^2\, M^3} - \frac{\mu^2 + \Delta^2}{\mu^2\, M} \right] .
\end{align}
With $K = \mu^2 - \Delta^2 \left( 1 - \eta_z^2 \right)$, these simplify to Eq.~\eqref{eq:epsb-ztilt}. For $\Delta \to 0$ only the term $\cos^2 \gamma/E_F$ survives in $\epsilon_{b, z}$, with $1/E_F = \left( 1 + \eta_z \sin \gamma \right)/\mu$. The odd term integrates to zero and $\int_0^{2\pi} d\gamma\, \cos^2 \gamma = \pi$. This gives $\epsilon_{b, z} = 1 + g\, e^2\, k_0/(4\, \mu)$, independent of $\eta_z$, and $\epsilon_{b, x} = 1 + g\, e^2\, k_0/(8\, \mu)$ follows in the same way.

For the imaginary part, $\Im \left[ 1/(\omega - 2\, E_k + i\, 0^+) \right] = -\, \pi\, \delta (\omega - 2\, E_k)$ in Eq.~\eqref{eq:polinter}. The same steps as above give, with $\Im \varepsilon^{\rm RPA} = -\, v (q)\, \Im \Pi$,
\begin{align}
\label{eq:app-imeps}
\Im \varepsilon^{\rm RPA} (q\, \bs{\hat z}, \omega) &= \frac{g\, e^2\, k_0}{4\, \omega} \int_{\rm arc} \! d\gamma \left( \cos^2 \gamma + \frac{4\, \Delta^2}{\omega^2} \sin^2 \gamma \right) , \nn
\Im \varepsilon^{\rm RPA} (q\, \bs{\hat x}, \omega) &= \frac{g\, e^2\, k_0}{8\, \omega} \int_{\rm arc} \! d\gamma \left( \sin^2 \gamma + \frac{4\, \Delta^2}{\omega^2} \cos^2 \gamma \right) .
\end{align}
The delta function sets $E_k = \omega/2$, which is the circle $v_0\, \kappa = \Upsilon \equiv \sqrt{\omega^2/4 - \Delta^2}$. A point of this circle is unblocked if $\varepsilon_2 > \mu$. On the circle $\varepsilon_2 = \omega/2 + \eta_z\, \Upsilon \sin \gamma$. For $\eta_z > 0$ the unblocked arc is therefore $\sin \gamma > \tau $ with $ \tau = \left( \mu - \omega/2 \right)/(\eta_z\, \Upsilon)$. For the PTNR, $\Upsilon = \omega/2$ and $ \tau = \left( 2\, \mu/\omega - 1 \right)/\eta_z$. For $|\tau| < 1$ the arc is $\gamma_1 < \gamma < \pi - \gamma_1$ with $\gamma_1 = \arcsin \tau $, and
\begin{align}
\int_{\rm arc} d\gamma\, \cos^2 \gamma = \frac{\pi - 2\, \gamma_1}{2} - \tau \, \sqrt{1 - \tau^2} \,, \quad
\int_{\rm arc} d\gamma\, \sin^2 \gamma = \frac{\pi - 2\, \gamma_1}{2} + \tau\, \sqrt{1 - \tau^2} \,.
\end{align}
The first follows from the antiderivative $\int d\gamma\, \cos^2 \gamma = \gamma/2 + \sin (2\, \gamma)/4$ between $\gamma_1$ and $\pi - \gamma_1$, with $\sin (2\, \gamma_1) = 2\, \tau\, \sqrt{1 - \tau^2}$. The second follows from $\sin^2 \gamma = 1 - \cos^2 \gamma$ and the arc length $\pi - 2\, \gamma_1$. Inserting these into Eq.~\eqref{eq:app-imeps} and using $4\, \Delta^2 - \omega^2 = -\, 4\, \Upsilon^2$ gives, for $\bs q \parallel \bs{\hat z}$,
\begin{align}
\Im \varepsilon^{\rm RPA} (q\, \bs{\hat z}, \omega) = \frac{g\, e^2\, k_0}{4\, \omega} \left[ \frac{\pi - 2\, \gamma_1}{2} \left( 1 + \frac{4\, \Delta^2}{\omega^2} \right) - \frac{4\, \Upsilon^2}{\omega^2}\, \tau\, \sqrt{1 - \tau^2} \right] ,
\end{align}
which is the expression quoted in Sec.~\ref{sec:gaptiltz}. The $\bs{\hat x}$ and $\bs{\hat y}$ channels follow in the same way. They carry half the prefactor, and the sign of the term with $\tau$ is reversed. For $\Delta = 0$ the prefactors reduce to $\pi\, g\, e^2\, k_0/(4\, \omega)$ and $\pi\, g\, e^2\, k_0/(8\, \omega)$, and the brackets become $\mathcal{G}_+ (\tau )$ and $\mathcal{G}_- (\tau)$ of Sec.~\ref{sectiltz}. The arc is empty for $ \tau \geq 1$, and it is the whole circle, with $\gamma_1 = -\pi/2$, for $ \tau \leq -1$.

\subsection{Approach of the dielectric function to the interband onset}
\label{app:onsetz}

This appendix derives how $\Re \varepsilon^{\rm RPA}$ approaches the interband onset for the axially-tilted PTNR and GNR, which Appendix~\ref{app:onset} classifies. The interband part of the dielectric function is $\varepsilon_{\rm inter, i} = -\, 4\, \pi\, e^2\, \Pi_{\rm inter}/q^2$. Eq.~\eqref{eq:pinterztiltgnr} gives
\begin{align}
\varepsilon_{\rm inter, z} = \frac{g\, e^2\, k_0}{4\, \pi} \int_0^{2\pi} \! d\gamma\, \mathcal{W}_z \left( E_F (\gamma), \gamma \right) , \quad
\varepsilon_{\rm inter, x} = \varepsilon_{\rm inter, y} = \frac{g\, e^2\, k_0}{8\, \pi} \int_0^{2\pi} \! d\gamma\, \mathcal{W}_x \left( E_F (\gamma), \gamma \right) ,
\end{align}
where $E_F (\gamma)$ is the energy at the lower edge of the unblocked part of the ray. The only singular function in $\mathcal{W}$ is $P (E)$, and it is singular at $E = \omega/2$, where $P (E) = -\, \omega^{-1} \ln \!\left( E - \omega/2 \right) + \text{analytic}$. The term $-1/E$ in $\mathcal{W}$ is analytic. The weights of $P$ are $c_z = \cos^2\gamma + 4\, \Delta^2 \sin^2\gamma/\omega^2$ in $\mathcal{W}_z$ and $c_x = \sin^2\gamma + 4\, \Delta^2 \cos^2\gamma/\omega^2$ in $\mathcal{W}_x$. A non-analytic term can therefore arise only at those $\gamma$ where $E_F (\gamma)$ reaches $\omega_{\rm th}/2$.

We begin with the fully-conducting regime, $\mu > \Delta$, and take $\eta_z > 0$. Here $E_F (\gamma)$ is the energy of the point where the ray meets the ellipse. On the ellipse $E_k = \mu - v_0\, \eta_z\, w$ by Appendix~\ref{app:gnrztilt}, and the minimum of $E_F$ therefore sits at the largest $w$. This is the top of the ellipse, $u = 0$, which lies on the ray $\gamma_* = \pi/2$. On the circle map, $E_k = \left( \mu - \eta_z \sqrt{K} \sin\vartheta \right)/\nu^2$ and the top is at $\vartheta = \pi/2$. It gives $\omega_{\rm th} = 2\, E_{\min}$ with $E_{\min} = \left( \mu - \eta_z\, \sqrt{K} \right)/\nu^2$, as in the main text. For the curvature of $E_F$ we set $\vartheta = \pi/2 + t$. To leading order in $t$ we have $u \simeq -\, \tilde R\, t$ with $\tilde R = \sqrt{K}/(v_0\, \nu)$, and $w \simeq w_m$ with $w_m = \left( \sqrt{K} - \mu\, \eta_z \right)/(v_0\, \nu^2)$. The angle of the ray follows from $\cot\gamma = u/w$ as $\gamma - \pi/2 \simeq \tilde R\, t/w_m$, and the energy is $E_k \simeq E_{\min} + \eta_z \sqrt{K}\, t^2/(2\, \nu^2)$. Eliminating $t$ gives
\begin{align}
\label{eq:app-EF-curv}
E_F (\gamma) \simeq E_{\min} + \frac{\mathcal{E}}{2} \left( \gamma - \frac{\pi}{2} \right)^2 , \quad
\mathcal{E} = \frac{\eta_z \left( \sqrt{K} - \eta_z\, \mu \right)^2}{\nu^4\, \sqrt{K}} = \frac{\eta_z \left( \mu^2 - \Delta^2 \right)^2}{\sqrt{K} \left( \sqrt{K} + \eta_z\, \mu \right)^2} \,.
\end{align}
The second form uses $\left( \sqrt{K} - \eta_z\, \mu \right) \left( \sqrt{K} + \eta_z\, \mu \right) = \nu^2 \left( \mu^2 - \Delta^2 \right)$. Differentiating the explicit $E_F (\gamma)$ of Eq.~\eqref{eq:kappaFztiltgapped} twice at $\sin\gamma = 1$ gives the same $\mathcal{E}$.

Let $\check\delta \equiv E_{\min} - \omega/2 = (\omega_{\rm th} - \omega)/2$. Near $\gamma_*$ we have $E_F - \omega/2 = \check\delta + \mathcal{E}\, \hat x^2/2$ with $\hat x = \gamma - \pi/2$. As in Appendix~\ref{app:onset}, the non-analytic part of the integral of the logarithm follows from its derivative,
\begin{align}
\frac{d}{d\check\delta} \int d\hat x\, \ln \!\left( \check\delta + \frac{\mathcal{E}}{2}\, \hat x^2 \right) = \pi \sqrt{\frac{2}{\mathcal{E}\, \check\delta}} \,.
\end{align}
Integration over $\check\delta$ gives $2\,\pi\, \sqrt{(\omega_{\rm th} - \omega)/\mathcal{E}}$. The logarithm enters $P$ with the factor $-\, 1/\omega_{\rm th}$, and the weights at $\gamma_*$ are $c_z = 4\, \Delta^2/\omega_{\rm th}^2$ and $c_x = 1$, since $\cos\gamma_* = 0$. Hence
\begin{align}
\label{eq:onset-cusp-z}
\delta \varepsilon_{\rm inter, z} \simeq -\, \frac{2\, g\, e^2\, k_0\, \Delta^2}{\omega_{\rm th}^3} \sqrt{\frac{\omega_{\rm th} - \omega}{\mathcal{E}}} \,, \quad
\delta \varepsilon_{\rm inter, x} = \delta \varepsilon_{\rm inter, y} \simeq -\, \frac{g\, e^2\, k_0}{4\, \omega_{\rm th}} \sqrt{\frac{\omega_{\rm th} - \omega}{\mathcal{E}}} \,.
\end{align}
The quadratic form of $E_F$ must hold over the arc of half-width $\sqrt{2\, \check\delta/\mathcal{E}}$, and Eq.~\eqref{eq:onset-cusp-z} is therefore valid for $\omega_{\rm th} - \omega \ll \mathcal{E}$. In contrast to the in-plane tilt, the two channels differ at the cusp. At $\gamma_*$ the ray lies along $w$. A transfer along $\bs{\hat z}$ is then parallel to $\bs k$ in the local plane, and its form-factor is carried by the mass alone, whereas a transfer along $\bs{\hat x}$ is perpendicular to $\bs k$ and has the full weight. For $\eta_z \to 0$ we have $\mathcal{E} \simeq \eta_z \sqrt{\mu^2 - \Delta^2}$, and the coefficients diverge as $\eta_z^{-1/2}$, where the logarithm of the untilted case takes over. For $\mu \to \Delta^+$ we have $\mathcal{E} \simeq (\mu - \Delta)^2/(\eta_z^2\, \Delta)$. The window of validity then closes quadratically, and the cusp gives way to the logarithm of the displaced-torus regime treated below.

For the PTNR, $\Delta = 0$, we have $\sqrt{K} = \mu$, $E_{\min} = \mu/(1 + \eta_z)$, $\omega_{\rm th} = 2\, \mu/(1 + \eta_z)$, and $\mathcal{E} = \eta_z\, \mu/(1 + \eta_z)^2$. The weight $c_z = \cos^2\gamma \simeq \hat x^2$ now vanishes at $\gamma_*$. For this weight the derivative above is replaced by $\int dx\, x^2/(\delta + \mathcal{E}\, x^2/2) = (2/\mathcal{E}) \int dx \left[ 1 - \delta/(\delta + \mathcal{E}\, x^2/2) \right]$. Its non-analytic part is $-\, \pi\, \sqrt{\delta}\, (2/\mathcal{E})^{3/2}$, and integration over $\delta$ gives $-\, (2\, \pi/3)\, (2\, \delta/\mathcal{E})^{3/2}$ for the integral of $x^2 \ln \!\left( \delta + \mathcal{E}\, x^2/2 \right)$. We obtain
\begin{align}
\delta \varepsilon_{\rm inter, x} = \delta \varepsilon_{\rm inter, y} \simeq -\, \frac{g\, e^2\, k_0 \left( 1 + \eta_z \right)^2}{8\, \mu} \sqrt{\frac{\omega_{\rm th} - \omega}{\eta_z\, \mu}} \,, \quad
\delta \varepsilon_{\rm inter, z} \simeq \frac{g\, e^2\, k_0 \left( 1 + \eta_z \right)^4}{12\, \mu} \left( \frac{\omega_{\rm th} - \omega}{\eta_z\, \mu} \right)^{3/2} .
\end{align}
The first expression is Eq.~\eqref{eq:onset-cusp-z} at $\Delta = 0$. The $\bs{\hat z}$ channel carries only the exponent $3/2$, as does the $\bs{\hat y}$ channel of the in-plane tilt.

In the displaced-torus regime, $\Delta_{\rm eff} < \mu < \Delta$, the onset is $\omega_{\rm th} = 2\, \Delta$, and the lower edge of every unblocked ray is $E = \Delta$ at $\kappa = 0$. The integrand of Eq.~\eqref{eq:pinterztiltgnr} is $\mathcal{W} (\Delta, \gamma) - \mathcal{W} (E_1, \gamma) + \mathcal{W} (E_2, \gamma)$ inside the window and $\mathcal{W} (\Delta, \gamma)$ outside it. Since $E_{1,2} > \Delta$, the terms with $E_1$ and $E_2$ stay analytic at $\omega = 2\, \Delta$. The term $P (\Delta) \to (2\, \Delta)^{-1} \ln \!\left[ 4\, \Delta/(\omega_{\rm th} - \omega) \right]$ is singular for every $\gamma$, and both weights equal $\cos^2\gamma + \sin^2\gamma = 1$ there. The $\gamma$-integral is $2\, \pi$, and we obtain
\begin{align}
\delta \varepsilon_{\rm inter, z} \simeq \frac{g\, e^2\, k_0}{4\, \Delta} \ln \!\left(  \frac{4\, \Delta}{\omega_{\rm th} - \omega} \right), \quad
\delta \varepsilon_{\rm inter, x} = \delta \varepsilon_{\rm inter, y} \simeq \frac{g\, e^2\, k_0}{8\, \Delta} \ln \!\left(  \frac{4\, \Delta}{\omega_{\rm th} - \omega} \right).
\end{align}
This is Eq.~\eqref{eq:onset-log} with the gapped arc replaced by the whole ring, $\mathcal{A}_z = 2\, \pi$ and $\mathcal{A}_x = \pi$. It does not depend on $\eta_z$, and it equals the result of the untilted doped GNR at $\mu = \Delta$. The real part diverges to $+\infty$, and the Drude term gives $-\infty$ at $\omega \to 0$. Since the interband part increases with $\omega$, the zero of $\Re \varepsilon^{\rm RPA}$ is unique and lies below $\omega_{\rm th}$ at every coupling.


\section{Tilted Weyl node: calculation details}
\label{app:weyl}

Here, we give the derivations behind the results of Sec.~\ref{secweyl}.

\subsection{Fermi surface and density of states}

For Eq.~\eqref{eq:hweyl}, the energies are $\varepsilon_\pm (\bs k) = \pm v_0\, k + v_0\, \bs{\eta} \cdot \bs{k}$. They satisfy $\varepsilon_\mp (- \bs k) = - \varepsilon_\pm (\bs k)$, and all results are therefore even in $\mu$. We take $\mu > 0$. The conduction band is $\varepsilon_+ = v_0\, k \left( 1 + \eta \cos\varpi \right)$, where $\varpi$ is the angle between $\bs k$ and $\bs{\hat n}$, and its velocity is $\bs{V} = v_0 \left( \bs{\hat k} + \bs{\eta} \right)$. Setting $\Xi = \cos\varpi$, the FS is $k_F = \mu/[v_0 \left( 1 + \eta\, \Xi \right)]$, and the radial derivative of the energy at the FS is $v_0 \left( 1 + \eta\, \Xi \right)$. Throughout, $\nu^2 = 1 - \eta^2$. The DOS follows from
\begin{align}
\rho = g \int \frac{d^3\bs k}{(2\pi)^3}\, \delta (\varepsilon_+ - \mu) = \frac{g}{(2\pi)^3} \int d\Omega\, \frac{k_F^2}{v_0 \left( 1 + \eta\, \Xi \right)} = \frac{g\, \mu^2}{4\, \pi^2\, v_0^3} \int_{-1}^{1} \frac{d\Xi}{\left( 1 + \eta\, \Xi \right)^3} \,.
\end{align}
Using $\int_{-1}^{1} d\Xi \left( 1 + \eta\, \Xi \right)^{-3} = 2/\nu^4$ gives Eq.~\eqref{eq:rhoweyl}.

\subsection{Drude weights}

From Eq.~\eqref{eqdrude}, $D_{ij} = \pi\, g\, e^2 \int d^3\bs k/(2\pi)^3\, \delta (\varepsilon_+ - \mu)\, V_i\, V_j$. The components of the velocity along $\bs{\hat n}$ and perpendicular to it are $V_\parallel = v_0 \left( \Xi + \eta \right)$ and $V_\perp = v_0 \sin\varpi \cos\varphi$, where $\varphi$ is the azimuthal angle around $\bs{\hat n}$. The off-diagonal components vanish after the $\varphi$ integral. In the same way as for the DOS,
\begin{align}
D_\parallel = \frac{g\, e^2\, \mu^2}{4\, \pi\, v_0}\, \check J_\parallel \,, \quad
D_\perp = \frac{g\, e^2\, \mu^2}{8\, \pi\, v_0}\, \check J_\perp \,, \quad
\check J_\parallel = \int_{-1}^{1} \frac{\left( \Xi + \eta \right)^2\, d\Xi}{\left( 1 + \eta\, \Xi \right)^3} \,, \quad
\check J_\perp = \int_{-1}^{1} \frac{\left( 1 - \Xi^2 \right) d\Xi}{\left( 1 + \eta\, \Xi \right)^3} \,.
\end{align}
With the substitution $\Psi = 1 + \eta\, \Xi$, these integrals are elementary and give
\begin{align}
\check J_\parallel = \frac{2 \left( \text{arctanh}\, \eta - \eta \right)}{\eta^3} \,, \quad
\check J_\perp = \frac{2}{\eta^3} \left( \frac{\eta}{\nu^2} - \text{arctanh}\, \eta \right) ,
\end{align}
which yields Eq.~\eqref{eq:dweyl}. Expanding in $\eta$ gives $\check J_\parallel = 2/3 + 2\, \eta^2/5 + \mathcal{O} (\eta^4)$ and $\check J_\perp = 4/3 + 8\, \eta^2/5 + \mathcal{O} (\eta^4)$, from which Eq.~\eqref{eq:dweylsmall} follows. As a check, $\check J_\parallel + \check J_\perp = \int_{-1}^{1} d\Xi \left( 1 + \eta^2 + 2\, \eta\, \Xi \right)/(1 + \eta\, \Xi)^3 = 2/\nu^2$, which gives the sum rule. As $\eta \to 1$, $\check J_\parallel$ diverges only logarithmically, whereas $\check J_\perp$ diverges as $\nu^{-2}$.

\subsection{Interband onset and background dielectric constant}

At $q \to 0$ only vertical transitions contribute, with energy $2\, v_0\, k$ because the tilt term cancels. A transition at $\bs k$ is allowed when the lower-band state is occupied and the upper-band state is empty, which requires $k > k_F (\varpi)$. The lowest allowed energy is therefore $2\, v_0 \min_\varpi k_F (\varpi) = 2\, \mu/(1 + \eta)$, reached at $\varpi = 0$. This proves Eq.~\eqref{eq:thweyl}.

For the background dielectric constant, we use Eq.~\eqref{eq:polinterstatic}. The interband matrix element between the states $\bs k$ and $\bs k + \bs q$ of opposite bands is $|\langle +, \bs k + \bs q | -, \bs k \rangle|^2 = q_i\, q_j\, g_{ij} (\bs k)$ to leading order in $\bs q$. Here, $g_{ij} = \left( \delta_{ij} - \hat k_i\, \hat k_j \right)/(4\, k^2)$ is the quantum metric, which follows from the eigenstates of $\bs k \cdot \bs \sigma$ and is independent of the tilt. Both orderings of the bands contribute $-1/(2\, v_0\, k)$ in the unblocked region $k > k_F (\varpi)$, where the lower band is filled and the upper band is empty. The static polarisability is therefore
\begin{align}
\Pi_{\rm inter} (\bs q, 0) = -\, \frac{g\, q_i\, q_j}{4\, v_0 \left( 2\, \pi \right)^3} \int d\Omega \left( \delta_{ij} - \hat k_i\, \hat k_j \right) \int_{k_F (\varpi)}^{\Lambda_{\rm c}} \frac{dk}{k} \,,
\end{align}
and the background dielectric constant is $\epsilon_{b} = 1 - 4\, \pi\, e^2\, \Pi_{\rm inter}/q^2$. For $\bs q \parallel \bs{\hat n}$, the angular weight is $1 - \Xi^2$. For $\bs q \perp \bs{\hat n}$ it averages over $\varphi$ to $\left( 1 + \Xi^2 \right)/2$. The radial integral is $\int_{k_F (\varpi)}^{\Lambda_{\rm c}} dk/k = \ln \!\left( v_0\, \Lambda_{\rm c}/\mu \right) + \ln \!\left( 1 + \eta\, \Xi \right)$, where the second term is the tilt dependence of the lower limit. The first term, after the angular integral $\int_{-1}^{1} d\Xi \left( 1 - \Xi^2 \right) = \int_{-1}^{1} d\Xi \left( 1 + \Xi^2 \right)/2 = 4/3$, gives the logarithm in Eq.~\eqref{eq:epsbweyl}. The second term gives $\mathcal{B}_\parallel$ and $\mathcal{B}_\perp$. Their expansions follow from $\ln \!\left( 1 + \eta\, \Xi \right) = \eta\, \Xi - \eta^2 \Xi^2/2 + \dots$. The odd term vanishes upon integration, and $\int_{-1}^{1} d\Xi \left( \Xi^2 - \Xi^4 \right) = 4/15$ and $\int_{-1}^{1} d\Xi \left( \Xi^2 + \Xi^4 \right)/2 = 8/15$ give the values $-2\, \eta^2/15$ and $-4\, \eta^2/15$. The constant term depends on how the cutoff is implemented, and only the dependence on $\mu$ is regularisation-independent.

For the imaginary part, we use Eq.~\eqref{eq:polinter} at $\omega \gg v_0\, q$, where the denominators are evaluated at $q = 0$. The two orderings of the bands combine into $4\, v_0\, k/[(\omega + i\, 0^+)^2 - 4\, v_0^2\, k^2]$. With the matrix element $q_i\, q_j\, g_{ij}$ and the measure $k^2\, dk\, d\Omega$, the factors of $k$ combine as $k^2 \times 1/(4\, k^2) \times 4\, v_0\, k = v_0\, k$, and
\begin{align}
\Pi_{\rm inter} (\bs q, \omega) = \frac{g\, v_0\, q_i\, q_j}{(2\, \pi)^3} \int d\Omega \left( \delta_{ij} - \hat k_i\, \hat k_j \right) \int_{k_F (\varpi)}^{\Lambda_{\rm c}} \frac{k\, dk}{(\omega + i\, 0^+)^2 - 4\, v_0^2\, k^2} \,.
\end{align}
The imaginary part of the integrand is $-\, \pi\, \delta (\omega^2 - 4\, v_0^2\, k^2)$, which has its support at $k = \omega/(2\, v_0)$ and gives $\int k\, dk\, \delta (\omega^2 - 4\, v_0^2\, k^2) = 1/(8\, v_0^2)$. The delta function contributes only if $\omega/(2\, v_0) > k_F (\varpi)$, which is the condition $1 + \eta\, \Xi > 2\, \mu/\omega$. Hence
\begin{align}
\Im \Pi_{\rm inter} (\bs q, \omega) = -\, \frac{g\, q_i\, q_j}{64\, \pi^2\, v_0} \int d\Omega \left( \delta_{ij} - \hat k_i\, \hat k_j \right) \Theta \left( \omega - \frac{2\, \mu}{1 + \eta\, \Xi} \right).
\end{align}
For $\bs q \parallel \bs{\hat n}$ and $\bs q \perp \bs{\hat n}$, the angular weights are again $1 - \Xi^2$ and $\left( 1 + \Xi^2 \right)/2$, and $\int d\Omega = 2\, \pi \int d\Xi$. We define $\Xi_\omega = \max \left[ -1, \left( 2\, \mu/\omega - 1 \right)/\eta \right]$ for $\omega \geq 2\, \mu/(1 + \eta)$. The two elementary integrals $\int_{\Xi_\omega}^{1} d\Xi \left( 1 - \Xi^2 \right) = 2/3 - \Xi_\omega + \Xi_\omega^3/3$ and $\int_{\Xi_\omega}^{1} d\Xi \left( 1 + \Xi^2 \right) = 4/3 - \Xi_\omega - \Xi_\omega^3/3$ then give, with $\Im \varepsilon^{\rm RPA} = -\, 4\, \pi\, e^2\, \Im \Pi_{\rm inter}/q^2$,
\begin{align}
\Im \varepsilon^{\rm RPA}_\parallel = \frac{g\, e^2}{8\, v_0} \left( \frac{2}{3} - \Xi_\omega + \frac{\Xi_\omega^3}{3} \right), \quad
\Im \varepsilon^{\rm RPA}_\perp = \frac{g\, e^2}{16\, v_0} \left( \frac{4}{3} - \Xi_\omega - \frac{\Xi_\omega^3}{3} \right),
\end{align}
and both vanish for $\omega < 2\, \mu/(1 + \eta)$. For $\omega \geq 2\, \mu/(1 - \eta)$, we have $\Xi_\omega = -\, 1$, and both expressions equal $g\, e^2/(6\, v_0)$. This is also the result for $\eta \to 0$ at $\omega > 2\, \mu$, and it is Eq.~\eqref{eq:imweyl}.


\bibliography{ref_pl}

@article{nielsen,
title = {A no-go theorem for regularizing chiral fermions},
journal = {Physics Letters B},
volume = {105},
number = {2},
pages = {219-223},
year = {1981},
issn = {0370-2693},
doi = {https://doi.org/10.1016/0370-2693(81)91026-1},
url = {https://www.sciencedirect.com/science/article/pii/0370269381910261},
author = {H.B. Nielsen and M. Ninomiya}
}

@article{lv_weyl,
title="{Experimental Discovery of Weyl Semimetal TaAs}",
  author = {Lv, B. Q. and Weng, H. M. and Fu, B. B. and Wang, X. P. and Miao, H. and Ma, J. and Richard, P. and Huang, X. C. and Zhao, L. X. and Chen, G. F. and Fang, Z. and Dai, X. and Qian, T. and Ding, H.},
  journal = {Phys. Rev. X},
  volume = {5},
  issue = {3},
  pages = {031013},
  numpages = {8},
  year = {2015},
  month = {Jul},
  publisher = {American Physical Society},
  doi = {10.1103/PhysRevX.5.031013},
  url = {https://link.aps.org/doi/10.1103/PhysRevX.5.031013}
}

@article{yang_weyl,
title="{Discovery of a Weyl fermion semimetal and topological Fermi arcs}",
author = {Su-Yang Xu  and Ilya Belopolski  and Nasser Alidoust  and Madhab Neupane  and Guang Bian  and Chenglong Zhang  and Raman Sankar  and Guoqing Chang  and Zhujun Yuan  and Chi-Cheng Lee  and Shin-Ming Huang  and Hao Zheng  and Jie Ma  and Daniel S. Sanchez  and BaoKai Wang  and Arun Bansil  and Fangcheng Chou  and Pavel P. Shibayev  and Hsin Lin  and Shuang Jia  and M. Zahid Hasan },
journal = {Science},
volume = {349},
number = {6248},
pages = {613-617},
year = {2015},
doi = {10.1126/science.aaa9297},
URL = {https://www.science.org/doi/abs/10.1126/science.aaa9297}
}

@article{ruan_weyl,
title="{Symmetry-protected ideal Weyl semimetal in HgTe-class materials}",
	volume = {7},
	copyright = {2016 The Author(s)},
	issn = {2041-1723},
	url = {https://www.nature.com/articles/ncomms11136},
	doi = {10.1038/ncomms11136},
	number = {1},
	urldate = {2022-11-16},
	journal = {Nature Communications},
	author = {Ruan, Jiawei and Jian, Shao-Kai and Yao, Hong and Zhang, Haijun and Zhang, Shou-Cheng and Xing, Dingyu},
	month = apr,
	year = {2016},
	pages = {11136}
}

@article{ips-plasmon-luttinger-epjb,
title = "{Plasmon modes in Luttinger semimetals}",
author = {Mandal, Ipsita},
journal = {Eur. Phys. J. B},
volume = {96},
number = {10},
pages = {150},
year = {2023},
month = {Oct},
doi = {10.1140/epjb/s10051-023-00596-x},
url = {https://doi.org/10.1140/epjb/s10051-023-00596-x}
}

@article{ips-plasmon-luttinger-aop,
title = "{Search for plasmons in isotropic Luttinger semimetals}",
author = {Mandal, Ipsita},
journal = {Ann. Phys.},
volume = {406},
pages = {173--185},
year = {2019},
month = {Jul},
doi = {10.1016/j.aop.2019.04.002},
url = {https://doi.org/10.1016/j.aop.2019.04.002}
}

@article{hwang-dassarma-graphene,
  title = {Dielectric function, screening, and plasmons in two-dimensional graphene},
  author = {Hwang, E. H. and Das Sarma, S.},
  journal = {Phys. Rev. B},
  volume = {75},
  issue = {20},
  pages = {205418},
  numpages = {6},
  year = {2007},
  month = {May},
  publisher = {American Physical Society},
  doi = {10.1103/PhysRevB.75.205418},
  url = {https://link.aps.org/doi/10.1103/PhysRevB.75.205418}
}

@article{pyatkovskiy-gapped-graphene,
  author  = {Pyatkovskiy, P. K.},
  title   = {Density of states and screening in gapped graphene},
  journal = {J. Phys.: Condens. Matter},
  volume  = {21},
  number  = {2},
  pages   = {025506},
  year    = {2009},
  doi     = {10.1088/0953-8984/21/2/025506}
}

@ARTICLE{ips-gnr-strain,
author = {{Haidar}, Firdous and {Jaffar A.}, Muhammed and {Mandal}, Ipsita},
title = "{Interplay of strain-induced axial gauge fields and intrinsic band-topology in the magnetoelectric conductivity of gapped nodal rings}",
journal={The European Physical Journal B},
year={2026},
month={Sep},
day={24},
volume={99},
number={9},
pages={147},
issn={1434-6036},
doi={10.1140/epjb/s10051-026-01238-8},
url={https://doi.org/10.1140/epjb/s10051-026-01238-8}
}

@ARTICLE{ips-sanskar-2dgnr,
author = {{Mandal}, Ipsita and {Sharma}, Sanskar},
title = "{Magnetoconductivity of two-dimensional Dirac cones and gapped nodal-rings under impurity-potentials in the ultraquantum limit}",
journal = {arXiv e-prints},
         year = 2026,
        month = jul,
archivePrefix = {arXiv},
       eprint = {2607.18769},
 primaryClass = {cond-mat.mes-hall},
      url = {https://arxiv.org/abs/2607.18769}
}

@ARTICLE{ips-firdous-vnr,
       author = {{Haidar}, Firdous and {Mandal}, Ipsita},
        title = "{Direction-dependent magnetoelectric conductivity from dipolar topological semimetals}",
      journal = {arXiv e-prints},
         year = 2026,
        month = sep,
archivePrefix = {arXiv},
       eprint = {2609.26676},
 primaryClass = {cond-mat.mes-hall},
url = {https://arxiv.org/abs/2609.26676}
}

@ARTICLE{ips-vnr-plasmon,
      title= "{Plasmon modes in tilted three-dimensional nodal-ring semimetals. II. Vortex nodal ring}", 
      author={Ipsita Mandal},
      journal = {arXiv e-prints},
      year={2026},
 month = oct,
      archivePrefix={arXiv},
      eprint = {2610.03188},
      primaryClass={cond-mat.mes-hall},
      url={https://arxiv.org/abs/2610.03188}, 
}

@article{lpge-ptnr,
  title = "{Linear photogalvanic effect induced by quantum metric in $\mathcal{P}T$-symmetric tilted nodal-line semimetals}",
  author = {Huang, Yi-Bei and Huang, Meng-Rou and Zheng, Shi-Han and Duan, Hou-Jian and Deng, Ming-Xun and Wang, Rui-Qiang},
  journal = {Phys. Rev. B},
  volume = {111},
  issue = {11},
  pages = {115116},
  numpages = {9},
  year = {2025},
  month = {Mar},
  publisher = {American Physical Society},
  doi = {10.1103/PhysRevB.111.115116},
  url = {https://link.aps.org/doi/10.1103/PhysRevB.111.115116}
}

@article{gnr-hall-ah,
  title = {Anomalous Hall optical conductivity in tilted topological nodal-line semimetals},
  author = {Wang, Chen and Xu, Wen-Hui and Zhu, Chang-Yong and Chen, Jin-Na and Zhou, Yong-Long and Deng, Ming-Xun and Duan, Hou-Jian and Wang, Rui-Qiang},
  journal = {Phys. Rev. B},
  volume = {103},
  issue = {16},
  pages = {165104},
  numpages = {7},
  year = {2021},
  month = {Apr},
  publisher = {American Physical Society},
  doi = {10.1103/PhysRevB.103.165104},
  url = {https://link.aps.org/doi/10.1103/PhysRevB.103.165104}
}

@ARTICLE{nlr-acoustic,
       author = {{Cheng}, Zheyu and {Guan}, Yi-Jun and {Xue}, Haoran and {Ge}, Yong and {Jia}, Ding and {Long}, Yang and {Yuan}, Shou-Qi and {Sun}, Hong-Xiang and {Chong}, Yidong and {Zhang}, Baile},
        title = "{Three-dimensional flat Landau levels in an inhomogeneous acoustic crystal}",
      journal = {Nature Communications},
         year = 2024,
        month = mar,
       volume = {15},
          eid = {2174},
        pages = {2174},
          doi = {10.1038/s41467-024-46517-z},
       adsurl = {https://ui.adsabs.harvard.edu/abs/2024NatCo..15.2174C}
}

@ARTICLE{ips-dipole-vnr,
       author = {{Mandal}, Ipsita},
        title = "{Unconventional magnetoelectric conductivity and electrochemical response from dipole-like sources of Berry curvature}",
      journal = {arXiv e-prints},
         year = 2026,
        month = feb,
archivePrefix = {arXiv},
       eprint = {2602.08844},
 primaryClass = {cond-mat.mes-hall},
url = {https://arxiv.org/abs/2602.08844}
}

@article{marcus-tilt_magneto,
  title = {Magneto-optical conductivity in generic {W}eyl semimetals},
  author = {St\aa{}lhammar, Marcus and Larana-Aragon, Jorge and Knolle, Johannes and Bergholtz, Emil J.},
  journal = {Phys. Rev. B},
  volume = {102},
  issue = {23},
  pages = {235134},
  numpages = {14},
  year = {2020},
  doi = {10.1103/PhysRevB.102.235134}
}

@article{ahn-tilted-nl,
  title = {Electrodynamics on {F}ermi Cyclides in Nodal Line Semimetals},
  author = {Ahn, Seongjin and Mele, E. J. and Min, Hongki},
  journal = {Phys. Rev. Lett.},
  volume = {119},
  issue = {14},
  pages = {147402},
  numpages = {5},
  year = {2017},
  month = {Oct},
  publisher = {American Physical Society},
  doi = {10.1103/PhysRevLett.119.147402},
  url = {https://link.aps.org/doi/10.1103/PhysRevLett.119.147402}
}

@ARTICLE{ips-nlsm-ph,
title = "{Direction-dependent linear response for gapped nodal-line semimetals in planar-Hall configurations}",
author =  {{Rather}, Fasil Hussain and {Haidar}, Firdous and {Jaffar A.}, Muhammed and {Mandal}, Ipsita},
journal={Eur. Phys. J. B},
year={2026},
month={Mar},
day={17},
volume={99},
number={3},
pages={41},
issn={1434-6036},
doi={10.1140/epjb/s10051-026-01131-4},
url={https://doi.org/10.1140/epjb/s10051-026-01131-4}
}

@Article{arpes-nlsm,
title= "{Experimental observation of drumhead surface states in SrAs$_3$}",
author={Hosen, M. Mofazzel and Dhakal, Gyanendra and Wang, Baokai and Poudel, Narayan and Dimitri, Klauss and Kabir, Firoza
and Sims, Christopher and Regmi, Sabin and Gofryk, Krzysztof and Kaczorowski, Dariusz and Bansil, Arun and Neupane, Madhab},
journal={Scientific Reports},
year={2020},
month={Feb},
day={17},
volume={10},
number={1},
pages={2776},
issn={2045-2322},
doi={10.1038/s41598-020-59200-2},
url={https://doi.org/10.1038/s41598-020-59200-2}
}

@article{fu_nlsm,
  title = {Topological nodal line semimetals with and without spin-orbital coupling},
  author = {Fang, Chen and Chen, Yige and Kee, Hae-Young and Fu, Liang},
  journal = {Phys. Rev. B},
  volume = {92},
  issue = {8},
  pages = {081201},
  numpages = {5},
  year = {2015},
  month = {Aug},
  publisher = {American Physical Society},
  doi = {10.1103/PhysRevB.92.081201},
  url = {https://link.aps.org/doi/10.1103/PhysRevB.92.081201}
}

@article{phe_nlsm,
  title = "{Planar Hall effect in topological Weyl and nodal-line semimetals}",
  author = {Li, Lei and Cao, Jin and Cui, Chaoxi and Yu, Zhi-Ming and Yao, Yugui},
  journal = {Phys. Rev. B},
  volume = {108},
  issue = {8},
  pages = {085120},
  numpages = {8},
  year = {2023},
  month = {Aug},
  publisher = {American Physical Society},
  doi = {10.1103/PhysRevB.108.085120},
  url = {https://link.aps.org/doi/10.1103/PhysRevB.108.085120}
}

@article{yang1,
  title = "{Sign reversal of magnetoresistivity in massive nodal-line semimetals due to the Lifshitz transition of the Fermi surface}",
  author = {Yang, Min-Xue and Geng, Hao and Luo, Wei and Sheng, Li and Chen, Wei and Xing, D. Y.},
  journal = {Phys. Rev. B},
  volume = {104},
  issue = {16},
  pages = {165149},
  numpages = {8},
  year = {2021},
  month = {Oct},
  publisher = {American Physical Society},
  doi = {10.1103/PhysRevB.104.165149},
  url = {https://link.aps.org/doi/10.1103/PhysRevB.104.165149}
}

@article{yang_review_nlsm,
author = {Min-Xue Yang, Wei Luo and Wei Chen},
title = {Quantum transport in topological nodal-line semimetals},
journal = {Advances in Physics: X},
volume = {7},
number = {1},
pages = {2065216},
year = {2022},
publisher = {Taylor \& Francis},
doi = {10.1080/23746149.2022.2065216},
URL = {https://doi.org/10.1080/23746149.2022.2065216}
}

@article{schnyder_nodal,
  title = {Topological transport in {D}irac nodal-line semimetals},
  author = {Rui, W. B. and Zhao, Y. X. and Schnyder, Andreas P.},
  journal = {Phys. Rev. B},
  volume = {97},
  issue = {16},
  pages = {161113},
  numpages = {6},
  year = {2018},
  month = {Apr},
  publisher = {American Physical Society},
  doi = {10.1103/PhysRevB.97.161113},
  url = {https://link.aps.org/doi/10.1103/PhysRevB.97.161113}
}

@Article{biao_nodal,
author={Yang, Biao and Bi, Yangang and Zhang, Rui-Xing and Zhang, Ruo-Yang and You, Oubo and Zhu, Zhihong and Feng, Jing and Sun, Hongbo and Chan, C. T. and Liu, Chao-Xing and Zhang, Shuang},
title={Momentum space toroidal moment in a photonic metamaterial},
journal={Nature Communications},
year={2021},
month={Mar},
day={19},
volume={12},
number={1},
pages={1784},
issn={2041-1723},
doi={10.1038/s41467-021-22063-w},
url={https://doi.org/10.1038/s41467-021-22063-w}
}

@article{linearize-nlsm,
  title = {Weak Localization and Antilocalization in Nodal-Line Semimetals: {D}imensionality and Topological Effects},
  author = {Chen, Wei and Lu, Hai-Zhou and Zilberberg, Oded},
  journal = {Phys. Rev. Lett.},
  volume = {122},
  issue = {19},
  pages = {196603},
  numpages = {6},
  year = {2019},
  month = {May},
  publisher = {American Physical Society},
  doi = {10.1103/PhysRevLett.122.196603},
  url = {https://link.aps.org/doi/10.1103/PhysRevLett.122.196603}
}

@article{balents-nodal,
  title = {Topological nodal semimetals},
  author = {Burkov, A. A. and Hook, M. D. and Balents, Leon},
  journal = {Phys. Rev. B},
  volume = {84},
  issue = {23},
  pages = {235126},
  numpages = {14},
  year = {2011},
  month = {Dec},
  publisher = {American Physical Society},
  doi = {10.1103/PhysRevB.84.235126},
  url = {https://link.aps.org/doi/10.1103/PhysRevB.84.235126}
}

@article{expt2_nlsm,
title = {Low Carrier Density Metal Realized in Candidate Line-Node {D}irac Semimetals {C}a{A}g{P} and {C}a{A}g{A}s},
author = {{Okamoto}, Yoshihiko and {Inohara}, Takumi and {Yamakage}, Ai and {Yamakawa}, Youichi and {Takenaka}, Koshi},
journal = {Journal of the Physical Society of Japan},
year = 2016,
        month = dec,
       volume = {85},
       number = {12},
        pages = {123701},
doi = {10.7566/JPSJ.85.123701},
URL = {https://doi.org/10.7566/JPSJ.85.123701}
}

@article{expt1_nlsm,
    author = {Xie, Lilia S. and Schoop, Leslie M. and Seibel, Elizabeth M. and Gibson, Quinn D. and Xie, Weiwei and Cava, Robert J.},
    title = "{A new form of Ca$_3$P$_2$ with a ring of Dirac nodes}",
    journal = {APL Materials},
    volume = {3},
    number = {8},
    pages = {083602},
    year = {2015},
    month = {07},
    issn = {2166-532X},
    doi = {10.1063/1.4926545},
    url = {https://doi.org/10.1063/1.4926545}
    }

@ARTICLE{ips-tilted,
       author = {{Mandal}, Ipsita},
        title = "{Anisotropic conductivity for the type-I and type-II phases of Weyl/multi-Weyl semimetals in planar Hall set-ups}",
      journal = {Annals of Physics},
volume = {482},
pages = {170181},
year = {2025},
issn = {0003-4916},
doi = {https://doi.org/10.1016/j.aop.2025.170181},
url = {https://www.sciencedirect.com/science/article/pii/S0003491625002635},
author = {Ipsita Mandal}
}

@ARTICLE{ips-sandip-fano,
       author = {{Bera}, Sandip and {Mandal}, Ipsita},
        title = "{Floquet scattering and Fano resonances in nodal-ring and multi-Weyl semimetals: Role of propagating and evanescent modes}",
      journal = {arXiv e-prints},
         year = 2026,
        month = sep,
archivePrefix = {arXiv},
       eprint = {2609.00631},
 primaryClass = {cond-mat.mes-hall},
      url = {https://arxiv.org/abs/2609.00631}
}

@article{ips-magnus,
	title={Magnus {H}all effect in three-dimensional topological semimetals},
	author = {Sekh, Sajid and Mandal, Ipsita},
	DOI= "10.1140/epjp/s13360-022-02840-2",
	url= "https://doi.org/10.1140/epjp/s13360-022-02840-2",
	journal = {Eur. Phys. J.  Plus},
	year = 2022,
	volume = 137,
	number = 6,
	pages = 736
}

@article{ahn_mWeyl,
title={Collective modes in multi-{W}eyl semimetals},
	volume = {6},
	issn = {2045-2322},
	url = {https://www.nature.com/articles/srep34023},
	doi = {10.1038/srep34023},
	number = {1},
	urldate = {2022-11-16},
	journal = {Scientific Reports},
	author = {Ahn, Seongjin and Hwang, E. H. and Min, Hongki},
	month = sep,
	year = {2016},
	pages = {34023}
}

@article{yan17_topological,
  title={Topological materials: {W}eyl semimetals},
  author={Yan, Binghai and Felser, Claudia},
  journal={Annual Rev. of Condensed Matter Phys.},
  volume={8},
  pages={337--354},
  year={2017},
  doi={10.1146/annurev-conmatphys-031016-025458}
}

@article{burkov11_Weyl,
  title = "{{W}eyl semimetal in a topological insulator multilayer}",
  author = {Burkov, A. A. and Balents, Leon},
  journal = {Phys. Rev. Lett.},
  volume = {107},
  issue = {12},
  pages = {127205},
  numpages = {4},
  year = {2011},
  doi = {10.1103/PhysRevLett.107.127205}
}

@article{huang15_observation,
  title = {Observation of the Chiral-Anomaly-Induced Negative Magnetoresistance in 3D {W}eyl Semimetal {T}a{A}s},
  author = {Huang, Xiaochun and Zhao, Lingxiao and Long, Yujia and Wang, Peipei and Chen, Dong and Yang, Zhanhai and Liang, Hui and Xue, Mianqi and Weng, Hongming and Fang, Zhong and Dai, Xi and Chen, Genfu},
  journal = {Phys. Rev. X},
  volume = {5},
  issue = {3},
  pages = {031023},
  numpages = {9},
  year = {2015},
  doi = {10.1103/PhysRevX.5.031023}
}

@article{yadav23_magneto,
  title={Magneto-optical conductivity in the type-{I} and type-{II} phases of {W}eyl/multi-{W}eyl semimetals},
  author={Yadav, Shivam and Sekh, Sajid and Mandal, Ipsita},
  journal={Physica B: Condensed Matter},
  volume={656},
  pages={414765},
  year={2023},
  doi={10.1016/j.physb.2023.414765}
}

@article{cuteo_nlsm,
  title = "{Almost ideal nodal-loop semimetal in monoclinic CuTeO$_3$ material}",
  author = {Li, Si and Liu, Ying and Fu, Botao and Yu, Zhi-Ming and Yang, Shengyuan A. and Yao, Yugui},
  journal = {Phys. Rev. B},
  volume = {97},
  issue = {24},
  pages = {245148},
  numpages = {8},
  year = {2018},
  month = {Jun},
  publisher = {American Physical Society},
  doi = {10.1103/PhysRevB.97.245148},
  url = {https://link.aps.org/doi/10.1103/PhysRevB.97.245148}
}

@ARTICLE{alkaline_nlsm,
       author = {{Hirayama}, Motoaki and {Okugawa}, Ryo and {Miyake}, Takashi and {Murakami}, Shuichi},
        title = "{Topological Dirac nodal lines and surface charges in fcc alkaline earth metals}",
      journal = {Nature Communications},
         year = 2017,
        month = jan,
       volume = {8},
        pages = {14022},
          doi = {10.1038/ncomms14022},
       adsurl = {https://ui.adsabs.harvard.edu/abs/2017NatCo...814022H}
}

@Article{enke,
author={Liu, Enke
and Sun, Yan
and Kumar, Nitesh
and Muechler, Lukas
and Sun, Aili
and Jiao, Lin
and Yang, Shuo-Ying
and Liu, Defa
and Liang, Aiji
and Xu, Qiunan
and Kroder, Johannes
and S{\"u}{\ss}, Vicky
and Borrmann, Horst
and Shekhar, Chandra
and Wang, Zhaosheng
and Xi, Chuanying
and Wang, Wenhong
and Schnelle, Walter
and Wirth, Steffen
and Chen, Yulin
and Goennenwein, Sebastian T. B.
and Felser, Claudia},
title= "{Giant anomalous Hall effect in a ferromagnetic Kagome-lattice semimetal}",
journal={Nature Physics},
year={2018},
month={Nov},
day={01},
volume={14},
number={11},
pages={1125-1131},
issn={1745-2481},
doi={10.1038/s41567-018-0234-5},
url={https://doi.org/10.1038/s41567-018-0234-5}
}

@article{claudia_nlsm,
  title = "{Large anomalous Hall and Nernst effects from nodal line symmetry breaking in Fe$_2$MnX (X = P, As, Sb)}",
  author = {Noky, Jonathan and Xu, Qiunan and Felser, Claudia and Sun, Yan},
  journal = {Phys. Rev. B},
  volume = {99},
  issue = {16},
  pages = {165117},
  numpages = {5},
  year = {2019},
  month = {Apr},
  publisher = {American Physical Society},
  doi = {10.1103/PhysRevB.99.165117},
  url = {https://link.aps.org/doi/10.1103/PhysRevB.99.165117}
}

@article{rahmipoor-2d-nlsm,
  title = {Collective excitations and screening in two-dimensional tilted nodal-line semimetals},
  author = {Rahimpoor, Hamid and Abedinpour, Saeed H.},
  journal = {Phys. Rev. B},
  volume = {109},
  issue = {4},
  pages = {045120},
  numpages = {9},
  year = {2024},
  month = jan,
  doi = {10.1103/PhysRevB.109.045120}
}

@article{rhim-kim-friedel,
  title = {Anisotropic density fluctuations, plasmons, and Friedel oscillations in nodal line semimetal},
  author = {Rhim, Jun-Won and Kim, Yong Baek},
  journal = {New J. Phys.},
  volume = {18},
  pages = {043010},
  numpages = {12},
  year = {2016},
  doi = {10.1088/1367-2630/18/4/043010}
}

@book{giuliani-vignale,
  title={Quantum Theory of the Electron Liquid},
  author={Giuliani, Gabriele and Vignale, Giovanni},
  isbn={9780521821124},
  url={https://books.google.com/books?id=cVYzGDsC3GoC},
  year={2005},
  publisher={Cambridge University Press}
}

\end{document}